\documentclass[10pt,twoside]{article}
\usepackage[english]{babel}
\usepackage{droid}
\usepackage{famems}
\usepackage{braket}
\usepackage{citeref}
\usepackage{cmap}
\usepackage{graphics}
\usepackage{graphicx}
\usepackage{graphpap}
\usepackage{amsmath,amstext,amsfonts,amssymb,amsthm}
\usepackage{mathtools}
\usepackage{dsfont}
\usepackage[mathscr]{eucal}
\usepackage{array}
\usepackage{multicol}
\usepackage{multirow}
\usepackage{multicol}
\usepackage{tabularx}
\usepackage{xcolor}
\usepackage{color}
\usepackage{colortbl}
\usepackage{ifthen}
\usepackage{pdflscape}
\usepackage{epstopdf}
\usepackage[normalem]{ulem}
\usepackage{rotating}
\usepackage{hhline}
\usepackage{fontawesome}
\usepackage{scrextend}
\usepackage{arydshln}
\usepackage{ctable}
\usepackage{url}
\usepackage[breaklinks]{hyperref}
\usepackage{hyperref}
\hypersetup{
    colorlinks,
    citecolor=black,
    filecolor=black,
    linkcolor=black,
    urlcolor=
teal
}
\usepackage{cleveref}
\definecolor{MyYellow50}{RGB}{255,255,128}
\definecolor{MyYellow0}{RGB}{255,255,0}
\definecolor{MyYellow75}{RGB}{255,255,192}
\definecolor{MyYellow85}{RGB}{255,255,240}
\definecolor{MyAqua50}{RGB}{127,255,255}
\definecolor{MyAqua0}{RGB}{0,255,255}
\definecolor{MyAqua75}{RGB}{192,255,255}
\definecolor{MyAqua85}{RGB}{224,255,255}
\definecolor{MyGreen0}{RGB}{0,255,0}
\definecolor{MyGreen50}{RGB}{128,255,128}
\definecolor{MyYellow}{rgb}{1,1,0}
\definecolor{MyYellow75}{rgb}{1,1,0.75}
\definecolor{MyRed95}{rgb}{1,0.95,0.95}
\definecolor{MyRed85}{rgb}{1,0.85,0.85}
\definecolor{MyRed75}{rgb}{1,0.75,0.75}
\definecolor{MyRed50}{rgb}{1,0.50,0.50}
\definecolor{MyRed25}{rgb}{1,0.25,0.25}
\definecolor{MyRed0}{rgb}{1,0,0}
\definecolor{MyRed}{rgb}{1,0,0}
\definecolor{MyBlue}{rgb}{0,0,1}
\definecolor{MyGreen}{rgb}{0.75,1,0.75}
\definecolor{MyGreen35}{rgb}{0.35,1,0.35}
\definecolor{MyGreen75}{rgb}{0.75,1,0.75}
\definecolor{MyBlue40}{rgb}{0.40,0.40,1}
\definecolor{MyBlue60}{rgb}{0.60,0.60,1}
\definecolor{MyBlue75}{rgb}{0.75,0.75,1}
\definecolor{MyBlue85}{rgb}{0.85,0.85,1}
\definecolor{MyGray}{rgb}{0.95,0.95,0.95}
\definecolor{MyWhite}{rgb}{1,1,1}
\definecolor{MyBlack}{rgb}{0,0,0}
\definecolor{MyProbability}{rgb}{1.00,0.25,0.25}
\definecolor{MyConfidency}{rgb}{1.00,0.25,1.00}
\definecolor{MyEventology}{rgb}{0.25,1.00,1.00}
\definecolor{MyOrange}{rgb}{1,0.56,0.25}
\definecolor{MyOrangeB}{rgb}{1,0.65,0.45}
\definecolor{MyMagentaP}{rgb}{1.00,0.45,0.65}
\definecolor{MyMagenta}{rgb}{1.00,0,1.00}
\definecolor{MyAqua}{rgb}{0,0.75,0.75}
\makeatletter
\renewcommand{\paragraph}{\@startsection{paragraph}{4}{0ex}%
   {-3.25ex plus -1ex minus -0.2ex}%
   {1.5ex plus 0.2ex}%
   {\normalfont\normalsize\tt}}
\makeatother

\begin{document}
\baselineskip=11pt

\newcounter{ctrwar}\setcounter{ctrwar}{0} 
\newcounter{ctrdef}\setcounter{ctrdef}{0}
\newcounter{ctrdefpre}\setcounter{ctrdefpre}{0}
\newcounter{ctrTh}\setcounter{ctrTh}{0}
\newcounter{ctrnot}\setcounter{ctrnot}{0}
\newcounter{ctrATT}\setcounter{ctrATT}{0} 
\newcounter{ctrcor}\setcounter{ctrcor}{0}
\newcounter{ctrAx}\setcounter{ctrAx}{0}
\newcounter{ctrexa}\setcounter{ctrexa}{0}
\newcounter{ctrlem}\setcounter{ctrlem}{0}
\newcounter{ctrPRO}\setcounter{ctrPRO}{0}
\newcounter{ctrrem}\setcounter{ctrrem}{0}
\newcounter{ctrass}\setcounter{ctrass}{0}
\newcounter{ctrmem}\setcounter{ctrmem}{0}

\newcommand{\bfPB}{\mbox{\protect\reflectbox{\bf P}\hspace{-0.4em}{\bf B}}}
\newcommand{\bfEl}{\mbox{\protect\reflectbox{\bf E}}}
\newcommand{\bfEr}{\mathbf{E}}
\newcommand{\bfEE}{\mbox{\protect\reflectbox{\bf E}\hspace{-0.4em}{\bf E}}}
\newcommand{\bfphi}{\mbox{\boldmath$\varphi$}}
\newcommand{\bfPhi}{\mbox{\boldmath$\Phi$}}
\newcommand{\bfPhii}{\mbox{\scriptsize\boldmath$\Phi$}}
\newcommand{\rS}{\reflectbox{\bf S}}
\newcommand{\rsS}{\,\reflectbox{\scriptsize\bf S}}
\newcommand{\rssS}{\,\reflectbox{\tiny\bf S}}
\newcommand{\reS}{\reflectbox{S}}
\newcommand{\resS}{\,\reflectbox{\scriptsize S}}
\newcommand{\ressS}{\,\reflectbox{\tiny S}}
\newcommand{\scrER}{\mathscr E\!\mathscr R}

\newcommand{\ER}{\mathscr E\!\mathscr R}

\numberwithin{equation}{section}

\renewcommand{\figurename}{\scriptsize  Figure}
\renewcommand{\tablename}{\scriptsize  Table}
\renewcommand\refname{\scriptsize References}
\renewcommand\contentsname{\scriptsize Contents}
\setcounter{page}{78}

\numberwithin{equation}{section}
\titleOneAuthoren{
Elements of the Kopula (eventological copula) theory
}{Oleg Yu. Vorobyev}{
Institute of Mathematics and Computer Science\\
Siberian Federal University\\
Krasnoyarsk\\
\tiny
\url{mailto:oleg.yu.vorobyev@gmail.com}\\
\url{http://www.sfu-kras.academia.edu/OlegVorobyev}
}

\label{vorobyev-13}

\colontitleen{Vorobyev}
\footavten{Oleg Yu. Vorobyev}
\setcounter{footnote}{0}
\setcounter{equation}{0}
\setcounter{figure}{0}
\setcounter{table}{0}
\setcounter{section}{0}

\vspace{-22pt}

\begin{abstracten}
\textit{New in the probability theory and eventology theory, the concept of Kopula (eventological copula) is introduced\footnote{\label{Kopulas-footnote}To distinguish the quite differently defined notion of an \emph{eventological copula} from the classical concept of copula in the sense of Sklar (1959), the following radical terminology with capital ``K'' is used:
\emph{Kopula} = eventological copula;
\emph{$N$-Kopula} = eventological $N$-copula.}.
The theorem on the characterization of the sets of events by Kopula is proved, which serves as the eventological pre-image of the well-known Sclar's theorem on copulas (1959). The Kopulas of doublets and triplets of events are given, as well as of some $N$-sets of events.}
\end{abstracten}\hspace*{-8pt}\footnote[777]{Editing the text of November 12, 2017.}\\[-21pt]
\begin{keywordsen}
\emph{Eventology, probability, Kolmogorov event, event, set of events, Kopula (eventological copula), Kopula characterizing a set of events.}
\end{keywordsen}

\section{Introduction\label{intro}}

Long time ago and little by little, the incentive for this work materialized in the theory of sets of events, eventology \cite{Vorobyev2007}, where the need to locate the classes of event-probability distributions (e.p.d.) of the sets of events (s.e.), which were so arbitrary and spacious to be able without let or hindrance to deal with the relationships between pairs, triples, quadruplets, etc., of events, in other words, to understand the structure of statistical dependencies and generalities between events from some s.e. A similar need is perhaps the only one that has always fueled the development of the probability theory and statistics, which in one way or another are theories of studying and evaluating the structures of statistical dependencies and generalities in the distributions of sets of events.

The \emph{classical copula theory} \cite{Sklar1959, Nelsen1999, Nelsen2003, Alsina2006}, existing since the 50s of the last century, allows us to construct classes of joint distribution functions that have given marginal distribution functions.
In eventology, the theory of sets of events, proposed in the paper the \emph{theory of Kopula (eventological copula)} allows us to solve a similar problem --- to build classes of e.p.d's of sets of events whose events happen with given \emph{probabilities of marginal events}.

\subsection{General statement of the problem of the Kopula theory}

We formulate the general statement of the problem of the $N$-Kopula theory for $N$-sets of events.
If in the classical theory the copula is the tool for selecting some \emph{family of joint d.f.'s of a set of random variables} from the set of all d.f.'s with given marginal d.f.'s, then in the eventological theory the Kopula is the tool for selecting a family of \emph{e.p.d.'s of the 1st kind of the set events} from the set of all e.p.d.'s with the given probabilities of marginal events.

However, unlike the classical d.f.'s the functions of e.p.d.'s of the 1st kind of the $N$-s.e. $\frak{X}$ are functions that are defined as the sets
\begin{equation}\label{Nepd0}
\begin{split}
\{ p(X /\!\!/ \frak{X}), X \subseteq \frak{X} \}
\end{split}
\end{equation}
of all its $2^N$ values, probabilities of the 1st kind $p(X /\!\!/\frak{X})$, on the set of all subsets of this $N$ -s.e.

So, let's clarify, Kopula is the tool for selecting a family of sets of the form (\ref{Nepd0}) from the set of all sets with given probabilities of marginal events.

To specify a family of sets of $2^N$ probabilities of the 1st kind (\ref{Nepd0}), it is necessary and sufficient to specify a family of sets from the $2^N-1$ parameters, since all the probabilities in each set must be nonnegative and give in the sum of one. And to specify a family of sets of $2^N$ probability values of the 1st kind (\ref{Nepd0}) with given probabilities of marginal events that form the $\frak{X}$-set
\begin{equation}
\breve{p} = \{p_x, x \in \frak{X}\},
\end{equation}
it is necessary and sufficient to specify a family of sets from the $2^N-N-1$ parameters, since for each collection there must be another $N$ constraints for events $x\in\frak{X}$:
\begin{equation}
\sum_{x \in X \subseteq \frak{X}} p(X /\!\!/ \frak{X}) = p_x.
\end{equation}

Therefore, ``to define the family of functions of e.p.d.'s of the 1st kind with given probabilities of marginal events'' means ``to define a family of sets from the $2^N-N-1$ parameters'' as sets of functions of these probabilities. Eventological theory should solve this problem with the help of a convenient tool, the Kopula, which allows us to define a family of sets from the $2^N-N-1$ parameters as sets of functions from marginal probabilities, which in turn can be made dependent on a number of auxiliary parameters.

In general, the $N$-Kopula in the eventological theory is an instrument for defining the family of probability distributions of the 1st kind of the $N$-s.e., in the form of a family of $2^N$-set of functions of the probabilities of their $N$ marginal events.

\subsection{Prolegomena of the Kopula theory}

The main results of this paper are presented in a rather rigorous mathematical manner. And although the definitions, statements and proofs are provided with examples and illustrations, in order to visualize the ideas underlying the Kopula theory, in my opinion, a number of preliminary explanations in a less strict context, which are collected in several prolegomena, may be necessary.

If in some set of events $\frak{X}$, some events from the subset $X\subseteq\frak{X}$ are replaced by their complements, then we get a new set of events $\frak{X}^{(c|X)}=X+(\frak{X}-X)^{(c)}$, which is called the $X$-phenomenon of s.e. $\frak{X}$. The set of all such $X$-phenomena for $X\subseteq\frak{X}$ is called the $2^\frak{X}$-phenomenon-dom of s.e. $\frak{X}$.
In \cite{Vorobyev2015famems12} a rather distinct theory of set-phenomena and the phenomenon-dom of some s.e.

Similarly, the theory of set-phenomena \cite{Vorobyev2015famems12} defines the phenomena and phenomenon-dom of the set $\breve{p}=\Set{p_x, \, x \in \frak{X}}$ of the probabilities of marginal events $x\in\frak{ X}$: the $X$-phenomenon $\breve{p}^{(c|X)}=\Set{p_x, x \in X}+\Set{1-p_x, x \in \frak{X}-X}$ of the set of marginal probabilities $\breve{p}$ is obtained by replacing the marginal probabilities $p_x$ by complementary marginal probabilities $1-p_x$ when $x \in \frak{X}-X$.

\texttt{Prolegomenon \!\refstepcounter{ctrPRO}\arabic{ctrPRO}\,\label{pro-1}\itshape\enskip\scriptsize (set-phenomenon of a set of events and a set of probabilities of events).}
The main conclusion of the above theory is obvious: \emph{probability distribution of the s.e. $\frak{X}$ characterizes the probability distribution and the set of marginal probabilities of each set-phenomenon from its $2^\frak{X}$-phenomenon-dom}.

\texttt{Prolegomenon \!\refstepcounter{ctrPRO}\arabic{ctrPRO}\,\label{pro-2}\itshape\enskip\scriptsize (set-phenomenal transformation).}
\emph{1) For each pair $\frak{X}^{(c|X)}, \frak{X}^{(c|Y)}$ set-phenomena of the set of events $\frak{X}$ their probability distributions are related to each other set-phenomenal transformations}.
\emph{2) In each pair $\breve{p}^{(c|X)}, \breve{p}^{(c|Y)}$ set-phenomena of the set of marginal probabilities $\breve{p}$ events from $\frak{X}$ are also interconnected by set-phenomenal transformations}.

The event $x \in \frak{X}$ is called half-rare \cite{Vorobyev2015famems12} if the probability $p_x=\mathbf{P}(x)$ with which it happens is not more than half: $p_x \leqslant 1/2$. If all events from the s.e. $\frak{X}$ are half-rare, we speak of a set of half-rare events, or a half-rare s.e.

\texttt{Prolegomenon \!\refstepcounter{ctrPRO}\arabic{ctrPRO}\,\label{pro-3}\itshape\enskip\scriptsize (sets of half-rare events and its Kopulas).}
\emph{1) It is not difficult to guess that for any s.e. $\frak{X}$, in the $2^\frak{X}$ phenomenon-dom of the sets of events $\frak{X}^{(c|X)}, X\subseteq\frak{X}$, and in the $\frak{X}$-phenomenon-dom of the sets of its marginal probabilities $\breve{p}^{(c|X)}, X\subseteq\frak{X}$, there is always a half-rare set-phenomenon. If, in addition, there are no events in $\frak{X}$ happening with probability $1/2$, then such a half-rare set-phenomenon is unique.}
\emph{2) A Kopula of some family of half-rare $N$-s.e.'s is generated by $2^N$ functions from half-rare variables defined on the half-hypercube $[0,1/2]^N$ with values from $[0,1]$ that are continued by the set-phenomenal transformations of the half-rare variables to the corresponding half-hypercubes, all together completely filling the unit hypercube}.

\texttt{Prolegomenon \!\refstepcounter{ctrPRO}\arabic{ctrPRO}\,\label{pro-4}\itshape\enskip\scriptsize (an invariance of the copula with respect to the order of half-rare events).}
Our task is to construct a Kopula of a family of arbitrary (unordered) sets of half-rare events, i.e. a 1-function, the arguments of which form an unordered set of probabilities of their marginal events. Therefore, it is natural to require such a function to be invariant with respect to the order of its arguments; with respect to the order of events in these sets. In other words, it is natural to consider this 1-function as a function of a set of arguments, rather than a vector of arguments with ordered components, as it is usually assumed.

\texttt{Prolegomenon \!\refstepcounter{ctrPRO}\arabic{ctrPRO}\,\label{pro-5}\itshape\enskip\scriptsize (insertable sets of half-rare events and a frame half-rare event).}
\emph{Two insertable sets of half-rare events} for a given set of half-rare events $\frak{X}=\{x_0\}+\mathcal{X}$ with the \emph{frame half-rare event} $x_0\in \frak{X}$, happening with the highest probability among all events from $\frak{X}$, are two sets of half-rare events $ \mathcal {X}'= \{x_0\} (\cap) \mathcal {X}$ and $\mathcal {X}'\!' = \{x_0^c\} (\cap) \mathcal {X}$ that partition the set of other events $ \mathcal {X} = \frak { X} - \{x_0 \} $ into two: $ \mathcal {X} = \mathcal {X} '(+) \mathcal {X}' \! '$. The events of one of them, $ \mathcal {X} '$, are contained in the \emph {frame half-rare event} $ x_0 $, and the events of the other, $ \mathcal {X}' \! '$, are contained in its complement $ x_0 ^ c = \Omega-x_0 $.

\texttt{Prolegomenon \!\refstepcounter{ctrPRO}\arabic{ctrPRO}\,\label{pro-6}\itshape\enskip\scriptsize (the insertable sets of events and conditional e.p.d.'s of a set of events with respect to the frame event and its complement).}
Conditional e.p.d.'s of the 1st kind of one s.e. $ \frak {X} $ with respect to the other s.e. $ \frak {Y} $ are defined in the traditional way \cite{Vorobyev2007}. However, until now attempts to define such a ``conditional'' s.e., which would have given a conditional e.p.d. of the 1st kind, turned out to be completely impractical \cite{Vorobyev2006aa}. The concept of \emph{two insertable s.e.'s} in a frame event is a well-defined ``ersatz'' of such ``conditional'' s.e.'s. The e.p.d.'s of this ``ersatz'', although they do not coincide with two conditional e.p.d.'s of the 1st kind with respect to the frame event and its complement, but they are fully characterized by them. The converse is also true: the e.p.d.'s of two frame s.e.'s characterize the corresponding two conditioned e.p.d.'s of the 1st kind.

\texttt{Prolegomenon \!\refstepcounter{ctrPRO}\arabic{ctrPRO}\,\label{pro-7}\itshape\enskip\scriptsize (a frame method of constructing a Kopula of an arbitrary set of half-rare events).}
A \emph{frame method} constructs a Kopula of a family of arbitrary sets of half-rare events on the basis of a conditional scheme by means of a recurrence formula via conditional e.p.d.'s concerning the frame event and its complement. A recurrent formula associates this Kopula with two Kopulas of families of their insertable sets of half-rare events of smaller dimension, which are characterized by the corresponding conditional e.p.d.'s of the 1st kind (see Prolegomenon \ref{pro-6}).

\texttt{Prolegomenon \!\refstepcounter{ctrPRO}\arabic{ctrPRO}\,\label{pro-8}\itshape\enskip\scriptsize (a set-phenomenal transformation of a half-rare Copula to an arbitrary one).}
\emph{To construct the Kopula of a family of arbitrary s.e.'s it is enough to construct the Kopula of the family of their half-rare set-phenomena and apply a set-phenomenal transformation to this Kopula.}

\texttt{Prolegomenon \!\refstepcounter{ctrPRO}\arabic{ctrPRO}\,\label{pro-9}\itshape\enskip\scriptsize (Cartesian representation of the $N$-Kopula in $\mathbb{R}^N$).}
\emph{It follows from the Prolegomenon \ref{pro-4} that the Cartesian representation of the $N$-Kopula in $\mathbb{R}^N$ should be a symmetric function of $N$ ordered variables, marginal probabilities of events from $N$-s.e. $\frak{X}$, which is defined on the $N$-dimensional unit hypercube $[0,1]^N$. The Cartesian representation of the $N$-Kopula is based on the fact that its symmetric image takes the same values on all permutations of its arguments, that is, is defined by the permutation of $N$ events group.
Moreover, the value of such a symmetric function on an arbitrary $N$-vector $\bar{w}=\{w_1,...,w_N\} \in [0,1]^N$ is equal to the value of the $N$-Kopula on an ordered $\frak{X}$-set of marginal probabilities of half-rare events $\breve{p}=\{p_x, x \in \frak{X}\}$, the ordered half-rare projection of the $N$-vector $\bar{w}$, the order of the variables in which is defined by an $N$-permutation $\pi_{\bar{w}}$ that has the components of the half-rare projection $\bar{w}^*$ in decreasing order, where
\begin{equation}\label{half-rare-projection}
\begin{split}
w^*_n =
\begin{cases}
w_n, & w_n \leqslant 1/2,\\
1-w_n, & w_n > 1/2
\end{cases}
\end{split}
\end{equation}
are components of the $N$-vector $\bar{w}^*$ of the half-rare projection of the $N$-vector $\bar{w}$, $n = 1, ..., N$. As a result, the ordered half-rare $\frak{X}$ is the set of marginal probabilities $\breve{p}=\breve{p}(\bar{w})$, on which the $N$-Kopula takes the same value as a symmetric function on $\bar{w}$, is given by the formula
\begin{equation}\label{ordered-half-rare-projection}
\begin{split}
\breve{p}(\bar{w}) = \pi_{\bar{w}}(\bar{w}^*),
\end{split}
\end{equation}
which defines the Cartesian representation of the $N$-Kopula in $\mathbb{R}^N$ for each $\bar{w} \in [0,1]^N$.}

\section{The Kopula: definition, theorem and the simplest Kopulas\label{def-Th-Kopula}}

We consider the \emph{general probability space of Kolmogorov events} $(\Omega, \mathcal {A}^\mho, \mathbf {P})$, some \emph{particular probability space of events}$ (\Omega, \mathcal{A}, \mathbf{P})$ and the $N$-\emph{set of events ($N$-s.e.)} $\frak {X} \subset \mathcal {A}$ with the event-probability distribution (e.p.d.\footnote{The abbreviations: \emph{e.p.d.} and \emph{e.c.d.} are used for the \emph{event-probability distribution} and for the \emph{event-covariance distribution}.}) of the 1st kind
$$
{\mbox{\boldmath$p$}}(\frak{X}) = \{ p(X /\!\!/ \frak{X}) : X \subseteq \frak{X} \},
$$
and of the second kind
$$
{\mbox{\boldmath$p$}}_\frak{X} = \{ p_{X /\!\!/ \frak{X}} : X \subseteq \frak{X} \},
$$
which, recall, are related to each other by the Mobius inversion formulas:
$$
p_{X /\!\!/ \frak{X}} = \sum_{X \subseteq Y} p(Y /\!\!/ \frak{X}),
$$
$$
p(X /\!\!/ \frak{X}) = \sum_{X \subseteq Y} (-1)^{|Y|-|X|} p_{Y /\!\!/ \frak{X}}.
$$

\texttt{Definition \!\refstepcounter{ctrdef}\arabic{ctrdef}\,\label{def-PHness}\itshape\enskip\scriptsize (set-phenomena of a s.e. and its phenomenon-dom).}
Every $ N $-s.e. $\frak {X} \subset \mathcal {A} $ generates its own \emph{$ 2^ \frak {X} $-phenomenon-dom}, defined as a $ 2 ^ N $-family
\begin{equation}\label{phenomenness}
\begin{split}
2^{(c|\frak{X})} = \left\{ \frak{X}^{(c|X)}, X \subseteq \frak{X} \right\},
\end{split}
\end{equation}
composed of $ N $-s.e. in the form
$$
\frak{X}^{(c|X /\!\!/ \frak{X})} = \frak{X}^{(c|X)} = X + (\frak{X}-X)^{(c)} \subset \mathcal{A},
$$
which for each $ X \subseteq \frak {X} $ is called its \emph {set-phenomen \cite{Vorobyev2015famems12}, \emph{more precisely}, $ X $-phenomen}, where
$$
X^{(c)} = \{ x^c : x \in X  \}
$$
is an \emph{М-complement of the s.e.} $X \subseteq \frak{X}$.

We also recall that probabilities of the second kind
$$
p_x = p_{\{x\}} = \mathbf{P}\left( \bigcap_{x \in \{x\} \subseteq \frak{X}} x \right) = \mathbf{P}(x)
$$
are \emph{probabilities of marginal events} from $\{x\}\subseteq\frak{X}$ (\emph{marginal probabilities}), probabilities of the second kind
$$
p_{xy} = p_{\{x,y\}} = \mathbf{P}\left( \bigcap_{z \in \{x,y\} \subseteq \frak{X}} z \right) = \mathbf{P}(x \cap y)
$$
are probabilities of double intersections of events from $\{x,y\}\subseteq\frak{X}$, and probabilities of the second kind
$$
p_{Z_n} = \mathbf{P}\left( \bigcap_{x \in Z_n \subseteq \frak{X}} x \right)
$$
are probabilities of $n$-intersections of events from $Z_n\subseteq\frak{X}$, where $|Z_n|=n$;

\texttt{Definition \!\refstepcounter{ctrdef}\arabic{ctrdef}\,\label{def-p-PHness}\itshape\enskip\scriptsize (set-phenomena of the set of probabilities of events from a s.e. and its phenomen-dom).}
The $N$-set of probabilities of events from $\frak{X}$
$$
\breve{p} = \{ p_x : x \in \frak{X} \}
$$
also generates its \emph{$2^{\frak{X}}$-phenomen-dom}, the $2^N$-totality
\begin{equation}\label{p-phenomenness}
\begin{split}
2^{(c|\breve{p})} = \left\{ \breve{p}^{(c|X/\!\!/\frak{X})}, X \subseteq \frak{X} \right\},
\end{split}
\end{equation}
composed of $N$-sets in the form
$$
\breve{p}^{(c|X/\!\!/\frak{X})} = \left\{ p_z : z \in \frak{X}^{(c|X)} \right\},
$$
and defined for $X \subseteq \frak{X}$ as the \emph{$N$-set of probabilities of events from $X$-phenomenon} $\frak{X}^{(c|X)}$ of the s.e. $\frak{X}$ where for $p_z \in \breve{p}^{(c|X/\!\!/\frak{X})}$
$$
p_z = \begin{cases}
p_x, & z=x \in X,\cr
1-p_x& z=x^c \in X^{(c)}.
\end{cases}
$$
In particular, for $X=\frak{X}$
$$
\breve{p}^{(c|\frak{X}/\!\!/\frak{X})} = \left\{ p_x : x \in \frak{X} \right\} = \breve{p}.
$$

We denote by
\begin{equation}\label{def-Th-Kopula1}
\begin{split}
\psi :  \bigotimes_{x \in \frak{X}} [0,1]^x \to \mathbb{R}_0^+
\end{split}
\end{equation}
a nonnegative bounded numerical function defined on the \emph{set-product} \cite{Vorobyev2014famems9}, \emph{$\frak{X}$-hypercube}
$$
[0,1]^{\otimes \frak{X}} = \bigotimes_{x \in \frak{X}} [0,1]^x.
$$
Arguments of $\psi$ form the $N$-set
$$
\breve{w} = \{ w_x : x \in \frak{X} \} \in [0,1]^{\otimes \frak{X}}
$$
which generates its own \emph{$2^\frak{X}$-phenomenon-dom}, the $2^N$-totality
\begin{equation}\label{def-w-PHness}
\begin{split}
2^{(c|\breve{w})}=\left\{ \breve{w}^{(c|X/\!\!/\frak{X})}, X \subseteq \frak{X} \right\}
\end{split}
\end{equation}
of $N$-sets of arguments:
\begin{equation}\label{def-Th-Kopula2}
\begin{split}
\breve{w}^{(c|X/\!\!/\frak{X})} = \left\{ w_z : z \in \frak{X}^{(c|X)} \right\}
\end{split}
\end{equation}
where for $w_z \in \breve{w}^{(c|X/\!\!/\frak{X})}$
$$
w_z = \begin{cases}
w_x, & z=x \in X,\cr
1-w_x& z=x^c \in X^{(c)}.
\end{cases}
$$

Let
\begin{equation}\label{def-Th-Kopula3}
\begin{split}
\Psi_\frak{X} = \left\{ \psi \left| \ \psi :  [0,1]^{\otimes \frak{X}} \to \mathbb{R}_0^+ \right. \right\}
\end{split}
\end{equation}
be the family of all the nonnegative bounded numerical functions on the $\frak{X}$-hypercube.

\texttt{Definition \!\refstepcounter{ctrdef}\arabic{ctrdef}\,\label{norm}({\itshape\footnotesize normalized function on the $\frak{X}$\!-hypercube})\!.}
A function $\psi \in \Psi_\frak{X}$ is called \emph{normalized} on the $\frak{X}$-hypercube if for each $\breve{w} \in [0,1]^{\otimes \frak{X}}$
\begin{equation}\label{def-Th-Kopula4}
\begin{split}
\sum_{X \subseteq \frak{X}} \psi\left(\breve{w}^{(c|X/\!\!/\frak{X})}\right) = 1,
\end{split}
\end{equation}
i.e., the sum of its values on all the $N$-sets of arguments from $2^\frak{X}$-phenomenon-dom $2^{(c|\breve{w})}$ is one.

\texttt{Definition \!\refstepcounter{ctrdef}\arabic{ctrdef}\,\label{one}({\itshape\footnotesize a 1-function on the $\frak{X}$-hypercube}\,).}
A function $\psi \in \Psi_\frak{X}$ is called a \emph{1-function} on the $\frak{X}$-hypercube if for all $\breve{w} \in [0,1]^{\otimes \frak{X}}$ $x$-\emph{marginal equalities} are satisfied for all $x\in\frak X$:
\begin{equation}\label{def-Th-Kopula5}
\begin{split}
\sum_{x \in X \subseteq \frak{X}} \psi\left(\breve{w}^{(c|X/\!\!/\frak{X})}\right) = w_x,
\end{split}
\end{equation}
i.e., the sum of its values on $x$-halves of $N$-sets of arguments from the $2^\frak{X}$-phenomenon-dom $2^{(c|\breve{w})}$ is $w_x$.

Denote by
$$
\Psi_\frak{X}^0 = \left\{ \psi \in \Psi_\frak{X} : \sum_{X \subseteq \frak{X}} \psi\left(\breve{w}^{(c|X/\!\!/\frak{X})}\right) = 1;
\breve{w} \in [0,1]^{\otimes \frak{X}} \right\}
$$
the family of functions, normalized on the $\frak{X}$-hypercube; and by
$$
\Psi_\frak{X}^1 = \left\{\! \psi \in \Psi_\frak{X}\!: \hspace{-8pt}\sum_{x\in X \subseteq \frak{X}} \!\psi\!\left(\breve{w}^{(c|X/\!\!/\frak{X})}\right) = w_x;
\breve{w} \in [0,1]^{\otimes \frak{X}} \!\right\}
$$
the family of 1-functions on the $\frak{X}$-hypercube.

\texttt{Лемма \!\refstepcounter{ctrlem}\arabic{ctrlem}\,\label{lem-normings}({\itshape\footnotesize properties of 1-functions on the $\{x,y\}$-square}\,).} \emph{A strict inclusion is fair:}
$$
\Psi_{\{x,y\}}^1 \subset \Psi_{\{x,y\}}^0.
$$

\texttt{Proof}. In other words, the lemma states: 1) if $\psi \in \Psi_{\{x,y\}}^1$ is a 1-function on the $\{x,y\}$-square then $\psi \in \Psi_{\{x,y\}}^0$ is a normalized function on the $\{x,y\}$-square; 2) among the normalized functions from $\Psi_{\{x,y\}}^0$ there is one which is not a 1-function. But this is obvious, as it is confirmed by the following simple examples.

\begin{figure}[h!]
\centering
\includegraphics[width=2.65in]{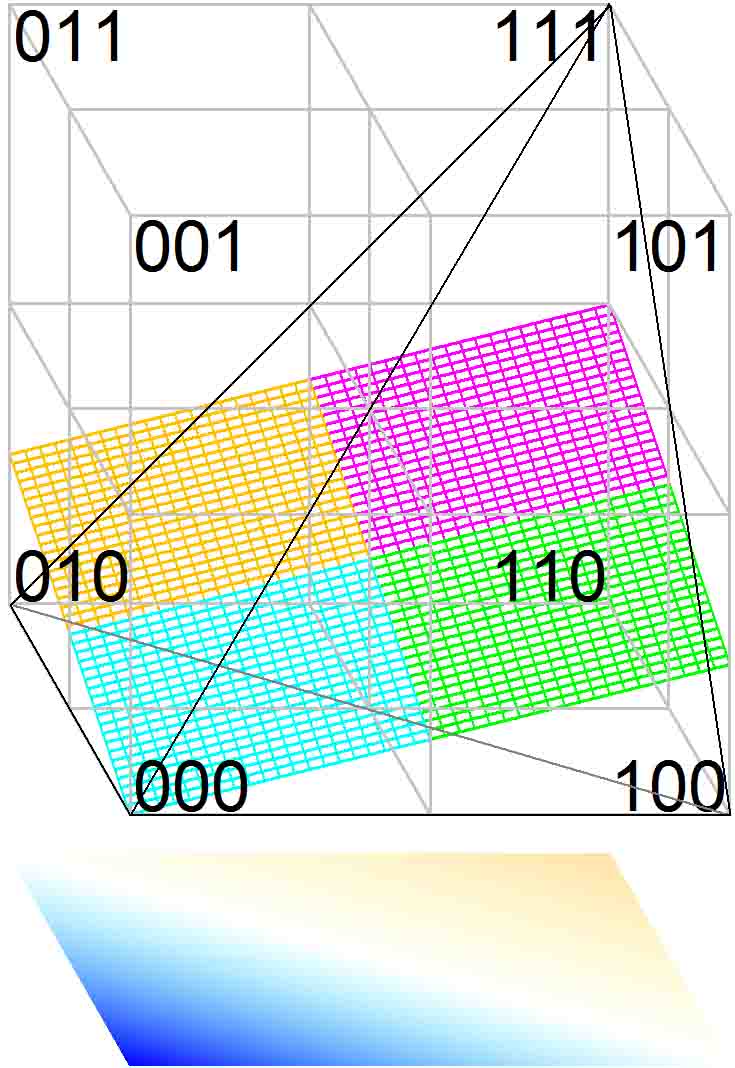}

\vspace{-5pt}

\caption{The graph of the Cartesian representation of the normalized function $\psi(w_x,w_y)=(w_x+w_y)/4$ on the $\{x,y\}$-square from $\Psi_{\{x,y\}}^0$ which is not a 1-function.} %
\label{normalizing-func}
\end{figure}

First, indeed, for the doublet of events $\frak{X}=\{x,y\}$ by the definition of a 1-function, we have
\begin{equation}\label{def-Th-Kopula6}
\begin{split}
\psi(w_x,w_y)+\psi(w_x,1-w_y)=w_x,
\end{split}
\end{equation}
\begin{equation}\label{def-Th-Kopula7}
\begin{split}
\psi(w_x,w_y)+\psi(1-w_x,w_y)=w_y,
\end{split}
\end{equation}
\begin{equation}\label{def-Th-Kopula8}
\begin{split}
\psi(1-w_x,1-w_y)+\psi(w_x,1-w_y)=1-w_y,
\end{split}
\end{equation}
\begin{equation}\label{def-Th-Kopula9}
\begin{split}
\psi(1-w_x,1-w_y)+\psi(1-w_x,w_y)=1-w_x.
\end{split}
\end{equation}
The sums (\ref{def-Th-Kopula6}) and (\ref{def-Th-Kopula9}) as well as the sums (\ref{def-Th-Kopula7}) and (\ref{def-Th-Kopula8}) as a result give
\begin{equation}\label{def-Th-Kopula10}
\begin{split}
\psi(w_x,w_y)+\psi(w_x,1-w_y)+\\
+\psi(1-w_x,1-w_y)+\psi(w_x,1-w_y)=1,
\end{split}
\end{equation}
i.e., $\psi \in \Psi_{\{x,y\}}^0$ is a normalized function on the $\{x,y\}$-square.

Second, the function (see its graph in Fig. \ref{normalizing-func}\footnote{In this figure and others, which illustrate the doublets of events, the map of this function on a unit square is shown under the graph in conditional colors where the white color corresponds to the level 1/4.})
$$
\psi(w_x,w_y)=(w_x+w_y)/4
$$
is normalized on the $\{x,y\}$-square, since
\begin{equation}\nonumber
\begin{split}
\psi(w_x,w_y)&+\psi(w_x,1-w_y)+\\
\psi(1-w_x,w_y)&+\psi(1-w_x,1-w_y)=\\
=(w_x+w_y)/4&+(w_x+1-w_y)/4+\\
(1-w_x+w_y)/4&+(1-w_x+1-w_y)/4 = 1.
\end{split}
\end{equation}
However, it is not a 1-function, since
\begin{equation}\nonumber
\begin{split}
\psi(w_x,w_y)+\psi(w_x,1-w_y)&=(w_x+w_y)/4+\\
(w_x+1-w_y)/4 &= w_x/2+1/4 \not= w_x,\\
\\
\psi(w_x,w_y)+\psi(1-w_x,w_y)&=(w_x+w_y)/4+\\
(1-w_x+w_y)/4 &= w_y/2+1/4 \not= w_y.
\end{split}
\end{equation}
The lemma is proved.

Of course, in the general case, for an arbitrary s.e. $\frak{X}$ the same lemma is fulfilled.

\texttt{Lemma \!\!\refstepcounter{ctrlem}\arabic{ctrlem}\,\label{lem-1-0}({\itshape\footnotesize properties of 1-functions on the $\frak{X}$-hypercube}\,).} \emph{A strict inclusion is fair:}
$$
\Psi_{\frak{X}}^1 \subset \Psi_{\frak{X}}^0.
$$

\texttt{Proof} is similar.

\texttt{Note \!\refstepcounter{ctrnot}\arabic{ctrnot}\,\label{not-representation-1-function}\itshape\footnotesize (a representation of a 1-function on the $\frak{X}$-hypercube in the form of $2^{|\frak{X}|}$-set of functions).} Any 1-function $\psi \in \Psi_{\frak{X}}^1$ on the $\frak{X}$-hypercube $[0,1]^{\otimes \frak{X}}$ for each $\breve{w} \in [0,1]^{\otimes \frak{X}}$ is represented in the form of $2^{|\frak{X}|}$-set of the following functions:
\begin{equation}\label{def-1-function-2N-functions}
\begin{split}
\psi(\breve{w}) &= \Big\{\psi_X(\breve{w}), \ X \subseteq \frak{X}\Big\}=\\
&= \Big\{\psi\left( \breve{w}^{(c|X/\!\!/\frak{X})} \right), \ X \subseteq \frak{X}\Big\}.
\end{split}
\end{equation}

\texttt{Definition \!\refstepcounter{ctrdef}\arabic{ctrdef}\,\label{Kopula}({\itshape\footnotesize Kopula}\,).}
The 1-functions ${\mbox{\boldmath$\mathscr{K}$}} \in \Psi_\frak{X}^1 \subset \Psi_\frak{X}$ is called \emph{$|\frak{X}|$-Kopulas\footnote{see the footnote \ref{Kopulas-footnote} on page \pageref{Kopulas-footnote}.}} of the s.e. $\frak{X}$.

As well as every 1-function (\ref{def-1-function-2N-functions}), any $|\frak{X}|$-Kopula of the s.e. $\frak{X}$ can be represented for $\breve{w} \in [0,1]^{\otimes \frak{X}}$ in the form of $2^{|\frak{X}|}$-set of the following functions:
\begin{equation}\label{def-Kopula-2N-Kopulas}
\begin{split}
{\mbox{\boldmath$\mathscr{K}$}}(\breve{w}) &= \Big\{{\mbox{\boldmath$\mathscr{K}$}}_X(\breve{w}), \ X \subseteq \frak{X}\Big\}=\\
&= \Big\{{\mbox{\boldmath$\mathscr{K}$}}\left( \breve{w}^{(c|X/\!\!/\frak{X})} \right), \ X \subseteq \frak{X}\Big\}.
\end{split}
\end{equation}

\texttt{Note \!\refstepcounter{ctrnot}\arabic{ctrnot}\,\label{not-character-properties}\itshape\footnotesize  (characteristic properties of Kopula).}
Each Kopula ${\mbox{\boldmath$\mathscr{K}$}}$ has two characteristic properties

1) Kopula is \emph{nonnegative}:
\begin{equation}\label{def-Kopula-positive}
\begin{split}
{\mbox{\boldmath$\mathscr{K}$}}\left(\breve{w}^{(c|X/\!\!/\frak{X})}\right) \geqslant 0
\end{split}
\end{equation}
for $X \subseteq \frak{X}$, since by definition ${\mbox{\boldmath$\mathscr{K}$}} \in \Psi_\frak{X}$;

2) Kopula ia satisfied $x$-\emph{marginal equalities}:
\begin{equation}\label{def-Kopula-wx}
\begin{split}
\sum_{x \in X \subseteq \frak{X}} {\mbox{\boldmath$\mathscr{K}$}}\left(\breve{w}^{(c|X/\!\!/\frak{X})}\right) = w_x
\end{split}
\end{equation}
for $x \in \frak{X}$, since by definition ${\mbox{\boldmath$\mathscr{K}$}} \in \Psi_\frak{X}^1$;

From (\ref{def-Kopula-wx}) by Lemma \ref{lem-1-0} a \emph{probabilistic normalization of the Kopula} follows:
\begin{equation}\label{def-Kopula-1}
\begin{split}
\sum_{X \subseteq \frak{X}} {\mbox{\boldmath$\mathscr{K}$}}\left(\breve{w}^{(c|X/\!\!/\frak{X})}\right) = 1.
\end{split}
\end{equation}
From (\ref{def-Kopula-positive}) and (\ref{def-Kopula-1}) \emph{terrace-by-terrace probabilistic normalization of the Kopula} follows:
\begin{equation}\label{def-Kopula-pX}
\begin{split}
0 \leqslant {\mbox{\boldmath$\mathscr{K}$}}\left(\breve{w}^{(c|X/\!\!/\frak{X})}\right) \leqslant 1
\end{split}
\end{equation}
for $X \subseteq \frak{X}$.

\subsection{Characterization of a set of events by Kopula}

The eventological analogue and the preimage of the well-known Sklar theorem on copulas \cite{Sklar1959} is the following theorem.

\texttt{Theorem \!\refstepcounter{ctrTh}\arabic{ctrTh}\,\label{Th-CharByK}({\itshape\footnotesize characterization of a s.e. by Kopula}\,).}
\emph{Let ${\mbox{\boldmath$p$}} = \{p(X /\!\!/ \frak{X}) : X \subseteq \frak{X}\}$
be the e.p.d. of the 1st kind of the s.e. $\frak{X}$ with $\frak{X}$-set of probabilities of marginal events
$
\breve{p} = \{ p_x : x \in \frak{X} \} \in [0,1]^{\otimes \frak{X}}.
$
Then there is a $|\frak{X}|$-Kopula ${\mbox{\boldmath$\mathscr{K}$}} \in \Psi_\frak{X}^1$ that defines a family of e.p.d.'s of the 1st find of the s.e. $\frak{X}$. This family contains the e.p.d. ${\mbox{\boldmath$p$}}$, when Kopula's arguments coincide with $\breve{p}$. In other words, the such Kopula that foe all $X \subseteq \frak{X}$
\begin{equation}\label{CharByK}
\begin{split}
p(X /\!\!/ \frak{X}) = {\mbox{\boldmath$\mathscr{K}$}}_X\left(\breve{p}\right) = {\mbox{\boldmath$\mathscr{K}$}}\left(\breve{p}^{(c|X/\!\!/\frak{X})}\right).
\end{split}
\end{equation}
\uwave{Conversely}, for any $\frak{X}$-set of probabilities of marginal events $\breve{p} \in [0,1]^{\otimes \frak{X}}$ and any $|\frak{X}|$-Kopula ${\mbox{\boldmath$\mathscr{K}$}} \in \Psi_\frak{X}^1$, the function ${\mbox{\boldmath$p$}}\!=\!\{p(X /\!\!/ \frak{X}):X\subseteq~\frak{X}\}$, defined by formulas (\ref{CharByK}) for $X \subseteq \frak{X}$, is an e.p.d. of the 1st kind, which characterizes the s.e. $\frak{X}$ with given $\frak{X}$-set of the probabilities of marginal events~$\breve{p}$.}

\texttt{Proof} is a direct consequence of the properties of e.p.d. of the 1st kind of the s.e. $\frak{X}$ and the $|\frak{X}|$-Kopula. First, if the e.p.d. of the 1st kind  ${\mbox{\boldmath$p$}} = \{p(X /\!\!/ \frak{X}) : X \subseteq \frak{X}\}$ of some s.e. $\frak{X}$ with the $\frak{X}$-set of marginal probabilities $\breve{p} = \{ p_x : x \in \frak{X} \}$ is defined, then from properties of probabilities of the 1st kind it follows that for $x \in \frak{X}$
\begin{equation}\label{xinX}
\begin{split}
p_x = \sum_{x \in X \subseteq \frak{X}} p(X /\!\!/ \frak{X}),
\end{split}
\end{equation}
i.e., the function ${\mbox{\boldmath$\mathscr{K}$}}$, defined by the e.p.d. of the 1st kind ${\mbox{\boldmath$p$}}$ and formulas (\ref{def-Th-Kopula1}), satisfies $x$-marginal equalities for $x \in \frak{X}$:
\begin{equation}\label{xinK}
\begin{split}
p_x = \sum_{x \in X \subseteq \frak{X}} {\mbox{\boldmath$\mathscr{K}$}}(\breve{p}^{(c|X /\!\!/ \frak{X})})
\end{split}
\end{equation}
(required for being a 1-function:: ${\mbox{\boldmath$\mathscr{K}$}} \in \Psi_\frak{X}^1$) and serves as the $|\frak{X}|$-Kopula.

Second, if the function ${\mbox{\boldmath$\mathscr{K}$}}$ is the $|\frak{X}|$-Kopula, then by Lemma \ref{lem-normings}: ${\mbox{\boldmath$\mathscr{K}$}} \in \Psi_\frak{X}^1 \subset \Psi_\frak{X}^0$, i.e., it is normalized and, consequently, by (\ref{CharByK}) the function ${\mbox{\boldmath$p$}}$ is normalized too:
\begin{equation}\label{p-norming}
\begin{split}
1 = \sum_{X \subseteq \frak{X}} p(X /\!\!/ \frak{X}).
\end{split}
\end{equation}
In addition, from (\ref{CharByK}) and from the fact that the $|\frak{X}|$-Kopula is a 1-function, (\ref{xinX}) follows for all $x \in \frak{X}$. Therefore, the function ${\mbox{\boldmath$p$}}$ is a e.p.d. of the 1st kind of the s.e. $\frak{X}$ with the $\frak{X}$-set of marginal probabilities $\breve{p}$. The theorem is proved.

\subsection{Convex combination of Kopulas\label{Kopulas-convex-combination}}

\texttt{Lemma \!\refstepcounter{ctrlem}\arabic{ctrlem}\,\label{lem-Kopulas-convex-combination}({\itshape\footnotesize convex combination of Kopulas}\,).} A \emph{convex combination of an arbitrary set of Kopulas of one and the same s.e. is its Kopula too.}

\texttt{Proof} without tricks. Let $\frak{X}$ be a s.e., and
\begin{equation}\label{set-of-Kopulas}
\begin{split}
{\mbox{\boldmath$\mathscr{K}$}}_1,\ldots,{\mbox{\boldmath$\mathscr{K}$}}_n
\end{split}
\end{equation}
be some set of its Kopulas. Let us prove that their convex combination
\begin{equation}\label{Kopulas-convex-combination-proof1}
\begin{split}
{\mbox{\boldmath$\mathscr{K}$}}=\sum_{i=1}^n \alpha_i{\mbox{\boldmath$\mathscr{K}$}}_i
\end{split}
\end{equation}
(where, of course, $\alpha_1+\ldots+\alpha_n=1, \ \alpha_i \geqslant 0, i=1,\ldots,n$) is also a Kopula. For this it suffices to prove that ${\mbox{\boldmath$\mathscr{K}$}}$ is a 1-function. In other words, that for $x \in \frak{X}$
\begin{equation}\label{Kopulas-convex-combination-proof2}
\begin{split}
\sum_{x \in X \subseteq \frak{X}} {\mbox{\boldmath$\mathscr{K}$}}(\breve{p}^{(c|X /\!\!/ \frak{X})})=p_x.
\end{split}
\end{equation}
Since each Kopula from the set (\ref{set-of-Kopulas}) has a property of a 1-function, then for $x \in \frak{X}$ we get what is required:
\begin{equation}\label{Kopulas-convex-combination-proof2}
\begin{split}
&\sum_{x \in X \subseteq \frak{X}} {\mbox{\boldmath$\mathscr{K}$}}(\breve{p}^{(c|X /\!\!/ \frak{X})})=\\
&=\sum_{i=1}^n \alpha_i\sum_{x \in X \subseteq \frak{X}} {\mbox{\boldmath$\mathscr{K}$}}_i(\breve{p}^{(c|X /\!\!/ \frak{X})}) =\\
&=\sum_{i=1}^n \alpha_i p_x = p_x.
\end{split}
\end{equation}

\texttt{Corollary \!\refstepcounter{ctrcor}\arabic{ctrcor}\,\label{cor-Kopulas-convex-space}({\itshape\footnotesize convex combination of Kopulas}\,).} \emph{For every set of events $\frak{X}$ the space of 1-functions $\Psi_\frak{X}^1$, as well as the space of its Kopulas, is a convex manifold.}

\subsection{Kopula of free variables: A~computational aspect\label{Kopula-free}}

Without set-phenomenon transformations and variable transformations, analytic work on sets of half-rare events (s.h-r.e.'s) (see \cite{Vorobyev2015famems12}) and sets of their marginal probabilities, is unlikely to be effective. However, in specific calculations at first, because of their unaccustomedness, these compulsory wisdoms can cause misunderstandings, leading to errors. Therefore, it is useful, in order to avoid unnecessary stumbling during calculations, to introduce separate notation for \emph{half-rare marginal probabilities events} from s.h-r.e's. $\frak{X}$ and its set-phenomena, that is, probabilities that are not greater than half, in order to distinguish them from \emph{free marginal probabilities}, to the values of which there are no restrictions.

So, we will talk about \emph{half-rare variables (h-r variables)} and \emph{free variables}, assigning special notation to them\footnote{Just remember \cite{Vorobyev2015famems12}, that the formula of $X$-renumbering any $\frak{X}$-set of probabilities of events has the form for $X \subseteq \frak{X}$: $\breve{p}^{(c|X /\!\!/ \frak{X})}=\{p_x, x \in X\}+\{1-p_x, x \in \frak{X}-X\}$.}:
\begin{equation}\label{St-fv-denotations}
\begin{split}
&\breve{p}=\{p_x, x \in \frak{X}\} \in [0,1/2]^{\otimes \frak{X}}\\
-&\mbox{$\frak{X}$-set of half-rare variables},\\
&\breve{p}^{(c|X /\!\!/ \frak{X})}\in [0,1/2]^{\otimes X} \otimes (1/2,1]^{\otimes \frak{X}-X}\\
-&\mbox{$X$-renumbering $\breve{p}$},\\
&\breve{w}=\{w_x, x \in \frak{X}\} \in [0,1]^{\otimes \frak{X}}\\
-&\mbox{$\frak{X}$-set of free variables},\\
&\breve{w}^{(c|X /\!\!/ \frak{X})}\in [0,1]^{\otimes \frak{X}}\\
-&\mbox{$X$-renumbering $\breve{w}$},\\
\end{split}
\end{equation}
and always interpreting them as probabilities of events.
In particular, for the half-rare doublet $\frak{X}=\{x,y\}$ we have:
\begin{equation}\label{St-fv-denotations-doublet}
\begin{split}
&\breve{p}=\{p_x,p_y\} \in [0,1/2]^x \otimes [0,1/2]^y\\
-&\mbox{$\frak{X}$-set of half-rare variables},\\
&p_{xy}(p_x,p_y) \in [0,\min\{p_x,p_y\}]\\
-&\mbox{half-rare function of half-rare variables},\\
&\\
&\breve{w}=\{w_x,w_y\} \in [0,1]^x \otimes [0,1]^y\\
-&\mbox{$\frak{X}$-set of free variables},\\
&w_{xy}(w_x,w_y) \in [0,\min\{w_x,w_y\}]\\
-&\mbox{free function of free variables}.
\end{split}\hspace{-30pt}
\end{equation}

\texttt{Note \!\refstepcounter{ctrnot}\arabic{ctrnot}\,\label{not-replacing}\itshape\enskip\scriptsize (phenomenon replacement between half-rare and free variables).}
For every $X \subset \frak{X}$ \emph{phenomenon replacement} of half-rare variables $\breve{p} \in [0,1/2]^{\otimes \frak{X}}$ by free variables $\breve{w}\in [0,1]^{\otimes \frak{X}}$ and vise-versa is defined for $X \subseteq \frak{X}$ by mutually inverse formulas of the set-phenomenon transformation of the form:
\begin{equation}\label{replacing}
\begin{split}
&\breve{p}=
\breve{p}^{(c|\frak{X} /\!\!/ \frak{X})} =\\
&=\begin{cases}
\breve{w}^{(c|\emptyset /\!\!/ \frak{X})}, & w_x>1/2, x \in \frak{X},\\
\ldots, & \ldots\\
\breve{w}^{(c|X /\!\!/ \frak{X})}, & w_x\leqslant 1/2, x \in X,\\
                                   & w_x>1/2, x \in \frak{X}-X,\\
\ldots, & \ldots \\
\breve{w}^{(c|\frak{X} /\!\!/ \frak{X})}, & w_x \leqslant 1/2, x \in \frak{X}
\end{cases}
\end{split}
\end{equation}
where for $x \in \frak{X}$ the following agreement is always accepted (see, for example, paragraph \ref{N-Venn-diagram}):
\begin{equation}\label{replacing-x}
\begin{split}
p_x =
\begin{cases}
w_x, & w_x \leqslant 1/2,\\
1-w_x, & w_x > 1/2.
\end{cases}
\end{split}
\end{equation}

\subsection{Kopula for a monoplet of events\label{1-Kopulas}}

Let's construct the \emph{$1$-Kopula} ${\mbox{\boldmath$\mathscr{K}$}} \in \Psi^1_\frak{X}$ of a family of e.p.d.'s of monoplet of events $\frak{X}=\{x\}$ with the e.p.d. of the 1st kind
$$
\Big\{p(X /\!\!/ \{x\}), X \subseteq \{x\}\Big\} = \{p(\emptyset /\!\!/ \{x\}),p(x /\!\!/ \{x\})\}
$$
and $\{x\}$-monoplet of marginal probabilities $\{p_x\}$, where
$$
p_x = \mathbf{P}(x) = p(x /\!\!/ \{x\}).
$$
In other words, let's construct a 1-function on the unit $\frak{X}$-segment, i.e., a such nonnegative bounded numerical function
$$
{\mbox{\boldmath$\mathscr{K}$}} : [0,1] \to [0,1],
$$
that for $x \in \frak{X}$
$$
\sum_{x \in X \subseteq \frak{X}} {\mbox{\boldmath$\mathscr{K}$}}\left(\breve{w}^{(c|X/\!\!/\frak{X})}\right) = w_x.
$$
Since for $X \subseteq \frak{X}=\{x\}$
$$
{\mbox{\boldmath$\mathscr{K}$}}\left(\breve{w}^{(c|X/\!\!/\{x\})}\right) =
\begin{cases}
{\mbox{\boldmath$\mathscr{K}$}}\left(1-w_x\right), & X=\emptyset,\cr
{\mbox{\boldmath$\mathscr{K}$}}\left(w_x\right), & X=\{x\},
\end{cases}
$$
then a marginal and global normalization of the function $\mbox{\boldmath${\mbox{\boldmath$\mathscr{K}$}}$}$ are written as:
\begin{equation}\label{1-normings}
\begin{split}
{\mbox{\boldmath$\mathscr{K}$}}(w_x) &= w_x,\\
{\mbox{\boldmath$\mathscr{K}$}}(w_x)+{\mbox{\boldmath$\mathscr{K}$}}(1-w_x) &= 1,
\end{split}
\end{equation}
and the global normalization obviously follows from the marginal one, which agrees with Lemma \ref{lem-1-0}; and from the marginal normalization it follows that the 1-copula ${\mbox{\boldmath$\mathscr{K}$}}$ of an arbitrary monoplet of events $\frak{X}=\{x\}$ is defined for free variables $\breve{w}=\{w_x\} \in [0,1]^x$ by a one formula:
\begin{equation}\label{1-Kopula-def}
\begin{split}
{\mbox{\boldmath$\mathscr{K}$}}(\breve{w})={\mbox{\boldmath$\mathscr{K}$}}(w_x) = w_x,
\end{split}
\end{equation}
which provides two values on each $2^{(c|\breve{w})}$-penomenon-dom by ``free'' formulas:
\begin{equation}\label{1-Kopula-formula}
\begin{split}
&{\mbox{\boldmath$\mathscr{K}$}}\left(\breve{w}^{(c|X/\!\!/\{x\})}\right) =\\
&=
\begin{cases}
{\mbox{\boldmath$\mathscr{K}$}}\left(1-w_x\right) = 1-w_x, & X = \emptyset,\\
{\mbox{\boldmath$\mathscr{K}$}}\left(w_x\right) = w_x, & X = \{x\}.
\end{cases}
\end{split}
\end{equation}
and the e.p.d. of the 1st kind of this monoplet with $\{x\}$-monoiplet of probabilities of events $\breve{p}=\{p_x\}\in [0,1]^x$ are defined for half-rare variables by the 1-Kopula (\ref{1-Kopula-def}) for $X \subseteq \{x\}$ by exactly the same ``half-rare'' formulas:
\begin{equation}\label{1-epd-from-Kopula}
\begin{split}
&p(X /\!\!/ \{x\})={\mbox{\boldmath$\mathscr{K}$}}\left(\breve{p}^{(c|X/\!\!/\{x\})}\right) =\\
&=
\begin{cases}
{\mbox{\boldmath$\mathscr{K}$}}\left(1-p_x\right) = 1-p_x, & X = \emptyset,\\
{\mbox{\boldmath$\mathscr{K}$}}\left(p_x\right) = p_x, & X = \{x\}.
\end{cases}
\end{split}
\end{equation}

\subsection{Kopulas for a doublet of events}

Let's construct an example of \emph{2-Kopulas} ${\mbox{\boldmath$\mathscr{K}$}} \in \Psi^1_\frak{X}$ of families of a doublet of events $\frak{X}=\{x,y\}$, in other words, let's construct on the unit $\{x,y\}$-square the such nonnegative bounded numerical functions
$$
{\mbox{\boldmath$\mathscr{K}$}} : [0,1]^{\otimes\{x,y\}} \to [0,1],
$$
that for all $z \in \{x,y\}$
$$
\sum_{z \in Z \subseteq \{x,y\}} {\mbox{\boldmath$\mathscr{K}$}}\left(\breve{w}^{(c|Z/\!\!/\{x,y\})}\right) = w_z.
$$
Since each 2-set of arguments $\breve{w} \in [0,1]^x \otimes [0,1]^y$ generates $2^{\{x,y\}}$-phenomenon-dom
\begin{equation}\label{2-PHness}
\begin{split}
2^{(c|\breve{w})}=\{\breve{w}, \breve{w}^{(c|\{x\})}, \breve{w}^{(c|\{y\})}, \breve{w}^{(c|\emptyset)}\},
\end{split}
\end{equation}
composed from forth its set-phenomena
\begin{equation}\label{2-setPHs}
\begin{split}
\breve{w}=\breve{w}^{(c|\{x,y\}/\!\!/\{x,y\})}&=\{w_x,w_y\},\\
\breve{w}^{(c|\{x\}/\!\!/\{x,y\})}&=\{w_x,1-w_y\},\\
\breve{w}^{(c|\{x\}/\!\!/\{x,y\})}&=\{1-w_x,w_y\},\\
\breve{w}^{(c|\emptyset/\!\!/\{x,y\})}&=\{1-w_x,1-w_y\},
\end{split}
\end{equation}
then
\begin{equation}\nonumber
\begin{split}
{\mbox{\boldmath$\mathscr{K}$}}\left(\breve{w}\right) &= {\mbox{\boldmath$\mathscr{K}$}}(w_x,w_y),\\
{\mbox{\boldmath$\mathscr{K}$}}\left(\breve{w}^{(c|\{x\}/\!\!/\{x,y\})}\right) &= {\mbox{\boldmath$\mathscr{K}$}}(w_x,1-w_y),\\
{\mbox{\boldmath$\mathscr{K}$}}\left(\breve{w}^{(c|\{y\}/\!\!/\{x,y\})}\right) &= {\mbox{\boldmath$\mathscr{K}$}}(1-w_x,w_y)\\
{\mbox{\boldmath$\mathscr{K}$}}\left(\breve{w}^{(\!c|\emptyset\!/\!\!/\{x,y\})}\right) &=  {\mbox{\boldmath$\mathscr{K}$}}(1-w_x,1-w_y),
\end{split}
\end{equation}
and normalizations for every $\breve{w} \in [0,1]^{\otimes \{x,y\}}$ are written as:
\begin{equation}\nonumber
\begin{split}
{\mbox{\boldmath$\mathscr{K}$}} (w_x,w_y)+
{\mbox{\boldmath$\mathscr{K}$}} (w_x,1-w_y)&= w_x,\\
{\mbox{\boldmath$\mathscr{K}$}} (w_x,w_y)+
{\mbox{\boldmath$\mathscr{K}$}} (1-w_x,w_y)&= w_y,\\
{\mbox{\boldmath$\mathscr{K}$}} (w_x,w_y)+
{\mbox{\boldmath$\mathscr{K}$}} (w_x,1-w_y)&+\\
+{\mbox{\boldmath$\mathscr{K}$}} (1-w_x,w_y)+
{\mbox{\boldmath$\mathscr{K}$}} (1-w_x,1-w_y) &= 1.
\end{split}
\end{equation}
The e.p.d. of the 1st kind of doublet of events is defined by the 2-Kopula for $X \subseteq \{x,y\}$ in half-rare variables by general formulas:
\begin{equation}\label{2-Kopula-rare-formulas}
\begin{split}
&p(X /\!\!/ \{x,y\})={\mbox{\boldmath$\mathscr{K}$}}\left(\breve{p}^{(c|X/\!\!/\{x,y\})}\right) =\\
&=
\begin{cases}
{\mbox{\boldmath$\mathscr{K}$}}\left(1-p_x,1-p_y\right), & X = \emptyset,\\
{\mbox{\boldmath$\mathscr{K}$}}\left(p_x,1-p_y\right), & X = \{x\}.\\
{\mbox{\boldmath$\mathscr{K}$}}\left(1-p_x,p_y\right), & X = \{y\},\\
{\mbox{\boldmath$\mathscr{K}$}}\left(p_x,p_y\right), & X = \{x,y\},
\end{cases}\\
&=
\begin{cases}
1-p_x-p_y+p_{xy}(\breve{p}), & X = \emptyset,\\
p_x-p_{xy}(\breve{p}), & X = \{x\}.\\
p_y-p_{xy}(\breve{p}), & X = \{y\},\\
p_{xy}(\breve{p}), & X = \{x,y\},
\end{cases}
\end{split}
\end{equation}
where $p_{xy}(\breve{p})$ is functional parameter that has a sense of probability of double intersection.

This e.p.d. of the 1st kind of doublet of events in the free functional parameters and variables (after replacement (\ref{replacing-x})) has the form:
\begin{equation}\label{2-Kopula-free-formulas}
\begin{split}
&p(X /\!\!/ \{x,y\})={\mbox{\boldmath$\mathscr{K}$}}\left(\breve{w}^{(c|X/\!\!/\{x,y\})}\right) =\\
&=
\begin{cases}
w_x+w_y-1+w_{xy}(1-w_x,1-w_y),&\\
w_x>1/2, w_y>1/2 \Leftrightarrow X=\emptyset,&\\
&\\
w_x-w_{xy}(w_x,1-w_y),        &\\
w_x\leqslant 1/2, w_y>1/2 \Leftrightarrow X=\{x\},&\\
&\\
w_y-w_{xy}(1-w_x,w_y),          &\\
w_x>1/2,w_y \leqslant 1/2 \Leftrightarrow X=\{y\},&\\
&\\
w_{xy}(w_x,w_y),                 &\\
w_x\leqslant 1/2, w_y \leqslant 1/2 \Leftrightarrow X=\{x,y\}.&
\end{cases}
\end{split}
\end{equation}

\subsection{Kopula for a doublet of independent events}

\begin{figure}[h!]
\centering
\includegraphics[width=2.65in]{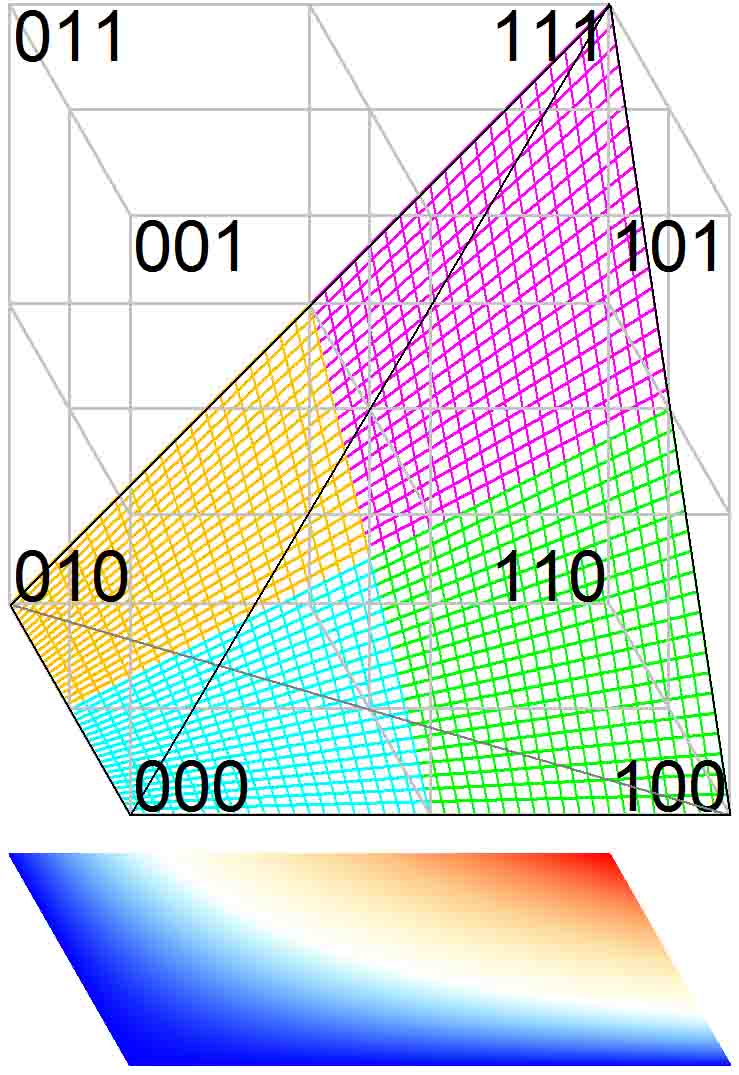}

\vspace{-5pt}

\caption{Graphs of Cartesian representation of the 2-Kopula of a family of e.p.d.'s of an independent half-rare doublet of events $\{x,y\}$;
probabilities of the 1st kind are marked by different colors : $p(xy)$ (aqua), $p(x)$ (lime), $p(y)$ (yellow) и $p(\emptyset)$ (fuchsia).} %
\label{ind-panKopula-1}
\end{figure}

The simplest example of a 1-function on a $\{x,y\}$-square is the so-called \emph{independent 2-Kopula}, which for free variables $\breve{w} \in [0,1]^{\otimes \{x,y\}}$ is defined by the formula:
\begin{equation}\label{ind-2-Kopula}
\begin{split}
{\mbox{\boldmath$\mathscr{K}$}}\left(\breve{w}\right) &= w_xw_y.
\end{split}
\end{equation}
This provides it on each $2^{(c|\breve{w})}$-phenomenon the following four values:
\begin{equation}\label{2-PHness}
\begin{split}
{\mbox{\boldmath$\mathscr{K}$}}\left(\breve{w}^{(c|\{x,y\}/\!\!/\{x,y\})}\right) &= w_xw_y,\\
{\mbox{\boldmath$\mathscr{K}$}}\left(\breve{w}^{(c|\{x\}/\!\!/\{x,y\})}\right)   &= w_x(1-w_y),\\
{\mbox{\boldmath$\mathscr{K}$}}\left(\breve{w}^{(c|\{y\}/\!\!/\{x,y\})}\right)   &= (1-w_x)w_y,\\
{\mbox{\boldmath$\mathscr{K}$}}\left(\breve{w}^{(\!c|\emptyset\!/\!\!/\{x,y\})}\right)&=  (1-w_x)(1-w_y).
\end{split}
\end{equation}
Indeed, the so-defined \emph{independent 2-Kopula} is a 1-function because
$$
\sum_{x \in X \subseteq \{x,y\}}\!\!\!\!\!\!\!\! {\mbox{\boldmath$\mathscr{K}$}}\left(\breve{w}^{(c|X/\!\!/\{x,y\})}\right)
= w_xw_y+w_x(1-w_y) = w_x,
$$
$$
\sum_{y \in X \subseteq \{x,y\}}\!\!\!\!\!\!\!\! {\mbox{\boldmath$\mathscr{K}$}}\left(\breve{w}^{(c|X/\!\!/\{x,y\})}\right)
= w_xw_y+(1-w_x)w_y = w_y.
$$
The e.p.d. of the 1st kind of doublet of independent events with the $\{x,y\}$-set of probabilities of events $\breve{p}$ is defined by four values of the independent 2-Kopula (\ref{ind-2-Kopula}) on its $2^{(c|\breve{p})}$-penomenon-dom by general formulas in half-rare variables (see Fig. \ref{ind-panKopula-1}), i.e., for $X \subseteq \{x,y\}$:
\begin{equation}\label{2-epd-from-ind-Kopula}
\begin{split}
&p(X /\!\!/ \{x,y\})={\mbox{\boldmath$\mathscr{K}$}}\left(\breve{p}^{(c|X/\!\!/\{x,y\})}\right) =\\
&=
\begin{cases}
(1-p_x)(1-p_y), & X = \emptyset,\\
p_x(1-p_y), & X = \{x\}.\\
(1-p_x)p_y, & X = \{y\},\\
p_xp_y, & X = \{x,y\}.
\end{cases}
\end{split}
\end{equation}

\subsection{2-Kopula of free variables: A~computational aspect\label{2-Kopula-free}}

With the phenomenal substitution (\ref{replacing}) \emph{half-rare 2-Kopula} as a function of the free variables takes the equivalent form:
\begin{equation}\label{2-KopulaSt-fvs}
\begin{split}
{\mbox{\boldmath$\mathscr{K}$}}\left(\breve{w}\right)
&=\begin{cases}
w_{xy}\left(\breve{w}^{(c|\frak{X}/\!\!/\frak{X})}\right), &\\
\hspace{20pt}\breve{w} \in [0,1/2]^x \otimes [0,1/2]^y,&\\[5pt]
w_x-w_{xy}\left(\breve{w}^{(c|\{x\}/\!\!/\frak{X})}\right), &\\
\hspace{20pt} \breve{w} \in [0,1/2]^x \otimes (1/2,1]^y,\\[5pt]
w_y-w_{xy}\left(\breve{w}^{(c|\{y\}/\!\!/\frak{X})}\right), &\\
\hspace{20pt} \breve{w} \in (1/2,1]^x \otimes [0,1/2]^y,\\[5pt]
w_x+w_y-1+w_{xy}\left(\breve{w}^{(c|\emptyset/\!\!/\frak{X})}\right), &\\
\hspace{20pt} \breve{w} \in (1/2,1]^x \otimes (1/2,1]^y.
\end{cases}\\
\end{split}
\end{equation}
We rewrite this formula a pair of times, in order to understand the properties of the phenomenon substitution of variables and not get confused in the calculations:
\begin{equation}\label{2-KopulaSt-fvs1}
\hspace{-2pt}\begin{split}
{\mbox{\boldmath$\mathscr{K}$}}\left(\breve{w}\right)
&=\begin{cases}
w_{xy}\left(w_x,w_y\right), &\\
\hspace{20pt}\breve{w} \in [0,1/2]^x \otimes [0,1/2]^y,&\\[5pt]
w_x-w_{xy}\left(w_x,1-w_y\right), &\\
\hspace{20pt} \breve{w} \in [0,1/2]^x \otimes (1/2,1]^y,\\[5pt]
w_y-w_{xy}\left(1-w_x,w_y\right), &\\
\hspace{20pt} \breve{w} \in (1/2,1]^x \otimes [0,1/2]^y,\\[5pt]
w_x+w_y-1+w_{xy}\left(1-w_x,1-w_y\right), &\\
\hspace{20pt} \breve{w} \in (1/2,1]^x \otimes (1/2,1]^y;
\end{cases}
\end{split}\hspace{-13pt}
\end{equation}
and again with restrictions in the form of the familiar ``human'' inequalities\footnote{probabilistic normalization: ``$0 \leqslant ... \leqslant 1$'' is assumed by default.}:
\begin{equation}\label{2-KopulaSt-fvs2}
\hspace{-2pt}\begin{split}
{\mbox{\boldmath$\mathscr{K}$}}\left(\breve{w}\right)
&=\begin{cases}
w_{xy}\left(w_x,w_y\right), &\\
\hspace{20pt} w_x \leqslant 1/2, w_y \leqslant 1/2,&\\[5pt]
w_x-w_{xy}\left(w_x,1-w_y\right), &\\
\hspace{20pt} w_x \leqslant 1/2, 1/2 < w_y,\\[5pt]
w_y-w_{xy}\left(1-w_x,w_y\right), &\\
\hspace{20pt} 1/2 < w_x, w_y \leqslant 1/2,\\[5pt]
w_x+w_y-1+w_{xy}\left(1-w_x,1-w_y\right), &\\
\hspace{20pt} 1/2 < w_x, 1/2 < w_y
\end{cases}
\end{split}\hspace{-13pt}
\end{equation}
where
\begin{equation}\label{2-KopulaSt-FBs}
\begin{split}
&0 \leqslant w_{xy}(w_x,w_y) \leqslant \min \{w_x,w_y\},\\
&0 \leqslant w_{xy}(w_x,1\!-\!w_y) \leqslant \min \{w_x,1\!-\!w_y\},\\
&0 \leqslant w_{xy}(1\!-\!w_x,w_y) \leqslant \min \{1\!-\!w_x,w_y\},\\
&0 \leqslant w_{xy}(1\!-\!w_x,1\!-\!w_y) \leqslant \min \{1\!-\!w_x,1\!-\!w_y\}\\
\end{split}
\end{equation}
are the Fr\'echet inequalities for $w_{xy}$ as the half-rare probability of double intersection of half-rare events: either half-rare events $x$ or $y$, or their half-rare complements, when events $x$ or $y$ are not half-rare.

Note that the conditional formulas (\ref{2-KopulaSt-fvs}), (\ref{2-KopulaSt-fvs1}) and (\ref{2-KopulaSt-fvs2}) can not be rewritten as four unconditional formulas, because these conditions are in the right, and not in the left. This is explained exclusively by the properties of the phenomenon replacement of half-rare variables by free ones (\ref{replacing}), which, for this reason, leads to formulas that are convenient for calculations.

\texttt{Note \!\refstepcounter{ctrnot}\arabic{ctrnot}\,\label{not1}\itshape\enskip\scriptsize (half-rare 2-Kopula of free variables of an independent doublet of events).}
With the functional parameter $w_{xy}(w_x,w_y)=w_xw_y$, which corresponds to the probability of double intersection of independent events $x$ and $y$ happening with probabilities $w_x$ and $w_y$, and means, of course, that the all the following four equations are satisfied:
\begin{equation}\nonumber
\begin{split}
w_{xy}(w_x,w_y)&=w_xw_y,\\
w_{xy}(w_x,1-w_y)&=w_x(1-w_y),\\
w_{xy}(1-w_x,w_y)&=(1-w_x)w_y,\\
w_{xy}(1-w_x,1-w_y)&=(1-w_x)(1-w_y);
\end{split}
\end{equation}
from (\ref{2-KopulaSt-fvs2}) it follows that a half-rare 2-Kopula from free variables of the family of e.p.d.'s of the independent doublet of events $\frak{X}=\{x,y\}$ with $\frak{X}$-sets of free marginal probabilities $\breve{w}=\{w_x,w_y\} \in [0,1]^{\otimes \frak{X}}$ has the \emph{same} view on all $2^{(c|\breve{w})}$-phenomenon-doms:
\begin{equation}\label{2-KopulaSt-fvs3}
\begin{split}
{\mbox{\boldmath$\mathscr{K}$}}\left(\breve{w}\right) = w_xw_y.
\end{split}
\end{equation}

\subsection{Upper 2-Kopula of Fr\'echet}

An example of a 1-function on a $\{x,y\}$-square is the so-called \emph{upper 2-Kopula of Fr\'echet}, which suggests the probabilities of a double intersection to be its upper Fr\'echet boundary. In other words, the only functional free parameter in (\ref{2-Kopula-free-formulas}) is: \begin{equation}\label{up-FRECHET-free}
\begin{split}
w_{xy}(\breve{w}) = w_{xy}^+(\breve{w}) = \min \{ w_x, w_y \}.
\end{split}
\end{equation}
The upper 2-Kopula of Fr\'echet from free variables $\breve{w} \in [0,1]^{\otimes \{x,y\}}$ is defined by the formulas:
\begin{equation}\label{2-Kopula-free-up-FRECHET0}
\begin{split}
&p(X /\!\!/ \{x,y\})={\mbox{\boldmath$\mathscr{K}$}}\left(\breve{w}^{(c|X/\!\!/\{x,y\})}\right) =\\
&=
\begin{cases}
w_x+w_y-1+\min \{ 1-w_x, 1-w_y \},&\\
w_x>1/2, w_y>1/2 \Leftrightarrow X=\emptyset,&\\
&\\[-10pt]
w_x-\min \{ w_x, 1-w_y \},        &\\
w_x\leqslant 1/2, w_y>1/2 \Leftrightarrow X=\{x\},&\\
&\\[-10pt]
w_y-\min \{ 1-w_x, w_y \},          &\\
w_x>1/2,w_y \leqslant 1/2 \Leftrightarrow X=\{y\},&\\
&\\[-10pt]
\min \{ w_x, w_y \},                 &\\
w_x\leqslant 1/2, w_y \leqslant 1/2 \Leftrightarrow X=\{x,y\}.&
\end{cases}
\end{split}
\end{equation}
After simple transformations, these formulas provide the upper 2-Kopula of Fr\'echet on each $2^{(c|\breve{w})}$-phenomenon-dom the following four values of free variables:
\begin{equation}\label{2-Kopula-free-up-FRECHET}
\begin{split}
&p(X /\!\!/ \{x,y\})={\mbox{\boldmath$\mathscr{K}$}}\left(\breve{w}^{(c|X/\!\!/\{x,y\})}\right) =\\
&=
\begin{cases}
\min \{ w_x, w_y \},&\\
w_x>1/2, w_y>1/2 \Leftrightarrow X=\emptyset,&\\
&\\[-10pt]
\max \{ 0, w_x+w_y-1 \},        &\\
w_x\leqslant 1/2, w_y>1/2 \Leftrightarrow X=\{x\},&\\
&\\[-10pt]
\max \{ 0, w_x+w_y-1 \},          &\\
w_x>1/2,w_y \leqslant 1/2 \Leftrightarrow X=\{y\},&\\
&\\[-10pt]
\min \{ w_x, w_y \},                 &\\
w_x\leqslant 1/2, w_y \leqslant 1/2 \Leftrightarrow X=\{x,y\},&\\
\end{cases}\\
&=
\begin{cases}
w_{xy}^+(w_x, w_y),&\\
w_x>1/2, w_y>1/2 \Leftrightarrow X=\emptyset,&\\
&\\[-10pt]
w_{xy}^-(w_x, w_y),        &\\
w_x\leqslant 1/2, w_y>1/2 \Leftrightarrow X=\{x\},&\\
&\\[-10pt]
w_{xy}^-(w_x, w_y),          &\\
w_x>1/2,w_y \leqslant 1/2 \Leftrightarrow X=\{y\},&\\
&\\[-10pt]
w_{xy}^+(w_x, w_y),                 &\\
w_x\leqslant 1/2, w_y \leqslant 1/2 \Leftrightarrow X=\{x,y\},&
\end{cases}
\end{split}
\end{equation}
which, as can be seen, are defined by the upper and lower Fr\'echet boundaries of the probability of double intersecyion, depending on the combination of the values of the free variables.

The same four values from the half-rare variables have the form:
\begin{equation}\label{2-Kopula-free-up-FRECHET}
\begin{split}
&p(X /\!\!/ \{x,y\})={\mbox{\boldmath$\mathscr{K}$}}\left(\breve{p}^{(c|X/\!\!/\{x,y\})}\right) =\\
&=
\begin{cases}
\min \{ 1-p_x, 1-p_y \},& X=\emptyset,\\
&\\[-10pt]
\max \{ 0, p_x-p_y \},        & X=\{x\},\\
&\\[-10pt]
\max \{ 0, p_y-p_x \},          & X=\{y\},\\
&\\[-10pt]
\min \{ p_x, p_y \},                 & X=\{x,y\}.\\
\end{cases}\\
\end{split}
\end{equation}
If $p_x \geqslant p_y$, this formula takes the form:
\begin{equation}\label{2-Kopula-free-up-FRECHETpxpy}
\begin{split}
&p(X /\!\!/ \{x,y\})={\mbox{\boldmath$\mathscr{K}$}}\left(\breve{p}^{(c|X/\!\!/\{x,y\})}\right) =\\
&=
\begin{cases}
1-p_x,& X=\emptyset,\\
&\\[-10pt]
p_x-p_y,        & X=\{x\},\\
&\\[-10pt]
0,          & X=\{y\},\\
&\\[-10pt]
p_y,                 & X=\{x,y\}.\\
\end{cases}\\
\end{split}
\end{equation}
And if $p_x < p_y$, then this formula takes the form:
\begin{equation}\label{2-Kopula-free-up-FRECHETpypx}
\begin{split}
&p(X /\!\!/ \{x,y\})={\mbox{\boldmath$\mathscr{K}$}}\left(\breve{p}^{(c|X/\!\!/\{x,y\})}\right) =\\
&=
\begin{cases}
1-p_y,& X=\emptyset,\\
&\\[-10pt]
0,        & X=\{x\},\\
&\\[-10pt]
p_y-p_x,          & X=\{y\},\\
&\\[-10pt]
p_y,                 & X=\{x,y\}.\\
\end{cases}\\
\end{split}
\end{equation}

So the definite \emph{upper 2-Kopula of Fr\'echet} is indeed a 1-function, due to the fact that when $w_x \geqslant w_y$
\begin{equation}\label{2-Kopula-up-FRECHET-1function-geq}
\hspace*{-3pt}
\begin{split}
\sum_{x \in X \subseteq \{x,y\}}\!\!\!\!\!\!\!\! {\mbox{\boldmath$\mathscr{K}$}}\left(\breve{w}^{(c|X/\!\!/\{x,y\})}\right)
&= w_y+(w_x-w_y) = w_x,\\
\sum_{y \in X \subseteq \{x,y\}}\!\!\!\!\!\!\!\! {\mbox{\boldmath$\mathscr{K}$}}\left(\breve{w}^{(c|X/\!\!/\{x,y\})}\right)
&= w_y+0 = w_y,
\end{split}\hspace*{-13pt}
\end{equation}
and when $w_x < w_y$
\begin{equation}\label{2-Kopula-up-FRECHET-1function-leq}
\hspace*{-3pt}
\begin{split}
\sum_{x \in X \subseteq \{x,y\}}\!\!\!\!\!\!\!\! {\mbox{\boldmath$\mathscr{K}$}}\left(\breve{w}^{(c|X/\!\!/\{x,y\})}\right)
&= w_x+0 = w_x,\\
\sum_{y \in X \subseteq \{x,y\}}\!\!\!\!\!\!\!\! {\mbox{\boldmath$\mathscr{K}$}}\left(\breve{w}^{(c|X/\!\!/\{x,y\})}\right)
&= w_x+(w_y-w_x) = w_y.
\end{split}\hspace*{-13pt}
\end{equation}

\begin{figure}[h!]
\centering
\includegraphics[width=2.65in]{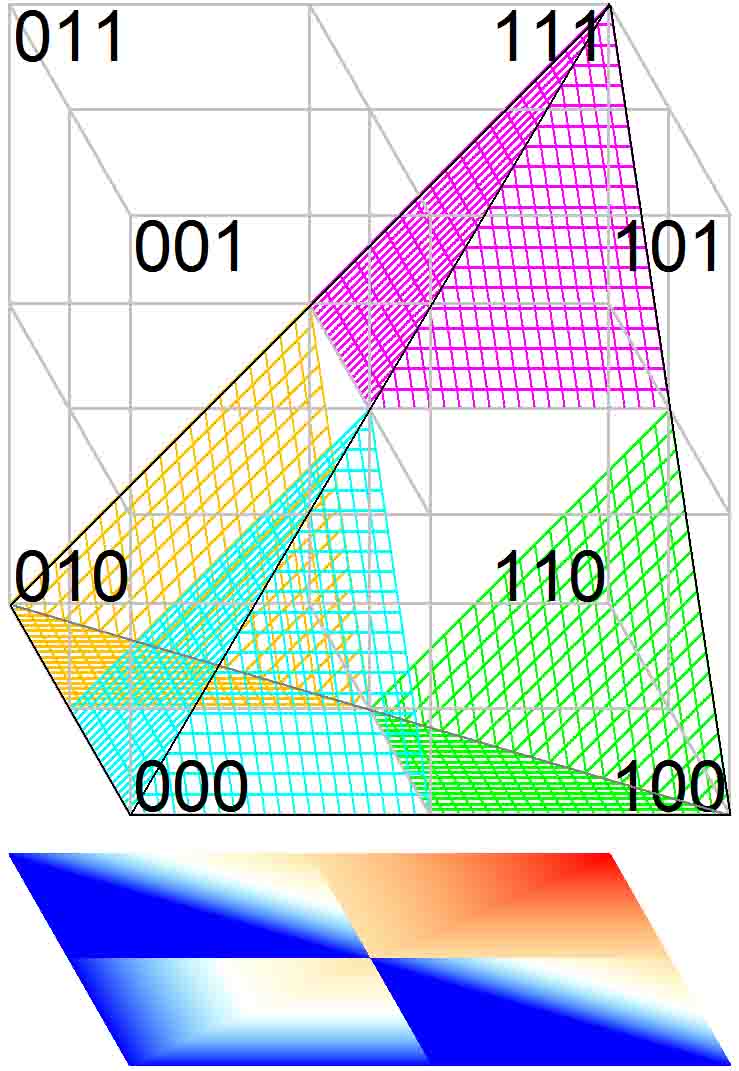}

\vspace{-5pt}

\caption{Graphs of the Cartesian representation of the upper 2-Kopula of Fr\'echet of a family of e.p.d.'s of half-rare doublet of events $\{x,y\}$; probabilities of the 1st kind are marked by different colors: $p(xy)$ (aqua), $p(x)$ (lime), $p(y)$ (yellow) и $p(\emptyset)$ (fuchsia).} %
\label{up-2-Kopula}
\end{figure}

\begin{figure}[h!]
\centering
\includegraphics[width=2.65in]{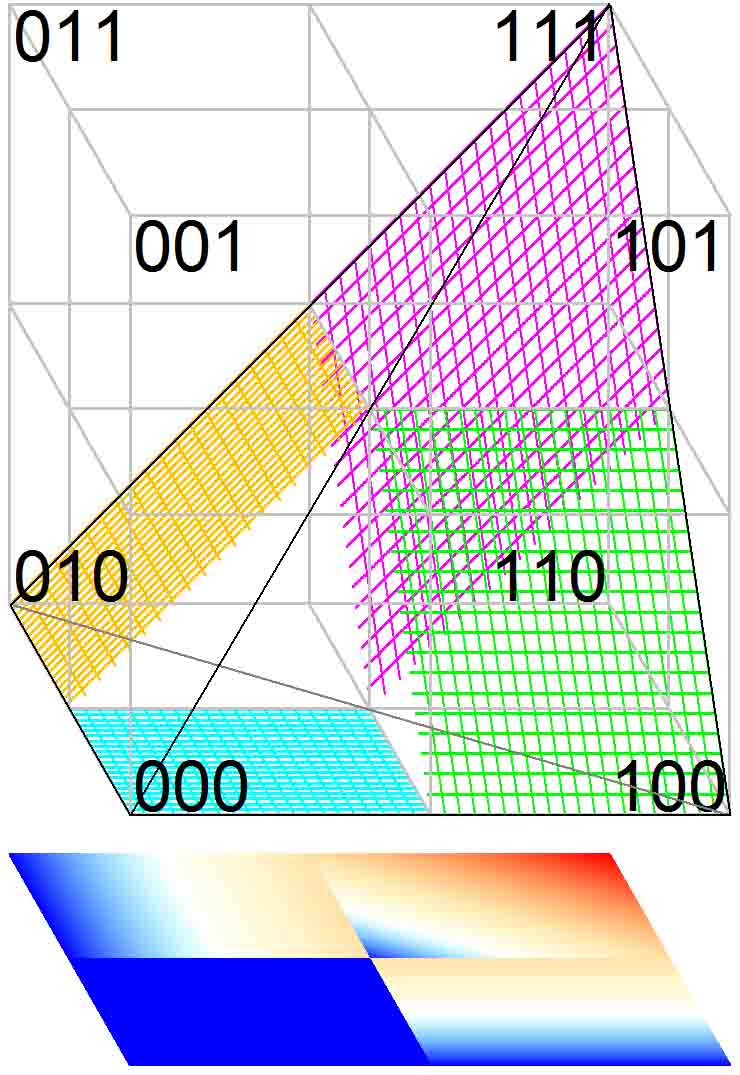}

\vspace{-5pt}

\caption{Graphs of the Cartesian representation of the lower 2-Kopula of Fr\'echet of a family of e.p.d.'s of half-rare doublet of events $\{x,y\}$; probabilities of the 1st kind are marked by different colors: $p(xy)$ (aqua), $p(x)$ (lime), $p(y)$ (yellow) и $p(\emptyset)$ (fuchsia).} %
\label{down-2-Kopula}
\end{figure}

\subsection{Lower 2-Kopula of Fr\'echet}

A once more example of a 1-function on a $\{x,y\}$-square is the so-called \emph{lower 2-Kopula of Fr\'echet}, which suggests the probabilities of a double intersection to be its upper Fr\'echet boundary. In other words, the only functional free parameter in (\ref{2-Kopula-free-formulas}) is:
\begin{equation}\label{up-FRECHET-free}
\begin{split}
w_{xy}(\breve{w}) = w_{xy}^-(\breve{w}) = \max \{ 0, w_x+w_y-1 \}.
\end{split}
\end{equation}
The lower 2-Kopula of Fr\'echet from free variables $\breve{w} \in [0,1]^{\otimes \{x,y\}}$ is defined by the formulas:
\begin{equation}\label{2-Kopula-free-up-FRECHET0}
\begin{split}
&p(X /\!\!/ \{x,y\})={\mbox{\boldmath$\mathscr{K}$}}\left(\breve{w}^{(c|X/\!\!/\{x,y\})}\right) =\\
&=
\begin{cases}
w_x+w_y-1+\max \{ 0, 1-w_x-w_y \},&\\
w_x>1/2, w_y>1/2 \Leftrightarrow X=\emptyset,&\\
&\\[-10pt]
w_x-\max \{ 0, w_x-w_y \},        &\\
w_x\leqslant 1/2, w_y>1/2 \Leftrightarrow X=\{x\},&\\
&\\[-10pt]
w_y-\max \{ 0, w_y-w_x \},          &\\
w_x>1/2,w_y \leqslant 1/2 \Leftrightarrow X=\{y\},&\\
&\\[-10pt]
\max \{ 0, w_x+w_y-1 \},                 &\\
w_x\leqslant 1/2, w_y \leqslant 1/2 \Leftrightarrow X=\{x,y\}.&
\end{cases}
\end{split}
\end{equation}
After simple transformations, these formulas provide the lower 2-Kopula of Fr\'echet on each $2^{(c|\breve{w})}$-phenomenon-dom the following four values of free variables:
\begin{equation}\label{2-Kopula-free-up-FRECHET}
\begin{split}
&p(X /\!\!/ \{x,y\})={\mbox{\boldmath$\mathscr{K}$}}\left(\breve{w}^{(c|X/\!\!/\{x,y\})}\right) =\\
&=
\begin{cases}
\max \{ 0, w_x+w_y-1 \},&\\
w_x>1/2, w_y>1/2 \Leftrightarrow X=\emptyset,&\\
&\\[-10pt]
\min \{ w_x, w_y \},        &\\
w_x\leqslant 1/2, w_y>1/2 \Leftrightarrow X=\{x\},&\\
&\\[-10pt]
\min \{ w_x, w_y \},          &\\
w_x>1/2,w_y \leqslant 1/2 \Leftrightarrow X=\{y\},&\\
&\\[-10pt]
\max \{ 0, w_x+w_y-1 \},                 &\\
w_x\leqslant 1/2, w_y \leqslant 1/2 \Leftrightarrow X=\{x,y\},&\\
\end{cases}\\
&=
\begin{cases}
w_{xy}^-(w_x, w_y),&\\
w_x>1/2, w_y>1/2 \Leftrightarrow X=\emptyset,&\\
&\\[-10pt]
w_{xy}^+(w_x, w_y),        &\\
w_x\leqslant 1/2, w_y>1/2 \Leftrightarrow X=\{x\},&\\
&\\[-10pt]
w_{xy}^+(w_x, w_y),          &\\
w_x>1/2,w_y \leqslant 1/2 \Leftrightarrow X=\{y\},&\\
&\\[-10pt]
w_{xy}^-(w_x, w_y),                 &\\
w_x\leqslant 1/2, w_y \leqslant 1/2 \Leftrightarrow X=\{x,y\},&
\end{cases}
\end{split}
\end{equation}
which, as can be seen, are also defined by the upper and lower Fr\'echet boundaries of the probability of double intersection only in other combinations of the values of the free variables.

The same four values from the half-rare variables have the more simple form:
\begin{equation}\label{2-Kopula-free-up-FRECHET}
\hspace*{-3pt}
\begin{split}
&p(X /\!\!/ \{x,y\})={\mbox{\boldmath$\mathscr{K}$}}\left(\breve{p}^{(c|X/\!\!/\{x,y\})}\right) =\\
&\!=\!
\begin{cases}
\max \{ 0, 1\!-\!p_x\!-\!p_y \},& X\!=\!\emptyset,\\
&\\[-10pt]
\min \{ p_x, 1\!-\!p_y \},        & X\!=\!\{x\},\\
&\\[-10pt]
\min \{ 1\!-\!p_x, p_y \},          & X\!=\!\{y\},\\
&\\[-10pt]
\max \{ 0, p_x\!+\!p_y\!-\!1 \},                 & X\!=\!\{x,y\},\\
\end{cases}\\
&\!=\!
\begin{cases}
1\!-\!p_x\!-\!p_y,& X\!=\!\emptyset,\\
&\\[-10pt]
p_x,              & X\!=\!\{x\},\\
&\\[-10pt]
p_y,              & X\!=\!\{y\},\\
&\\[-10pt]
0,                & X\!=\!\{x,y\},
\end{cases}
\end{split}\hspace*{-23pt}
\end{equation}

So the definite \emph{lower 2-Kopula of Fr\'echet} is indeed a 1-function, due to the fact that for all half-rare variables
\begin{equation}\label{2-Kopula-up-FRECHET-1function-geq}
\hspace*{-3pt}
\begin{split}
\sum_{x \in X \subseteq \{x,y\}}\!\!\!\!\!\!\!\! {\mbox{\boldmath$\mathscr{K}$}}\left(\breve{p}^{(c|X/\!\!/\{x,y\})}\right)
&= p_x+0 = p_x,\\
\sum_{y \in X \subseteq \{x,y\}}\!\!\!\!\!\!\!\! {\mbox{\boldmath$\mathscr{K}$}}\left(\breve{w}^{(c|X/\!\!/\{x,y\})}\right)
&= p_y+0 = p_y.
\end{split}\hspace*{-13pt}
\end{equation}

\begin{figure}[h!]
\centering
\includegraphics[width=2.65in]{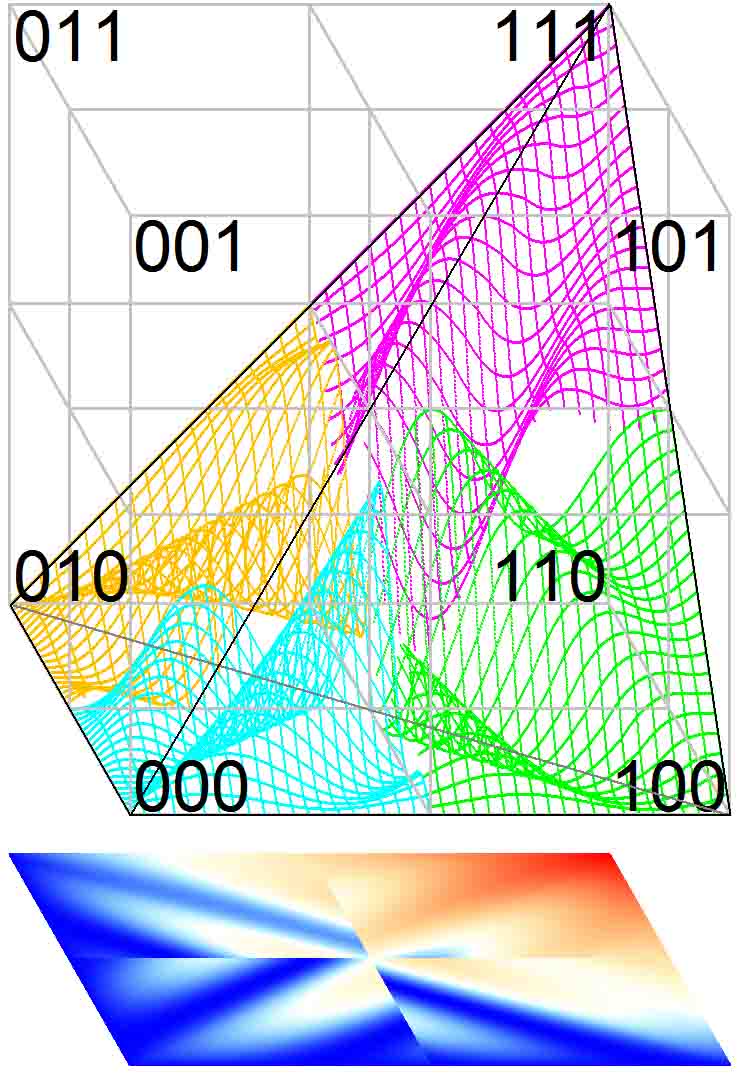}

\vspace{-5pt}
\caption{Graphs of the Cartesian representation of the convex combination of upper and lower 2-Kopulas of Fr\'echet of a family of e.p.d.'s of half-rare doublet of events $\{x,y\}$ with functional weight parameter (\ref{alpha-up-down}) in the formula (\ref{up-down-FRECHET-convex}); probabilities of the 1st kind are marked by different colors: $p(xy)$ (aqua), $p(x)$ (lime), $p(y)$ (yellow) и $p(\emptyset)$ (fuchsia).} %
\label{fig-up-down-FRECHET-sin15}
\end{figure}

\subsection{Convex combinations of the lower, upper and independent 2-Kopulas of Fr\'echet}

A rather general example of a 1-function on a $\{x,y\}$-square is the \emph{convex combinations of the upper, lower, and independent 2-Kopulas ofwise Fr\'echet}, which propose the probabilities of a pair intersection to become a convex combination (this is allowed by the lemma \ref{lem-Kopulas-convex-combination}) of its upper and lower Fr\'echet boundaries, as well as the probability of double intersection of independent events with some functional weighting parameter.

\subsubsection{Convex combination of the lower and upper 2-Kopulas of Fr\'echet}

A convex combination of the lower and upper 2-Kopula of Fr\'echet can be ensured by the unique functional free parameter $w_{xy}(\breve{w})$ in (\ref{2-Kopula-free-formulas}), in which the probability of double intersection is computed by the following formula:
\begin{equation}\label{up-down-FRECHET-convex}
\begin{split}
&w_{xy}(\breve{w}) =\\
&\!=\!
(1\!-\!\alpha)/2w_{xy}^-(\breve{w})+(1\!+\!\alpha)/2 w_{xy}^+(\breve{w})
\end{split}
\end{equation}
where $\alpha=\alpha(\breve{w}) \in [-1,1]$ is an arbitrary function on $[0,1]^{\otimes \{x,y\}}$ with values from $[-1,1]$, and
\begin{equation}\label{FRECHET-up-down}
\begin{split}
w_{xy}^-(\breve{w})&=\max\{0,w_x+w_y-1\},\\
w_{xy}^+(\breve{w})&=\min\{\breve{w}\}
\end{split}
\end{equation}
are the lower and upper Fr\'echet-boundaries of probability of double intersection.

\begin{figure}[h!]
\centering
\includegraphics[width=2.65in]{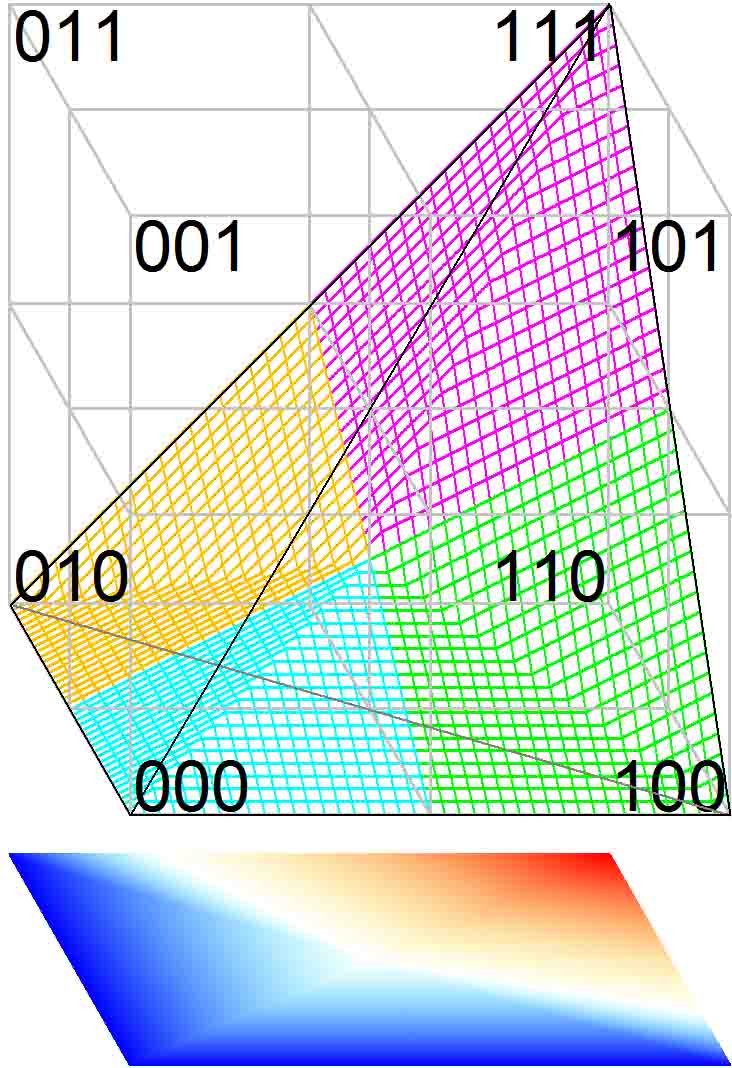}

\vspace{-5pt}

\caption{Graphs of the Cartesian representation of the convex combination of upper and lower 2-Kopulas of Fr\'echet of a family of e.p.d.'s of half-rare doublet of events $\{x,y\}$ with the constant functional weight parameter $\alpha(\breve{w})=0$ in the formula (\ref{up-down-FRECHET-convex}); probabilities of the 1st kind are marked by different colors: $p(xy)$ (aqua), $p(x)$ (lime), $p(y)$ (yellow) и $p(\emptyset)$ (fuchsia).} %
\label{fig-up-down-FRECHET-const}
\end{figure}

For $\alpha = -1 $, the probability $ w_{xy}(\breve{w}) = w_{xy}^-(\breve{w}) $ coincides with the lower Fr\'echet boundary of half-rare marginal probabilities; for $\alpha = 1 $, the probability $ w_{xy}(\breve{w}) = w_{xy}^+(\breve{w}) $ coincides with the upper Fr\'echet boundary of marginal probabilities. Unfortunately, these are the properties of a convex combination such that for $\alpha = 0 $ the probability of double intersection is equal to half of the sum of its lower and upper Fr\'echet boundaries: $ w_{xy}(\breve{w}) =(w_{xy}^-(\breve{w}) + w_{xy}^+(\breve{w})) / 2 $ (see Figure\ref{fig-up-down-FRECHET-const}), and not an independent 2-Kopula, no matter how much we want it. This ``blunder'' of the convex combination can easily be corrected by conjugation of two convex combinations, as done below.

In Fig. \ref{fig-up-down-FRECHET-sin15} it is a graph of this 2-Kopula for a deliberately intricate weight function with values from $[-1,1]$:
\begin{equation}\label{alpha-up-down}
\begin{split}
\alpha=\alpha(\breve{w}) = \sin(15(w_x-w_y)).
\end{split}
\end{equation}

\subsubsection{The conjugation of two convex combinations of the independent one with the lower and upper 2-Kopula of Fr\'echet}

We construct two convex combinations of the the independent 2-copula and the lower and upper 2-Kopulas of Fr\'echet.
The conjugation of these two convex combinations can be ensured by the unique functional free parameter $w_{xy}(\breve{w})$ in (\ref{2-Kopula-free-formulas}) by the following conjugation formula for two convex combinations:
\begin{equation}\label{up-down-FRECHET-conting}
\begin{split}
&w_{xy}(\breve{w}) =\\
&\!=\!\begin{cases}
w_{xy}^+(\breve{w})\max\{\breve{w}\} (1\!+\!\alpha),&\\
\alpha\leqslant 0;&\\
&\\[-6pt]
w_{xy}^+(\breve{w}) \Big(\max\{\breve{w}\} (1\!-\!\alpha)\!+\!\alpha\Big),&\\
\alpha>0,&
\end{cases}
\end{split}
\end{equation}
where $\alpha=\alpha(\breve{w}) \in [-1,1]$ is an arbitrary function on $[0,1]^{\otimes \{x,y\}}$ with values from $[-1,1]$, and
\begin{equation}\label{FRECHET-up}
\begin{split}
w_{xy}^+(\breve{w})&=\min\{\breve{w}\}
\end{split}
\end{equation}
is the upper Fr\'echet-boundary of probability of double intersection. For $\alpha=0$, the probability $w_{xy}(\breve{w})=\min\{\breve{w}\}\max\{\breve{w}\}=w_xw_y$ coincides with the probability of double intersection of independent events; for $\alpha=-1$, the probability  $w_{xy}(\breve{w})=0$ coincides with the lower Fr\'echet-boundary of half-rare marginal probabilities; for $\alpha=1$, the probability $w_{xy}(\breve{w})=w_{xy}^+(\breve{w})$ coincides with the upper Fr\'echet-boundary of marginal probabilities.

In Fig. \ref{fig-up-down-FRECHET-sin15} it is a graph of this 2-Kopula for the same weight function with values from $[-1,1]$ as in the previous example.
\begin{equation}\label{alpha-sin15}
\begin{split}
\alpha=\alpha(\breve{w}) = \sin(15(w_x-w_y)).
\end{split}
\end{equation}

\section{The frame method of construction of Kopula\label{frame-method}}

\subsection{Inserted sets of events and conditional e.p.d.'s\label{inserting-sets}}

\texttt{Definition \!\refstepcounter{ctrdef}\arabic{ctrdef}\,\label{def-inserting-set}\itshape\enskip\scriptsize (inserted s.e.'s).} For each pais of s.e.'s $\mathcal{X}$ and $\mathcal{Y}$ with the joint e.p.d.
\begin{equation}\label{inserting-set-joint-epd}
\begin{split}
\{ p(X+Y /\!\!/ \mathcal{X}+\mathcal{Y})\}, X \subseteq \mathcal{X}, Y \subseteq \mathcal{Y} \}
\end{split}
\end{equation}
for every $Y \subseteq \mathcal{Y}$ the \emph{$Y$-inserted s.e.'s, generated by $\mathcal{X}$, in the frame s.e. $\mathcal{Y}$} are s.e.'s, which are denoted by $\mathcal{X}^{(\cap Y /\!\!/ \mathcal{Y})}$, and defined as the following M-intersection\footnote{M-intersection is an intersection by Minkowski.}:
\begin{equation}\label{inserting-set}
\begin{split}
\mathcal{X}^{(\cap Y /\!\!/ \mathcal{Y})} &= \mathcal{X} \, (\cap) \, \{\textsf{ter}(Y /\!\!/ \mathcal{Y})\} =\\
&=\{ x \cap \textsf{ter}(Y /\!\!/ \mathcal{Y}), \, x \in \mathcal{X} \}
\end{split}
\end{equation}
and have the e.p.d., which coincides with the projection of the joint e.p.d. (\ref{inserting-set-joint-epd}) for fixed $Y \subseteq \mathcal{Y}$ and every $X \subseteq \mathcal{X}$:
\begin{equation}\label{epd-inserting-set}
\begin{split}
p(X \, (\cap) \, \{\textsf{ter}(Y /\!\!/ \mathcal{Y})\}) = p(X+Y /\!\!/ \mathcal{X}+\mathcal{Y})\}.
\end{split}
\end{equation}

\texttt{Definition \!\refstepcounter{ctrdef}\arabic{ctrdef}\,\label{def-inserting-set-pseudo-epd}\itshape\enskip\scriptsize (event-probabilistic pseudo-distribution of an inserted s.e.).} For each $Y \subseteq \mathcal{Y}$ the $Y$-inserted s.e.
\begin{equation}\label{inserting-set1}
\begin{split}
\mathcal{X}^{(\cap Y /\!\!/ \mathcal{Y})} &= \mathcal{X} \, (\cap) \, \{\textsf{ter}(Y /\!\!/ \mathcal{Y})\}
\end{split}
\end{equation}
with the e.p.d. (\ref{epd-inserting-set}) has the \emph{event-probabilistic $Y$-pseudo-distribution}, which is defined as a set of probabilitieso of terraced events that coincide with probabilities from the e.p.d. (\ref{epd-inserting-set}) for all $X \subseteq \mathcal{X}$ excepting $X=\emptyset$:
\begin{equation}\label{inserting-set-Yepd}
\begin{split}
&p^{(Y)}(X+Y /\!\!/ \mathcal{X}+\mathcal{Y}) =\\
&=\begin{cases}
p(X+Y /\!\!/ \mathcal{X}+\mathcal{Y}), & X \ne \emptyset,\\
p(Y /\!\!/ \mathcal{X}+\mathcal{Y})-1+p(Y /\!\!/ \mathcal{Y}), & X=\emptyset,
\end{cases}\\[3pt]
&=\begin{cases}
p(X+Y /\!\!/ \mathcal{X}+\mathcal{Y}), & X \ne \emptyset,\\
\displaystyle p(Y /\!\!/ \mathcal{Y})-\sum_{{X \ne \emptyset}\atop{X \subseteq \mathcal{X}}} p(X+Y /\!\!/ \mathcal{X}+\mathcal{Y}), & X=\emptyset.
\end{cases}
\end{split}\hspace*{-13pt}
\end{equation}
The sum of all probabilities from every $Y$-pseudo-distribution (\ref{inserting-set-Yepd}) is $p(Y /\!\!/ \mathcal{Y})=\mathbf{P}(\textsf{ter}(Y /\!\!/ \mathcal{Y}))$, the probability of a terraced event, generated by the \emph{frame s.e.} $\mathcal{Y}$, in which the given s.e. $\mathcal{X}^{(\cap Y /\!\!/ \mathcal{Y})}$ is inserted.

Thus, the only difference of e.p.d.'s of $Y$-inserted s.e.'s from their event-probabilistic $Y$-pseudo-distributions, lies in the fact that the sums of the probabilities of the terraced events, from which the $Y$-pseudo-distributions are composed, are normalized not by unity, but by the probabilities of the corresponding frame terraces events $p(Y /\!\!/ \mathcal{Y})$. And the sum of the normalizing constants by $Y \subseteq \mathcal{Y}$ is obviously equal to one.

\texttt{Note \!\refstepcounter{ctrnot}\arabic{ctrnot}\,\label{not-inserting-set-symmetry}\itshape\enskip\scriptsize (symmetry of inserted and frame s.e.'s).} In Definition \ref{def-inserting-set} the s.e. $\mathcal{X}$ and $\mathcal{Y}$ can always be swapped, i.e., to take the s.e. $\mathcal{X}$ on a role of the \emph{frame} one, and the s.e. $\mathcal{Y}$ to take on a role of s.e., that generates \emph{$X$-inserted s.e.'s} for every $X \subseteq \mathcal{X}$:
\begin{equation}\label{inserting-set-symmetry}
\begin{split}
\mathcal{Y}^{(\cap X /\!\!/ \mathcal{X})} &= \mathcal{Y} \, (\cap) \, \{\textsf{ter}(X /\!\!/ \mathcal{X})\} =\\
&=\{ y \cap \textsf{ter}(X /\!\!/ \mathcal{X}), \, y \in \mathcal{Y} \}.
\end{split}
\end{equation}

\texttt{Note \!\refstepcounter{ctrnot}\arabic{ctrnot}\,\label{not-inserting-set}\itshape\enskip\scriptsize (M-sum of the all inserted s.e.'s).} The M-sum\footnote{M-sum is a sum by Minkowski.} of the all $Y$-inserted s.e.'s $\mathcal{X}^{(\cap Y /\!\!/ \mathcal{Y})}$ for $Y \subseteq \mathcal{Y}$ forms the given s.e. $\mathcal{X}$:
\begin{equation}\label{inserting-set-1}
\begin{split}
\mathcal{X} &= \mathop{\left(\sum\right)}_{Y \subseteq \mathcal{Y}} \mathcal{X}^{(\cap Y /\!\!/ \mathcal{Y})}=\\
&= \underbrace{\mathcal{X}^{(\cap \emptyset /\!\!/ \mathcal{Y})}\, (+) \,... \,(+) \,\mathcal{X}^{(\cap \mathcal{Y} /\!\!/ \mathcal{Y})}}_{2^{|\mathcal{Y}|}}.
\end{split}
\end{equation}

\texttt{Note \!\refstepcounter{ctrnot}\arabic{ctrnot}\,\label{not-inserting-from-conditional}\itshape\enskip\scriptsize (charcterization of $Y$-inserted s.e.'s by conditional e.p.d.'s of the 1st kind).}
The e.p.d. of $Y$-inserted s.e. $\mathcal{X}^{(\cap Y /\!\!/ \mathcal{Y})}$ with every $Y \subseteq \mathcal{Y}$ has a form for $X \subseteq \mathcal{X}$:
\begin{equation}\label{inserting-from-conditional}
\hspace{-3pt}\begin{split}
&p\left(X (\cap) \{\textsf{ter}(Y /\!\!/ \mathcal{Y})\} /\!\!/ \mathcal{X}^{(\cap Y /\!\!/ \mathcal{Y})}\right) =\\
&=\begin{cases}
p(X+Y /\!\!/ \mathcal{X}+\mathcal{Y}), &\!\!\!\!\!\emptyset \ne X \subseteq \mathcal{X},\\
p(Y /\!\!/ \mathcal{X}+\mathcal{Y})+1-p(Y /\!\!/ \mathcal{Y}), &\!\!\!\!\!X=\emptyset,\\
\end{cases}\\
&=\begin{cases}
p(X /\!\!/ \mathcal{X} \mid Y /\!\!/ \mathcal{Y})p(Y /\!\!/ \mathcal{Y}), &\!\!\!\!\!\emptyset \ne X \subseteq \mathcal{X},\\
1-(1-p(\emptyset /\!\!/ \mathcal{X} \mid Y /\!\!/ \mathcal{Y}))p(Y /\!\!/ \mathcal{Y}), &\!\!\!\!\!X=\emptyset
\end{cases}\hspace{-30pt}
\end{split}
\end{equation}
where for every $Y \subseteq \mathcal{Y}$
\begin{equation}\label{inserting-set-2}
\begin{split}
p(X /\!\!/ \mathcal{X} \mid Y /\!\!/ \mathcal{Y}) = \frac{p(X+Y /\!\!/  \mathcal{X}+\mathcal{Y})}{p(Y /\!\!/ \mathcal{Y})},
\end{split}
\end{equation}
i.e., the probabilities of the 1st kind, forming for $X \subseteq \mathcal{X}$ the \emph{$Y$-conditional e.p.d. of the 1st kind} of the s.e. $\mathcal{X}$ with respect to terraced event $\textsf{ter}(Y /\!\!/ \mathcal{Y})$ generated by the s.e. $\mathcal{Y}$.

In other words, $Y$-inserted s.e. $\mathcal{X}^{(\cap Y /\!\!/ \mathcal{Y})}$ for $Y \subseteq \mathcal{Y}$ are characterized by formulas (\ref{inserting-from-conditional}) and $Y$-conditional e.p.d.'s of the 1st kind of the s.e. $\mathcal{X}$ with respect to the terraced event  $\textsf{ter}(Y /\!\!/ \mathcal{Y})$, generated by the s.e. $\mathcal{Y}$.

\texttt{Note \!\refstepcounter{ctrnot}\arabic{ctrnot}\,\label{not-Yinserting-and-conditional}\itshape\enskip\scriptsize (mutual characterization of conditional e.p.d.'s of the 1st kind and pseudo-distributions of inserted s.e.'s).}
The connection between each $Y$-pseudo-distribution of the $Y$-inserted s.e. with the corresponding $Y$-conditional e.p.d. looks simpler. It is sufficient for each fixed $Y \subseteq \mathcal{Y}$ to normalize all its probabilities of ``inserted'' terraced events by the probability of a terraced event $p(Y /\!\!/ \mathcal{Y})$ in order to obtain corresponding to the $Y$-conditional probabilities regarding the fact that the corresponding frame terraces event $\textsf{ter}(Y /\!\!/ \mathcal{Y})$ happened. As a result, we have the following obvious inversion formulas:
\begin{equation}\label{conditional-from-inserting-pseudo}
\begin{split}
p(X /\!\!/ \mathcal{X} \mid Y /\!\!/ \mathcal{Y}) &= \frac{p^{(Y)}(X+Y /\!\!/  \mathcal{X}+\mathcal{Y})}{p(Y /\!\!/ \mathcal{Y})},\\
p^{(Y)}(X+Y /\!\!/  \mathcal{X}+\mathcal{Y}) &= p(X /\!\!/ \mathcal{X} \mid Y /\!\!/ \mathcal{Y})p(Y /\!\!/ \mathcal{Y}).
\end{split}
\end{equation}

\texttt{Note \!\refstepcounter{ctrnot}\arabic{ctrnot}\,\label{not-practicability}\itshape\footnotesize (about the appropriateness of the concept of inserted s.e.'s).}
It would seem, why develop a theory of inserted s.e.'s, pseudo-distributions of which are simply characterized by conditional e.p.d.'s. Is not it better to instead practice the theory of conditional e.p.d., especially since this theory has long had excellent recommendations in many areas. However, in eventology, as the theory of events, which prefers to work directly with sets of events, there is one rather serious objection. The fact is that conditional e.p.d., as any e.p.d. in eventology, there must be a set of some events, in this case, a set of well-defined ``conditional events''. But until now it has not been possible to give a satisfactory definition of the ``conditional event'', except for my impractical definition in \cite{Vorobyev2006aa}. So, the inserted s.e.'s are a completely satisfactory ``surrogate'' definition of the sets of ``conditional events''. Such that e.p.d.'s of inserted s.e.'s, although they do not coincide with the desired conditional e.p.d.'s, but are associated with them by well-defined mutual-inverse transformations. As a result, inserted s.e.'s play the role of a convenient eventological tool for working with conditional e.p.d.'s of a one set of events regarding terrace events generated by another set of events.

\texttt{Example \!\refstepcounter{ctrexa}\arabic{ctrexa}\,\label{exa-inserting-set}\itshape\enskip\scriptsize (two inserted s.e.'s in a frame monoplet).} Let in formulas (\ref{inserting-set-joint-epd}) the s.e. $\mathcal{X}$ is an arbitrary set, and the s.e. $\mathcal{Y}=\{y\}$ is a frame monoplet of events, which have the joint e.p.d. in a form:
\begin{equation}\label{example-inserting-set-joint-epd}
\begin{split}
\{ p(X+Y /\!\!/ \mathcal{X}+\{y\}), X \subseteq \mathcal{X}, Y \subseteq \{y\} \}.
\end{split}
\end{equation}
Then there is the $\{y\}$-inserted s.e. and the $\emptyset$-inserted s.e.:
\begin{equation}\label{exa-inserting-set1}
\begin{split}
\mathcal{X}^{(\cap \{y\} /\!\!/ \{y\})}&=\{x \cap y, x \in \mathcal{X}\},\\
\mathcal{X}^{(\cap \emptyset /\!\!/ \{y\})}&=\{x \cap y^c, x \in \mathcal{X}\}.
\end{split}
\end{equation}
These inserted s.e.'s are characterized for every of two subsets of the monoplet $Y=\{y\} \subseteq \{y\}$ and $Y = \emptyset \subseteq \{y\}$ by formulas (\ref{epd-inserting-set}) and by two corresponding e.p.d.'s
\begin{equation}\label{edps-inserting-set}
\begin{split}
\{ p(X /\!\!/ \mathcal{X}+\{y\})\}, & \ X \subseteq \mathcal{X} \},\\
\{ p(X+\{y\} /\!\!/ \mathcal{X}+\{y\})\}, & \ X \subseteq \mathcal{X} \}.
\end{split}
\end{equation}
which by formulas (\ref{inserting-set-Yepd}) define two \emph{$Y$-pseudo-distributions} for $X \subseteq \mathcal{X}$:
\begin{equation}\label{inserting-set-Yepd-y}
\begin{split}
&p^{(\{y\})}(X+\{y\} /\!\!/ \mathcal{X}+\{y\}) =\\
&=\begin{cases}
p(X+\{y\} /\!\!/ \mathcal{X}+\{y\}), &\hspace{-10pt}  X \ne \emptyset,\\
p(\{y\} /\!\!/ \mathcal{X}+\{y\})-1+p(\{y\} /\!\!/ \{y\}), &\hspace{-10pt} X=\emptyset,
\end{cases}\\
&=\begin{cases}
p(X+\{y\} /\!\!/ \mathcal{X}+\{y\}), & X \ne \emptyset,\\
p(\{y\} /\!\!/ \mathcal{X}+\{y\})-1+p_y, & X=\emptyset.
\end{cases}
\end{split}\hspace*{-13pt}
\end{equation}
\begin{equation}\label{inserting-set-Yepd-emptyset}
\begin{split}
&p^{(\emptyset)}(X /\!\!/ \mathcal{X}+\{y\}) =\\
&=\begin{cases}
p(X /\!\!/ \mathcal{X}+\{y\}), & X \ne \emptyset,\\
p(\emptyset /\!\!/ \mathcal{X}+\{y\})-1+p(\emptyset /\!\!/ \{y\}), & X=\emptyset,
\end{cases}\\
&=\begin{cases}
p(X /\!\!/ \mathcal{X}+\{y\}), & X \ne \emptyset,\\
p(\emptyset /\!\!/ \mathcal{X}+\{y\})-p_y, & X=\emptyset,
\end{cases}\end{split}\hspace*{-13pt}
\end{equation}
where $p_y=\mathbf{P}(y)$ is a probability of the \emph{frame event}  $y~\in~\{y\}$.

First of all, note that the sum of the probabilities of terraced events from the $\{y\}$-pseudo-distribution (\ref{inserting-set-Yepd-y}) is $ p_y $, and the probabilities of the $\emptyset$-pseudo-distribution (\ref{inserting-set-Yepd-emptyset}) is $ 1-p_y $; and secondly, that
these two pseudo-distributions define a joint e.p.d. of the s.e. $ \mathcal {X} $ and the monoplet $ \{y \} $, i.e., e.p.d. of the s.e. $\mathcal{X} + \{y \}$, which is related to them by fairly obvious formulas for $ Z \subseteq \mathcal {X} + \{y \} $:
\begin{equation}\label{inserting-recurrent1}
\hspace*{-3pt}
\begin{split}
&p(Z /\!\!/ \mathcal{X}+\{y\})\!=\!\begin{cases}
p^{(\{y\})}(Z /\!\!/ \mathcal{X}+\{y\}),       & \hspace*{-7pt} y \in Z,\\
p^{(\emptyset)}(Z /\!\!/ \mathcal{X}+\{y\}),   & \hspace*{-7pt} y \not\in Z.
\end{cases}
\end{split}\hspace*{-13pt}
\end{equation}
The formulas (\ref{inserting-recurrent1}) are recurrent, connecting the e.p.d. of s.e. $ \mathcal {X} + \{y \} $ with two pseudo-distributions of the inserted s.e. $\mathcal{X}'=\mathcal{X}^{(\cap \{y\} /\!\!/ \mathcal{Y})}$ and $\mathcal{X}'\!'=\mathcal{X}^{(\cap \emptyset /\!\!/ \mathcal{Y})}$ whose power is less by one. The inversion formulas (\ref{conditional-from-inserting-pseudo}) allow recurrence formulas (\ref{inserting-recurrent1}) to express the e.p.d. of $ \mathcal {X} + \{y \} $ via the conditional e.p.d. with respect to one of its events $ y \in \mathcal {X} + \{y \} $ and its complements $y^c=\Omega-y$:
\begin{equation}\label{inserting-recurrentC}
\hspace*{-3pt}
\begin{split}
&p(Z /\!\!/ \mathcal{X}+\{y\})=\\
&=\begin{cases}
p(X /\!\!/ \mathcal{X} \mid \{y\} /\!\!/ \{y\})p_y,         & \hspace*{-7pt} Z=X+\{y\},\\
                                                         & \hspace*{-7pt} X \subseteq \mathcal{X};\\
p(X /\!\!/ \mathcal{X} \mid \emptyset /\!\!/ \{y\})(1-p_y), & \hspace*{-7pt} Z=X,\\
                                                         & \hspace*{-7pt} X \subseteq \mathcal{X},
\end{cases}
\end{split}\hspace*{-13pt}
\end{equation}
where $ Z \subseteq \mathcal {X} + \{y \} $.
Note that these formulas, like (\ref{inserting-recurrent1}), can be used recursively to express the e.p.d. of s.e. $ \mathcal {X} + \{y \} $ through two conditional e.p.d.'s of the s.e. $ \mathcal {X} $ whose power is less by one.

\subsection{Inserted and conditional Kopulas of a family of sets of events with respect to the set of events\label{Kopula-inserting-conditional}}

\texttt{Definition \!\refstepcounter{ctrdef}\arabic{ctrdef}\,\label{def-inserting-Kopulas}\itshape\footnotesize (inserted Kopulas).}
The $N$-Kopulas of $Y$-inserted $N$-s.e.'s
\begin{equation}\label{inserting-Kopula}
\begin{split}
\mathcal{X}^{(\cap Y /\!\!/ \mathcal{Y})} &= \mathcal{X} \, (\cap) \, \{\textsf{ter}(Y /\!\!/ \mathcal{Y})\}=\\
&=\Set{x \cap \textsf{ter}(Y /\!\!/ \mathcal{Y}), x \in \mathcal{X}},
\end{split}
\end{equation}
which for each $Y \subseteq \mathcal{Y}$ are defined (see Definition \ref{def-inserting-set}) as intersections by Minkowsi of the s.e. $\mathcal{X}$ with terraced events $\textsf{ter}(Y /\!\!/ \mathcal{Y})$, generated by the s.e. $\mathcal{Y}$, are called the \emph{$Y$-inserted $N$-Kopulas} with respect to the s.e. $\mathcal{Y}$. Such $Y$-inserted $N$-Kopulas characterizes e.p.d.'s of the 1st kind of $Y$-inserted $N$-s.e.'s by formulas for $X \subseteq \mathcal{X}$
\begin{equation}\label{inserting-Kopulas}
\hspace*{-10pt}
\begin{split}
&p(X (\cap) \{\textsf{ter}(Y /\!\!/ \mathcal{Y})\})=\\
&={\mbox{\boldmath$\mathscr{K}$}}^{(Y)}\left(\breve{p}^{\left(c|X^{(\cap Y /\!\!/ \mathcal{Y})} /\!\!/ \mathcal{X}^{(\cap Y /\!\!/ \mathcal{Y})}\right)}\!\right)\!,
\end{split}\hspace*{-3pt}
\end{equation}
where
\begin{equation}\label{inserting-Kopula-marginal-probabilities}
\hspace*{-10pt}
\begin{split}
&\breve{p}^{\left(c|\mathcal{X}^{(\cap Y /\!\!/ \mathcal{Y})} /\!\!/ \mathcal{X}^{(\cap Y /\!\!/ \mathcal{Y})}\right)}\!=\!\breve{p}^{(Y)} =\\
&\!=\!\Set{\!p_x^{(Y)}, x \in \mathcal{X}\!}
\!=\!\Big\{\mathbf{P}(x \cap \textsf{ter}(Y /\!\!/ \mathcal{Y})), x \in \mathcal{X}\Big\}
\end{split}\hspace*{-13pt}
\end{equation}
is the set of probabilities of ``inserted'' marginal events from the $\mathcal{X}^{(\cap Y /\!\!/ \mathcal{Y})}$, and
\begin{equation}\label{inserting-Kopula-marginal-probabilitiesPH}
\hspace*{-10pt}
\begin{split}
&\breve{p}^{\left(c|X^{(\cap Y /\!\!/ \mathcal{Y})} /\!\!/ \mathcal{X}^{(\cap Y /\!\!/ \mathcal{Y})}\right)} =
\Set{p_x^{(Y)}, x \in X}+\\
&+\Set{p(Y /\!\!/ \mathcal{X})-p_x^{(Y)}, x \in \mathcal{X}-X}
\end{split}\hspace*{-13pt}
\end{equation}
are $X$-phenomena of the set of ``inserted'' marginal probabilities $\breve{p}^{(Y)}$.

We also need to define an inserted $Y$-pseudo-Kopula with respect to the s.e. $\mathcal{Y}$, which characterizes the $Y$-pseudo-distribution of the $Y$-inserted s.e. $\mathcal{X}^{(\cap Y /\!\!/ \mathcal{Y})}$, inserted into the terraces event $\textsf{ter}(Y /\!\!/ \mathcal{Y})$ generated by the s.e. $\mathcal{Y}$. Although the $Y$-pseudo-Kopula is not a Kopula, i.e., is not a 1-function, it has properties very reminiscent of the Kopula properties.

\texttt{Definition \!\refstepcounter{ctrdef}\arabic{ctrdef}\,\label{def-inserting-pseudoKopulas}\itshape\footnotesize (inserted pseudo-Kopulas).}
The \emph{$Y$-pseudo-Kopula} of the $Y$-pseudo-distribution of $Y$-inserted s.e.
\begin{equation}\label{inserting-pseudoKopula}
\begin{split}
\mathcal{X}^{(\cap Y)} = \mathcal{X}^{(\cap Y /\!\!/ \mathcal{Y})} = \mathcal{X} \, (\cap) \, \{\textsf{ter}(Y /\!\!/ \mathcal{Y})\}
\end{split}
\end{equation}
with respect to the s.e. $\mathcal{Y}$
is a such function $\mathscr{K}^{(Y)}$ on $\mathcal{X}$-hypercube with sides $[0,p(Y/\!\!/ \mathcal{Y})]$ that

1) \emph{is non-negative}:
\begin{equation}\label{pseudoKopula-positive}
\begin{split}
\mathscr{K}^{(Y)}\left(\breve{w}^{\left(c|X^{(\cap Y)}/\!\!/\frak{X}^{(\cap Y)}\right)}\right) \geqslant 0
\end{split}
\end{equation}
for $\breve{w}^{\left(c|X^{(\cap Y)}/\!\!/\frak{X}^{(\cap Y)}\right)} \in [0,p(Y/\!\!/ \mathcal{Y})]^{\otimes \mathcal{X}}$ для $X \subseteq \frak{X}$;

2) satisfies the \emph{$Y$-marginal equalities} for $x \in \frak{X}$:
\begin{equation}\label{pseudoKopula-wx}
\begin{split}
\sum_{x \in X \subseteq \frak{X}} \mathscr{K}^{(Y)}\left(\breve{w}^{\left(c|X^{(\cap Y)}/\!\!/\frak{X}^{(\cap Y)}\right)}\right) = w_{x \cap \textsf{ter}(Y /\!\!/ \mathcal{X})}^{\left(c|X^{(\cap Y)}/\!\!/\frak{X}^{(\cap Y)}\right)}
\end{split}
\end{equation}
where
\begin{equation}\label{pseudoKopula-PH}
\begin{split}
\breve{w}^{\left(c\mid X^{(\cap Y)}/\!\!/\mathcal{X}^{(\cap Y)}\right)} =
\left\{
w_{x \cap \textsf{ter}(Y /\!\!/ \mathcal{X})}^{\left(
c \mid X^{(\cap Y)} /\!\!/ \mathcal{X}^{(\cap Y)}
\right)},
x \in \mathcal{X}
\right\}
\end{split}
\end{equation}
is a $X$-phenomenon of the $\mathcal{X}$-set of marginal probabilities of the $Y$-pseudo-distribution of $Y$-inserted s.e. $\mathcal{X}^{(\cap Y)}$, i.e.,
\begin{equation}\label{pseudoKopula-wPH}
\begin{split}
&w_{x \cap \textsf{ter}(Y /\!\!/ \mathcal{X})}^{\left(c \mid X^{(\cap Y)} /\!\!/ \mathcal{X}^{(\cap Y)}\right)} =\\
&=\begin{cases}
w_{x \cap \textsf{ter}(Y /\!\!/ \mathcal{X})}, & x \in X,\\
p(Y /\!\!/ \mathcal{Y})-w_{x \cap \textsf{ter}(Y /\!\!/ \mathcal{X})}, & x \in \mathcal{X}-X.
\end{cases}
\end{split}
\end{equation}

From (\ref{pseudoKopula-wx}) and (\ref{pseudoKopula-wPH}) it follows the \emph{probabilistic $Y$-normalization of pseudo-Kopula}:
\begin{equation}\label{pseudoKopula-pY}
\begin{split}
\sum_{X \subseteq \mathcal{X}} \mathscr{K}^{(Y)}\left(\breve{w}^{\left(c \mid X^{(\cap Y)} /\!\!/ \mathcal{X}^{(\cap Y)}\right)}\right) = p(Y /\!\!/ \mathcal{Y}).
\end{split}
\end{equation}
And from (\ref{pseudoKopula-positive}) and (\ref{pseudoKopula-pY}) it follows the \emph{terrace-by-terrace probabilistic $Y$-normalization of pseudo-Kopula}:
\begin{equation}\label{pseudoKopula-pX}
\begin{split}
0 \leqslant \mathscr{K}^{(Y)}\left(\breve{w}^{\left(c \mid X^{(\cap Y)} /\!\!/ \mathcal{X}^{(\cap Y)}\right)}\right) \leqslant p(Y /\!\!/ \mathcal{Y})
\end{split}
\end{equation}
for $X \subseteq \frak{X}$.

Such $Y$-pseudo-Kopulas characterize the $Y$-pseudo-distribution (\ref{inserting-set-Yepd}) of $Y$-inserted s.e.'s  $\mathcal{X}^{(\cap Y)}$ by formulas for $X \subseteq \mathcal{X}$
\begin{equation}\label{inserting-pseudoKopula}
\hspace*{-10pt}
\begin{split}
&p^{(Y)}(X+Y /\!\!/ \mathcal{X}+\mathcal{Y})=\\
&=\mathscr{K}^{(Y)}\left(\breve{p}^{\left(c|X^{(\cap Y /\!\!/ \mathcal{Y})} /\!\!/ \mathcal{X}^{(\cap Y /\!\!/ \mathcal{Y})}\right)}\!\right)\!,
\end{split}\hspace*{-3pt}
\end{equation}
where
\begin{equation}\label{inserting-pseudoKopula-marginal-probabilities}
\hspace*{-10pt}
\begin{split}
&\breve{p}^{\left(c|\mathcal{X}^{(\cap Y /\!\!/ \mathcal{Y})} /\!\!/ \mathcal{X}^{(\cap Y /\!\!/ \mathcal{Y})}\right)} = \breve{p}^{(Y)} =\\
&=\Set{p_x^{(Y)}, x \in \mathcal{X}}
=\Big\{\mathbf{P}(x \cap \textsf{ter}(Y /\!\!/ \mathcal{Y})), x \in \mathcal{X}\Big\}
\end{split}\hspace*{-8pt}
\end{equation}
is a set of $Y$-marginal probabilities, coinciding with the set of marginal probabilities of $Y$-inserted s.e.'s $\mathcal{X}^{(\cap Y)}$, and
\begin{equation}\label{inserting-pseudoKopula-marginal-probabilitiesPH}
\hspace*{-10pt}
\begin{split}
&\breve{p}^{\left(c|X^{(\cap Y /\!\!/ \mathcal{Y})} /\!\!/ \mathcal{X}^{(\cap Y /\!\!/ \mathcal{Y})}\right)} =
\Set{p_x^{(Y)}, x \in X}+\\
&+\Set{p(Y /\!\!/ \mathcal{X})-p_x^{(Y)}, x \in \mathcal{X}-X}
\end{split}\hspace*{-13pt}
\end{equation}
are $X$-phenomena of the set $Y$-marginal probabilities $\breve{p}^{(Y)}$.

\texttt{Definition \!\refstepcounter{ctrdef}\arabic{ctrdef}\,\label{def-conditional-Kopulas}\itshape\footnotesize (conditional Kopulas).}
The $N$-Kopulas, characterizing $Y$-inserted e.p.d.'s of the 1st kind of the $N$-s.e. $\mathcal{X}$ with respect to the terraced event $\textsf{ter}(Y /\!\!/ \mathcal{Y})$, generated by the s.e. $\mathcal{Y}$, i.e., e.p.d.'s of the 1st kind, defined by joint e.p.d. $\mathcal{X}$ and $\mathcal{Y}$ by formulas with fixed $Y \subseteq \mathcal{Y}$ for $X \subseteq \mathcal{X}$:
\begin{equation}\label{epd-conditional}
\begin{split}
p(X  /\!\!/ \mathcal{X} \mid Y /\!\!/ \mathcal{Y}) = \frac{p(X+Y /\!\!/ \mathcal{X}+\mathcal{Y})}{p(Y /\!\!/ \mathcal{Y})},
\end{split}
\end{equation}
are called the \emph{$Y$-conditional $N$-Kopulas} of the $N$-s.e. $\mathcal{X}$ with respect to the terraced event $\textsf{ter}(Y /\!\!/ \mathcal{Y})$, generated by the s.e. $\mathcal{Y}$.

Such $Y$-cvonditional $N$-Kopulas characterize the $Y$-conditional e.p.d. of the 1st kind (\ref{epd-conditional}) by formulas for $X \subseteq \mathcal{X}$:
\begin{equation}\label{conditional-Kopula}
\hspace*{-10pt}
\begin{split}
p(X  /\!\!/ \mathcal{X} \mid Y /\!\!/ \mathcal{Y})
\!=\!{\mbox{\boldmath$\mathscr{K}$}}^{\mid Y}
\left(
\breve{p}^{\left(c|X  /\!\!/ \mathcal{X} \mid Y /\!\!/ \mathcal{Y}\right)}
\!\right)\!,
\end{split}\hspace*{-3pt}
\end{equation}
where
\begin{equation}\label{conditional-Kopula-marginal-probabilities}
\hspace*{-10pt}
\begin{split}
&\breve{p}^{\left(c|\mathcal{X}  /\!\!/ \mathcal{X} \mid Y /\!\!/ \mathcal{Y}\right)} = \breve{p}^{\mid Y} = \Set{p_x^{\mid Y}, x \in \mathcal{X}} =\\
&=\Big\{\mathbf{P}(x \cap \textsf{ter}(Y /\!\!/ \mathcal{Y}))/p(Y /\!\!/ \mathcal{Y}), x \in \mathcal{X}\Big\}
\end{split}\hspace*{-13pt}
\end{equation}
is a set of conditional marginal probabilities of events $x \in \mathcal{X}$ with respect to the terraced event $\textsf{ter}(Y /\!\!/ \mathcal{Y})$, and
\begin{equation}\label{conditional-Kopula-marginal-probabilitiesPH}
\hspace*{-10pt}
\begin{split}
&\breve{p}^{\left(c|X  /\!\!/ \mathcal{X} \mid Y /\!\!/ \mathcal{Y}\right)} =\\
&=\Big\{p_x^{\mid Y}, x \in X\Big\}+\Big\{1-p_x^{\mid Y}, x \in \mathcal{X}-X\Big\}
\end{split}\hspace*{-13pt}
\end{equation}
are $X$-phenomenon of the set of conditional marginal probabilities $\breve{p}^{\mid Y}$.

\texttt{Note \!\refstepcounter{ctrnot}\arabic{ctrnot}\,\label{not-conditional-inserting-marginal}\itshape\footnotesize (connection between conditional and ``inserted'' marginal probaabilities).}
Conditional marginal probabilities are connected with ``inserted'' marginal probabilities for $x \in \mathcal{X}$ by the formula of conditional probability:
\begin{equation}\label{conditional-inserting-marginal-probabilities}
\hspace*{-10pt}
\begin{split}
p_x^{\mid Y} = \frac{1}{p(Y /\!\!/ \mathcal{Y})} p_x^{(Y)},
\end{split}\hspace*{-13pt}
\end{equation}
since ``inserted'' marginal probabilities (\ref{inserting-Kopula-marginal-probabilities}) are probabilities of intersections of events $x \in \mathcal{X}$ with the terraced event $\textsf{ter}(Y /\!\!/ \mathcal{Y})$. The connection between the corresponding set of conditional  ``inserted'' marginal probabilities we shall write in the similar way:
\begin{equation}\label{conditional-inserting-set-marginal-probabilities}
\hspace*{-10pt}
\begin{split}
\breve{p}^{\mid Y} &= \frac{1}{p(Y /\!\!/ \mathcal{Y})} \breve{p}^{(Y)},\\
\breve{p}^{\left(c|X  /\!\!/ \mathcal{X} \mid Y /\!\!/ \mathcal{Y}\right)} &= \frac{1}{p(Y /\!\!/ \mathcal{Y})} \breve{p}^{\left(c|X^{(\cap Y /\!\!/ \mathcal{Y})} /\!\!/ \mathcal{X}^{(\cap Y /\!\!/ \mathcal{Y})}\right)}.
\end{split}\hspace*{-13pt}
\end{equation}

\texttt{Note \!\refstepcounter{ctrnot}\arabic{ctrnot}\,\label{not-conditional-pseudo-Kopulas}\itshape\footnotesize (connection between conditional Kopulas and inserted pseudo-Kopulas).} From Definition \ref{def-conditional-Kopulas} of conditional Kopula and Definition \ref{def-inserting-pseudoKopulas} of inserted pseudo-Kopula with respect to the s.e. $\mathcal{Y}$, and also from the formula (\ref{conditional-inserting-set-marginal-probabilities}) it follows the simple inversion formulas that connect conditional Kopulas and inserted Pseudo-Kopulas of the family of sets of events $\mathcal{X}$ for $X \subseteq \mathcal{X}$:
\begin{equation}\label{conditional-Kopula-from-pseudoKopula}
\hspace*{-10pt}
\begin{split}
&{\mbox{\boldmath$\mathscr{K}$}}^{\mid Y}\!\!\!
\left(
\breve{p}^{\left(c|X  /\!\!/ \mathcal{X} \mid Y /\!\!/ \mathcal{Y}\right)}
\!\right)\!=\\
&=\frac{1}{p(Y /\!\!/ \mathcal{Y})} \, \mathscr{K}^{(Y)}\!\!\left(
p(Y /\!\!/ \mathcal{Y})\breve{p}^{\left(c|X  /\!\!/ \mathcal{X} \mid Y /\!\!/ \mathcal{Y}\right)}
\!\right)\!,\\
&\\
&\mathscr{K}^{(Y)}\!\!\left(
\breve{p}^{\left(c|X^{(\cap Y /\!\!/ \mathcal{Y})} /\!\!/ \mathcal{X}^{(\cap Y /\!\!/ \mathcal{Y})}\right)}
\!\right)\!=\\
&=p(Y /\!\!/ \mathcal{Y}) \, {\mbox{\boldmath$\mathscr{K}$}}^{\mid Y}\!\!\!
\left(
\frac{1}{p(Y /\!\!/ \mathcal{Y})}\breve{p}^{\left(c|X^{(\cap Y /\!\!/ \mathcal{Y})} /\!\!/ \mathcal{X}^{(\cap Y /\!\!/ \mathcal{Y})}\right)}
\!\right)\!.\\
\end{split}\hspace*{-13pt}
\end{equation}

\texttt{Note \!\refstepcounter{ctrnot}\arabic{ctrnot}\,\label{not1}\itshape\footnotesize (two formulas of full probability for a Kopula).} The Kopula ${\mbox{\boldmath$\mathscr{K}$}}$ of s.e. $\mathcal{X}$ is expressed through $Y$-conditional Kpulas ${\mbox{\boldmath$\mathscr{K}$}}^{\mid Y}$ for $Y \subseteq \mathcal{Y}$ by the usual formula of full probability:
\begin{equation}\label{Kopula-full-probability-conditionalKopulas}
\hspace*{-3pt}
\begin{split}
{\mbox{\boldmath$\mathscr{K}$}}\!\left(\breve{p}^{\left( c \mid X /\!\!/ \mathcal{X}\right)}\!\right)\!=\!\!\sum_{Y \subseteq \mathcal{Y}}
{\mbox{\boldmath$\mathscr{K}$}}^{\mid Y}\!\!\!\left(\breve{p}^{\left( c \mid X /\!\!/ \mathcal{X} \mid Y /\!\!/ \mathcal{Y}\right)}\right) p(Y /\!\!/ \mathcal{Y}).
\end{split}\hspace*{-2pt}
\end{equation}
From (\ref{Kopula-full-probability-conditionalKopulas}) and (\ref{conditional-Kopula-from-pseudoKopula}) we obtain an analogue of the formula of total probability --- the representation of the Kopula of s.e. $\mathcal{X}$ in the form of sum of $Y$-pseudo-Kopulas by $Y \subseteq \mathcal{Y}$:
\begin{equation}\label{Kopula-full-probability-pseudoKopulas}
\hspace*{-7pt}
\begin{split}
{\mbox{\boldmath$\mathscr{K}$}}\!\left(\breve{p}^{\left( c \mid X /\!\!/ \mathcal{X}\right)}\!\right)\!=\!\!\sum_{Y \subseteq \mathcal{Y}}
\mathscr{K}^{(Y\!)}\!\!\left(\breve{p}^{\left( c \mid X^{(\cap Y /\!\!/ \mathcal{Y})} /\!\!/ \mathcal{X}^{(\cap Y /\!\!/ \mathcal{Y})}\right)}\!\right)\!.
\end{split}\hspace*{-2pt}
\end{equation}

\texttt{Note \!\refstepcounter{ctrnot}\arabic{ctrnot}\,\label{not-Kopula-sum-sets}\itshape\footnotesize (Kopula of a sum of sets).} A Kopula of sum $\mathcal{X}+\mathcal{Y}$ of two s.e.'s $\mathcal{X}$ and $\mathcal{Y}$ characterizes their joint e.p.d. of the 1st kind and by definition has the form
\begin{equation}\label{Kopula-sum-sets}
\hspace*{-10pt}
\begin{split}
p(X+Y /\!\!/ \mathcal{X}+\mathcal{Y})={\mbox{\boldmath$\mathscr{K}$}}\!\left(\breve{p}^{\left( c \mid X+Y /\!\!/ \mathcal{X}+\mathcal{Y}\right)}\!\right),
\end{split}\hspace*{-13pt}
\end{equation}
where
\begin{equation}\label{Kopula-XPH}
\hspace*{-10pt}
\begin{split}
\breve{p}^{\left( c \mid X+Y /\!\!/ \mathcal{X}+\mathcal{Y}\right)} = \{p_x,x\in X\}+\{p_y,y\in Y\}+\\
+\{1-p_x, x \in \mathcal{X}-X\}+\{1-p_y, y \in \mathcal{Y}-Y\}
\end{split}\hspace*{-13pt}
\end{equation}
is the $(X+Y)$-phenomenon of the set of marginal probabilities
\begin{equation}\label{Kopula-XPH}
\hspace*{-10pt}
\begin{split}
\breve{p}^{\left( c \mid \mathcal{X}+\mathcal{Y} /\!\!/ \mathcal{X}+\mathcal{Y}\right)} = \{p_x,x\in \mathcal{X}\}+\{p_y,y\in \mathcal{Y}\}
\end{split}\hspace*{-13pt}
\end{equation}
for the sum $\mathcal{X}+\mathcal{Y}$.

From previous formulas (\ref{epd-conditional}), (\ref{conditional-Kopula}), and (\ref{inserting-pseudoKopula}) for a inserted pseudo-Kopula and conditional Kopula we obtain formulas
\begin{equation}\label{Kopula-sum-sets-inserting}
\hspace*{-13pt}
\begin{split}
&{\mbox{\boldmath$\mathscr{K}$}}\!\left(\breve{p}^{\left( c \mid X+Y /\!\!/ \mathcal{X}+\mathcal{Y}\right)}\!\right)=\\
&=\mathscr{K}^{(Y\!)}\!\!\left(\breve{p}^{\left( c \mid X^{(\cap Y /\!\!/ \mathcal{Y})} /\!\!/ \mathcal{X}^{(\cap Y /\!\!/ \mathcal{Y})}\right)}\!\right)\!,
\end{split}\hspace*{-10pt}
\end{equation}
\begin{equation}\label{Kopula-sum-sets-conditional}
\hspace*{-10pt}
\begin{split}
&{\mbox{\boldmath$\mathscr{K}$}}\!\left(\breve{p}^{\left( c \mid X+Y /\!\!/ \mathcal{X}+\mathcal{Y}\right)}\!\right)=\\
&={\mbox{\boldmath$\mathscr{K}$}}^{|Y}\!\!\!\left(\breve{p}^{\left( c \mid X /\!\!/ \mathcal{X} \mid Y /\!\!/ \mathcal{Y}\right)}\!\right) {\mbox{\boldmath$\mathscr{K}$}}\!\left(\breve{p}^{\left( c \mid Y /\!\!/ \mathcal{Y}\right)}\!\right),
\end{split}\hspace*{-13pt}
\end{equation}
that for each $Y \subseteq \mathcal{Y}$ connect the Kopula of sum $\mathcal{X}+\mathcal{Y}$ with the product of $Y$-conditional Kopula $\mathcal{X}$ with respect to $\mathcal{Y}$ and the value of Kopula $\mathcal{Y}$ at $Y$-phenomenon; and also with the $Y$-inserted pseudo-Kopula of $\mathcal{X}$ which is inserted in the terraced event $\textsf{ter}(Y /\!\!/ \mathcal{Y})$, generated by $\mathcal{Y}$.

\subsection{Theory of the frame method for constructing Kopula\label{theory-framing-method}}

The basis of the \emph{frame method} of constructing Kopula is a rather simple idea of composing an arbitrary $N$-s.e. $\frak{X}$ using the recurrence \emph{frame} formula:
\begin{equation}\label{framing-cup2}
\begin{split}
\frak{X} = \{x_0,x_1,...,x_{N-1}\} = \{x_0\} + \mathcal{X},
\end{split}
\end{equation}
where $(N\!-\!1)$-s.e.'s
\begin{equation}\label{framing-cup1}
\begin{split}
\mathcal{X} = \frak{X}-\{x_0\}=\{x_1,...,x_{N-1}\} = \Big(\mathcal{X}'\,(+)\,\mathcal{X}'\!'\Big)
\end{split}
\end{equation}
are composed from two $(N\!-\!1)$-s.e.'s $\mathcal{X}'$ and $\mathcal{X}'\!'$ by set-theoretic operation of $M$-union\footnote{\emph{М-intersection} and \emph{М-union} are an intersection and union od sets by Minkowski (see details in \cite{Vorobyev2007}).} and defined as the \emph{inserted s.e.'s in the frame monoplet} $\{x_0\}$ by the following formulas:
\begin{equation}\label{Mcap}
\begin{split}
\mathcal{X}' &= \mathcal{X}^{(\cap \{x_0\} /\!\!/ \{x_0\})}=\\
&=\{x_0\} \,(\cap)\, \mathcal{X} = \{x_0 \cap x_1,...,x_0 \cap x_{N-1}\},\\
&\\[-6pt]
\mathcal{X}'\!' &= \mathcal{X}^{(\cap \emptyset /\!\!/ \{x_0\})}=\\
&=\{x_0^c\} \,(\cap)\, \mathcal{X} = \{x_0^c \cap x_1,...,x_0^c \cap x_{N-1}\}.
\end{split}\hspace{-30pt}
\end{equation}

This simple idea allows us to find the recurrent frame formulas for the $N$-Kopula of s.e. $\frak{X}$ as functions of the set of marginal probabilities $\breve{p}=\{p_0,p_1,\ldots,p_{N-1}\}$.

The \emph{frame method}  relies on formulas (\ref{inserting-recurrent1}) and (\ref{inserting-recurrentC}) and also correspondingly on (\ref{Kopula-sum-sets-inserting}) and (\ref{Kopula-sum-sets-conditional}), and constructs two recurrent formulas:
\begin{equation}\label{frame-method-pseudo}
\begin{split}
{\mbox{\boldmath$\mathscr{K}$}}_\frak{X}(\breve{p}) &= \textrm{Recursion}_1\Big( \mathscr{K}^{(\{x_0\})}_{\mathcal{X}'}(\breve{p}), \mathscr{K}^{(\emptyset)}_{\mathcal{X}'\!'}(\breve{p})\Big),
\end{split}
\end{equation}
\begin{equation}\label{frame-method-conditional}
\begin{split}
{\mbox{\boldmath$\mathscr{K}$}}_\frak{X}(\breve{p}) &= \textrm{Recursion}_2\Big( {\mbox{\boldmath$\mathscr{K}$}}^{\mid \{x_0\}}_{\mathcal{X}'}(\breve{p}), {\mbox{\boldmath$\mathscr{K}$}}^{\mid \emptyset}_{\mathcal{X}'\!'}(\breve{p}),p_0\Big),
\end{split}
\end{equation}
for the $N$-Kopula of $N$-s.e. $\frak{X}$ through known probability $p_0$ of the event $x_0$ and together with it through two known \emph{inserted pseudo-$(N\!-\!1)$-Kopulas} (see Definition \ref{def-inserting-pseudoKopulas}), i.e., through pseudo-$(N\!-\!1)$-Kopulas of inserted $(N\!-\!1)$-s.e.'s $\mathcal{X}'$ and $\mathcal{X}'\!'$ in the frame monoplet $\{x_0\}$, either through two known \emph{conditional $(N\!-\!1)$-Kopulas} (see Definition \ref{def-conditional-Kopulas}) with respect to the frame monoplet $\{x_0\}$ of the same  $\mathcal{X}'$ and $\mathcal{X}'\!'$.

Note, that for the sake of brevity in the formulas (\ref{frame-method-pseudo}) and (\ref{frame-method-conditional}) we use the following abbreviations, of course, given that $\frak{X}=\mathcal{X}+\{x_0\}$:
\begin{equation}\label{abbreviations}
\begin{split}
{\mbox{\boldmath$\mathscr{K}$}}_\frak{X}(\breve{p})
&={\mbox{\boldmath$\mathscr{K}$}}\!\left(\breve{p}^{\left( c \mid X+Y /\!\!/ \mathcal{X}+\{x_0\}\right)}\!\right),\\
\mathscr{K}^{(\{x_0\})}_{\mathcal{X}'}(\breve{p})
&=\mathscr{K}^{(\{x_0\})}\!\!\left(\breve{p}^{\left( c \mid X^{(\cap \{x_0\} /\!\!/ \{x_0\})} /\!\!/ \mathcal{X}'\right)}\!\right),\\
\mathscr{K}^{(\emptyset)}_{\mathcal{X}'\!'}(\breve{p})
&=\mathscr{K}^{(\emptyset)}\!\!\left(\breve{p}^{\left( c \mid X^{(\cap \emptyset /\!\!/ \{x_0\})} /\!\!/ \mathcal{X}'\!'\right)}\!\right),\\
{\mbox{\boldmath$\mathscr{K}$}}^{\mid \{x_0\}}_{\mathcal{X}'}(\breve{p})
&={\mbox{\boldmath$\mathscr{K}$}}^{|\{x_0\}}\!\!\!\left(\breve{p}^{\left( c \mid X /\!\!/ \mathcal{X} \mid \{x_0\} /\!\!/ \{x_0\}\right)}\!\right),\\
{\mbox{\boldmath$\mathscr{K}$}}^{\mid \emptyset}_{\mathcal{X}'\!'}(\breve{p})
&={\mbox{\boldmath$\mathscr{K}$}}^{|\emptyset}\!\!\!\left(\breve{p}^{\left( c \mid X /\!\!/ \mathcal{X} \mid \emptyset /\!\!/ \{x_0\}\right)}\!\right).
\end{split}
\end{equation}

\texttt{Note \!\refstepcounter{ctrnot}\arabic{ctrnot}\,\label{not-term}\itshape\enskip\scriptsize (about term ``frame'').} Although in formulas (\ref{framing-cup2}) and (\ref{framing-cup1}) only monoplet $\{x_0\}$ is a \emph{frame} set, we shall call \emph{frame (with respect to this monoplet)} the s.e. $\frak{X}$ itself, construcyed from two inserted s.e.'s $\mathcal{X}'=\mathcal{X}^{(\cap \{x_0\} /\!\!/ \{x_0\})}$ and $\mathcal{X}'\!'=\mathcal{X}^{(\cap \emptyset /\!\!/ \{x_0\})}$, more hoping to clarify understanding than to cause misunderstandings.

\texttt{Note \!\refstepcounter{ctrnot}\arabic{ctrnot}\,\label{not-frame-recurrent-formulas}\itshape\footnotesize (recurrent formulas of the frame method).}
Getting rid of abbreviations (\ref{abbreviations}) and using (\ref{Kopula-sum-sets-inserting}) and (\ref{Kopula-sum-sets-conditional}), we write the recurrent formulas of the \emph{frame method} (\ref{frame-method-pseudo}) and (\ref{frame-method-conditional}) in the expanded form:
\begin{equation}\label{frame-recurrent-pseudo}
\hspace*{-13pt}
\begin{split}
&{\mbox{\boldmath$\mathscr{K}$}}\!\left(\breve{p}^{\left( c \mid X+Y /\!\!/ \mathcal{X}+\{x_0\}\right)}\!\right)=\\
&=\begin{cases}
\mathscr{K}^{(\{x_0\})}\!\!\left(\breve{p}^{\left( c \mid X^{(\cap \{x_0\} /\!\!/ \{x_0\})} /\!\!/ \mathcal{X}'\right)}\!\right)\!,& \hspace*{-8pt} Y=\{x_0\},\\
&\\[-5pt]
\mathscr{K}^{(\emptyset)}\!\!\left(\breve{p}^{\left( c \mid X^{(\cap \emptyset /\!\!/ \{x_0\})} /\!\!/ \mathcal{X}'\!'\right)}\!\right)\!,& \hspace*{-8pt} Y=\emptyset,
\end{cases}\\
\end{split}\hspace*{-13pt}
\end{equation}
\begin{equation}\label{frame-recurrent-conditional}
\hspace*{-13pt}
\begin{split}
&{\mbox{\boldmath$\mathscr{K}$}}\!\left(\breve{p}^{\left( c \mid X+Y /\!\!/ \mathcal{X}+\{x_0\}\right)}\!\right)=\\
&=\begin{cases}
{\mbox{\boldmath$\mathscr{K}$}}^{|\{x_0\}}\!\Big(\breve{p}^{\left( c \mid X /\!\!/ \mathcal{X} \mid \{x_0\} /\!\!/ \{x_0\}\right)}\!\Big)p_0, & \hspace*{-8pt} Y=\{x_0\},\\
&\\[-5pt]
{\mbox{\boldmath$\mathscr{K}$}}^{|\emptyset}\!\Big(\breve{p}^{\left( c \mid X /\!\!/ \mathcal{X} \mid \emptyset /\!\!/ \{x_0\}\right)}\!\Big)(1-p_0), & \hspace*{-8pt} Y=\emptyset.
\end{cases}
\end{split}\hspace*{-13pt}
\end{equation}

Although formulas (\ref{framing-cup2}) and (\ref{framing-cup1}) are satisfied for any s.e., but in the proposed \emph{frame method} (\ref{frame-method-pseudo}) and (\ref{frame-method-conditional}) we use only \emph{half-rare s.e. (s.h-r.e.)} \cite{Vorobyev2015famems12}.
This, however, does not detract from the generality of its application, since the set-phenomenon transformations are any s.e. can be obtained from its half-rare projection \cite{Vorobyev2015famems12}.

We will make the following useful

\texttt{Note \!\refstepcounter{ctrnot}\arabic{ctrnot}\label{not-universal-half-rare}\itshape\enskip\scriptsize (any \emph{half-rare s.e.} is composed by the frame method from two inserted s.e.'s which are always half-rare).} If the \emph{frame s.e.} $\frak{X}$ in (\ref{framing-cup2}) ans (\ref{framing-cup1}) is half-rare, i.e., its marginal probabilities from $\breve{p}=\{p_0,p_1,...,p_{N-1}\}$ are not more than half, for example:
\begin{equation}\label{...}
\begin{split}
1/2 \geqslant p_0 \geqslant p_1  \geqslant ...  \geqslant p_{N-1},
\end{split}
\end{equation}
then the \emph{both inserted s.e.'s} $\mathcal{X}'=\{x'_1,...,x'_{N-1}\}$ and $\mathcal{X}'\!'=\{x'\!'_1,...,x'\!'_{N-1}\}$, and together with them and the s.e. $\mathcal{X}$ are also half-rare by its Definition (\ref{Mcap}). In other words, their marginal probabilities from $\breve{p}'=\{p'_1,...,p'_{N-1}\}$ и $\breve{p}'\!'=\{p'\!'_1,...,p'\!'_{N-1}\}$
do not exceed the corresponding marginal probabilities events from the frame s.e. $\frak{X}$:
\begin{equation}\label{...}
\begin{split}
p_1 \geqslant p'_1, ... , p_{N-2} \geqslant p'_{N-1},\\
p_1 \geqslant p'\!'_1, ... , p_{N-2} \geqslant p'\!'_{N-1},
\end{split}
\end{equation}
and marginal probabilities from $\breve{p}_{N-1}=\{p_1,...,p_{N-1}\}$ are half-rare by definition.
Thus, any half-rare $N$-s.e. is composed by the frame method with the formula (\ref{framing-cup1}) from two inserted $(N\!-\!1)$-s.e.'s $\mathcal{X}'$ and $\mathcal{X}'\!'$, which are required to be half-rare.

\texttt{Lemma \!\refstepcounter{ctrlem}\arabic{ctrlem}\label{lem-independent-half-rare}\itshape\enskip\scriptsize (about \emph{independent half-rare s.e.'s}, constructed by the frame method from two inserted \emph{half-rare s.e.'s}).} That in the family of half-rare s.e.'s $\frak{X}$ with sets of marginal probabilitiues $\breve{p}$, constructed by the frame method from two inserted half-rare s.e.'s $\mathcal{X}'$ and $\mathcal{X}'\!'$, there was an \emph{independent half-rare s.e.}, it is \emph{necessary} so that the sets of marginal probabilities are related to the marginal probabilities of the frame s.e. $\frak{X}=\{x_0\}+\mathcal{X}=\{x_0\}+(\mathcal{X}'(+)\mathcal{X}'\!')$ by the following way:
\begin{equation}\label{inserting-marginals}
\begin{split}
\breve{p}'&=\Set{p'_1,\ldots,p'_{N-1}}\\
&=\Set{p_1p_0,\ldots,p_{N-1}p_0},\\
\breve{p}'\!'&=\Set{p'\!'_1,\ldots,p'\!'_{N-1}}\\
&=\Set{p_1(1-p_0),\ldots,p_{N-1}(1-p_0)};
\end{split}
\end{equation}
and \emph{sufficient} so that the e.p.d. of the 1st kind of inserted s.e.'s $\mathcal{X}'$ and $\mathcal{X}'\!'$
to be calculated from the formulas for $X \subseteq \mathcal{X}$:
\begin{equation}\label{inserting-marginals1}
\begin{split}
&p(X' /\!\!/ \mathcal{X}')=\mathbf{P}\left( \bigcap_{x' \in X'} x' \bigcap_{x' \in \mathcal{X}'-X'} x'^c \right)=\\
&=\begin{cases}
\displaystyle p_0 \prod_{x \in X} p_x \prod_{x \in \mathcal{X}-X} (1-p_x), & X \ne \emptyset,\\
\displaystyle p_0 \prod_{x \in \mathcal{X}} (1-p_x)+1-p_0, & X = \emptyset,\\
\end{cases}\\
&\\[-8pt]
&p(X'\!' /\!\!/ \mathcal{X}'\!')=\mathbf{P}\left( \bigcap_{x'\!' \in X'\!'} x'\!' \bigcap_{x'\!' \in \mathcal{X}'\!'-X'\!'} x'\!'^c \right)=\\
&=\begin{cases}
\displaystyle (1-p_0) \prod_{x \in X} p_x \prod_{x \in \mathcal{X}-X} (1-p_x), & X \ne \emptyset,\\
\displaystyle (1-p_0) \prod_{x \in \mathcal{X}} (1-p_x)+p_0, & X = \emptyset,
\end{cases}
\end{split}
\end{equation}
where
\begin{equation}\label{inserting-marginals2}
\begin{split}
X'&=\Set{x', x \in X}=\Set{x_0 \cap x, x \in X} \subseteq \mathcal{X}',\\
X'\!'&=\Set{x'\!', x \in X}=\Set{x_0^c \cap x, x \in X} \subseteq \mathcal{X}'\!'.
\end{split}
\end{equation}

\textsf{Proof}. The \emph{necessity} is obvious, since the inserted marginal probabilities of the independent s.e. $\frak{X}$ are probabilities of double intersections of independent events which have the required form for $n=1,...,N-1$:
\begin{equation}\label{inserting-marginals3}
\begin{split}
p'_n = \mathbf{P}(x_0 \cap x_n) &= p_n p_0,\\
p'\!'_n = \mathbf{P}(x_0^c \cap x_n) &= p_n (1-p_0).
\end{split}
\end{equation}
The \emph{sfficiency} follows from (\ref{inserting-marginals1})\footnote{By the way, the necessary condition also follows from (\ref{inserting-marginals1}).} and formulas that connect the e.p.d. of the 1st kind of frame s.e. $\frak{X}$ with the e.p.d. of the 1st kind of inserted s.e.'s $\mathcal{X}'$ and $\mathcal{X}'\!'$, which have the form for $X \subseteq \mathcal{X}$:
\begin{equation}\label{inserting-marginals4}
\begin{split}
&p(X+Y /\!\!/ \mathcal{X}+\{x_0\}) =\\
&=\begin{cases}
p(X' /\!\!/ \mathcal{X}'), & Y=\{x_0\}, X \ne \emptyset,\\
p(\emptyset /\!\!/ \mathcal{X}')-1+p_0, & Y=\{x_0\}, X = \emptyset,\\
&\\
p(X'\!' /\!\!/ \mathcal{X}'\!'), & Y=\emptyset, X \ne \emptyset,\\
p(\emptyset /\!\!/ \mathcal{X}'\!')-p_0, & Y=\emptyset, X = \emptyset.
\end{cases}
\end{split}
\end{equation}
Demanding (\ref{inserting-marginals4}) to perform sufficient conditions (\ref{inserting-marginals1}), we get
\begin{equation}\label{inserting-marginals5}
\hspace*{-8pt}
\begin{split}
&p(X+Y /\!\!/ \mathcal{X}+\{x_0\}) =\\
&=\begin{cases}
\displaystyle p_0 \prod_{x \in X} p_x \!\!\! \prod_{x \in \mathcal{X}\!-\!X}\!\!\!(1\!-\!p_x), &\hspace{-10pt} Y=\{x_0\}, X \subseteq \mathcal{X},\\
&\\
\displaystyle (1\!-\!p_0) \prod_{x \in X} p_x \!\!\! \prod_{x \in \mathcal{X}\!-\!X}\!\!\!(1\!-\!p_x), &\hspace{-10pt} Y=\emptyset, X \subseteq \mathcal{X}.
\end{cases}
\end{split}\hspace*{-13pt}
\end{equation}
As a result, for the s.e. $\frak{X}$ we have the e.p.d. of the 1st kind of independent events:
\begin{equation}\label{inserting-marginals6}
\begin{split}
&p(X+Y /\!\!/ \mathcal{X}+\{x_0\}) =\\
&= p(Z /\!\!/ \frak{X}) = \prod_{x \in Z} p_x \prod_{x \in \frak{X}-Z} (1-p_x),
\end{split}
\end{equation}
where
\begin{equation}\label{inserting-marginals7}
\begin{split}
Z=\begin{cases}
X+\{x_0\}, & Y=\{x_0\}, X \subseteq \mathcal{X},\\
X, & Y=\emptyset, X \subseteq \mathcal{X}.
\end{cases}
\end{split}
\end{equation}
The lemma is proved.

\section{The Kopula theory for monoplets of events\label{1-Kopula-theory}}

Theory of the Kopula of monoplets of events (1-Kopula) seemed to be completed by the formula (\ref{1-epd-from-Kopula}). This formula defines the 1-Kopula of an arbitrary monoplet of events $\{x\}$ with $\{x\}$-monoplet of marginal probabilities $\breve{p}=\{p_x\} \in [0,1]^x$ in the unique form:
\begin{equation}\label{1-Kopula-formula-0}
\begin{split}
{\mbox{\boldmath$\mathscr{K}$}}\left(\breve{p}\right) = {\mbox{\boldmath$\mathscr{K}$}}\left(p_x\right)=p_x,
\end{split}
\end{equation}
which
which provides 2 values on each $2^{(c|\breve{p})}$-phenomenon-dom by general formulas for $X \subseteq \{x\}$:
\begin{equation}\label{1-Kopula-formula-1}
\begin{split}
&{\mbox{\boldmath$\mathscr{K}$}}\left(\breve{p}^{(c|X/\!\!/\{x\})}\right) =\\
&=
\begin{cases}
{\mbox{\boldmath$\mathscr{K}$}}\left(1-p_x\right) = 1-p_x, & X = \emptyset,\\
{\mbox{\boldmath$\mathscr{K}$}}\left(p_x\right) = p_x, & X = \{x\}.
\end{cases}
\end{split}
\end{equation}

However, the formula (\ref{1-Kopula-formula-1}) can be generalized in the following simple way:
\begin{equation}\label{1-Kopula-formula-general}
\begin{split}
&{\mbox{\boldmath$\mathscr{K}$}}\left(\breve{p}^{(c|X/\!\!/\{x\})}\right) =\\
&=
\begin{cases}
{\mbox{\boldmath$\mathscr{K}$}}^{\emptyset}\left(1-p_x\right) = 1-{\mbox{\boldmath$\mathscr{K}$}}^{\{x\}}\left(p_x\right), & X = \emptyset,\\
{\mbox{\boldmath$\mathscr{K}$}}^{\{x\}}\left(p_x\right), & X = \{x\},
\end{cases}
\end{split}
\end{equation}
where ${\mbox{\boldmath$\mathscr{K}$}}^{\{x\}}$ is any function such that in half-rare variables:
\begin{equation}\label{1-Kopula-formula-general-rare}
\begin{split}
{\mbox{\boldmath$\mathscr{K}$}}^{\{x\}} : [0,1/2] \to [0,1/2],
\end{split}
\end{equation}
and in free variables:
\begin{equation}\label{1-Kopula-formula-any}
\begin{split}
{\mbox{\boldmath$\mathscr{K}$}}^{\{x\}} : [0,1] \to [0,1].
\end{split}
\end{equation}

In this case, the 1-Kopula (\ref{1-Kopula-formula-1}) is an important special case of 1-Kopula (\ref{1-Kopula-formula-general}) when ${\mbox{\boldmath$\mathscr{K}$}}^{\{x\}}(p_x)=p_x$. This case corresponds to a uniform marginal distribution function on the unit interval in the theory of the classical copula \cite{Sklar1959}.

\section{The Kopula theory for doublets of events}

\subsection{The frame method for constructing a half-rare doublet of events\label{2-framing-method}}

In order to construct by the frame method the $\breve{p}$-\uwave{ordered} \emph{frame half-rare doublet of events}
\begin{equation}\label{...}
\begin{split}
\frak{X}=\{x,y\}=\{x\}+\mathcal{X}=\{x\}+\Big(\mathcal{X}'(+)\mathcal{X}'\!'\Big)
\end{split}
\end{equation}
with the $\frak{X}$-set of marginal probabilities $\breve{p}=\{p_x,p_y\}$, where
\begin{equation}\label{2order}
\begin{split}
1/2 \geqslant p_x \geqslant p_y \geqslant 0,
\end{split}
\end{equation}
let's suppose that we have at our disposal two half-rare \emph{inserted monoplets of events}
\begin{equation}\label{inserting-2-monoplets}
\begin{split}
\mathcal{X}'=\{x \cap y\}=\{s'\} \ \mbox{и} \ \mathcal{X}'\!'=\{x^c \cap y\}=\{s'\!'\},
\end{split}
\end{equation}
with known 1-Kopulas:
\begin{equation}\label{1-Kopula-formula-s1}
\begin{split}
&{\mbox{\boldmath$\mathscr{K}$}}'\left(\breve{p}^{(c|S/\!\!/\{s'\})}\right) =\\
&=
\begin{cases}
{\mbox{\boldmath$\mathscr{K}$}}'\left(1-p_{s'}\right) = 1-p_{s'}, & S = \emptyset,\\
{\mbox{\boldmath$\mathscr{K}$}}'\left(p_{s'}\right) = p_{s'}, & S = \{s'\},
\end{cases}
\end{split}
\end{equation}
\begin{equation}\label{1-Kopula-formula-s11}
\begin{split}
&{\mbox{\boldmath$\mathscr{K}$}}'\!'\left(\breve{p}^{(c|S/\!\!/\{s'\!'\})}\right) =\\
&=
\begin{cases}
{\mbox{\boldmath$\mathscr{K}$}}'\!'\left(1-p_{s'\!'}\right) = 1-p_{s'\!'}, & S = \emptyset,\\
{\mbox{\boldmath$\mathscr{K}$}}'\!'\left(p_{s'\!'}\right) = p_{s'\!'}, & S = \{s'\!'\},
\end{cases}
\end{split}
\end{equation}

\begin{figure}[ht!]
\vspace*{-15pt}
\centering
\includegraphics[width=2.33in]{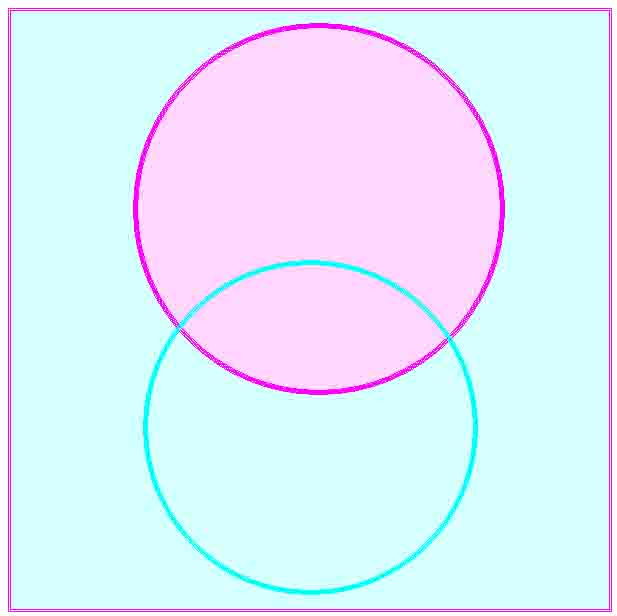}

\vspace{-165pt}

\hspace{-181pt} {\Large$\Omega$}

\vspace{-10pt}

\hspace{83pt} {\Large${\mbox{\boldmath$x$}}$}

\vspace{122pt}

\hspace{83pt} {\Large${\mbox{\boldmath$y$}}$}

\vspace{-116pt}

\hspace{0pt} {\scriptsize$s'^c \!\cap x=y^c\!\cap x$}

\vspace{34pt}

\hspace{0pt} {\Large${\mbox{\boldmath$s'$}}$}{\scriptsize$=x\cap y$}

\vspace{32pt}

\hspace{0pt} {\Large${\mbox{\boldmath$s'\!'$}}$}{\scriptsize$=x^c\!\cap y$}

\vspace{13pt}

\hspace{-97pt} {\scriptsize$s'\!'^c \!\cap x^c=y^c\!\cap x^c$}

\vspace{1pt}

\includegraphics[width=3.3in]{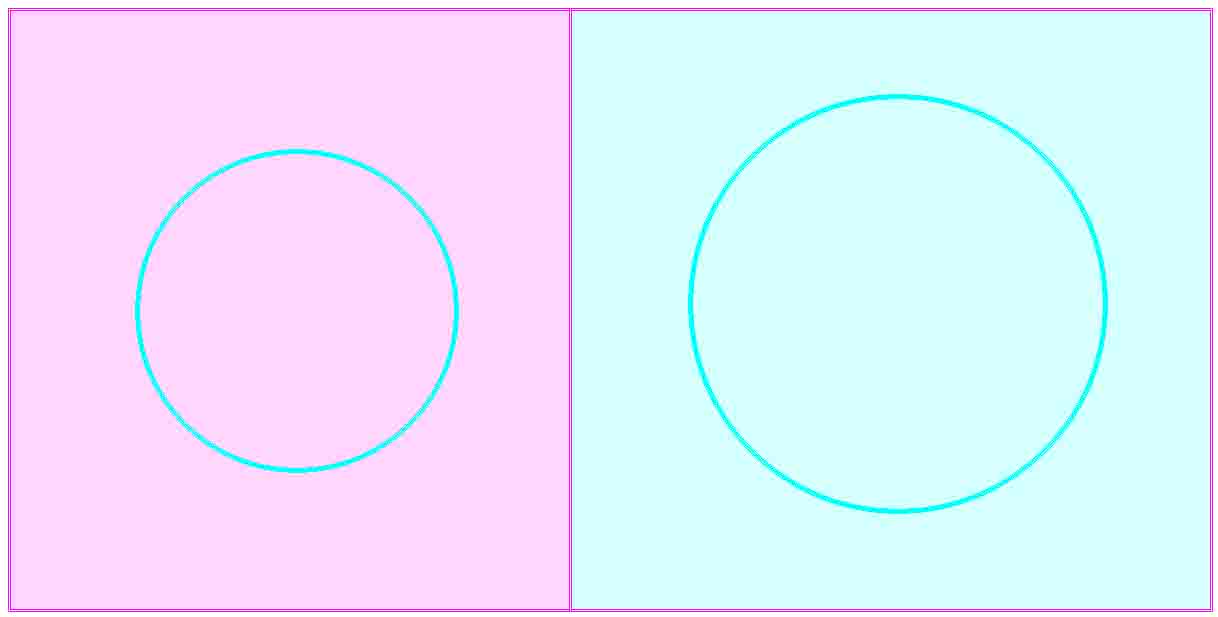}

\vspace{-134pt}

\hspace{-225pt}  {\Large$\Omega$}

\vspace{0pt}

\hspace{-180pt}  {\scriptsize$s'^c \!\cap x=y^c\!\cap x$}

\vspace{41pt}

\hspace{-5pt} ${\Large {\mbox{\boldmath$s'$}}}${\scriptsize$=x \cap y$} \hspace{70pt} ${\Large {\mbox{\boldmath$s'\!'$}}}${\scriptsize$=x^c\!\cap y$}

\vspace{38pt}

\hspace{167pt} {\scriptsize$s'\!'^c \!\cap x^c=y^c\!\cap x^c$}

\vspace{-7pt}

$\underbrace{\hspace{108pt}}_{\Large {\mbox{\boldmath$x$}}}
\underbrace{\hspace{124pt}}_{\Large {\mbox{\boldmath$x^c$}}}$

\vspace{-10pt}

\caption{Venn diagrams of the frame half-rare doublet of events $\frak{X}=\{x,y\}, \ 1/2 \geqslant p_x \geqslant p_y$ (up), and two inserted monoplets $\mathcal{X}'=\{s'\}$ and $\mathcal{X}'\!'=\{s'\!'\}$ (down) agreed with the frame doublet $\frak{X}$ in the following sense\!:\! $y\!=\!s'\!+s'\!'\!$ и $s' \subseteq x, s'\!' \subseteq x^c$.\label{fig-Venn2-1-xy}}
\end{figure}

By Definition of inserted monoplets (\ref{inserting-2-monoplets})
(see Fig. \ref{fig-Venn2-1-xy})
\begin{equation}\label{monoplet}
\begin{split}
s' &= x \cap y \subseteq x,\\
s'\!' &= x^c \cap y \subseteq x^c,
\end{split}
\end{equation}
and also because of the $\breve{p}$-ordering assumption (\ref{2order}), we get that
\begin{equation}\label{1-Kopula-2-monoplets-bounds-0}
\begin{split}
p_{s'}+p_{s'\!'}=p_y \leqslant p_x \leqslant 1/2 \leqslant 1-p_x.
\end{split}
\end{equation}
Consequently, the 1-Kopulas of inserted monoplanes (\ref{1-Kopula-formula-s1}) and (\ref{1-Kopula-formula-s11}) are bound by the sum of their marginal probabilities:

Consequently, the 1-Kopulas of inserted monoplets (\ref{1-Kopula-formula-s1}) and (\ref{1-Kopula-formula-s11})
are bound by the restriction on the sum of their marginal probabilities:
\begin{equation}\label{1-Kopula-2-monoplets-bounds-1}
\begin{split}
p_{s'}+p_{s'\!'}=p_y,
\end{split}
\end{equation}
and depend on only one parameter:
\begin{equation}\label{1-Kopula-2-monoplets-bounds}
\begin{split}
p_{s'} \in [0,p_y].
\end{split}
\end{equation}

We get the following formulas:
\begin{equation}\label{epd-doublet}
\hspace*{-3pt}
\begin{split}
p(xy /\!\!/ \{x,y\}) &\!=\! p(s' /\!\!/ \mathcal{X}') = p_{s'},\\
p(x /\!\!/ \{x,y\}) &\!=\!p(\emptyset /\!\!/ \mathcal{X}')-1+p_x=p_x-p_{s'},\\
p(y /\!\!/ \{x,y\}) &\!=\!p(s'\!' /\!\!/ \mathcal{X}'\!') = p_y-p_{s'},\\
p(\emptyset /\!\!/ \{x,y\}) &\!=\!p(\emptyset /\!\!/ \mathcal{X}'\!')-p_x=1-p_y-p_x+p_{s'}.
\end{split}\hspace*{-13pt}
\end{equation}
These formulas express the e.p.d. of the 1st kind of the $\breve{p}$-ordered half-rare \emph{frame doublet of events} $\frak{X}=\{x,y\}$ through the e.p.d. of the 1st kind of \emph{inserted monoplets} $\mathcal{X}'$ and $\mathcal{X}'\!'$, and the probability of \emph{frame event} $x$, and, in the final result, through their marginal probabilities $p_x$ and $p_y$, and marginal probability $p_{s'}$ of the inserted monoplet $\mathcal{X}'=\{s'\} = \{x \cap y\}$.

The formulas (\ref{epd-doublet}) express values of the 2-Kopula of $\breve{p}$-ordered doublet $\frak{X}=\{x,y\}$ through 1-Kopulas of inserted monoplets $\mathcal{X}'=\{s'\}$ and $\mathcal{X}'\!'=\{s'\!'\}$. Rewrite this in a form of an explicit recurrent formula:
\begin{equation}\label{2-Kopula-recurrent-formula}
\begin{split}
p(X /\!\!/ \{x,y\})={\mbox{\boldmath$\mathscr{K}$}}_{\frak{X}}\Big(\breve{p}^{(c|X /\!\!/ \{x,y\})} \Big)=\\
=\begin{cases}
{\mbox{\boldmath$\mathscr{K}$}}_{\mathcal{X}'}(p_{s'}), & X=\{x,y\},\\
{\mbox{\boldmath$\mathscr{K}$}}_{\mathcal{X}'}(1-p_{s'})-1+p_x, & X=\{x\},\\
&\\[-8pt]
{\mbox{\boldmath$\mathscr{K}$}}_{\mathcal{X}'\!'}(p_{s'\!'}), & X=\{y\},\\
{\mbox{\boldmath$\mathscr{K}$}}_{\mathcal{X}'\!'}(1-p_{s'\!'})-p_x, & X=\emptyset.
\end{cases}
\end{split}
\end{equation}
Considering (\ref{1-Kopula-formula}) and (\ref{1-Kopula-2-monoplets-bounds}), we will continue:
\begin{equation}\label{2-Kopula-recurrent-formula2}
\begin{split}
p(X /\!\!/ \{x,y\})={\mbox{\boldmath$\mathscr{K}$}}_{\frak{X}}\Big(\breve{p}^{(c|X /\!\!/ \{x,y\})} \Big)=\\
=\begin{cases}
p_{s'}, & X=\{x,y\},\\
p_x-p_{s'}, & X=\{x\},\\
&\\[-8pt]
p_{s'\!'}, & X=\{y\},\\
1-p_x-p_{s'\!'}, & X=\emptyset,
\end{cases}
\end{split}
\end{equation}
\begin{equation}\label{2-Kopula-recurrent-formula3}
\begin{split}
=\begin{cases}
p_{s'}, & X=\{x,y\},\\
p_x-p_{s'}, & X=\{x\},\\
&\\[-8pt]
p_y-p_{s'}, & X=\{y\},\\
1-p_x-p_y+p_{s'}, & X=\emptyset.
\end{cases}
\end{split}
\end{equation}
We note, by the way, that the restriction (\ref{1-Kopula-2-monoplets-bounds}) by the assumption of $\breve{p}$-ordered (\ref{2order}) is a special case of Fr\'echet-inequalities:
\begin{equation}\label{FRECHET-bounds-xy}
\begin{split}
0  \leqslant p_{s'} \leqslant p_{xy}^+=\min\{p_x,p_y\} = p_y.
\end{split}
\end{equation}

\begin{figure}[ht!]
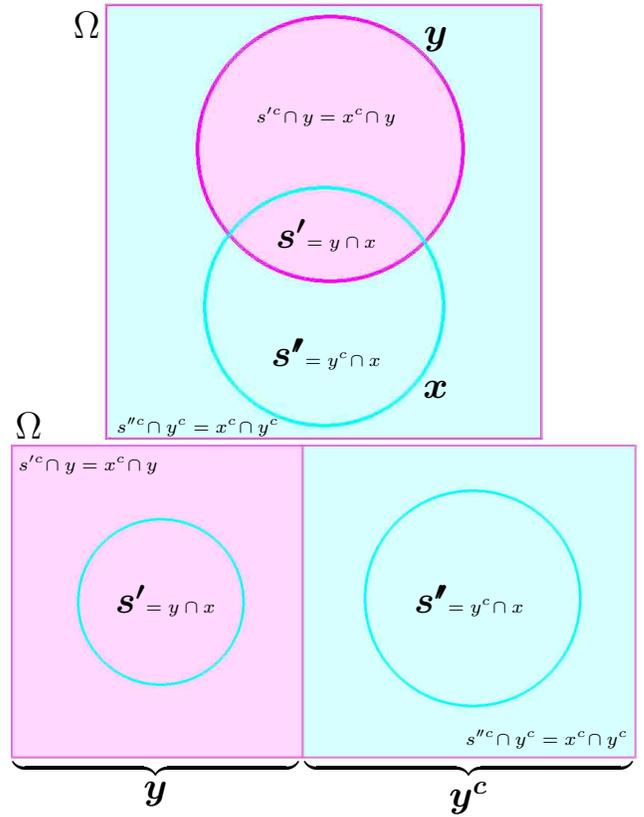

\vspace*{-15pt}
\centering
\includegraphics[width=2.33in]{xyRoseBluealoneSt.jpg}

\vspace{-165pt}

\hspace{-181pt} {\Large$\Omega$}

\vspace{-10pt}

\hspace{83pt} {\Large${\mbox{\boldmath$y$}}$}

\vspace{122pt}

\hspace{83pt} {\Large${\mbox{\boldmath$x$}}$}

\vspace{-116pt}

\hspace{0pt} {\scriptsize$s'^c \!\cap y=x^c\!\cap y$}

\vspace{34pt}

\hspace{0pt} {\Large${\mbox{\boldmath$s'$}}$}{\scriptsize$=y\cap x$}

\vspace{32pt}

\hspace{0pt} {\Large${\mbox{\boldmath$s'\!'$}}$}{\scriptsize$=y^c\!\cap x$}

\vspace{13pt}

\hspace{-97pt} {\scriptsize$s'\!'^c \!\cap y^c=x^c\!\cap y^c$}

\vspace{1pt}

\includegraphics[width=3.3in]{xRoseBlueSt.jpg}

\vspace{-134pt}

\hspace{-225pt}  {\Large$\Omega$}

\vspace{0pt}

\hspace{-180pt}  {\scriptsize$s'^c \!\cap y=x^c\!\cap y$}

\vspace{41pt}

\hspace{-5pt} ${\Large {\mbox{\boldmath$s'$}}}${\scriptsize$=y \cap x$} \hspace{70pt} ${\Large {\mbox{\boldmath$s'\!'$}}}${\scriptsize$=y^c\!\cap x$}

\vspace{38pt}

\hspace{167pt} {\scriptsize$s'\!'^c \!\cap y^c=x^c\!\cap y^c$}

\vspace{-7pt}

$\underbrace{\hspace{108pt}}_{\Large {\mbox{\boldmath$y$}}}
\underbrace{\hspace{124pt}}_{\Large {\mbox{\boldmath$y^c$}}}$

\vspace{-10pt}

\caption{Venn diagrams of the frame, otherwise ordered half-rare doublet events $\frak{X}=\{y,x\}, \ 1/2 \geqslant p_y \geqslant p_x$ (up), and two inserted monoplets $\mathcal{X}'=\{s'\}$ and $\mathcal{X}'\!'=\{s'\!'\}$ (down), agreed with the frame doublet $\frak{X}$ in the following sense:\! $x\!=\!s'\!+s'\!'\!$ и $s' \subseteq y, s'\!' \subseteq y^c$.\label{fig-Venn2-1-yx}}
\end{figure}

\texttt{Note \!\refstepcounter{ctrnot}\arabic{ctrnot}\,\label{not-another-order}\itshape\enskip\scriptsize (frame mathod for otherwise $\breve{p}$-ordered half-rare doublet of events).}
For \emph{otherwise $\breve{p}$-ordered} half-rare doublet of events
\begin{equation}\label{...}
\begin{split}
\frak{X}=\{y,x\} = \{y\}+\mathcal{X}=\{y\}+\Big( \mathcal{X}'(+)\mathcal{X}'\!'\Big),
\end{split}
\end{equation}
where
\begin{equation}\label{inserting-2-monoplets-yx}
\begin{split}
\mathcal{X}'=\{y \cap x\}=\{s'\} \ \mbox{и} \ \mathcal{X}'\!'=\{y^c \cap x\}=\{s'\!'\},
\end{split}
\end{equation}
with the $\frak{X}$-set of marginal probabilities $\breve{p}=\{p_y,p_x\}$, где $1/2 \geqslant p_y \geqslant p_x$, the assumptions (\ref{monoplet}) take symmetrical form:
\begin{equation}\label{monoplet-another}
\begin{split}
s' &= y \cap x \subseteq y,\\
s'\!' &= y^c \cap x \subseteq y^c.
\end{split}
\end{equation}

By Definition of inserted monoplets (\ref{inserting-2-monoplets-yx})
(см. Рис. \ref{fig-Venn2-1-yx})
\begin{equation}\label{monoplet}
\begin{split}
s' &= x \cap y \subseteq y,\\
s'\!' &= x \cap y^c \subseteq y^c,
\end{split}
\end{equation}
and also because of the assumption
and also because of the assumption of another $\breve{p}$-ordering, we get that
\begin{equation}\label{1-Kopula-yx-2-monoplets-bounds-0}
\begin{split}
p_{s'}+p_{s'\!'}=p_x \leqslant p_y \leqslant 1/2 \leqslant 1-p_y.
\end{split}
\end{equation}
Consequently, 1-copulas of inserted monoplanes are connected by a restriction on the sum of their marginal probabilities:
\begin{equation}\label{1-Kopula-yx-2-monoplets-bounds-1}
\begin{split}
p_{s'}+p_{s'\!'}=p_x,
\end{split}
\end{equation}
and depend on only one parameter:
\begin{equation}\label{1-Kopula-yx-2-monoplets-bounds}
\begin{split}
p_{s'} \in [0,p_x].
\end{split}
\end{equation}

By the assumptions, the following formulas are valid:
\begin{equation}\label{epd-doublet-yx}
\hspace*{-3pt}
\begin{split}
p(xy /\!\!/ \{y,x\}) &\!=\! p(s' /\!\!/ \mathcal{X}')=p_{s'},\\
p(y /\!\!/ \{y,x\}) &\!=\!p(\emptyset /\!\!/ \mathcal{X}')-1+p_y=p_y-p_{s'},\\
p(x /\!\!/ \{y,x\}) &\!=\!p(s'\!' /\!\!/ \mathcal{X}'\!')=p_x-p_{s'},\\
p(\emptyset /\!\!/ \{y,x\}) &\!=\!p(\emptyset /\!\!/ \mathcal{X}'\!')-p_y=1-p_x-p_y-p_{s'}.
\end{split}\hspace*{-13pt}
\end{equation}
These formulas express the e.p.d. of the 1st kind otherwise $\breve{p}$-ordered half-rare \emph{frame doublet of events} $\frak{X}$ through the e.p.d. of the 1st kind of \emph{inserted monoplets} $\mathcal{X}'$ and $\mathcal{X}'\!'$, and the probability of \emph{frame event} $y$, and, in the final result, through own marginal probabilities $p_x$ и $p_y$, and the marginal probabilities of inserted monoplet $\mathcal{X}'=\{s'\}=\{x \cap y\}$ (see Fig. \ref{fig-Venn2-1-yx}).

The formulas (\ref{epd-doublet-yx}) express values of the 2-Kopula of $\breve{p}$-ordered doublet $\frak{X}=\{y,x\}$ through 1-Kopulas of inserted monoplets $\mathcal{X}'=\{s'\}$ and $\mathcal{X}'\!'=\{s'\!'\}$. Rewrite this in the form of explicit recurrent formula:
\begin{equation}\label{2-Kopula-recurrent-formula-yx}
\begin{split}
p(X /\!\!/ \{y,x\})={\mbox{\boldmath$\mathscr{K}$}}_{\frak{X}}\Big(\breve{p}^{(c|X /\!\!/ \{y,x\})} \Big)=\\
=\begin{cases}
{\mbox{\boldmath$\mathscr{K}$}}_{\mathcal{X}'}(p_{s'}), & X=\{x,y\},\\
{\mbox{\boldmath$\mathscr{K}$}}_{\mathcal{X}'}(1-p_{s'})-1+p_y, & X=\{y\},\\
{\mbox{\boldmath$\mathscr{K}$}}_{\mathcal{X}'\!'}(p_{s'\!'}), & X=\{x\},\\
{\mbox{\boldmath$\mathscr{K}$}}_{\mathcal{X}'\!'}(1-p_{s'\!'})-p_y, & X=\emptyset.
\end{cases}
\end{split}
\end{equation}
Continue:
\begin{equation}\label{2-Kopula-recurrent-formula-yx2}
\begin{split}
p(X /\!\!/ \{x,y\})={\mbox{\boldmath$\mathscr{K}$}}_{\frak{X}}\Big(\breve{p}^{(c|X /\!\!/ \{x,y\})} \Big)=\\
=\begin{cases}
p_{s'}, & X=\{x,y\},\\
p_y-p_{s'}, & X=\{y\},\\
p_{s'\!'}, & X=\{x\},\\
1-p_x-p_{s'\!'}, & X=\emptyset,
\end{cases}
\end{split}
\end{equation}
\begin{equation}\label{2-Kopula-recurrent-formula-yx3}
\begin{split}
=\begin{cases}
p_{s'}, & X=\{x,y\},\\
p_y-p_{s'}, & X=\{y\},\\
p_x-p_{s'}, & X=\{x\},\\
1-p_x-p_y+p_{s'}, & X=\emptyset.
\end{cases}
\end{split}
\end{equation}
We note, as above, that the restriction (\ref{1-Kopula-yx-2-monoplets-bounds}) by the assumption of another $\breve{p}$-ordering is a special case of Fr\'echet inequalities:
\begin{equation}\label{ps-inequality-yx}
\begin{split}
0  \leqslant p_{s'} \leqslant p_{xy}^+=\min\{p_x,p_y\} = p_x.
\end{split}
\end{equation}

\subsection{The frame method: recurrent formulas for a half-rare doublet of events\label{2-framing-method-recurrent-formulas}}

The formulas (\ref{2-Kopula-recurrent-formula}), as well as formulas (\ref{2-Kopula-recurrent-formula3}), can be rewrite in the form of special cases of recurrent formulas (\ref{frame-recurrent-pseudo}) and (\ref{frame-recurrent-conditional}) from Note \ref{not-frame-recurrent-formulas} for the doublet $\frak{X}=\{x,y\}=\{x\}+\{y\}=\{x\}+\mathcal{X}$:
\begin{equation}\label{frame-recurrent-pseudo-doublet}
\hspace*{-13pt}
\begin{split}
&{\mbox{\boldmath$\mathscr{K}$}}\!\left(\breve{p}^{\left( c \mid X+Y /\!\!/ \mathcal{X}+\{x\}\right)}\!\right)=\\
&=\begin{cases}
\mathscr{K}^{(\{x\})}\!\!\left(\breve{p}^{\left( c \mid X^{(\cap \{x\} /\!\!/ \{x\})} /\!\!/ \mathcal{X}'\right)}\!\right)\!,& \hspace*{-8pt} Y=\{x\},\\
&\\[-5pt]
\mathscr{K}^{(\emptyset)}\!\!\left(\breve{p}^{\left( c \mid X^{(\cap \emptyset /\!\!/ \{x\})} /\!\!/ \mathcal{X}'\!'\right)}\!\right)\!,& \hspace*{-8pt} Y=\emptyset,
\end{cases}\\
\end{split}\hspace*{-13pt}
\end{equation}
\begin{equation}\label{frame-recurrent-conditional-doublet}
\hspace*{-13pt}
\begin{split}
&{\mbox{\boldmath$\mathscr{K}$}}\!\left(\breve{p}^{\left( c \mid X+Y /\!\!/ \mathcal{X}+\{x\}\right)}\!\right)=\\
&=\begin{cases}
{\mbox{\boldmath$\mathscr{K}$}}^{|\{x\}}\!\Big(\breve{p}^{\left( c \mid X /\!\!/ \mathcal{X} \mid \{x\} /\!\!/ \{x\}\right)}\!\Big)p_x, & \hspace*{-8pt} Y=\{x\},\\
&\\[-5pt]
{\mbox{\boldmath$\mathscr{K}$}}^{|\emptyset}\!\Big(\breve{p}^{\left( c \mid X /\!\!/ \mathcal{X} \mid \emptyset /\!\!/ \{x\}\right)}\!\Big)(1-p_x), & \hspace*{-8pt} Y=\emptyset.
\end{cases}
\end{split}\hspace*{-13pt}
\end{equation}
In the formulas (\ref{frame-recurrent-pseudo-doublet}) the pseudo-Kopulas $\mathscr{K}^{(\{x\})}$ and $\mathscr{K}^{(\emptyset)}$ of inserted monoplets $\mathcal{X}'$ and $\mathcal{X}'\!'$, correspondingly, are defined by the first and the second pairs of probabilities from (\ref{2-Kopula-recurrent-formula3}) correspondingly, i.e., by formulas:
\begin{equation}\label{epd-doublet-4-x-pseudoKopula}
\hspace*{-3pt}
\begin{split}
&\mathscr{K}^{(\{x\})}\!\!\left(\breve{p}^{\left( c \mid X^{(\cap \{x\} /\!\!/ \{x\})} /\!\!/ \mathcal{X}'\right)}\!\right)\!=\\
&=\begin{cases}
p_{s'}, & X=\{x,y\},\\
p_x-p_{s'}, & X=\{x\},\\
\end{cases}
\end{split}\hspace*{-13pt}
\end{equation}
\begin{equation}\label{epd--doublet-4-0-pseudoKopula}
\hspace*{-3pt}
\begin{split}
&\mathscr{K}^{(\emptyset)}\!\!\left(\breve{p}^{\left( c \mid X^{(\cap \emptyset /\!\!/ \{x\})} /\!\!/ \mathcal{X}'\!'\right)}\!\right)\!=\\
&=\begin{cases}
p_y-p_{s'}, & X=\{y\},\\
1-p_x-p_y+p_{s'}, & X=\emptyset,
\end{cases}
\end{split}\hspace*{-13pt}
\end{equation}
where, for example, for $X=\{y\}$
\begin{equation}\label{2-explanations1}
\hspace*{-13pt}
\begin{split}
&\{y\}^{(\cap \{x\} /\!\!/ \{x\})}\!=\!\{y \cap x\} \subseteq \mathcal{X}'\!=\!\{s'\},\\
&\{y\}^{(\cap \emptyset /\!\!/ \{x\})}\!=\!\{y \cap x^c\} \subseteq \mathcal{X}'\!'\!=\!\{y-s'\},
\end{split}\hspace*{-13pt}
\end{equation}
and the corresponding sets of marginal probabilities of inserted monoplets $\mathcal{X}'$ and $\mathcal{X}'\!'$ have the form
\begin{equation}\label{2-explanations2}
\hspace*{-13pt}
\begin{split}
\breve{p}^{\left( c \mid \{y\}^{(\cap \{x\} /\!\!/ \{x\})} /\!\!/ \mathcal{X}'\right)}\!&=\!\{p_{s'}\},\\
\breve{p}^{\left( c \mid \{y\}^{(\cap \emptyset /\!\!/ \{x\})} /\!\!/ \mathcal{X}'\!'\right)}\!&=\!\{p_y-p_{s'}\}.
\end{split}\hspace*{-13pt}
\end{equation}

In the formulas (\ref{frame-recurrent-conditional-doublet}) the conditional Kopulas
${\mbox{\boldmath$\mathscr{K}$}}^{|\{x\}}$ and ${\mbox{\boldmath$\mathscr{K}$}}^{|\emptyset}$
are defined by the first and the second pairs of probabilities from (\ref{2-Kopula-recurrent-formula3}), normalized by $p_x$ and by $1-p_x$ correspondingly, i.e. by the formulas:
\begin{equation}\label{epd-doublet-4-x-conditionalKopula}
\hspace*{-3pt}
\begin{split}
&{\mbox{\boldmath$\mathscr{K}$}}^{|\{x\}}\!\Big(\breve{p}^{\left( c \mid X /\!\!/ \mathcal{X} \mid \{x\} /\!\!/ \{x\}\right)}\!\Big)\!=\\
&=\begin{cases}
\frac{1}{p_x}p_{s'}, & X=\{x,y\},\\
\frac{1}{p_x}(p_x-p_{s'}), & X=\{x\},\\
\end{cases}
\end{split}\hspace*{-13pt}
\end{equation}
\begin{equation}\label{epd-doublet-4-0-conditionalKopula}
\hspace*{-3pt}
\begin{split}
&{\mbox{\boldmath$\mathscr{K}$}}^{|\emptyset}\!\Big(\breve{p}^{\left( c \mid X /\!\!/ \mathcal{X} \mid \emptyset /\!\!/ \{x\}\right)}\!\Big)\!=\\
&=\begin{cases}
\frac{1}{1-p_x}(p_y-p_{s'}), & X=\{y\},\\
\frac{1}{1-p_x}(1-p_x-p_y+p_{s'}), & X=\emptyset.
\end{cases}
\end{split}\hspace*{-13pt}
\end{equation}
The corresponding sets of marginal conditional probabilities of events $y \in \mathcal{X}$ with respect to the frame terraced events $\textsf{ter}(\{x\} /\!\!/ \{x\})=x$ и $\textsf{ter}(\emptyset /\!\!/ \{x\})=x^c$ correspondingly have the form:
\begin{equation}\label{2-conditional-marginals}
\hspace*{-13pt}
\begin{split}
\breve{p}^{\left( c \mid \mathcal{X} /\!\!/ \mathcal{X} \mid \{x\} /\!\!/ \{x\}\right)}\!&=\!\left\{\frac{p_{s'}}{p_x}\right\},\\
\breve{p}^{\left( c \mid \mathcal{X} /\!\!/ \mathcal{X} \mid \emptyset /\!\!/ \{x\}\right)}\!&=\!\left\{\frac{p_y-p_{s'}}{1-p_x}\right\}.
\end{split}\hspace*{-13pt}
\end{equation}
Remind, that Fr\'echet restrictions on the functional parameter $p_{s'}=p_{s'}(p_x,p_y)$ for $\breve{p}$-ordered half-rare doublet of events $\frak{X}=\{x,y\}$ have the form:
\begin{equation}\label{FRECHET-bounds-2xy}
\begin{split}
0  \leqslant p_{s'} \leqslant p_{xy}^+=\min\{p_x,p_y\} = p_y,
\end{split}
\end{equation}
and for otherwise $\breve{p}$-ordered half-rare doublet of events $\frak{X}=\{y,x\}$ have the form:
\begin{equation}\label{FRECHET-bounds-2yx}
\begin{split}
0  \leqslant p_{s'} \leqslant p_{xy}^+=\min\{p_x,p_y\} = p_x.
\end{split}
\end{equation}

\section{The Kopula theory for triplets of events\label{3-Kopulas-theory}}
\subsection{Independent $3$-Kopula}

First, without the frame method, which is not required here, consider the simplest example of a \emph{$3$-Kopula} ${\mbox{\boldmath$\mathscr{K}$}} \in \Psi^1_\frak{X}$ of the $(N\!-\!1)$-set of events $\frak{X}=\{x,y,z\}$, i.e., a 1-function on the unit $\frak{X}$-cube. In other words, construct such a nonnegative bounded numerical function
$$
{\mbox{\boldmath$\mathscr{K}$}} : [0,1]^{\otimes \frak{X}} \to [0,1],
$$
that for all $z \in \frak{X}$
$$
\sum_{x \in X \subseteq \frak{X}} {\mbox{\boldmath$\mathscr{K}$}}\left(\breve{w}^{(c|X/\!\!/\frak{X})}\right) = w_x.
$$
Such a simple example of a 1-function on $\frak{X}$-cube is so-called \emph{independent $(N\!-\!1)$-Kopula}, which for all free variables $\breve{w}=\{w_x,w_y,w_z\} \in [0,1]^x \otimes [0,1]^y \otimes [0,1]^z =[0,1]^{\otimes \frak{X}}$ is defined by the formula:
\begin{equation}\label{ind-3-Kopula}
\begin{split}
{\mbox{\boldmath$\mathscr{K}$}}\left(\breve{w}\right) &=  w_xw_yw_z,
\end{split}
\end{equation}
that provides it on each $2^{(c|\breve{w})}$-phenomenon-dom the following $2^3$ values:
\begin{equation}\label{ind-3-Kopulas}
\begin{split}
{\mbox{\boldmath$\mathscr{K}$}}\left(\breve{w}^{(c|X/\!\!/\frak{X})}\right) = \prod_{x \in X} w_x \prod_{x \in \frak{X}-X} (1-w_x)
\end{split}
\end{equation}
for $X \subseteq \frak{X}$.
Indeed, as in the case of the doublet of events this function is a 1-function, since for all $x \in \frak{X}$
$$
\sum_{x \in X \subseteq \frak{X}} \left( \prod_{z \in X} w_z \prod_{z \in \frak{X}-X} (1-w_z) \right) = w_x.
$$
The e.p.d. of the 1st kind of independent triplet of events $\frak{X}$ with the $\frak{X}$-set of probabilities of events $\breve{p}$ is defined by $2^3$ values of the independent 3-Kopula (\ref{ind-3-Kopula}) on $2^{(c|\breve{p})}$-phenomenon-dom by the general formulas of half-rare variables, i.e., for $X \subseteq \{x,y\}$:
\begin{equation}\label{3-epd-from-ind-Kopula}
\begin{split}
&p(X/\!\!/\frak{X})\!=\!{\mbox{\boldmath$\mathscr{K}$}}\left(\breve{p}^{(c|X/\!\!/\frak{X})}\right) \!=\!\prod_{x \in X} p_x \!\!\!\!\prod_{x \in \frak{X}-X} (1-p_x).
\end{split}
\end{equation}

\subsection{Three-dimensional maps of the independent 3-Kopula}

\begin{figure}[h!]
\centering
\hspace{-5pt}
\includegraphics[width=3.35in]{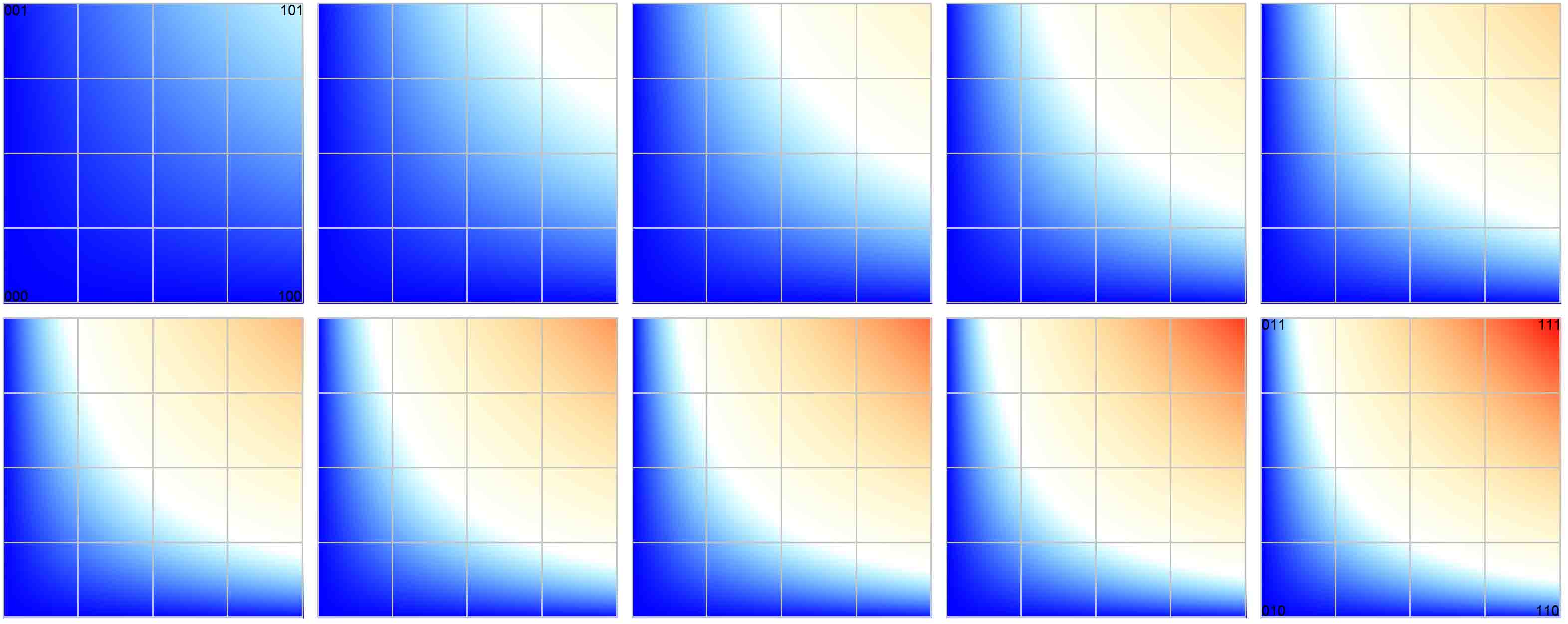}

\vspace{-10pt}

\caption{The visualization of projections of the same three-dimensional map of Cartesian representation of independent 3-Kopula of the triplet $\frak{X}=\{x,y,z\}$ on the unit cube in conditional colors with values of marginal probaboility $p_y=0.1, ..., 0.9, 1.0$, where the white color corresponds to points in which probabilities of all terraced events are 1/8. The orientation of axes: $(p_x,p_z)=$ (horizontal, vertical).} %
\label{3-Kopula-figs-maps}
\end{figure}

In Fig. \ref{3-Kopula-figs-maps} it is shown the results of visualization of the three-dimensional graph of independent 3-Kopula (\ref{ind-N-Kopula}) of the triplet $\frak{X}=\{x,y,z\}$, defined on the cube $[0,1]^3$, in projections on planes, which are orthogonal to the axis $p_y$.

\subsection{The frame method for constructing a half-rare triplet of events\label{3-framing-method}}

\begin{figure}[ht!]
\centering
\includegraphics[width=2.33in]{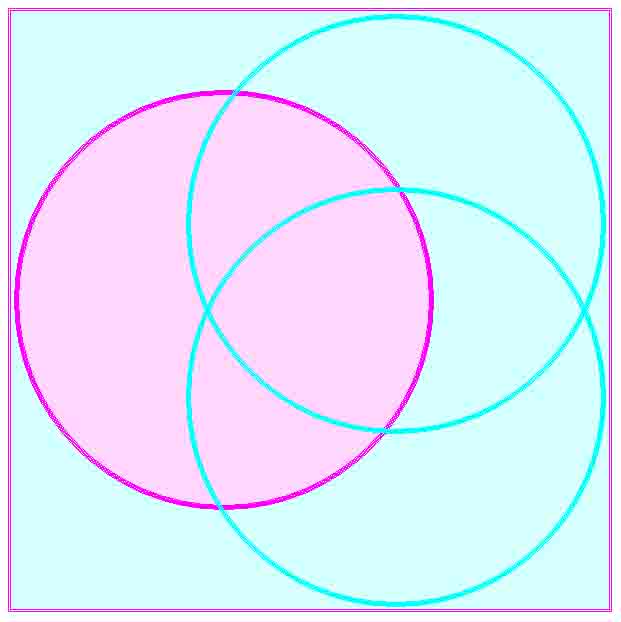}

\vspace{-165pt}

\hspace{-181pt} {\Large$\Omega$}

\vspace{-16pt}

\hspace{-115pt} {\scriptsize$ x^c \!\cap  s'\!'^c \!\cap t'\!'^c$}

\vspace{-6pt}

\hspace{133pt} {\Large${\mbox{\boldmath$y$}}$}

\vspace{95pt}

\hspace{-155pt} {\Large${\mbox{\boldmath$x$}}$}

\vspace{15pt}

\hspace{143pt} {\Large${\mbox{\boldmath$z$}}$}

\vspace{-130pt}

\hspace{75pt} {\scriptsize$s'\!' \!\cap t'\!'^c$}

\vspace{2pt}

\hspace{-19pt} {\scriptsize$s' \!\cap t'^c$}

\vspace{28pt}

\hspace{-13pt} {\scriptsize$x  \!\cap s'^c \!\cap t'^c$} \hspace{21pt} {\scriptsize$s' \!\cap t'$} \hspace{17pt} {\scriptsize$s'\!' \!\cap t'\!'$}

\vspace{24pt}

\hspace{-19pt} {\scriptsize$s'^c \!\cap t'$}

\vspace{2pt}

\hspace{80pt} {\scriptsize$ s'\!'^c \!\cap t'\!'$}

\vspace{29pt}

\includegraphics[width=3.3in]{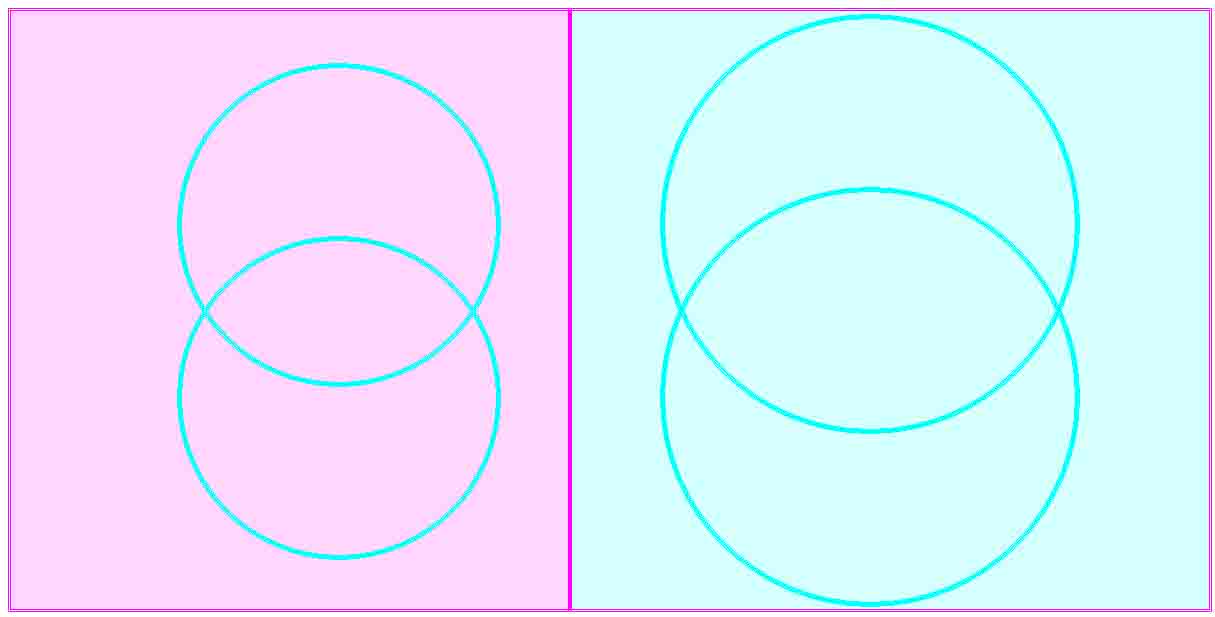}

\vspace{-134pt}

\hspace{-225pt}  {\Large$\Omega$}

\vspace{-1pt}

\hspace{70pt}  {\scriptsize$x \!\cap\! s'^c \!\!\cap\! t'^c$} \hspace{80pt}
{\scriptsize$x^c \!\!\cap\! s'\!'^c \!\!\cap\! t'\!'^c$}

\vspace{10pt}

{\scriptsize$s' \!\cap t'^c$} \hspace{75pt}
{\scriptsize$s'\!' \!\cap t'\!'^c$}

\vspace{18pt}

{\scriptsize$s' \!\cap t'$} \hspace{75pt}
{\scriptsize$s'\!' \!\cap t'\!'$}

\vspace{20pt}

{\scriptsize$s'^c \!\cap t'$} \hspace{75pt}
{\scriptsize$s'\!'^c \!\cap t'\!'$}

\vspace{-92pt}

\hspace{-45pt} ${\Large {\mbox{\boldmath$s'$}}}$ \hspace{70pt} ${\Large {\mbox{\boldmath$s'\!'$}}}$

\vspace{87pt}

\hspace{55pt} ${\Large {\mbox{\boldmath$t'$}}}$ \hspace{102pt} ${\Large {\mbox{\boldmath$t'\!'$}}}$

\vspace{-5pt}

$\underbrace{\hspace{109pt}}_{\Large {\mbox{\boldmath$x$}}}
\underbrace{\hspace{124pt}}_{\Large {\mbox{\boldmath$x^c$}}}$

\vspace{-10pt}

\caption{Venn diagrams of the frame half-rare trip0let of events $\frak{X}=\{x,y,z\}, \ 1/2 \geqslant p_x \geqslant p_y \geqslant p_z$ (up), and two inserted doublets $\mathcal{X}'=\{s',t'\}$ and $\mathcal{X}'\!'=\{s'\!',t'\!'\}$ (down), agreed with the frame triplet $\frak{X}$ in the following sense:\! $y\!=\!s'\!+s'\!'\!\!, z\!=\!t'\!+t'\!'\!$ и $s' \!\cup t' \subseteq x, s'\!' \!\cup t'\!' \subseteq x^c$.\label{fig-Venn3-2-zyx}} %
\end{figure}

In order to construct by the frame method the $\breve{p}$-\uwave{ordered} \emph{frame half-rare triplet of events} $\frak{X}=\{x,y,z\}$ with the $\frak{X}$-set of marginal probabilities $\breve{p}=\{p_x,p_y,p_z\}$, where
\begin{equation}\label{3order}
\begin{split}
1/2 \geqslant p_x \geqslant p_y \geqslant p_z \geqslant 0,
\end{split}
\end{equation}
let's suppose that
\begin{equation}\label{3order}
\begin{split}
\frak{X}=\{x\}+\{y,z\} = \{x\}+ (\mathcal{X}' (+) \mathcal{X}'\!')
\end{split}
\end{equation}
and in our disposal we have two \emph{inserted half-rare doublets of events}
$$
\mathcal{X}'=\{s',t'\} \ \mbox{и} \ \mathcal{X}'\!'=\{s'\!',t'\!'\},
$$
with the known  2-Kopulas (see Fig. \ref{fig-Venn3-2-zyx}) which by the definition satisfy the following inclusions:
\begin{equation}\label{3-coherence-xyz}
\hspace*{-3pt}
\begin{split}
&s' = x \cap y \subseteq x, \ \ \ \ t' = x \cap z \subseteq x, \ \ \ s' \cup t' \subseteq x,\\
&s'\!' = x^c \cap y \subseteq x^c, \ t'\!' = x^c \cap z \subseteq x^c\!, \  s'\!' \cup t'\!' \subseteq x^c.
\end{split}\hspace*{-13pt}
\end{equation}

In view of the assumptions made (see Fig. \ref{fig-Venn3-2-zyx})
\begin{equation}\label{triplet-6}
\begin{split}
\textsf{ter}(xyz /\!\!/ \frak{X}) &\!=\!x \!\cap y \cap z\!=\!s' \cap t' \!=\! \textsf{ter}(s't' /\!\!/ \mathcal{X}'),\\
\textsf{ter}(xy /\!\!/ \frak{X}) &\!=\!x \!\cap y \cap z^c\!=\!s' \cap t'^c\!=\! \textsf{ter}(s' /\!\!/ \mathcal{X}'),\\
\textsf{ter}(xz /\!\!/ \frak{X}) &\!=\!x \!\cap y^c \cap z\!=\!s'^c \cap t' \!=\! \textsf{ter}(t' /\!\!/ \mathcal{X}'),\\
\textsf{ter}(yz /\!\!/ \frak{X}) &\!=\!x^c \!\cap y \cap z\!=\!s'\!' \cap t'\!' \!=\! \textsf{ter}(s'\!'t'\!' /\!\!/ \mathcal{X}'\!'),\hspace{-50pt}\\
\textsf{ter}(y /\!\!/ \frak{X}) &\!=\!x^c \!\cap y \cap z^c\!=\!s'\!' \cap t'\!'^c \!=\! \textsf{ter}(s'\!' /\!\!/ \mathcal{X}'\!'),\hspace{-50pt}\\
\textsf{ter}(z /\!\!/ \frak{X}) &\!=\!x^c \!\cap y^c \cap z\!=\!s'\!'^c \cap t'\!' \!=\! \textsf{ter}(t'\!' /\!\!/ \mathcal{X}'\!'),\hspace{-50pt}
\end{split}
\end{equation}
these 6 terraced events are defined. All of them are generated by the \emph{frame half-rare triplet} $\frak{X}$, with the exception of two terraced events
\begin{equation}\label{triplet-2}
\begin{split}
\textsf{ter}(x /\!\!/ \frak{X}) &= x \cap y^c \cap z^c,\\
\textsf{ter}(\emptyset /\!\!/ \frak{X}) &= x^c \cap y^c \cap z^c,
\end{split}
\end{equation}
that are defined by the formulas:
\begin{equation}\label{doublet-2}
\begin{split}
\textsf{ter}(x /\!\!/ \frak{X}) = x - s' \cup t',\\
\textsf{ter}(\emptyset /\!\!/ \frak{X}) = x^c-s'\!' \cup t'\!',
\end{split}
\end{equation}
or, equivalently,
\begin{equation}\label{doublet-2a}
\begin{split}
\textsf{ter}(x /\!\!/ \frak{X}) = \textsf{ter}(\emptyset/\!\!/\mathcal{X}') - x^c,\\
\textsf{ter}(\emptyset /\!\!/ \frak{X}) = \textsf{ter}(\emptyset/\!\!/\mathcal{X}'\!') - x.
\end{split}
\end{equation}

In view of this, we obtain the formulas:
\begin{equation}\label{epd-triplet-8}
\begin{split}
p(xyz /\!\!/ \frak{X}) &\!=\!p(s't' /\!\!/ \mathcal{X}'),\\
p(xy /\!\!/ \frak{X}) &\!=\! p(s' /\!\!/ \mathcal{X}'),\\
p(xz /\!\!/ \frak{X}) &\!=\!p(t' /\!\!/ \mathcal{X}'),\\
p(x /\!\!/ \frak{X}) &\!=\!p(\emptyset /\!\!/ \mathcal{X}')-1+p_x,\\
&\\[-8pt]
p(yz /\!\!/ \frak{X}) &\!=\!p(s'\!'t'\!' /\!\!/ \mathcal{X}'\!'),\\
p(y /\!\!/ \frak{X}) &\!=\!p(s'\!' /\!\!/ \mathcal{X}'\!'),\\
p(z /\!\!/ \frak{X}) &\!=\!p(t'\!' /\!\!/ \mathcal{X}'\!'),\\
p(\emptyset /\!\!/ \frak{X}) &\!=\!p(\emptyset /\!\!/ \mathcal{X}'\!')-p_x,
\end{split}
\end{equation}
that express the e.p.d. of the 1st kind of \emph{frame half-rare triplet of events} $\frak{X}$ through the e.p.d. of the 1st kind of \emph{inserted half-rare doublets} $\mathcal{X}'$ and $\mathcal{X}'\!'$, and the probability of \emph{frame event} $x$.

In the language of e.p.d. of the 1st kind assumptions (\ref{3-coherence-xyz}) mean that
\begin{equation}\label{epd-triplet-y-z}
\begin{split}
p(s' /\!\!/ \mathcal{X}')&+p(s't' /\!\!/ \mathcal{X}')+\\
&+p(s'\!' /\!\!/ \mathcal{X}')+p(s'\!'t'\!' /\!\!/ \mathcal{X}')=p_y,\\
p(t' /\!\!/ \mathcal{X}')&+p(s't' /\!\!/ \mathcal{X}')+\\
&+p(t'\!' /\!\!/ \mathcal{X}')+p(s'\!'t'\!' /\!\!/ \mathcal{X}')=p_z,
\end{split}
\end{equation}
or the same in the language of probabilities events:
\begin{equation}\label{pxyz-triplet-y-z}
\begin{split}
p_{s'}+p_{s'\!'}=p_y,\\
p_{t'}+p_{t'\!'}=p_z,
\end{split}
\end{equation}
In addition, the third pair of inclusions under the assumptions (\ref{3-coherence-xyz}) means that
\begin{equation}\label{pxyz-triplet-cups}
\begin{split}
p_{s'}+p_{t'}-p_{s't'} &\leqslant p_x,\\
p_{s'\!'}+p_{t'\!'}-p_{s'\!'t'\!'} &\leqslant 1-p_x,
\end{split}
\end{equation}
where
\begin{equation}\label{pxyz-triplet-caps}
\begin{split}
p_{s't'} &= p(s't' /\!\!/ \mathcal{X}') = \mathbf{P}(s' \cap t'),\\
p_{s'\!'t'\!'} &= p(s'\!'t'\!' /\!\!/ \mathcal{X}'\!') = \mathbf{P}(s'\!' \cap t'\!')
\end{split}
\end{equation}
are probabilities of double intersections of events from the inserted doublets $\mathcal{X}'$ and $\mathcal{X}'\!'$.

Taking into account the Fr\'echet inequalities the restrictions (\ref{pxyz-triplet-y-z}) and (\ref{pxyz-triplet-cups}) are equivalent to the following inequalities for 4 parameters $p_{s'}, p_{t'}, p_{s't'}$ и $p_{s'\!'t'\!'}$ of inserted doublets $\mathcal{X}'$ and $\mathcal{X}'\!'$:
\begin{equation}\label{FRECHET-restraints-triplet}
\begin{split}
0 \leqslant & \ p_{s'} \leqslant p_y,\\
0 \leqslant & \ p_{t'} \leqslant p_z,\\
p_{s't'}^- \leqslant & \ p_{s't'} \leqslant p_{s't'}^+,\\
p_{s'\!'t'\!'}^- \leqslant & \ p_{s'\!'t'\!'} \leqslant p_{s'\!'t'\!'}^+,
\end{split}
\end{equation}
where
\begin{equation}\label{FRECHET-bounds-xyz}
\begin{split}
p_{s't'}^- &=\max \{0,p_{s'}+p_{t'}-p_x\},\\
p_{s't'}^+ &=\min \{p_{s'},p_{t'}\},\\
p_{s'\!'t'\!'}^-&=\max \{0,p_x+p_y+p_z-1-p_{s'}-p_{t'}\},\\
p_{s'\!'t'\!'}^+&=\min\{p_y-p_{s'},p_z-p_{t'}\},
\end{split}
\end{equation}
are the lower and upper Frechet-boundaries of probabilities of double intersections of inserted doublets  $\mathcal{X}'$ and $\mathcal{X}'\!'$ with respect to the frame monoplet $\{x\}$.

Let's write the formulas (\ref{epd-triplet-8}), using only these 4 parameters and remembering the restrictions (\ref{FRECHET-restraints-triplet}):
\begin{equation}\label{epd-triplet-8-4}
\hspace*{-3pt}
\begin{split}
p(xyz /\!\!/ \frak{X}) &\!=\!p_{s't'},\\
p(xy /\!\!/ \frak{X}) &\!=\! p_{s'}-p_{s't'},\\
p(xz /\!\!/ \frak{X}) &\!=\!p_{t'}-p_{s't'},\\
p(x /\!\!/ \frak{X}) &\!=\!p_x-p_{s'}-p_{t'}+p_{s't'},\\
&\\[-8pt]
p(yz /\!\!/ \frak{X}) &\!=\!p_{s'\!'t'\!'},\\
p(y /\!\!/ \frak{X}) &\!=\!p_y-p_{s'}-p_{s'\!'t'\!'},\\
p(z /\!\!/ \frak{X}) &\!=\!p_z-p_{t'}-p_{s'\!'t'\!'},\\
p(\emptyset /\!\!/ \frak{X}) &\!=\!1-p_x-p_y-p_z+p_{s'}+p_{t'}+p_{s'\!'t'\!'}.
\end{split}\hspace*{-13pt}
\end{equation}

\subsection{The frame method: recurrent formulas for a half-rare triplet of events\label{3-framing-method-recurrent-formulas}}

The formulas (\ref{epd-triplet-8-4}) as well as the formulas (\ref{epd-triplet-8}) can be written in the form of special cases of recurrence formulas (\ref{frame-recurrent-pseudo}) and (\ref{frame-recurrent-conditional}) from Note \ref{not-frame-recurrent-formulas} for the triplet $\frak{X}=\{x,y,z\}=\{x\}+\{y,z\}=\{x\}+\mathcal{X}$:
\begin{equation}\label{frame-recurrent-pseudo-triplet}
\hspace*{-13pt}
\begin{split}
&{\mbox{\boldmath$\mathscr{K}$}}\!\left(\breve{p}^{\left( c \mid X+Y /\!\!/ \mathcal{X}+\{x\}\right)}\!\right)=\\
&=\begin{cases}
\mathscr{K}^{(\{x\})}\!\!\left(\breve{p}^{\left( c \mid X^{(\cap \{x\} /\!\!/ \{x\})} /\!\!/ \mathcal{X}'\right)}\!\right)\!,& \hspace*{-8pt} Y=\{x\},\\
&\\[-5pt]
\mathscr{K}^{(\emptyset)}\!\!\left(\breve{p}^{\left( c \mid X^{(\cap \emptyset /\!\!/ \{x\})} /\!\!/ \mathcal{X}'\!'\right)}\!\right)\!,& \hspace*{-8pt} Y=\emptyset,
\end{cases}\\
\end{split}\hspace*{-13pt}
\end{equation}
\begin{equation}\label{frame-recurrent-conditional-triplet}
\hspace*{-13pt}
\begin{split}
&{\mbox{\boldmath$\mathscr{K}$}}\!\left(\breve{p}^{\left( c \mid X+Y /\!\!/ \mathcal{X}+\{x\}\right)}\!\right)=\\
&=\begin{cases}
{\mbox{\boldmath$\mathscr{K}$}}^{|\{x\}}\!\Big(\breve{p}^{\left( c \mid X /\!\!/ \mathcal{X} \mid \{x\} /\!\!/ \{x\}\right)}\!\Big)p_x, & \hspace*{-8pt} Y=\{x\},\\
&\\[-5pt]
{\mbox{\boldmath$\mathscr{K}$}}^{|\emptyset}\!\Big(\breve{p}^{\left( c \mid X /\!\!/ \mathcal{X} \mid \emptyset /\!\!/ \{x\}\right)}\!\Big)(1-p_x), & \hspace*{-8pt} Y=\emptyset.
\end{cases}
\end{split}\hspace*{-13pt}
\end{equation}
In the formulas (\ref{frame-recurrent-pseudo-triplet}) the inserted pseudo-Kopulas $\mathscr{K}^{(\{x\})}$ and $\mathscr{K}^{(\emptyset)}$ are defined by the first and the second four probabilities from (\ref{epd-triplet-8-4}) correspondingly, i.e., by the formulas:
\begin{equation}\label{epd-triplet-8-x-pseudoKopula}
\hspace*{-3pt}
\begin{split}
&\mathscr{K}^{(\{x\})}\!\!\left(\breve{p}^{\left( c \mid X^{(\cap \{x\} /\!\!/ \{x\})} /\!\!/ \mathcal{X}'\right)}\!\right)\!=\\
&=\begin{cases}
p_{s't'}, & X=\{y,z\},\\
p_{s'}-p_{s't'}, & X=\{y\},\\
p_{t'}-p_{s't'}, & X=\{z\},\\
p_x-p_{s'}-p_{t'}+p_{s't'}, & X=\emptyset,\\
\end{cases}
\end{split}\hspace*{-13pt}
\end{equation}
\begin{equation}\label{epd-triplet-8-0-pseudoKopula}
\hspace*{-3pt}
\begin{split}
&\mathscr{K}^{(\emptyset)}\!\!\left(\breve{p}^{\left( c \mid X^{(\cap \emptyset /\!\!/ \{x\})} /\!\!/ \mathcal{X}'\!'\right)}\!\right)\!=\\
&=\begin{cases}
p_{s'\!'t'\!'}, & X=\{y,z\},\\
p_y-p_{s'}-p_{s'\!'t'\!'}, & X=\{y\},\\
p_z-p_{t'}-p_{s'\!'t'\!'}, & X=\{z\},\\
1-p_x-p_y-p_z+&\\
+p_{s'}+p_{t'}+p_{s'\!'t'\!'}, & X=\emptyset,
\end{cases}
\end{split}\hspace*{-13pt}
\end{equation}
where, for example, for $X=\{y,z\}$
\begin{equation}\label{explanations1}
\hspace*{-13pt}
\begin{split}
&\{y,z\}^{(\cap \{x\} /\!\!/ \{x\})}\!=\!\{y \cap x, z \cap x\} \subseteq \mathcal{X}'\!=\!\{s',t'\},\\
&\{y,z\}^{(\cap \emptyset /\!\!/ \{x\})}\!=\!\{y \cap x^c, z \cap x^c\} \subseteq \mathcal{X}'\!'\!=\!\{s'\!',t'\!'\},
\end{split}\hspace*{-13pt}
\end{equation}
and the corresponding sets of marginal probabilities of inserted doublets $\mathcal{X}'$ и $\mathcal{X}'\!'$ have the form
\begin{equation}\label{explanations2}
\hspace*{-13pt}
\begin{split}
\breve{p}^{\left( c \mid \{y,z\}^{(\cap \{x\} /\!\!/ \{x\})} /\!\!/ \mathcal{X}'\!'\right)}\!&=\!\{p_{s'},p_{t'}\},\\
\breve{p}^{\left( c \mid \{y,z\}^{(\cap \emptyset /\!\!/ \{x\})} /\!\!/ \mathcal{X}'\!'\right)}\!&=\!\{p_y-p_{s'},p_z-p_{t'}\}.
\end{split}\hspace*{-13pt}
\end{equation}
In the formulas (\ref{frame-recurrent-conditional-triplet}) the conditional Kopulas
${\mbox{\boldmath$\mathscr{K}$}}^{|\{x\}}$ and ${\mbox{\boldmath$\mathscr{K}$}}^{|\emptyset}$
are defined by the first and the second four probabilities from (\ref{epd-triplet-8-4}), normalized by $p_x$ and by $1-p_x$ correspondingly, i.e., by the formulas:
\begin{equation}\label{epd-triplet-8-x-conditionalKopula}
\hspace*{-3pt}
\begin{split}
&{\mbox{\boldmath$\mathscr{K}$}}^{|\{x\}}\!\Big(\breve{p}^{\left( c \mid X /\!\!/ \mathcal{X} \mid \{x\} /\!\!/ \{x\}\right)}\!\Big)\!=\\
&=\begin{cases}
\frac{1}{p_x}p_{s't'}, & X=\{y,z\},\\
\frac{1}{p_x}(p_{s'}-p_{s't'}), & X=\{y\},\\
\frac{1}{p_x}(p_{t'}-p_{s't'}), & X=\{z\},\\
1-\frac{1}{p_x}(p_{s'}+p_{t'}-p_{s't'}), & X=\emptyset,\\
\end{cases}
\end{split}\hspace*{-13pt}
\end{equation}
\begin{equation}\label{epd-triplet-8-0-conditionalKopula}
\hspace*{-3pt}
\begin{split}
&{\mbox{\boldmath$\mathscr{K}$}}^{|\emptyset}\!\Big(\breve{p}^{\left( c \mid X /\!\!/ \mathcal{X} \mid \emptyset /\!\!/ \{x\}\right)}\!\Big)\!=\\
&=\begin{cases}
\frac{1}{1-p_x}p_{s'\!'t'\!'}, & X=\{y,z\},\\
\frac{1}{1-p_x}(p_y-p_{s'}-p_{s'\!'t'\!'}), & X=\{y\},\\
\frac{1}{1-p_x}(p_z-p_{t'}-p_{s'\!'t'\!'}), & X=\{z\},\\
\frac{1}{1-p_x}(1-p_x-p_y-p_z)+&\\
+\frac{1}{1-p_x}(p_{s'}+p_{t'}+p_{s'\!'t'\!'}), & X=\emptyset.
\end{cases}
\end{split}\hspace*{-13pt}
\end{equation}
The corresponding sets of marginal conditional probabilities of events $y, z \in \mathcal{X}$ with respect to the frame terraced  events $\textsf{ter}(\{x\} /\!\!/ \{x\})=x$ и $\textsf{ter}(\emptyset /\!\!/ \{x\})=x^c$ correspondingly have the from:
\begin{equation}\label{conditional-marginals}
\hspace*{-13pt}
\begin{split}
\breve{p}^{\left( c \mid \mathcal{X} /\!\!/ \mathcal{X} \mid \{x\} /\!\!/ \{x\}\right)}\!&=\!\left\{\frac{p_{s'}}{p_x},\frac{p_{t'}}{p_x}\right\},\\
\breve{p}^{\left( c \mid \mathcal{X} /\!\!/ \mathcal{X} \mid \emptyset /\!\!/ \{x\}\right)}\!&=\!\left\{\frac{p_y-p_{s'}}{1-p_x},\frac{p_z-p_{t'}}{1-p_x}\right\}.
\end{split}\hspace*{-13pt}
\end{equation}
Remind, that the four functional parameters $p_{s'},p_{t'},p_{s't'}$ and $p_{s'\!'t'\!'}$ in the recurrent formulas (\ref{frame-recurrent-pseudo-triplet}), (\ref{frame-recurrent-conditional-triplet}), and also in the formulas for pseudo-Kopulas (\ref{epd-triplet-8-x-pseudoKopula}), (\ref{epd-triplet-8-0-pseudoKopula}), and the conditional Kopulas (\ref{epd-triplet-8-x-conditionalKopula}), (\ref{epd-triplet-8-0-conditionalKopula}), obey the Frechet-constraints (\ref{FRECHET-restraints-triplet}).

\section{The Kopula theory for quadruplets of events\label{4-method}}

\subsection{The frame method for constructing a half-rare quadruplet of events\label{4-framing-method}}

\begin{figure}[ht!]
\centering
\includegraphics[width=3.3in]{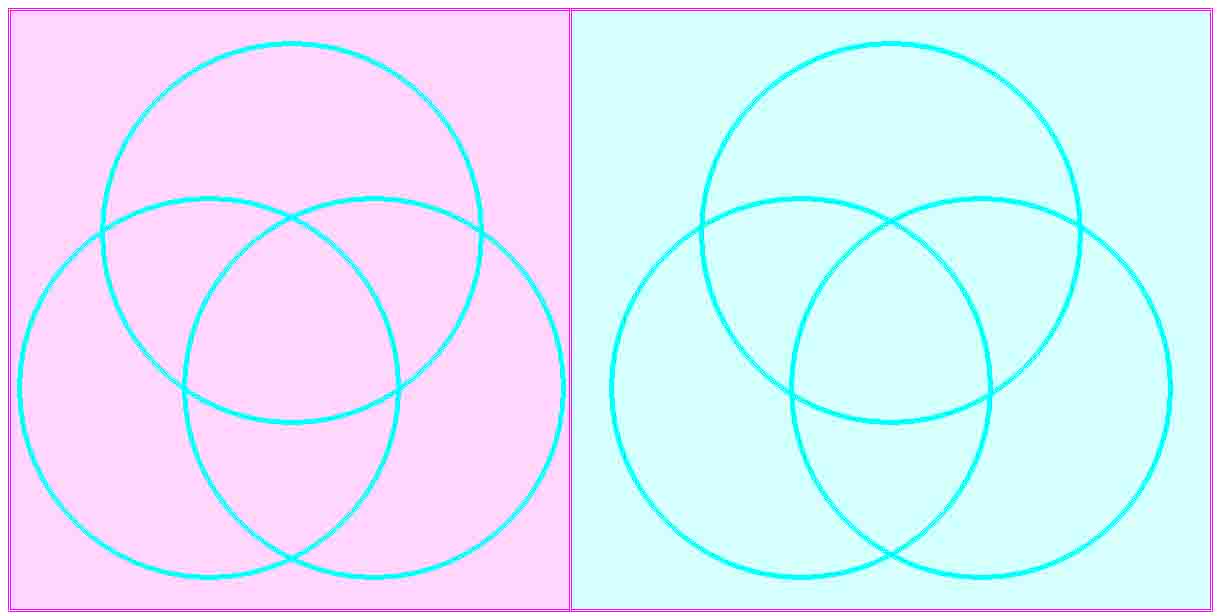}

\vspace{-134pt}

\hspace{-225pt}  {\Large$\Omega$}

\vspace{-4pt}

\hspace{-10pt}
{\tiny $\textsf{ter}(x \!/\!\!/\! \frak{X}\!)$}
\hspace{85pt}
{\tiny $\textsf{ter}(\emptyset \!/\!\!/\! \frak{X}\!)$}

\vspace{13pt}

\hspace{-12pt} {\tiny $\textsf{ter}(xy \!/\!\!/\! \frak{X}\!)$}
\hspace{84pt}
{\tiny $\textsf{ter}(y \!/\!\!/\! \frak{X}\!)$}

\vspace{3pt}

\hspace{-12pt}
{\tiny $\textsf{ter}(xyz \!/\!\!/\! \frak{X}\!)$}
\hspace{5pt}
{\tiny $\textsf{ter}(xyv \!/\!\!/\! \frak{X}\!)$}
\hspace{39pt}
{\tiny $\textsf{ter}(yz \!/\!\!/\! \frak{X}\!)$}
\hspace{8pt}
{\tiny $\textsf{ter}(yv \!/\!\!/\! \frak{X}\!)$}

\vspace{7pt}

\hspace{-12pt}
{\tiny $\textsf{ter}(xyzv \!/\!\!/\! \frak{X}\!)$}
\hspace{78pt}
{\tiny $\textsf{ter}(yzv \!/\!\!/\! \frak{X}\!)$}

\vspace{10pt}

\hspace{-10pt}
{\tiny $\textsf{ter}(xz \!/\!\!/\! \frak{X}\!)$}
\hspace{0pt}
{\tiny $\textsf{ter}(xzv \!/\!\!/\! \frak{X}\!)$}
\hspace{0pt}
{\tiny $\textsf{ter}(xv \!/\!\!/\! \frak{X}\!)$}
\hspace{14pt}
{\tiny $\textsf{ter}(z \!/\!\!/\! \frak{X}\!)$}
\hspace{5pt}
{\tiny $\textsf{ter}(zv \!/\!\!/\! \frak{X}\!)$}
\hspace{5pt}
{\tiny $\textsf{ter}(v \!/\!\!/\! \frak{X}\!)$}

\vspace{-77pt}

\hspace{-87pt} ${\small {\mbox{\boldmath$y\!\cap\! x$}}}$ \hspace{90pt} ${\small {\mbox{\boldmath$y\!\cap\! x^c$}}}$

\vspace{83pt}

${\small {\mbox{\boldmath$z\!\cap\! x$}}}$ \hspace{63pt} ${\small {\mbox{\boldmath$v\!\cap\! x$}}}$ \hspace{-6pt} ${\small {\mbox{\boldmath$z\!\cap\! x^c$}}}$ \hspace{70pt} ${\small {\mbox{\boldmath$v\!\cap\! x^c$}}}$

\vspace{-9pt}

$\underbrace{\hspace{109pt}}_{\Large {\mbox{\boldmath$x$}}}
\underbrace{\hspace{124pt}}_{\Large {\mbox{\boldmath$x^c$}}}$

\vspace{5pt}

\includegraphics[width=3.3in]{xyzvRoseBlueSt.jpg}

\vspace{-134pt}

\hspace{-225pt}  {\Large$\Omega$}

\vspace{-3pt}

\hspace{-5pt}
{\tiny $s\!'^{\!c}\!\!\cap\!\!t\!'^{\!c}\!\!\cap\!\!u\!'^{\!c} - x^c$}
\hspace{70pt}
{\tiny $s'\!'^{\!c}\!\!\cap\!\!t'\!'^{\!c}\!\!\cap\!\!u'\!'^{\!c} - x$}

\vspace{12pt}

\hspace{-12pt}
{\tiny $s'\!\!\cap\!\!t'^{\!c}\!\!\cap\!\!u'^{\!c}$}
\hspace{84pt}
{\tiny $s'\!'\!\!\cap\!\!t'\!'^{\!c}\!\!\cap\!\!u'\!'^{\!c}$}

\vspace{3pt}

\hspace{-5pt}
{\tiny $s'\!\!\cap\!\!t'\!\!\cap\!\!u'^{\!c}$}
\hspace{5pt}
{\tiny $s'\!\!\cap\!\!t'^{\!c}\!\!\cap\!\!u'$}
\hspace{43pt}
{\tiny $s'\!'\!\!\cap\!\!t'\!'\!\!\cap\!\!u'\!'^{\!c}$}
\hspace{2pt}
{\tiny $s'\!'\!\!\cap\!\!t'\!'^{\!c}\!\!\cap\!\!u'\!'$}

\vspace{7pt}

\hspace{-8pt}
{\tiny $s'\!\!\!\cap\!t'\!\!\!\cap\!\!u'$}
\hspace{87pt}
{\tiny $s'\!'\!\!\!\cap\!\!t'\!'\!\!\!\cap\!\!u'\!'$}

\vspace{12pt}

\hspace{-6pt}
{\tiny $s'^{\!c}\!\!\cap\!\!t'\!\!\cap\!\!u'^{\!c}$}
\hspace{1pt}
{\tiny $s'^{\!c}\!\!\cap\!\!t'\!\!\cap\!\!u'$}
\hspace{1pt}
{\tiny $s'^{\!c}\!\!\cap\!\!t'^{\!c}\!\!\cap\!\!u'$}
\hspace{9pt}
{\tiny $s'\!'^{\!c}\!\!\cap\!\!t'\!'\!\!\cap\!\!u'\!'^{\!c}$}
\hspace{-3pt}
{\tiny $s'\!'^{\!c}\!\!\cap\!\!t'\!'\!\!\cap\!\!u'\!'$}
\hspace{-3pt}
{\tiny $s'\!'^{\!c}\!\!\cap\!\!t'\!'^{\!c}\!\!\cap\!\!u'\!'$}

\vspace{-79pt}

\hspace{-75pt} ${\small {\mbox{\boldmath$s'$}}}$
\hspace{103pt} ${\small {\mbox{\boldmath$s'\!'$}}}$

\vspace{83pt}

\hspace{1pt}
${\small {\mbox{\boldmath$t'$}}}$
\hspace{80pt}
${\small {\mbox{\boldmath$u'$}}}$
\hspace{-1pt}
${\small {\mbox{\boldmath$t'\!'$}}}$
\hspace{96pt}
${\small {\mbox{\boldmath$u'\!'$}}}$

\vspace{-9pt}

$\underbrace{\hspace{109pt}}_{\Large {\mbox{\boldmath$x$}}}
\underbrace{\hspace{124pt}}_{\Large {\mbox{\boldmath$x^c$}}}$

\vspace{-10pt}

\caption{Venn diagrams of the frame half-rare quadruplet of events $\frak{X}=\{x,y,z,v\}, \ 1/2 \geqslant p_x \geqslant p_y \geqslant p_z \geqslant p_v$ (up), and two inserted triplets $\mathcal{X}'=\{s',t',u'\}$ and $\mathcal{X}'\!'=\{s'\!',t'\!',u'\!'\}$ (down), agreed with the frame quadruplet $\frak{X}$ in the following sense:\! $y\!=\!s'\!+s'\!'\!\!, z\!=\!t'\!+t'\!'\!, v\!=\!u'\!+u'\!'\!$ и $s' \!\cup t' \!\cup u' \subseteq x, s'\!' \!\cup t'\!' \!\cup u'\!' \subseteq x^c$.\label{fig-Venn4-3}} %
\end{figure}

In order by the frame method to construct the $\breve{p}$-\uwave{ordered} \emph{frame half-rare quadruplet of events} $\frak{X}=\{x,y,z,v\}$ with the $\frak{X}$-set of marginal probabilities $\breve{p}=\{p_x,p_y,p_z,p_v\}$, where
\begin{equation}\label{4order}
\begin{split}
1/2 \geqslant p_x \geqslant p_y \geqslant p_z \geqslant p_v \geqslant 0,
\end{split}
\end{equation}
let's suppose that
\begin{equation}\label{4order}
\begin{split}
\frak{X}=\{x\}+\{y,z,v\} = \{x\}+ (\mathcal{X}' (+) \mathcal{X}'\!')
\end{split}
\end{equation}
and we have two \emph{inserted half-rare triplets of events}
$$
\mathcal{X}'=\{s',t',u'\} \ \mbox{и} \ \mathcal{X}'\!'=\{s'\!',t'\!',u'\!'\},
$$
with the known 3-Kopulas, which by definition satisfy the following inclusions (see Fig. \ref{fig-Venn4-3}):
\begin{equation}\label{4-coherence-xyz}
\hspace*{-10pt}
\begin{split}
&s' = x \cap y \subseteq x, \ t' = x \cap z \subseteq x, \ u' = x \cap v \subseteq x,\\
&s' \cup t' \cup u' \subseteq x,\\
&\\[-8pt]
&s'\!' = x^c \cap y \subseteq x^c, \ t'\!' = x^c \cap z \subseteq x^c, \ u'\!' = x^c \cap v \subseteq x^c,\\
&s'\!' \cup t'\!' \cup u'\!' \subseteq x^c.
\end{split}\hspace*{-28pt}
\end{equation}

The recurrent formulas, which express the e.p.d. of the 1st kind of $\breve{p}$-ordered half-rare quadruplet of events $\frak{X}$ through the e.p.d. of the 1st kind of two inserted triplets $\mathcal{X}'$ and $\mathcal{X}'\!'$, follow from the general recurrent formulas (\ref{frame-recurrent-pseudo}) and (\ref{frame-recurrent-conditional}) in Note \ref{not-frame-recurrent-formulas} as well as in cases of a doublet and a triplet of events.
And therefore, and because of the cumbersomeness, these formulas are not represented here, but are only illustrated by Venn diagrams  (see Fig. \ref{fig-Venn4-3}).

\subsection{Recurrent Fr\'echet-restrictions in the frame method\label{FRECHET-reсurrent}}

Let us dwell in more detail on Fr\'echet-restrictions for $11=2^4-4-1$ functional parameters of a Kopula of quadruplet of events, to derive the recurrent sequence of such Fr\'echet-restrictions, which begins with Fr\'echet-restrictions for a doublet of events (\ref{FRECHET-bounds-2xy}), continues with Fr\'echet-restrictions for a triplet of events (\ref{FRECHET-restraints-triplet}), and should be supported by Fr\'echet-restrictions for parameters of a Kopula of quadruplet of events $\frak{X}=\{x,y,z,v\}$ and so on.

To this end, we first recall Fr\'echet-restrictions for parameters of Kopulas of a doublet and a triplet of events.

\subsubsection{Fr\'echet-restrictions for a doublet of events}

For a Kopula of doublet of events, the Fr\'echet-restrictions of a $1=2^2-2-1$ parameter of inserted monoplets $\mathcal{X}'$ and $\mathcal{X}'\!'$ have the form:
\begin{equation}\label{FRECHET-bounds-2UpDown}
\begin{split}
0  \leqslant p_{s'} \leqslant p_y.
\end{split}
\end{equation}

\subsubsection{Fr\'echet-restrictions for a triplet events}

For a Kopula of doublet of events, the Fr\'echet-restrictions of $4=2^3-3-1$ parameters of inserted doublets $\mathcal{X}'$ and $\mathcal{X}'\!'$ have the form:
\begin{equation}\label{FRECHET-restraints-triplet-bis}
\begin{split}
0 \leqslant & \ p_{s'} \leqslant p_y,\\
0 \leqslant & \ p_{t'} \leqslant p_z,\\
p_{s't'}^- \leqslant & \ p_{s't'} \leqslant p_{s't'}^+,\\
p_{s'\!'t'\!'}^- \leqslant & \ p_{s'\!'t'\!'} \leqslant p_{s'\!'t'\!'}^+,
\end{split}
\end{equation}
where
\begin{equation}\label{FRECHET-bounds-3UpDown}
\begin{split}
p_{s't'}^- &=\max \{0,p_{s'}+p_{t'}-p_x\},\\
p_{s't'}^+ &=\min \{p_{s'},p_{t'}\},\\
p_{s'\!'t'\!'}^-&=\max \{0,p_x+p_y+p_z-1-p_{s'}-p_{t'}\},\\
p_{s'\!'t'\!'}^+&=\min\{p_y-p_{s'},p_z-p_{t'}\},
\end{split}
\end{equation}
are the lower and upper Fr\'echet-boundaries probabilities of double intersections of events from inserted doublets  $\mathcal{X}'$ and $\mathcal{X}'\!'$ with respect to the frame monoplet $\{x\}$.

The case of triplet of events gives a new level of Fr\'echet-restrictions (the two last Fr\'echet-boundaries in (\ref{FRECHET-bounds-3UpDown})), when probabilities of double intersections of events from inserted doublets have Fr\'echet-boundaries that depend not only on marginal probabilities of the triplet, but and on inserted marginal probabilities on which, in turn, the usual Fr\'echet-restrictions mentioned above are imposed.

\subsubsection{Fr\'echet-restrictions for a quadruplet of events}

For a Kopula of quadruplet of events the Fr\'echet-restrictions of $11=2^4-4-1$ parameters of the inserted triplet $\mathcal{X}'$ and $\mathcal{X}'\!'$ have the form:
\begin{equation}\label{FRECHET-restraints-quadruplet}
\begin{split}
0 \leqslant & \ p_{s'} \leqslant p_y,\\
0 \leqslant & \ p_{t'} \leqslant p_z,\\
0 \leqslant & \ p_{u'} \leqslant p_v,\\
p_{s't'}^- \leqslant & \ p_{s't'} \leqslant p_{s't'}^+,\\
p_{s'u'}^- \leqslant & \ p_{s'u'} \leqslant p_{s'u'}^+,\\
p_{t'u'}^- \leqslant & \ p_{t'u'} \leqslant p_{t'u'}^+,\\
p_{s'\!'t'\!'}^- \leqslant & \ p_{s'\!'t'\!'} \leqslant p_{s'\!'t'\!'}^+,\\
p_{s'\!'u'\!'}^- \leqslant & \ p_{s'\!'u'\!'} \leqslant p_{s'\!'u'\!'}^+,\\
p_{t'\!'u'\!'}^- \leqslant & \ p_{t'\!'u'\!'} \leqslant p_{t'\!'u'\!'}^+,\\
p_{s't'u'}^- \leqslant & \ p_{s't'u'} \leqslant p_{s't'u'}^+,\\
p_{s'\!'t'\!'u'\!'}^- \leqslant & \ p_{s'\!'t'\!'u'\!'} \leqslant p_{s'\!'t'\!'u'\!'}^+,
\end{split}
\end{equation}
where
\begin{equation}\label{FRECHET-bounds-4UpDown}
\hspace*{-3pt}
\begin{split}
p_{s't'}^- &=\max \{0,p_{s'}+p_{t'}-p_x\},\\
p_{s't'}^+ &=\min \{p_{s'},p_{t'}\},\\
p_{s'\!'t'\!'}^-&=\max \{0,p_x+p_y+p_z-1-p_{s'}-p_{t'}\},\\
p_{s'\!'t'\!'}^+&=\min\{p_y-p_{s'},p_z-p_{t'}\},\\
p_{s'u'}^- &=\max \{0,p_{s'}+p_{u'}-p_x\},\\
p_{s'u'}^+ &=\min \{p_{s'},p_{u'}\},\\
p_{s'\!'u'\!'}^-&=\max \{0,p_x+p_y+p_v-1-p_{s'}-p_{u'}\},\\
p_{s'\!'u'\!'}^+&=\min\{p_y-p_{s'},p_v-p_{u'}\},\\
p_{t'u'}^- &=\max \{0,p_{t'}+p_{u'}-p_x\},\\
p_{t'u'}^+ &=\min \{p_{t'},p_{u'}\},\\
p_{t'\!'u'\!'}^-&=\max \{0,p_x+p_z+p_v-1-p_{t'}-p_{u'}\},\\
p_{t'\!'u'\!'}^+&=\min\{p_z-p_{t'},p_v-p_{u'}\},\\
p_{s't'u'}^- &=\max \{0,p_{s't'}+p_{s'u'}+p_{t'u'}-2p_x\},\\
p_{s't'u'}^+ &=\min \{p_{s't'},p_{s'u'},p_{t'u'}\},\\
p_{s'\!'t'\!'u'\!'}^-&=\max \{0,p_{s'\!'t'\!'}+p_{s'\!'u'\!'}+p_{t'\!'u'\!'}-2(1-p_x)\},\\
p_{s'\!'t'\!'u'\!'}^+&=\min\{p_{s'\!'t'\!'},p_{s'\!'u'\!'},p_{t'\!'u'\!'}\}
\end{split}\hspace*{-13pt}
\end{equation}
are the lower and upper Fr\'echet-boundaries of probabilities of double and triple intersections of events from the inserted triplets  $\mathcal{X}'$ and $\mathcal{X}'\!'$ with respect to the frame monoplet $\{x\}$.

The case of a quadruplet of events gives the following level of Fr\'echet-restrictions (the four last Fr\'echet-boundaries in (\ref{FRECHET-bounds-4UpDown})), when probabilities of triple intersections of events from inserted triplets have Fr\'echet-boundaries that depend directly not so much on marginal probabilities as on inserted probabilities of double intersections,
on which, in turn, Fr'echet-restrictions of the previous level, mentioned above, are imposed.

The all Fr\'echet-restrictions in the considered frame methods for a doublet, a triplet and a quadruplet of events differ from the usual Fr\'echet-restrictions, which are functions of only corresponding marginal probabilities. They differ in that they have a recurrent structure.
When, as the power of intersections of inserted events increases, the Fr\'echet-boundaries of their probabilities are functions of Fr\'echet-boundaries for probabilities of intersections of lower power.

The such Fr\'echet-restrictions and  Fr\'echet-boundaries for a doublet (\ref{FRECHET-bounds-2UpDown}), a triplet (\ref{FRECHET-restraints-triplet-bis},\ref{FRECHET-bounds-3UpDown}), a quadruplet (\ref{FRECHET-restraints-quadruplet},\ref{FRECHET-bounds-4UpDown}) of events and so on, will call the \emph{recurrent Fr\'echet-restrictions} and \emph{recurrent Fr\'echet-boundaries}.

\subsection{The frame method: recurrent formulas for a half-rare quadruplet of events\label{3-framing-method-recurrent-formulas}}

The recurrent formulas for Kopula of a quadruplet of events immediately can be written in the form of special cases of recurrence formulas (\ref{frame-recurrent-pseudo}) and (\ref{frame-recurrent-conditional}) from Note \ref{not-frame-recurrent-formulas} for the quadruplet $\frak{X}=\{x,y,z,v\}=\{x\}+\{y,z,v\}=\{x\}+\mathcal{X}$:
\begin{equation}\label{frame-recurrent-pseudo-quadruplet}
\hspace*{-13pt}
\begin{split}
&{\mbox{\boldmath$\mathscr{K}$}}\!\left(\breve{p}^{\left( c \mid X+Y /\!\!/ \mathcal{X}+\{x\}\right)}\!\right)=\\
&=\begin{cases}
\mathscr{K}^{(\{x\})}\!\!\left(\breve{p}^{\left( c \mid X^{(\cap \{x\} /\!\!/ \{x\})} /\!\!/ \mathcal{X}'\right)}\!\right)\!,& \hspace*{-8pt} Y=\{x\},\\
&\\[-5pt]
\mathscr{K}^{(\emptyset)}\!\!\left(\breve{p}^{\left( c \mid X^{(\cap \emptyset /\!\!/ \{x\})} /\!\!/ \mathcal{X}'\!'\right)}\!\right)\!,& \hspace*{-8pt} Y=\emptyset,
\end{cases}\\
\end{split}\hspace*{-13pt}
\end{equation}
\begin{equation}\label{frame-recurrent-conditional-quadruplet}
\hspace*{-13pt}
\begin{split}
&{\mbox{\boldmath$\mathscr{K}$}}\!\left(\breve{p}^{\left( c \mid X+Y /\!\!/ \mathcal{X}+\{x\}\right)}\!\right)=\\
&=\begin{cases}
{\mbox{\boldmath$\mathscr{K}$}}^{|\{x\}}\!\Big(\breve{p}^{\left( c \mid X /\!\!/ \mathcal{X} \mid \{x\} /\!\!/ \{x\}\right)}\!\Big)p_x, & \hspace*{-8pt} Y=\{x\},\\
&\\[-5pt]
{\mbox{\boldmath$\mathscr{K}$}}^{|\emptyset}\!\Big(\breve{p}^{\left( c \mid X /\!\!/ \mathcal{X} \mid \emptyset /\!\!/ \{x\}\right)}\!\Big)(1-p_x), & \hspace*{-8pt} Y=\emptyset.
\end{cases}
\end{split}\hspace*{-13pt}
\end{equation}
In the formulas (\ref{frame-recurrent-pseudo-quadruplet}) the inserted pseudo-Kopulas $\mathscr{K}^{(\{x\})}$ and $\mathscr{K}^{(\emptyset)}$ are defined by the octuples of probabilities, i.e., by the formulas:
\begin{equation}\label{epd-quadruplet-16-x-pseudoKopula}
\hspace*{-3pt}
\begin{split}
&\mathscr{K}^{(\{x\})}\!\!\left(\breve{p}^{\left( c \mid X^{(\cap \{x\} /\!\!/ \{x\})} /\!\!/ \mathcal{X}'\right)}\!\right)\!=\\
&=\begin{cases}
p_{s't'u'}, & \hspace{-8pt} X=\{y,z,v\},\\
p_{s't'}-p_{s't'u'}, & \hspace{-8pt} X=\{y,z\},\\
p_{s'u'}-p_{s't'u'}, & \hspace{-8pt} X=\{y,v\},\\
p_{t'u'}-p_{s't'u'}, & \hspace{-8pt} X=\{z,v\},\\
p_{s'}-p_{s't'}-p_{s'u'}+p_{s't'u'}, & \hspace{-8pt} X=\{z\},\\
p_{t'}-p_{s't'}-p_{t'u'}+p_{s't'u'}, & \hspace{-8pt} X=\{v\},\\
p_{u'}-p_{s'u'}-p_{t'u'}+p_{s't'u'}, & \hspace{-8pt} X=\{v\},\\
p_x-p_{s'}-p_{t'}-p_{u'}+&\\
+p_{s't'}+p_{s'u'}+p_{t'u'}-p_{s't'u'}, & \hspace{-8pt} X=\emptyset,
\end{cases}
\end{split}\hspace*{-13pt}
\end{equation}
\begin{equation}\label{epd-quadruplet-16-0-pseudoKopula}
\hspace*{-13pt}
\begin{split}
&\mathscr{K}^{(\emptyset)}\!\!\left(\breve{p}^{\left( c \mid X^{(\cap \emptyset /\!\!/ \{x\})} /\!\!/ \mathcal{X}'\!'\right)}\!\right)\!=\\
&=\begin{cases}
p_{s'\!'t'\!'u'\!'}, & \hspace{-8pt} X=\{y,z,v\},\\
p_{s'\!'t'\!'}-p_{s'\!'t'\!'u'\!'}, & \hspace{-8pt} X=\{y,z\},\\
p_{s'\!'u'\!'}-p_{s'\!'t'\!'u'\!'}, & \hspace{-8pt} X=\{y,v\},\\
p_{t'\!'u'\!'}-p_{s'\!'t'\!'u'\!'}, & \hspace{-8pt} X=\{z,v\},\\
p_y-p_{s'}-&\\
-p_{s'\!'t'\!'}-p_{s'\!'u'\!'}+p_{s'\!'t'\!'u'\!'}, & \hspace{-8pt} X=\{y\},\\
p_z-p_{t'}-&\\
-p_{s'\!'t'\!'}-p_{t'\!'u'\!'}+p_{s'\!'t'\!'u'\!'}, & \hspace{-8pt} X=\{z\},\\
p_v-p_{t'}-&\\
-p_{s'\!'u'\!'}-p_{t'\!'u'\!'}+p_{s'\!'t'\!'u'\!'}, & \hspace{-8pt} X=\{v\},\\
1-p_x-p_y-p_z-p_v+&\\
+p_{s'}+p_{t'}+p_{u'}-&\\
-p_{s'\!'t'\!'}-p_{s'\!'u'\!'}-p_{t'\!'u'\!'}+&\\
+p_{s'\!'t'\!'u'\!'}, & \hspace{-8pt} X=\emptyset,
\end{cases}
\end{split}\hspace*{-33pt}
\end{equation}
where, for example, for $X=\{y,z,v\}$
\begin{equation}\label{4-explanations1}
\hspace*{-13pt}
\begin{split}
&\{y,z,v\}^{(\cap \{x\} /\!\!/ \{x\})}\!=\!\{y \cap x, z \cap x, v \cap x\} \subseteq \mathcal{X}',\\
&\{y,z,v\}^{(\cap \emptyset /\!\!/ \{x\})}\!=\!\{y \cap x^c, z \cap x^c, v \cap x^c\} \subseteq \mathcal{X}'\!',
\end{split}\hspace*{-13pt}
\end{equation}
and the corresponding sets of marginal probabilities of inserted triplets $\mathcal{X}'$ and $\mathcal{X}'\!'$ have the form
\begin{equation}\label{4-explanations2}
\hspace*{-13pt}
\begin{split}
\breve{p}^{\left( c \mid \{y,z,v\}^{(\cap \{x\} /\!\!/ \{x\})} /\!\!/ \mathcal{X}'\!'\right)}\!&=\!\{p_{s'},p_{t'},p_{u'}\},\\
&\\
\breve{p}^{\left( c \mid \{y,z,v\}^{(\cap \emptyset /\!\!/ \{x\})} /\!\!/ \mathcal{X}'\!'\right)}\!&=\!\{p_y-p_{s'},p_z-p_{t'},p_v-p_{u'}\}.
\end{split}\hspace*{-33pt}
\end{equation}
In the formulas (\ref{frame-recurrent-conditional-quadruplet}) the condirional Kopulas
${\mbox{\boldmath$\mathscr{K}$}}^{|\{x\}}$ and ${\mbox{\boldmath$\mathscr{K}$}}^{|\emptyset}$
are defined by the octuples of probabilities that are normalized by $p_x$ and by $1-p_x$ correspondingly, i.e., by the formulas:
\begin{equation}\label{epd-quadruplet-16-x-conditionalKopula}
\hspace*{-13pt}
\begin{split}
&{\mbox{\boldmath$\mathscr{K}$}}^{|\{x\}}\!\Big(\breve{p}^{\left( c \mid X /\!\!/ \mathcal{X} \mid \{x\} /\!\!/ \{x\}\right)}\!\Big)\!=\\
&=\begin{cases}
\frac{1}{p_x}p_{s't'u'}, & \hspace{-9pt} X=\{y,z,v\},\\
\frac{1}{p_x}(p_{s't'}-p_{s't'u'}), & \hspace{-9pt} X=\{y,z\},\\
\frac{1}{p_x}(p_{s'u'}-p_{s't'u'}), & \hspace{-9pt} X=\{y,v\},\\
\frac{1}{p_x}(p_{t'u'}-p_{s't'u'}), & \hspace{-9pt} X=\{z,v\},\\
\frac{1}{p_x}(p_{s'}-p_{s't'}-p_{s'u'}+p_{s't'u'}), & \hspace{-9pt} X=\{z\},\\
\frac{1}{p_x}(p_{t'}-p_{s't'}-p_{t'u'}+p_{s't'u'}), & \hspace{-9pt} X=\{v\},\\
\frac{1}{p_x}(p_{u'}-p_{s'u'}-p_{t'u'}+p_{s't'u'}), & \hspace{-9pt} X=\{v\},\\
1-\frac{1}{p_x}(p_{s'}+p_{t'}+p_{u'})+&\\
+\frac{1}{p_x}(p_{s't'}+p_{s'u'}+p_{t'u'})-&\\
-\frac{1}{p_x}p_{s't'u'}, & \hspace{-9pt} X=\emptyset,
\end{cases}
\end{split}\hspace*{-33pt}
\end{equation}
\begin{equation}\label{epd-quadruplet-16-0-conditionalKopula}
\hspace*{-13pt}
\begin{split}
&{\mbox{\boldmath$\mathscr{K}$}}^{|\emptyset}\!\Big(\breve{p}^{\left( c \mid X /\!\!/ \mathcal{X} \mid \emptyset /\!\!/ \{x\}\right)}\!\Big)\!=\\
&=\begin{cases}
\frac{1}{1-p_x}p_{s'\!'t'\!'u'\!'}, & \hspace{-8pt} X=\{y,z,v\},\\
\frac{1}{1-p_x}(p_{s'\!'t'\!'}-p_{s'\!'t'\!'u'\!'}), & \hspace{-8pt} X=\{y,z\},\\
\frac{1}{1-p_x}(p_{s'\!'u'\!'}-p_{s'\!'t'\!'u'\!'}), & \hspace{-8pt} X=\{y,v\},\\
\frac{1}{1-p_x}(p_{t'\!'u'\!'}-p_{s'\!'t'\!'u'\!'}), & \hspace{-8pt} X=\{z,v\},\\
\frac{1}{1-p_x}(p_y-p_{s'})-&\\
-\frac{1}{1-p_x}(p_{s'\!'t'\!'}+p_{s'\!'u'\!'}-p_{s'\!'t'\!'u'\!'}), & \hspace{-8pt} X=\{y\},\\
\frac{1}{1-p_x}(p_z-p_{t'})-&\\
-\frac{1}{1-p_x}(p_{s'\!'t'\!'}+p_{t'\!'u'\!'}-p_{s'\!'t'\!'u'\!'}), & \hspace{-8pt} X=\{z\},\\
\frac{1}{1-p_x}(p_v-p_{t'})-&\\
-\frac{1}{1-p_x}(p_{s'\!'u'\!'}+p_{t'\!'u'\!'}-p_{s'\!'t'\!'u'\!'}), & \hspace{-8pt} X=\{v\},\\
1-\frac{1}{1-p_x}(p_y+p_z+p_v)+&\\
+\frac{1}{1-p_x}(p_{s'}+p_{t'}+p_{u'})-&\\
-\frac{1}{1-p_x}(p_{s'\!'t'\!'}+p_{s'\!'u'\!'}+p_{t'\!'u'\!'})+&\\
+\frac{1}{1-p_x}p_{s'\!'t'\!'u'\!'}, & \hspace{-8pt} X=\emptyset,
\end{cases}
\end{split}\hspace*{-33pt}
\end{equation}
The corresponding sets of marginal conditional probabilities of events $y, z \in \mathcal{X}$ with respect to the frame terraced  events $\textsf{ter}(\{x\} /\!\!/ \{x\})=x$ and $\textsf{ter}(\emptyset /\!\!/ \{x\})=x^c$ correspondingly have the form:
\begin{equation}\label{4-conditional-marginals}
\hspace*{-3pt}
\begin{split}
\breve{p}^{\left( c \mid \mathcal{X} /\!\!/ \mathcal{X} \mid \{x\} /\!\!/ \{x\}\right)}\!&=\!\left\{\frac{p_{s'}}{p_x},\frac{p_{t'}}{p_x},\frac{p_{u'}}{p_x}\right\},\\
\breve{p}^{\left( c \mid \mathcal{X} /\!\!/ \mathcal{X} \mid \emptyset /\!\!/ \{x\}\right)}\!&=\!\left\{\frac{p_y-p_{s'}}{1-p_x},\frac{p_z-p_{t'}}{1-p_x},\frac{p_v-p_{u'}}{1-p_x}\right\}.
\end{split}\hspace*{-13pt}
\end{equation}
Recall that 11 functional parameters
\begin{equation}\label{11-parameters}
\hspace*{-3pt}
\begin{split}
&p_{s'}, p_{t'}, p_{u'}\!,\\
&p_{s't'}, p_{s'u'}, p_{t'u'},\\
&p_{s'\!'t'\!'}, p_{s'\!'u'\!'}, p_{t'\!'u'\!'},\\
&p_{s't'u'}, p_{s'\!'t'\!'u'\!'}
\end{split}\hspace*{-13pt}
\end{equation}
in the recurrent formulas (\ref{epd-quadruplet-16-x-pseudoKopula}, \ref{epd-quadruplet-16-0-pseudoKopula}) and (\ref{epd-quadruplet-16-x-conditionalKopula}, \ref{epd-quadruplet-16-0-conditionalKopula}) obey the Fr\'echet-restrictions (\ref{FRECHET-restraints-quadruplet}) and Fr\'echet-boundaries (\ref{FRECHET-bounds-4UpDown}).

\section{The Kopula theory for a set of events\label{N-Kopulas-theory}}

\subsection{Independent $N$-Kopula}

First, without the frame method, which is not required here, let's consider the simplest example of the \emph{$N$-Kopula} ${\mbox{\boldmath$\mathscr{K}$}} \in \Psi^1_\frak{X}$ of an $N$-set of events $\frak{X}$, i.e., a 1-function on the unit $\frak{X}$-hypercube. In other words, we construct a nonnegative bounded numerical function
$$
{\mbox{\boldmath$\mathscr{K}$}} : [0,1]^{\otimes \frak{X}} \to [0,1],
$$
that for all $z \in \frak{X}$
$$
\sum_{x \in X \subseteq \frak{X}} {\mbox{\boldmath$\mathscr{K}$}}\left(\breve{w}^{(c|X/\!\!/\frak{X})}\right) = w_x.
$$
A such simplest example of a 1-function on the unit $\frak{X}$-hypercube is the so-called \emph{independent $N$-Kopula} which for all free variables $\breve{w} \in [0,1]^{\otimes \frak{X}}$ is defined by the formula:
\begin{equation}\label{ind-N-Kopula}
\begin{split}
{\mbox{\boldmath$\mathscr{K}$}}\left(\breve{w}\right) &= \prod_{x \in \frak{X}} w_x,
\end{split}
\end{equation}
that provides it on each $2^{(c|\breve{w})}$-phenomenon-dom the following $2^N$ values:
\begin{equation}\label{ind-N-Kopulas}
\begin{split}
{\mbox{\boldmath$\mathscr{K}$}}\left(\breve{w}^{(c|X/\!\!/\frak{X})}\right) = \prod_{x \in X} w_x \prod_{x \in \frak{X}-X} (1-w_x)
\end{split}
\end{equation}
for $X \subseteq \frak{X}$.
Indeed\footnote{Perhaps this statement deserves to be called a lemma, which, incidentally, is not difficult to prove.} as in the case of doublet of events this function is a 1-function, since for all $x \in \frak{X}$
$$
\sum_{x \in X \subseteq \frak{X}} \left( \prod_{z \in X} w_z \prod_{z \in \frak{X}-X} (1-w_z) \right) = w_x.
$$
The e.p.d. of the 1st kind of ondependent $N$-s.e. $\frak{X}$ with the $\frak{X}$-set of probabilities of events $\breve{p}$ is defined by $2^N$ values of the independent $N$-Kopula (\ref{ind-N-Kopula}) on the $2^{(c|\breve{p})}$-phenomenon-dom by the general formulas of half-rare variables, i.e., for $X \subseteq \{x,y\}$:
\begin{equation}\label{N-epd-from-ind-Kopula}
\begin{split}
&p(X/\!\!/\frak{X})\!=\!{\mbox{\boldmath$\mathscr{K}$}}\left(\breve{p}^{(c|X/\!\!/\frak{X})}\right) \!=\!\prod_{x \in X} p_x \!\!\!\!\prod_{x \in \frak{X}-X} (1-p_x).
\end{split}
\end{equation}

\subsection{The frame method for constructing a half-rare set of events\label{N-framing-method}}

The general recurrent formulas (\ref{frame-recurrent-pseudo}, \ref{frame-recurrent-conditional}) of the frame method for constructing a Kopula of a set of events are derived in Note \ref{not-frame-recurrent-formulas}. Recall these formulas:
\begin{equation}\label{frame-recurrent-pseudo2}
\hspace*{-13pt}
\begin{split}
&{\mbox{\boldmath$\mathscr{K}$}}\!\left(\breve{p}^{\left( c \mid X+Y /\!\!/ \mathcal{X}+\{x_0\}\right)}\!\right)=\\
&=\begin{cases}
\mathscr{K}^{(\{x_0\})}\!\!\left(\breve{p}^{\left( c \mid X^{(\cap \{x_0\} /\!\!/ \{x_0\})} /\!\!/ \mathcal{X}'\right)}\!\right)\!,& \hspace*{-8pt} Y=\{x_0\},\\
&\\[-5pt]
\mathscr{K}^{(\emptyset)}\!\!\left(\breve{p}^{\left( c \mid X^{(\cap \emptyset /\!\!/ \{x_0\})} /\!\!/ \mathcal{X}'\!'\right)}\!\right)\!,& \hspace*{-8pt} Y=\emptyset,
\end{cases}\\
\end{split}\hspace*{-13pt}
\end{equation}
\begin{equation}\label{frame-recurrent-conditional2}
\hspace*{-13pt}
\begin{split}
&{\mbox{\boldmath$\mathscr{K}$}}\!\left(\breve{p}^{\left( c \mid X+Y /\!\!/ \mathcal{X}+\{x_0\}\right)}\!\right)=\\
&=\begin{cases}
{\mbox{\boldmath$\mathscr{K}$}}^{|\{x_0\}}\!\Big(\breve{p}^{\left( c \mid X /\!\!/ \mathcal{X} \mid \{x_0\} /\!\!/ \{x_0\}\right)}\!\Big)p_0, & \hspace*{-8pt} Y=\{x_0\},\\
&\\[-5pt]
{\mbox{\boldmath$\mathscr{K}$}}^{|\emptyset}\!\Big(\breve{p}^{\left( c \mid X /\!\!/ \mathcal{X} \mid \emptyset /\!\!/ \{x_0\}\right)}\!\Big)(1-p_0), & \hspace*{-8pt} Y=\emptyset,
\end{cases}
\end{split}\hspace*{-13pt}
\end{equation}
which express the $N$-Kopula of $N$-s.e. $\frak{X}=\mathcal{X}+\{x_0\}$, where $\mathcal{X}=\mathcal{X}'(+)\mathcal{X}'\!'$, through the known probability $p_0$ of the event $x_0$ and together with it either two known \emph{inserted pseudo-$(N\!-\!1)$-Kopulas} (see Definition \ref{def-inserting-pseudoKopulas}), i.e., pseudo-$(N\!-\!1)$-Kopulas of inserted $(N\!-\!1)$-s.e.'s $\mathcal{X}'$ and $\mathcal{X}'\!'$ in the frame monoplet $\{x_0\}$, or through two known \emph{conditional $(N\!-\!1)$-Kopulas} (see Definition \ref{def-conditional-Kopulas}) with respect to the frame monoplet $\{x_0\}$ for the same $\mathcal{X}'$ and $\mathcal{X}'\!'$.

We write out more detailed formulas for corresponding pseudo-$(N\!-\!1)$-Kopulas:
\begin{equation}\label{pseudo-N-Kopulas}
\hspace*{-13pt}
\begin{split}
&\mathscr{K}^{(\{x_0\})}\!\!\left(\breve{p}^{\left( c \mid X^{(\cap \{x_0\} /\!\!/ \{x_0\})} /\!\!/ \mathcal{X}'\right)}\!\right)=
\mathscr{K}^{(\{x_0\})}\!\!\left(\breve{p}^{\left( c \mid X' /\!\!/ \mathcal{X}'\right)}\!\right)=\\
&=\begin{cases}
p(X' /\!\!/ \mathcal{X}'), & X' \ne \emptyset,\\
p(\emptyset /\!\!/ \mathcal{X}')-1+p_0, & X' = \emptyset,\\
\end{cases}\\
&\\
&\mathscr{K}^{(\emptyset)}\!\!\left(\breve{p}^{\left( c \mid X^{(\cap \emptyset /\!\!/ \{x_0\})} /\!\!/ \mathcal{X}'\!'\right)}\!\right)=
\mathscr{K}^{(\emptyset)}\!\!\left(\breve{p}^{\left( c \mid X'\!' /\!\!/ \mathcal{X}'\!'\right)}\!\right)=\\
&=\begin{cases}
p(X'\!' /\!\!/ \mathcal{X}'\!'), & X'\!' \ne \emptyset,\\
p(\emptyset /\!\!/ \mathcal{X}'\!')-p_0, & X'\!' = \emptyset,
\end{cases}\\
\end{split}\hspace*{-33pt}
\end{equation}
and for conditional $(N\!-\!1)$-Kopulas:
\begin{equation}\label{conditional-N-Kopulas}
\hspace*{-13pt}
\begin{split}
&{\mbox{\boldmath$\mathscr{K}$}}^{|\{x_0\}}\!\Big(\breve{p}^{\left( c \mid X /\!\!/ \mathcal{X} \mid \{x_0\} /\!\!/ \{x_0\}\right)}\!\Big)=\\
&=\begin{cases}
\frac{1}{p_0}p(X' /\!\!/ \mathcal{X}'), & X' \ne \emptyset,\\
\frac{1}{p_0}(p(\emptyset /\!\!/ \mathcal{X}')-1+p_0), & X' = \emptyset,\\
\end{cases}\\
&\\
&{\mbox{\boldmath$\mathscr{K}$}}^{|\emptyset}\!\Big(\breve{p}^{\left( c \mid X /\!\!/ \mathcal{X} \mid \emptyset /\!\!/ \{x_0\}\right)}\!\Big)=\\
&=\begin{cases}
\frac{1}{1-p_0}p(X'\!' /\!\!/ \mathcal{X}'\!'), & X'\!' \ne \emptyset,\\
\frac{1}{1-p_0}(p(\emptyset /\!\!/ \mathcal{X}'\!')-p_0), & X'\!' = \emptyset,
\end{cases}\\
\end{split}\hspace*{-13pt}
\end{equation}
where $X'=X^{(\cap \{x_0\} /\!\!/ \{x_0\})} = X (\cap) \{x_0\} = \{x \cap x_0, x \in X\}$ и $X'\!'=X^{(\cap \emptyset /\!\!/ \{x_0\})} = X (\cap) \{x_0^c\} = \{x \cap x_0^c, x \in X\}$ для $X \subseteq \mathcal{X}$.

\subsection{Recurrent formulas for Fr\'echet-boundaries and Fr\'echet-restrictions}

Now let us consider recurrent formulas for Fr\'echet-boundaries и Fr\'echet-restrictions and for the $2^N\!-\!N\!-\!1$ functional parameters of an $N$-Kopula of $N$-set of events
\begin{equation}\label{...}
\begin{split}
\frak{X} &= \{x_0,x_1,\ldots,x_{N\!-\!1}\}=\\
&=\{x_0\}+\{x_1,\ldots,x_{N\!-\!1}\}=\\
&=\{x_0\}+\mathcal{X} =\\
&=\{x_0\}+(\mathcal{X}'(+)\mathcal{X}'\!'),
\end{split}
\end{equation}
where
\begin{equation}\label{...}
\begin{split}
\mathcal{X}'&=\{x_0\cap x_1,\ldots,x_0\cap x_{N\!-\!1}\},\\
\mathcal{X}'\!'&=\{x_0^c\!\cap x_1,\ldots,x_0^c\!\cap x_{N\!-\!1}\}
\end{split}
\end{equation}
are inserted $(N\!-\!1)$-s.e.'s, and
\begin{equation}\label{...}
\begin{split}
\breve{p}^{(c \mid \frak{X} /\!\!/ \frak{X})} =
\{p_0,p_1,\ldots,p_{N\!-\!1}\}
\end{split}
\end{equation}
is the $\frak{X}$-set of probabilities of marginal events from $\frak{X}$, i.e., $p_n=\mathbf{P}(x_n), n=0,1,\ldots,N\!-\!1$.

Judging by the form of Fr\'echet-boundaries и Fr\'echet-restrictions for a doublet, a triplet and a quadruplet of events, collected in paragraph \ref{FRECHET-reсurrent}, these Fr\'echet-restrictions consists of two groups,
such that one of them, which refers to the parameters of the inserted $(N\!-\!1)$-s.e. $\mathcal{X}'$, consists of $2^{N\!-\!1}\!-\!1$ Fr\'echet-restrictions, and the other, which refers to the parameters of the inserted $(N\!-\!1)$-s.e. $\mathcal{X}'\!'$, consists of $2^{N\!-\!1}\!-\!(N\!-\!1)\!-\!1$ Fr\'echet-restrictions. And, as it should:
\begin{equation}\label{...}
\begin{split}
2^N\!-\!N-1\!=\!(2^{N\!-\!1}\!-\!1)\!+\!(2^{N\!-\!1}\!-\!(N\!-\!1)\!-\!1).
\end{split}
\end{equation}

The first group, related to the inserted $(N\!-\!1)$-s.e. $\mathcal{X}'$, contains Fr\'echet-restrictions for probabilities of the second kind
\begin{equation}\label{...}
\begin{split}
p_{X' /\!\!/ \mathcal{X}'} = \mathbf{P}\left(\bigcap_{x' \in X'} x'\right),
\end{split}
\end{equation}
that are numbered by nonempty subsets $X' \ne \emptyset$ of inserted $(N\!-\!1)$-s.e. $\mathcal{X}'$ (the number od such subsets: $2^{N\!-\!1}\!-\!1$); the second group, related to the inserted $(N\!-\!1)$-s.e. $\mathcal{X}'\!'$, contains Fr\'echet-restrictions for the such probabilities of the second kind:
\begin{equation}\label{...}
\begin{split}
p_{X'\!' /\!\!/ \mathcal{X}'\!'} = \mathbf{P}\left(\bigcap_{x'\!' \in X'\!'} x'\!'\right),
\end{split}
\end{equation}
that are numbered by subsets $X'\!' \subseteq \mathcal{X}'\!'$ with the power $|X'\!'| \geqslant 2$ (number of such subsets: $2^{N\!-\!1}\!-\!(N-1)\!-\!1$).

\texttt{Note \!\refstepcounter{ctrnot}\arabic{ctrnot}\,\label{not-subsets-n}\itshape\footnotesize (denotations for subsets of fixed power).}
To more conveniently represent the recurrent Fr\'echet-restrictions, agree to denote
\begin{equation}\label{...}
\begin{split}
X'_n \subseteq \mathcal{X}' & \ \Longleftrightarrow |X'_n|=n,\\
X'\!'_n \subseteq \mathcal{X}'\!' & \ \Longleftrightarrow |X'\!'_n|=n.
\end{split}
\end{equation}
the subsets consisting of $n$ events.
In this notation, for example, the $\frak{X}$--set of  marginal probabilities $\breve{p}^{(c \mid \frak{X} /\!\!/ \frak{X})}$ is written as the $\frak{X}$-set probabilities of the second kind that are numbered by monoplets of events $X_1=\{x\}, x \in \frak{X}$:
\begin{equation}\label{X1}
\begin{split}
\{p_0,p_1,\ldots,p_{N\!-\!1}\} = \{p_{X_1 /\!\!/ \frak{X}}, X_1 \subseteq \frak{X}\}.
\end{split}
\end{equation}
The set of probabilities of double intersection of events $x \in \frak{X}$, i.e., the set of probabilities of the second kind that are numbered by doublets, has the form:
\begin{equation}\label{X2}
\begin{split}
\{p_{\{x,y\} /\!\!/ \frak{X}}, \{x,y\} \subseteq \frak{X}\} =
\{p_{X_2 /\!\!/ \frak{X}}, X_2 \subseteq \frak{X}\}.
\end{split}
\end{equation}
And the set of probabilities of triple intersections of events $x \in \frak{X}$, i.e., the set of probabilities of the second kind that are numbered by triplets, has the form:
\begin{equation}\label{X3}
\begin{split}
\{p_{\{x,y,z\} /\!\!/ \frak{X}}, \{x,y,z\} \subseteq \frak{X}\} =
\{p_{X_3 /\!\!/ \frak{X}}, X_3 \subseteq \frak{X}\}
\end{split}
\end{equation}
and so on.

\texttt{Note \!\refstepcounter{ctrnot}\arabic{ctrnot}\,\label{not-FRECHET-recurrent-formulas}\itshape\footnotesize (recurrent formulas for Fr\'echet-boundaries and Fr\'echet-restrictions).}
Probabilities of $n$-intersections $(n=2,...,N\!-\!1)$ of events from the inserted s.e.'s $\mathcal{X}'$ and $\mathcal{X}'\!'$ have the \emph{recurrent Fr\'echet-restrictions} (see paragraph \ref{FRECHET-reсurrent}) that are written by denotations from Note \ref{not-subsets-n} by the following way with respect to $\mathcal{X}'$:
\begin{equation}\label{FRECHET-restraints1}
\begin{split}
p_{X'_n /\!\!/ \mathcal{X}'}^- \leqslant p_{X'_n /\!\!/ \mathcal{X}'} \leqslant p_{X'_n /\!\!/ \mathcal{X}'}^+,
\end{split}
\end{equation}
where
\begin{equation}\label{FRECHET-bounds1}
\hspace*{-13pt}
\begin{split}
p_{X'_n /\!\!/ \mathcal{X}'}^-&\!=\!\max\!\left\{\!\!0,p_x-\!\!\!\!\!\!\sum_{X'_{n\!-\!1} \subseteq X'_n}\!\!\!\!(p_x-p_{X'_{n\!-\!1} /\!\!/ \mathcal{X}'})\!\!\right\}\!,\\
p_{X'_n /\!\!/ \mathcal{X}'}^+&\!=\!\min_{X'_{n\!-\!1} \subseteq X'_n} \!\left\{\!p_{X'_{n\!-\!1} /\!\!/ \mathcal{X}'}, X'_{n\!-\!1} \subseteq \mathcal{X}'\!\right\}\!.
\end{split}\hspace*{-33pt}
\end{equation}
are \emph{recurrent the lower and upper Fr\'echet-boundaries}. And the lower Fr\'echet-boundary we can write somewhat differently after simple transformations:
\begin{equation}\label{FRECHET-bounds1-down}
\hspace*{-13pt}
\begin{split}
p_{X'_n /\!\!/ \mathcal{X}'}^-\!=\!\max\!\left\{\!0,\!\!\!\sum_{X'_{n\!-\!1} \subseteq X'_n}\!\!\!p_{X'_{n\!-\!1} /\!\!/ \mathcal{X}'}\!-\!(n\!-\!1)p_x \!\right\}\!.
\end{split}\hspace*{-13pt}
\end{equation}
Similar look the \emph{recurrent Fr\'echet-restrictions} with respect to the inserted s.e. $\mathcal{X}'\!'$:
\begin{equation}\label{FRECHET-restraints2}
\begin{split}
p_{X'\!'_n /\!\!/ \mathcal{X}'\!'}^- \leqslant p_{X'\!'_n /\!\!/ \mathcal{X}'\!'} \leqslant p_{X'\!'_n /\!\!/ \mathcal{X}'\!'}^+,
\end{split}
\end{equation}
where
\begin{equation}\label{FRECHET-bounds2}
\hspace*{-13pt}
\begin{split}
p_{X'\!'_n /\!\!/ \mathcal{X}'\!'}^-&\!=\!\max\!\left\{\!\!0,1\!-\!p_x\!-\!\!\!\!\!\!\!\sum_{X'\!'_{n\!-\!1} \subseteq X'\!'_n}\!\!\!\!(1\!-\!p_x\!-\!p_{X'\!'_{n\!-\!1} /\!\!/ \mathcal{X}'\!'})\!\!\right\}\!\!,\\[3pt]
&\\
p_{X'\!'_n /\!\!/ \mathcal{X}'\!'}^+&\!=\!\min_{X'\!'_{n\!-\!1} \subseteq X'\!'_n} \!\left\{\!p_{X'\!'_{n\!-\!1} /\!\!/ \mathcal{X}'\!'}, X'\!'_{n\!-\!1} \subseteq \mathcal{X}'\!'\!\right\}\!\!.
\end{split}\hspace*{-33pt}
\end{equation}
are \emph{recurrent the lower and upper Fr\'echet-boundaries}. And the lower Fr\'echet-boundary we can write somewhat differently after simple transformations:
\begin{equation}\label{FRECHET-bounds2-down}
\hspace*{-23pt}
\begin{split}
&p_{X'\!'_n /\!\!/ \mathcal{X}'\!'}^-=\\
&=\max\!\left\{\!0,\!\!\!\sum_{X'\!'_{n\!-\!1} \subseteq X'\!'_n}\!\!\!p_{X'\!'_{n\!-\!1} /\!\!/ \mathcal{X}'\!'}\!-\!(n\!-\!1)(1\!-\!p_x) \!\right\}\!.
\end{split}\hspace*{-13pt}
\end{equation}

It remains to write out more $N\!-\!1$ \emph{recurrent Fr\'echet-restrictions} on probabilities of marginal events from the inserted s.e. $\mathcal{X}'$, i.e., on probabilities of the second kind that are numbered by monoplets $X'_1 \subseteq \mathcal{X}'$:
\begin{equation}\label{FRECHET-marginals}
\begin{split}
0 \leqslant p_{X'_1 /\!\!/ \mathcal{X}'} \leqslant p_{X_1 /\!\!/ \mathcal{X}},
\end{split}
\end{equation}
which are restricted by marginal probabilities of events from the $(N\!-\!1)$-s.e. $\mathcal{X}$ and which together with the \emph{recurrent Fr\'echet-restrictions} (\ref{FRECHET-restraints1}, \ref{FRECHET-restraints2}) form the all totality of \emph{recurrent Fr\'echet-restrictions}.
This totality consists of $2^N\!-\!N\!-\!1$ restrictions. And \emph{recurrent the lower and upper Fr\'echet-boundaries} in these restrictions are defined by recurrent formulas (\ref{FRECHET-bounds1}, \ref{FRECHET-bounds2}).

\section{Parametrization of functional parameters of Kopula by Fr\'echet-correlations of inserted events\label{3Kopulas-parameters}}

Let's consider parametrization on an example of functional parameters $p_{s'}, p_{t'}, p_{s't'}$ и $p_{s'\!'t'\!'}$ of 3-Kopula of the $\breve{p}$-ordered half-rare triplet $\frak{X}=\{x,y,z\}$,
which in the frame method is constructed from two inserted pseudo-2-Kopulas.

\subsection{Parametrization of functional parameters $p_{s'}$ and $p_{t'}$}

The Fr\'echet-restriction of the functional parameter $p_{s'}=p_{s'}(p_x,p_y,p_z)$
\begin{equation}\label{2-Kopula-ps}
\begin{split}
0 \leqslant p_{s'} \leqslant p_y,
\end{split}
\end{equation}
that in the frame method has a sense of probability of double intersection of events $x$ and $y$:
\begin{equation}\label{2-Kopula-ps1}
\begin{split}
p_{s'}=p_{xy /\!\!/ \frak{X}} = \mathbf{P}(x \cap y),
\end{split}
\end{equation}
is baced on the notion of Fr\'echet-correlation \cite{Vorobyev2007}
\begin{equation}\label{Frechet-korrelation-2}
\begin{split}
\textsf{Kor}_{xy}=
\begin{cases}
\frac{\textsf{Kov}_{xy}}{|\textsf{Kov}_{xy}^-|}, & \textsf{Kov}_{xy} < 0,\\
&\\[-8pt]
\frac{\textsf{Kov}_{xy}}{\textsf{Kov}_{xy}^+}, & \textsf{Kov}_{xy} \geqslant 0,
\end{cases}
\end{split}
\end{equation}
where
\begin{equation}\label{Kovariance-2}
\begin{split}
\textsf{Kov}_{xy} = \mathbf{P}(x \cap y) - \mathbf{P}(x)\mathbf{P}(y)
\end{split}
\end{equation}
is a covariance of events $x$ and $y$, and
\begin{equation}\label{Kovariance-2-Frechet-bounds}
\begin{split}
\textsf{Kov}_{xy}^- &= \max \{ 0, p_x+p_y-1 \} - p_x p_y = -p_x p_y,\\
\textsf{Kov}_{xy}^+ &= \min \{ p_x, p_y \} - p_x p_y = p_y-p_xp_y
\end{split}
\end{equation}
are its the lower and upper Fr\'echet-boundaries.

From Definition (\ref{Frechet-korrelation-2}) we get the parametrization of functional parameter $p_{s'}$ by the double Fr\'echet-correlation on the following form:
\begin{equation}\label{parametrization-ps}
\begin{split}
&p_{s'}(p_x,p_y,p_z)=\\
&=\begin{cases}
p_xp_y-\textsf{Kor}_{xy}\textsf{Kov}_{xy}^-,& \textsf{Kor}_{xy} < 0,\\
p_xp_y+\textsf{Kor}_{xy}\textsf{Kov}_{xy}^+,& \textsf{Kor}_{xy} \geqslant 0
\end{cases}\\
&=\begin{cases}
p_xp_y+\textsf{Kor}_{xy}p_xp_y, & \textsf{Kor}_{xy} < 0,\\
p_xp_y+\textsf{Kor}_{xy}(p_y-p_xp_y), & \textsf{Kor}_{xy} \geqslant 0.
\end{cases}
\end{split}
\end{equation}
The parametrization of functional parameter $p_{t'}$ by the double Fr\'echet-correlation is similar:
\begin{equation}\label{parametrization-ps1}
\begin{split}
&p_{t'}(p_x,p_y,p_z)=\\
&=\begin{cases}
p_xp_z-\textsf{Kor}_{xz}\textsf{Kov}_{xz}^-,& \textsf{Kor}_{xz} < 0,\\
p_xp_z+\textsf{Kor}_{xz}\textsf{Kov}_{xz}^+,& \textsf{Kor}_{xz} \geqslant 0
\end{cases}\\
&=\begin{cases}
p_xp_z+\textsf{Kor}_{xz}p_xp_z, & \textsf{Kor}_{xz} < 0,\\
p_xp_z+\textsf{Kor}_{xz}(p_z-p_xp_z), & \textsf{Kor}_{xz} \geqslant 0.
\end{cases}
\end{split}
\end{equation}

\subsection{Inserted triple covariances and Fr\'echet-correlations\label{Kovariance3inserting}}

We recall first that an \emph{absolute triple Fr\'echet-correlation} \cite{Vorobyev2007} of three events $x,y$ and $z$ is defined similarly to the double one:
\begin{equation}\label{Frechet-korrelation-3}
\begin{split}
\textsf{Kor}_{xyz}=
\begin{cases}
\frac{\textsf{Kov}_{xyz}}{|\textsf{Kov}_{xyz}^-|}, & \textsf{Kov}_{xyz} < 0,\\
&\\[-8pt]
\frac{\textsf{Kov}_{xyz}}{\textsf{Kov}_{xyz}^+}, & \textsf{Kov}_{xyz} \geqslant 0,
\end{cases}
\end{split}
\end{equation}
where
\begin{equation}\label{Kovariance-3}
\begin{split}
\textsf{Kov}_{xyz} = \mathbf{P}(x \cap y \cap z) - \mathbf{P}(x)\mathbf{P}(y)\mathbf{P}(z)
\end{split}
\end{equation}
is the triple covariance of events $x,y$ and $z$, and
\begin{equation}\label{Kovariance-3-Frechet-bounds}
\begin{split}
\textsf{Kov}_{xyz}^- &= \max \{ 0, p_x+p_y+p_z-2 \} - p_x p_y p_z =\\
&= -p_x p_y p_z,\\
\textsf{Kov}_{xyz}^+ &= \min \{ p_x, p_y, p_z \} - p_x p_y p_z =\\
&= p_z-p_xp_yp_z
\end{split}
\end{equation}
are its \emph{absolute} the lower and upper Fr\'echet-boundaries.

The definition of the \emph{inserted} triple Fr\'echet-correlation differs of the definition of \emph{absolute} one (\ref{Frechet-korrelation-3}) in that its the lower and upper Fr\'echet-boundaries must depend on the e.p.d. of the inserted doublets $\mathcal{X}'$ and $\mathcal{X}'\!'$.
So they differ from \emph{absolute} Fr\'echet-boundaries (\ref{Kovariance-3-Frechet-bounds}) and have the form (\ref{2-Kopula-pst}), where
\begin{equation}\label{FRECHET-bounds-3UpDown-bis}
\begin{split}
p_{s't'}^-&=\max \{0,p_{s'}+p_{t'}-p_x\},\\
p_{s't'}^+&=\min\{p_{s'},p_{t'}\},\\
p_{s'\!'t'\!'}^-&=\max \{0,p_x+p_y+p_z-1-p_{s'}-p_{t'}\},\\
p_{s'\!'t'\!'}^+&=\min\{p_y-p_{s'},p_z-p_{t'}\},
\end{split}
\end{equation}
are the lower and upper Fr\'echet-boundaries of probabilities of double intersections of events from \emph{inserted} doublets  $\mathcal{X}'$ and $\mathcal{X}'\!'$ with respect to the frame monoplet $\{x\}$, which should serve \emph{inserted} the lower and upper Fr\'echet-boundaries of probabilities of triple intersections (\ref{2-Kopula-pst1}) of events from the triplet $\frak{X}=\{x\}+(\mathcal{X}'(+)\mathcal{X}'\!')$.
However, as might be expected, these Fr\'echet-boundaries are not always ready to serve as the lower and upper Fr\'echet-boundaries for probabilities of triple intersections.

For this reason, it is necessary to modify the definitions of two \emph{inserted} triple covariances and, respectively, --- \emph{inserted} the lower and upper Fr\'echet-boundaries of these covariances.

\textsf{The first modification of definitions (see Fig. \ref{fig-Kop3},\ref{fig-Kop3_minus},\ref{fig-Kop3_plus}
).}
For brevity, we denote $p^{(\{x\})}_\star = p_xp_yp_z, p^{(\emptyset)}_\star = (1-p_x)p_yp_z$. Two \emph{inserted} triple covariance are defined by the firmulas:
\begin{equation}\label{Kovariance-3-inserting}
\hspace*{-3pt}
\begin{split}
&\textsf{Kov}_{xyz}^{(\{x\})}\!\!=\!\!
\begin{cases}
p_{s't'}\!-\!p^{(\{x\})}_\star, & p^{(\{x\})}_\star \in \left[p_{s't'}^-, p_{s't'}^+\right],\\
p_{s't'}^-\!-\!p^{(\{x\})}_\star, & p^{(\{x\})}_\star < p_{s't'}^-,\\
p_{s't'}^+\!-\!p^{(\{x\})}_\star, & p_{s't'}^+ < p^{(\{x\})}_\star,\\
\end{cases}\\
&\\
&\textsf{Kov}_{xyz}^{(\emptyset)}\!\!=\!\!
\begin{cases}
p_{s'\!'t'\!'}\!-\!p^{(\emptyset)}_\star, &  p^{(\emptyset)}_\star \in \left[p_{s't'}^-,p_{s't'}^+\right],\\
p_{s'\!'t'\!'}^-\!-\!p^{(\emptyset)}_\star, &  p^{(\emptyset)}_\star < p_{s'\!'t'\!'}^-,\\
p_{s'\!'t'\!'}^+\!-\!p^{(\emptyset)}_\star, &  p_{s'\!'t'\!'}^+ < p^{(\emptyset)}_\star,\\
\end{cases}
\end{split}\hspace*{-33pt}
\end{equation}
and \emph{inserted} the lower and upper Fr\'echet-boundaries of these covariances --- by the formulas:
\begin{equation}\label{Kovariance-3-insertingFBUpDownX}
\hspace*{-23pt}
\begin{split}
&\textsf{Kov}_{xyz}^{-(\{x\})}\!\!=\!\!
\begin{cases}
p_{s't'}^-\!-\!p^{(\{x\})}_\star, & p^{(\{x\})}_\star \in \left[p_{s't'}^-, p_{s't'}^+\right],\\
p_{s't'}^-\!-\!p^{(\{x\})}_\star, & p^{(\{x\})}_\star < p_{s't'}^-,\\
p_{s't'}^+\!-\!p^{(\{x\})}_\star, & p_{s't'}^+ < p^{(\{x\})}_\star,\\
\end{cases}\\
&\\
&\textsf{Kov}_{xyz}^{+(\{x\})}\!\!=\!\!
\begin{cases}
p_{s't'}^+\!-\!p^{(\{x\})}_\star, & p^{(\{x\})}_\star \in \left[p_{s't'}^-, p_{s't'}^+\right],\\
p_{s't'}^-\!-\!p^{(\{x\})}_\star, & p^{(\{x\})}_\star < p_{s't'}^-,\\
p_{s't'}^+\!-\!p^{(\{x\})}_\star, & p_{s't'}^+ < p^{(\{x\})}_\star,\\
\end{cases}\\
\end{split}\hspace*{-33pt}
\end{equation}
\begin{equation}\label{Kovariance-3-insertingFBUpDownEmpty}
\hspace*{-13pt}
\begin{split}
&\textsf{Kov}_{xyz}^{-(\emptyset)}\!\!=\!\!
\begin{cases}
p_{s'\!'t'\!'}^-\!-\!p^{(\emptyset)}_\star, &  p^{(\emptyset)}_\star \in \left[p_{s't'}^-,p_{s't'}^+\right],\\
p_{s'\!'t'\!'}^-\!-\!p^{(\emptyset)}_\star, &  p^{(\emptyset)}_\star < p_{s'\!'t'\!'}^-,\\
p_{s'\!'t'\!'}^+\!-\!p^{(\emptyset)}_\star, &  p_{s'\!'t'\!'}^+ < p^{(\emptyset)}_\star,\\
\end{cases}\\
&\\
&\textsf{Kov}_{xyz}^{+(\emptyset)}\!\!=\!\!
\begin{cases}
p_{s'\!'t'\!'}^+\!-\!p^{(\emptyset)}_\star, &  p^{(\emptyset)}_\star \in \left[p_{s't'}^-,p_{s't'}^+\right],\\
p_{s'\!'t'\!'}^-\!-\!p^{(\emptyset)}_\star, &  p^{(\emptyset)}_\star < p_{s'\!'t'\!'}^-,\\
p_{s'\!'t'\!'}^+\!-\!p^{(\emptyset)}_\star, &  p_{s'\!'t'\!'}^+ < p^{(\emptyset)}_\star,\\
\end{cases}
\end{split}\hspace*{-43pt}
\end{equation}

We introduce some notation for brevity of the formulas:
\begin{equation}\label{p0}
\begin{split}
p^{(\{x\})}_0 &=\begin{cases}
p^{\{x\}}_\star, \ \ \mbox{if }  p^{\{x\}}_\star \in \left[ p_{s't'}^+, p_{s't'}^-\right],&\\
&\\
p_{s't'}^-+\frac{\left(p_{s't'}^--p^{\{x\}}_\star\right)\left(p_{s't'}^+-p_{s't'}^-\right)}{p_{s't'}^++p_{s't'}^--2p^{(\{x\})}_\star},&\\
\mbox{if } p^{\{x\}}_\star \not\in \left[ p_{s't'}^+, p_{s't'}^-\right],&
\end{cases}\\
&\\
p^{(\emptyset)}_0 &=\begin{cases}
p^{\emptyset}_\star, \ \ \mbox{если } p^{\emptyset}_\star \in \left[ p_{s't'}^+, p_{s't'}^-\right],&\\
&\\
p_{s'\!'t'\!'}^-+\frac{\left(p_{s'\!'t'\!'}^--p^{(\emptyset)}_\star\right)\left(p_{s't'}^+-p_{s't'}^-\right)}{p_{s'\!'t'\!'}^++p_{s'\!'t'\!'}^--2p^{(\emptyset)}_\star},&\\
\mbox{если } p^{\emptyset}_\star \not\in \left[ p_{s't'}^+, p_{s't'}^-\right],&
\end{cases}
\end{split}
\end{equation}
Two \emph{inserted} triple covariances are defined by the formulas:
\begin{equation}\label{Kovariance-3-inserting-2}
\hspace*{-3pt}
\begin{split}
\textsf{Kov}_{xyz}^{(\{x\})}&=p_{s't'}-p^{(\{x\})}_0\\
\textsf{Kov}_{xyz}^{(\emptyset)}&=p_{s''t''}-p^{(\emptyset)}_0,
\end{split}\hspace*{-33pt}
\end{equation}
and \emph{inserted} the lower and upper Fr\'echet-boundaries of these covariances --- by the formulas:
\begin{equation}\label{Kovariance-3-insertingFBUpDownX-2}
\hspace*{-23pt}
\begin{split}
\textsf{Kov}_{xyz}^{-(\{x\})}&=p_{s't'}^--p^{(\{x\})}_0,\\
\textsf{Kov}_{xyz}^{+(\{x\})}&=p_{s't'}^+-p^{(\{x\})}_0,\\
&\\
\textsf{Kov}_{xyz}^{-(\emptyset)}&=p_{s''t''}^--p^{(\emptyset)}_0,\\
\textsf{Kov}_{xyz}^{+(\emptyset)}&=p_{s''t''}^+-p^{(\emptyset)}_0.
\end{split}\hspace*{-43pt}
\end{equation}

$$
\star\star\star
$$

For any modification definitions of two \emph{inserted} Fr\'echet-correlations look in the usual way:
\begin{equation}\label{Frechet-korrelation-3-inserting}
\begin{split}
&\textsf{Kor}_{xyz}^{(\{x\})}=
\begin{cases}
\frac{\textsf{Kov}_{xyz}^{(\{x\})}}{\left|\textsf{Kov}_{xyz}^{-(\{x\})}\right|}, & \textsf{Kov}_{xyz}^{(\{x\})} < 0,\\
&\\[-8pt]
\frac{\textsf{Kov}_{xyz}^{(\{x\})}}{\textsf{Kov}_{xyz}^{+(\{x\})}}, & \textsf{Kov}_{xyz}^{(\{x\})} \geqslant 0,\\
\end{cases}\\
&\\
&\textsf{Kor}_{xyz}^{(\emptyset)}=
\begin{cases}
\frac{\textsf{Kov}_{xyz}^{(\emptyset)}}{\left|\textsf{Kov}_{xyz}^{-(\emptyset)}\right|}, & \textsf{Kov}_{xyz}^{(\emptyset)} < 0,\\
&\\[-8pt]
\frac{\textsf{Kov}_{xyz}^{(\emptyset)}}{\textsf{Kov}_{xyz}^{+(\emptyset)}}, & \textsf{Kov}_{xyz}^{(\emptyset)} \geqslant 0.
\end{cases}
\end{split}
\end{equation}

\subsection{Parametrization of functional parameters $p_{s't'}$ and $p_{s'\!'t'\!'}$}

The Fr\'echet-restriction of two functional parameters $p_{s't'}\!=\!p_{s't'}(p_x,p_y,p_z)$~и~$p_{s'\!'t'\!'}\!=\!p_{s'\!'t'\!'}(p_x,p_y,p_z)$
\begin{equation}\label{2-Kopula-pst}
\begin{split}
p_{s't'}^- \leqslant p_{s't'} \leqslant p_{s't'}^+,\\
p_{s'\!'t'\!'}^- \leqslant p_{s'\!'t'\!'} \leqslant p_{s'\!'t'\!'}^+,\\
\end{split}
\end{equation}
that in the frame method have a sense of probabilities of triple intersections of events:
\begin{equation}\label{2-Kopula-pst1}
\begin{split}
p_{s't'} = \mathbf{P}(x \cap y \cap z),\\
p_{s'\!'t'\!'} = \mathbf{P}(x^c \cap y \cap z),
\end{split}
\end{equation}
is based on the notion of the \emph{inserted} triple Fr\'echet-correlation.

From definitions (\ref{Frechet-korrelation-3-inserting}) and (\ref{FRECHET-bounds-3UpDown-bis}) we get the parametrization of functional parameter $p_{s't'}$ of the \emph{inserted} triple Fr\'echet-correlation $\textsf{Kor}_{xyz}^{(\{x\})}$ in the following form:
\begin{equation}\label{parametrization-pst1}
\begin{split}
&p_{s't'}(p_x,p_y,p_z)=\\
&=\begin{cases}
p_xp_yp_z-\textsf{Kor}_{xyz}^{(\{x\})}\textsf{Kov}_{xyz}^{-(\{x\})},&\\
\textsf{Kor}_{xyz}^{(\{x\})} < 0,&\\
&\\[-8pt]
p_xp_yp_z+\textsf{Kor}_{xyz}^{(\{x\})}\textsf{Kov}_{xyz}^{+(\{x\})},&\\
\textsf{Kor}_{xyz}^{(\{x\})} \geqslant 0,&
\end{cases}
\end{split}
\end{equation}
The parametrization of functional parameter $p_{s'\!'t'\!'}$ of the \emph{inserted} triple Fr\'echet-correlation $\textsf{Kor}_{xyz}^{(\emptyset)}$ follows from the same definitions (\ref{Frechet-korrelation-3-inserting}) and (\ref{FRECHET-bounds-3UpDown-bis}):
\begin{equation}\label{parametrization-pst2}
\begin{split}
&p_{s'\!'t'\!'}(p_x,p_y,p_z)=\\
&=\begin{cases}
p_xp_yp_z-\textsf{Kor}_{xyz}^{(\emptyset)}\textsf{Kov}_{xyz}^{-(\emptyset)},&\\
\textsf{Kor}_{xyz}^{(\{x\})} < 0,&\\
&\\[-8pt]
p_xp_yp_z+\textsf{Kor}_{xyz}^{(\emptyset)}\textsf{Kov}_{xyz}^{+(\emptyset)},&\\
\textsf{Kor}_{xyz}^{(\{x\})} \geqslant 0,&
\end{cases}
\end{split}
\end{equation}

\texttt{Note \!\refstepcounter{ctrnot}\arabic{ctrnot}\,\label{not1}\itshape\footnotesize (about parametrization of functional parameters of 3-Kopula by Fr\'echet-correlations).} The parametrization of the four functional parameters $p_{s'}, p_{t'}, p_{s't'}$ and $p_{s'\!'t'\!'}$ of 3-Kopula of the $\breve{p}$-ordered half-rare triplet $\frak{X}=\{x,y,z\}$ by two double Fr\'echet-correlations $\textsf{Kor}_{xy}$ and $\textsf{Kor}_{xz}$ (\ref{Frechet-korrelation-2}) and by two \emph{inserted} triple Fr\'echet-correlations $\textsf{Kor}_{xyz}^{(\{x\})}$ and $\textsf{Kor}_{xyz}^{(\emptyset)}$ (\ref{Frechet-korrelation-3-inserting}) has the following advantages. Each of four Fr\'echet-correlations is a numerical characteristics of dependency of events with values from fixed interval $[-1,+1]$. And these values clearly indicate the proximity to Fr\'echet-boundaries and to indepedent 3-Kopula. The value ``$-1$'' indicates to the lower Fr\'echet-boundary, the value ``$+1$'' --- to the upper Fr\'echet-boundary, and the value ``$0$'' --- to independent events.
For example, the equality of all these four Fr\'echet-correlations to zero determines a family of independent 3-Kopulas. Advantages of the proposed idea of parametrization of functional parameters of 3-Kopula are that
\begin{itemize}
\parskip=-3pt
\item
an each Fr\'echet-correlation can take arbitrary value from $[-1,+1]$ without any connection with the values of the other three Fr\'echet-correlations;
\item
the above parametrization algorithm for functional parameters of 3-Kopula extends to the parametrization of the functional parameters of $N$-Kopulas \emph{by inserted Fr\'echet-correlations of higher orders}.
\end{itemize}

\section{Examples of Kopulas of some families of sets of events\label{Kopulas-examples}}

\subsection{Examples of different 2-Kopulas with a functional parameter within Frechet boundaries}

\begin{figure}[h!]
\centering
\includegraphics[width=3.3in]{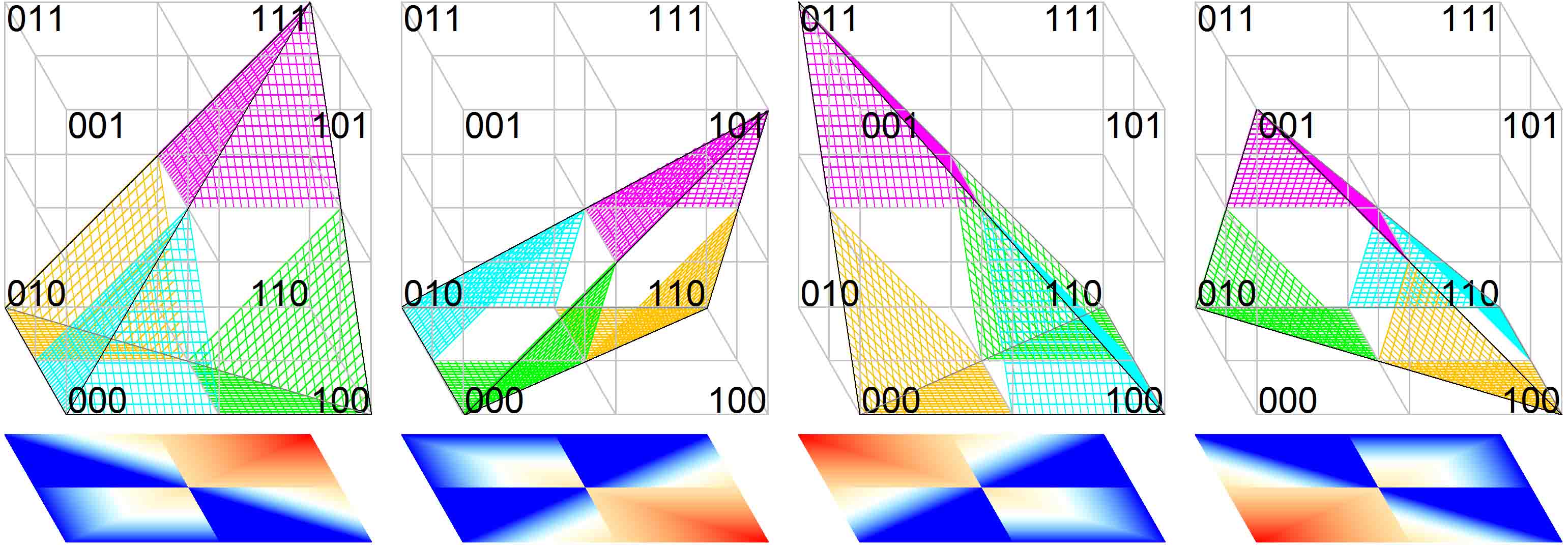}

\includegraphics[width=3.3in]{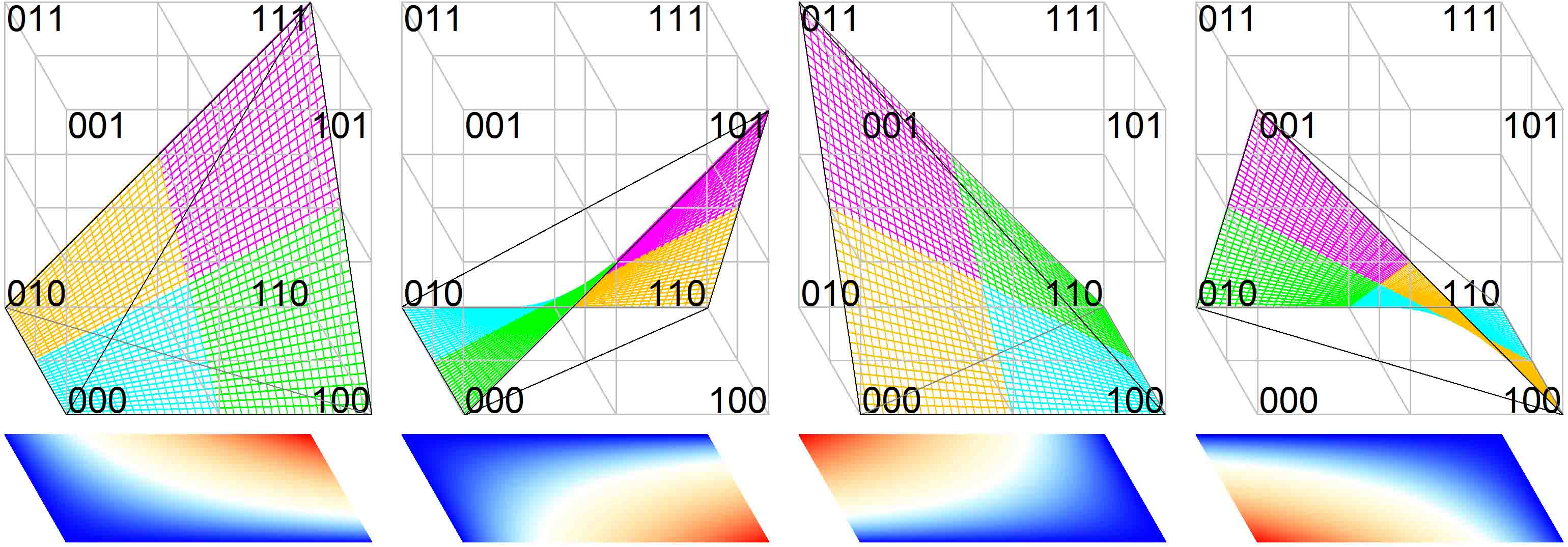}

\includegraphics[width=3.3in]{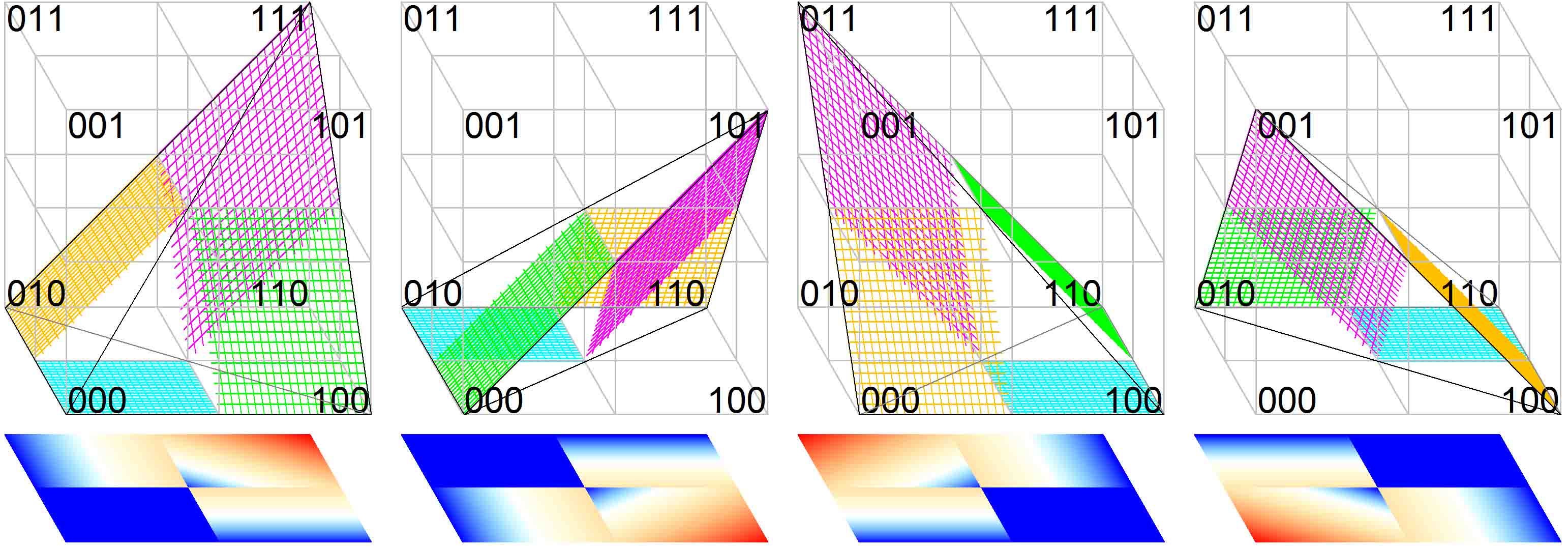}

\caption{Cartesian representations of 2-Kopulas of doublets of half-rare events $\frak{X}=\{x,y\}$ and its set-phenomena $\frak{X}^{(c|x)}$, $\frak{X}^{(c|y)}$ и $\frak{X}^{(c|xy)}$ corresponding to Frechet Kopula (from up to down): upper (embedded), independent, lower (minimum-intersected).} %
\label{fig-2-Kopulas-Frechet}
\end{figure}

Consider in Fig.\footnote{In each figure, below the graph, maps of these 2-Kopulas on unit squares in conditional colors are shown too, where the white color corresponds to the points at which the probabilities of terraced events are 1/4.} \ref{fig-2-Kopulas-Frechet}
a number of examples of 2-Kopulas of doublets of half-rare events $\frak{X}=\{x,y\}$ and its set-phenomena $\frak{X}^{(c|x)}=\{x,y^c\}$, $\frak{X}^{(c|y)}=\{x^c,y\}$, and $\frak{X}^{(c|xy)}=\{x^c,y^c\}$, each of which is characterized by its own functional parameter $\mathbf{P}(x \cap y) = p_{xy}(w_x,w_y)$, lying within the  Fr\'echet boundaries:
\begin{equation}\label{func-parameter}
\begin{split}
0 \leqslant p_{xy}(w_x,w_y) \leqslant \min \{w_x, w_y\}.
\end{split}
\end{equation}

Upper 2-Kopula of Fr\'echet (embedded):
\begin{equation}\label{2-Kopula-UP}
\begin{split}
p_{xy}(w_x,w_y) = \min\{w_x, w_y\}.
\end{split}
\end{equation}

Independent 2-Kopula of Fr\'echet:
\begin{equation}\label{2-Kopula-IND}
\begin{split}
p_{xy}(w_x,w_y) = w_x w_y.
\end{split}
\end{equation}

Lower 2-Kopula of Fr\'echet (minimum-intersected):
\begin{equation}\label{2-Kopula-DOWN}
\begin{split}
p_{xy}(w_x,w_y) = \max\{0,w_x+w_y-1\}.
\end{split}
\end{equation}

Half-independent 2-Kopula:
\begin{equation}\label{2-Kopula-ind}
\begin{split}
p_{xy}(w_x,w_y) = w_x w_y/2.
\end{split}
\end{equation}

Half-embedded 2-Kopula:
\begin{equation}\label{2-Kopula-ind}
\begin{split}
p_{xy}(w_x,w_y) = \min\{w_x, w_y\}/2.
\end{split}
\end{equation}

Arbitrary-embedded 2-Kopula:
\begin{equation}\label{2-Kopula-ind}
\begin{split}
&p_{xy}(w_x,w_y) =\\
&=\min\{w_x, w_y\}(1+\sin(15(w_x-w_y)))/2.
\end{split}
\end{equation}

Continuously-arbitrary-embedded 2-Kopula:
\begin{equation}\label{2-Kopula-ind}
\begin{split}
&p_{xy}(w_x,w_y) =\\
&= w_xw_y+(\alpha(w_x,w_y)-w_xw_y) \beta(w_x,w_y),
\end{split}
\end{equation}
where
\begin{equation}\label{2-Kopula-ind}
\begin{split}
&\alpha(w_x,w_y)=\\
&=\min\{w_x, w_y\}(1+\sin(15(w_x-w_y)))/2,\\
&\beta(w_x,w_y)=\sqrt[4]{(1/2-w_x)(1/2-w_y)}.
\end{split}
\end{equation}

\subsection{Examples of 2-Kopulas with a functional parameter corresponding to some classical copulas}

\begin{figure}[h!]
\centering
\includegraphics[width=3.3in]{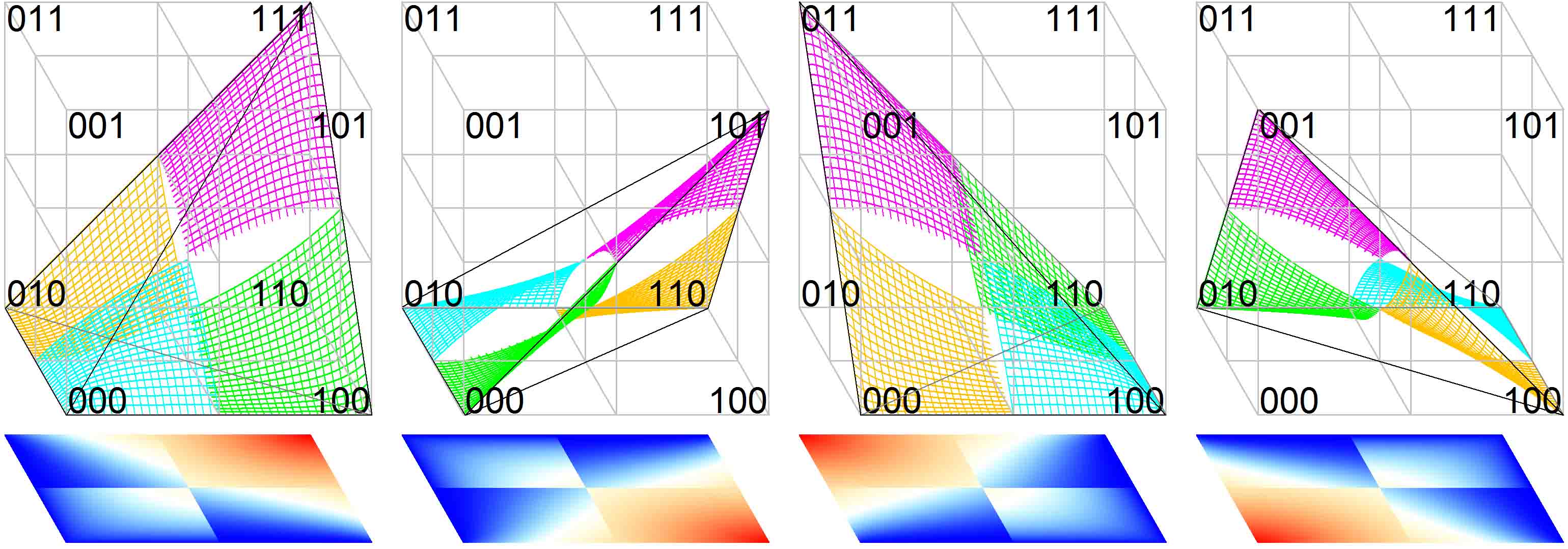}

\includegraphics[width=3.3in]{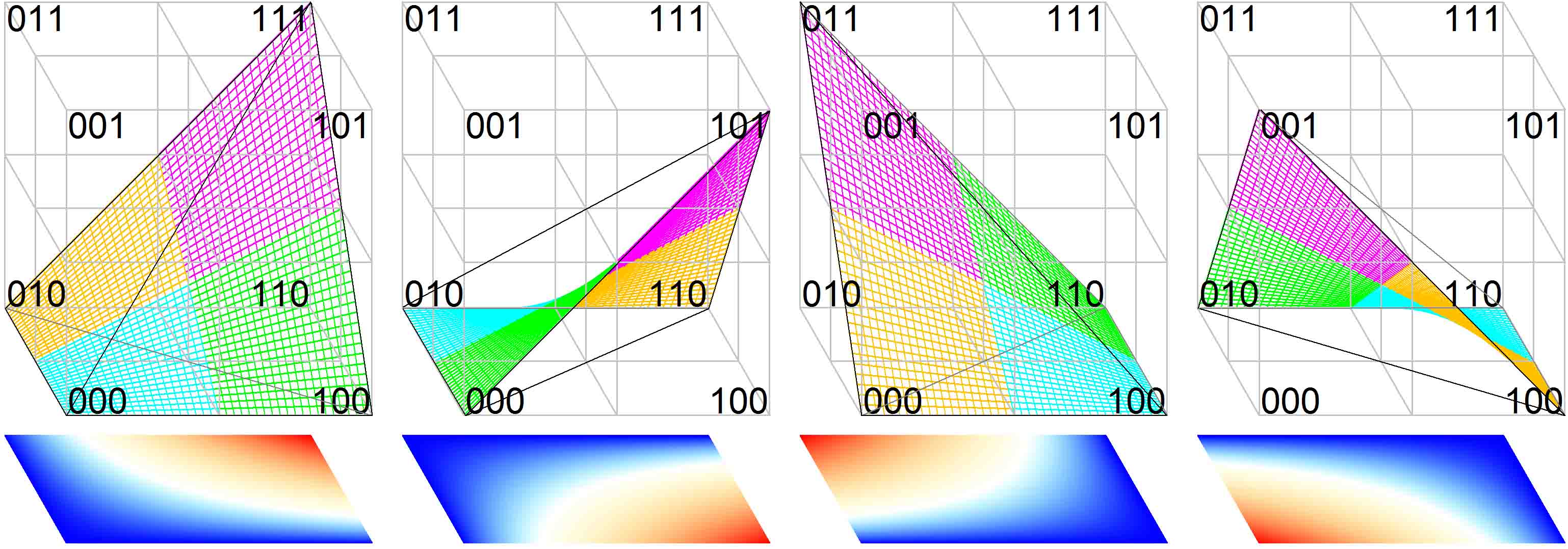}

\includegraphics[width=3.3in]{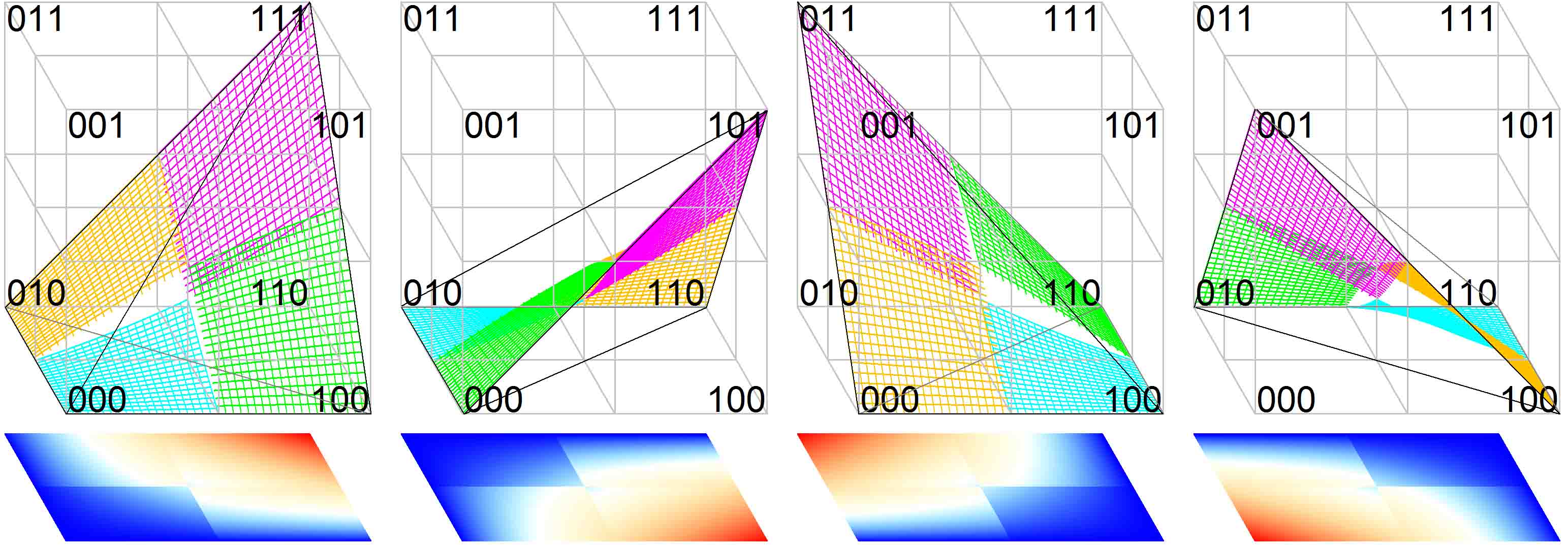}

\caption{Cartesian representations of 2-Kopulas of doublets of half-rare events $\frak{X}=\{x,y\}$ and its set-phenomena $\frak{X}^{(c|x)}$, $\frak{X}^{(c|y)}$ и $\frak{X}^{(c|xy)}$ corresponding to Ali-Mikhail-Haq Kopula (from up to down): from near-upper  ($\theta=0.999$) through independent ($\theta=0$) to lower ($\theta=-1.0$).} %
\label{fig-Kopulas-AMH}
\end{figure}

\begin{figure}[h!]
\centering
\includegraphics[width=3.3in]{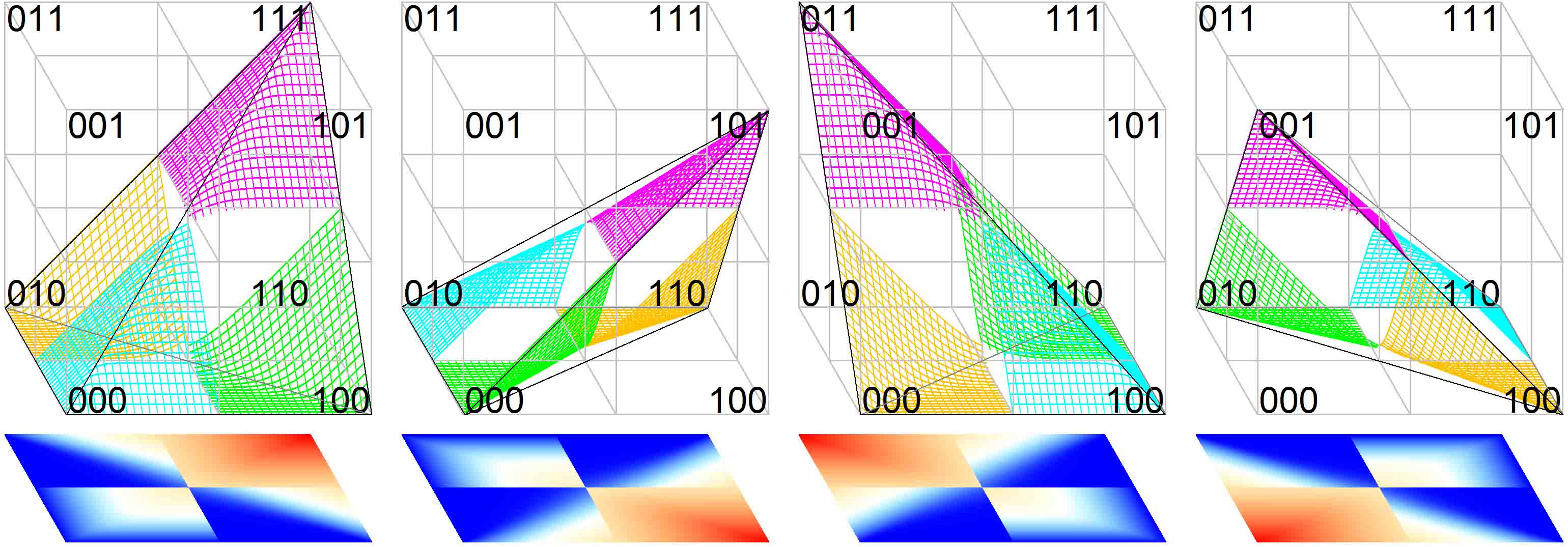}

\includegraphics[width=3.3in]{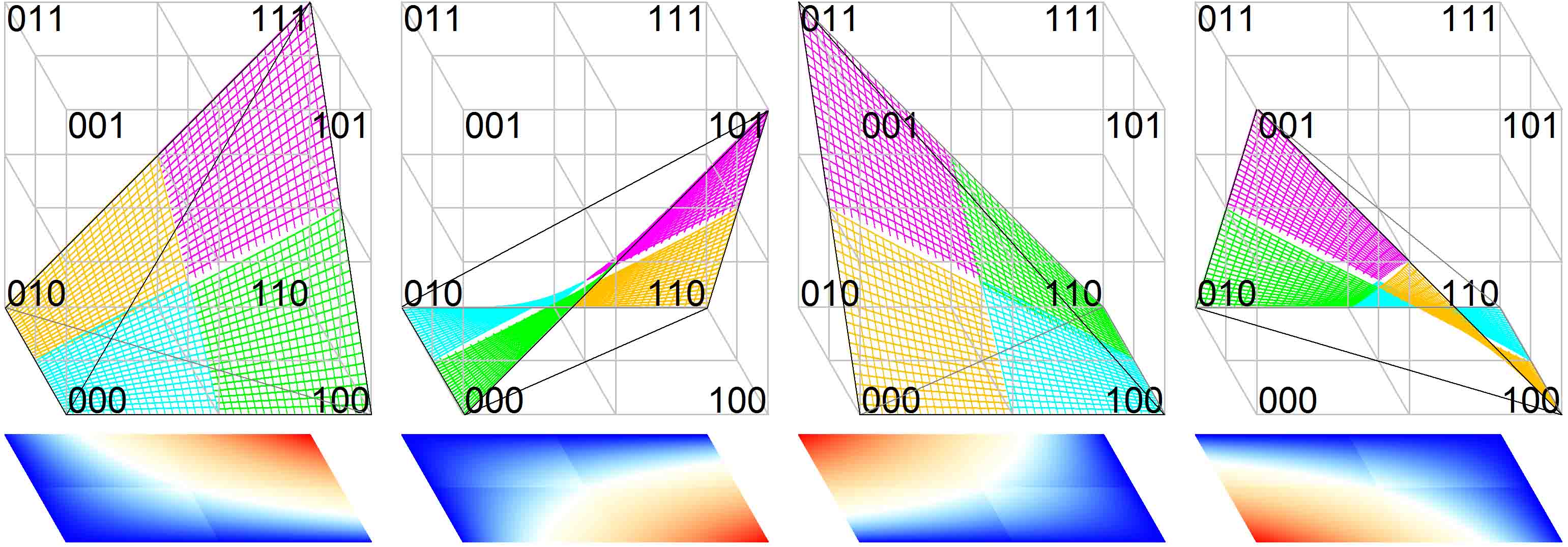}

\includegraphics[width=3.3in]{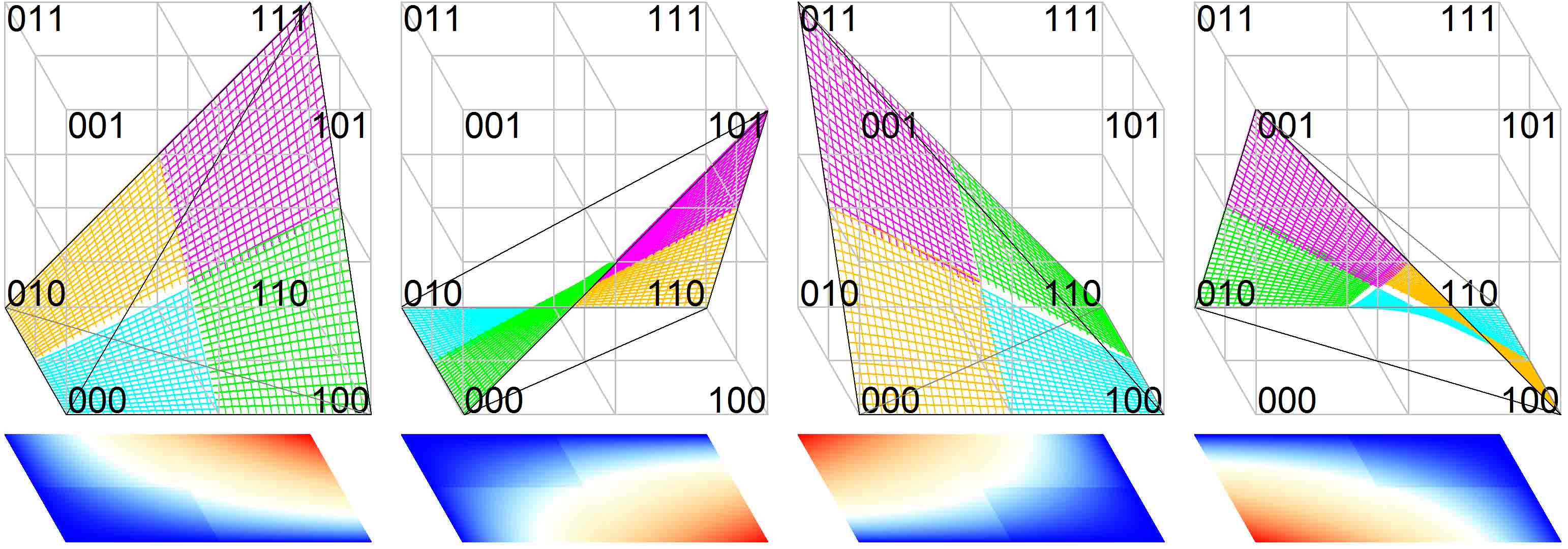}

\includegraphics[width=3.3in]{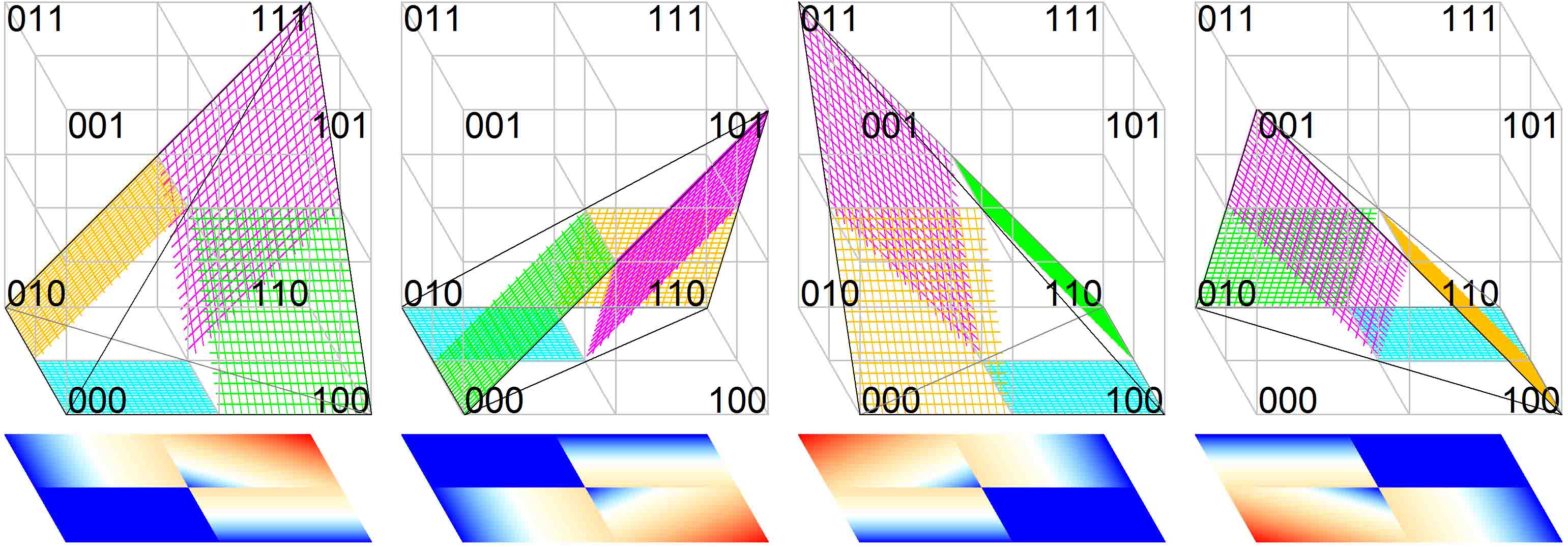}

\caption{Cartesian representations of 2-Kopulas of doublets of half-rare events $\frak{X}=\{x,y\}$ and its set-phenomena $\frak{X}^{(c|x)}$, $\frak{X}^{(c|y)}$ и $\frak{X}^{(c|xy)}$ corresponding to Clayton Kopula (from up to down): from near-upper  ($\theta=6.5$) around pinked-independent ($\theta=0.1, \, -0.1$) to lower ($\theta=-1.0$).} %
\label{fig-Kopulas-CLAYTON}
\end{figure}

\begin{figure}[h!]
\centering
\includegraphics[width=3.3in]{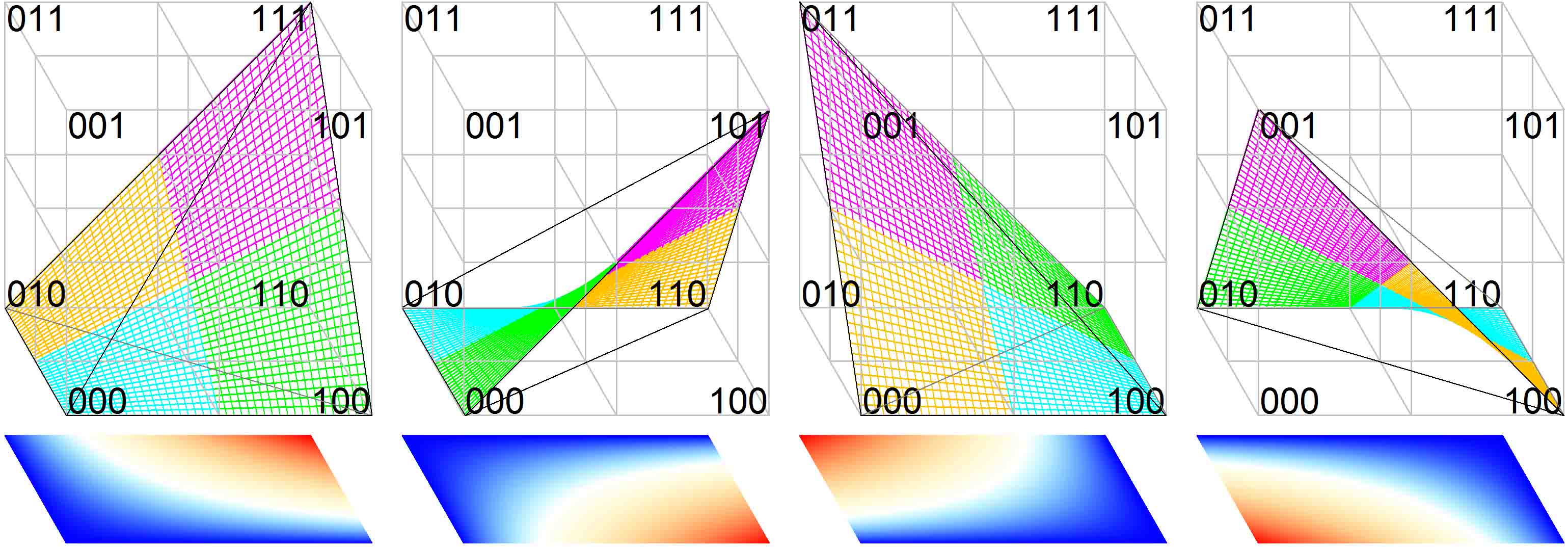}

\includegraphics[width=3.3in]{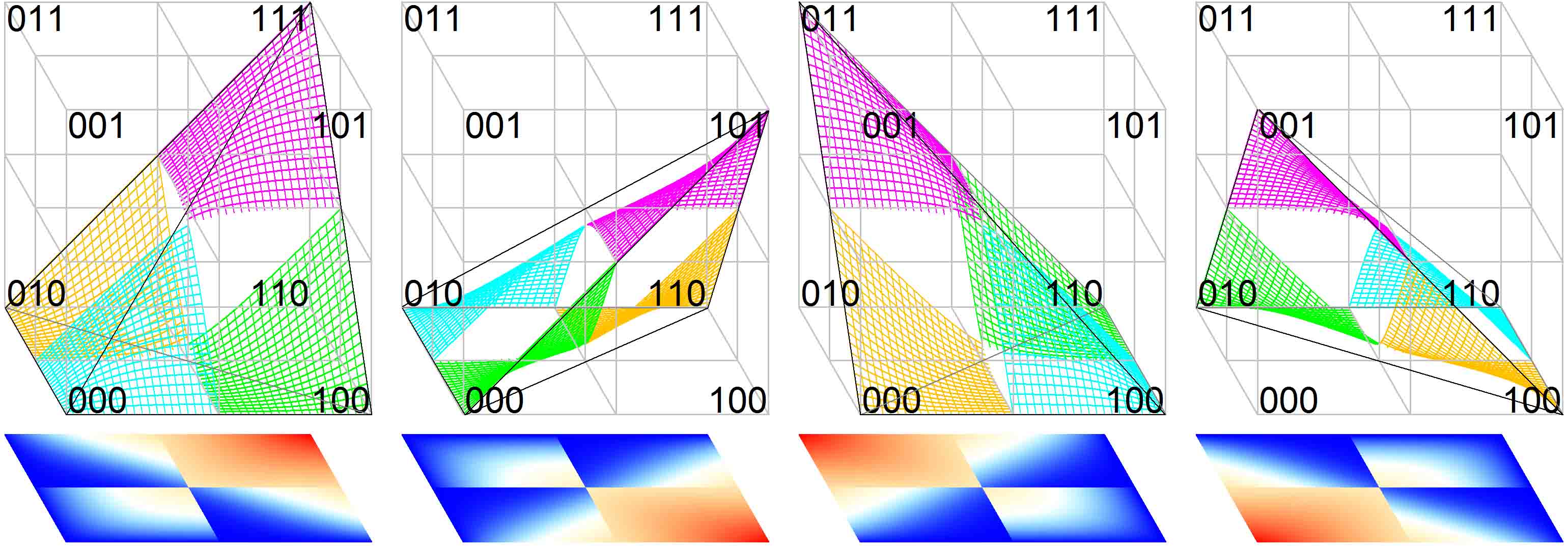}

\caption{Cartesian representations of 2-Kopulas of doublets of half-rare events $\frak{X}=\{x,y\}$ and its set-phenomena $\frak{X}^{(c|x)}$, $\frak{X}^{(c|y)}$ и $\frak{X}^{(c|xy)}$ corresponding to Joe Kopula (from up to down): from independent  ($\theta=1.0$) to near-lower ($\theta=6.5$).} %
\label{fig-Kopulas-JOE}
\end{figure}

\begin{figure}[h!]
\centering
\includegraphics[width=3.3in]{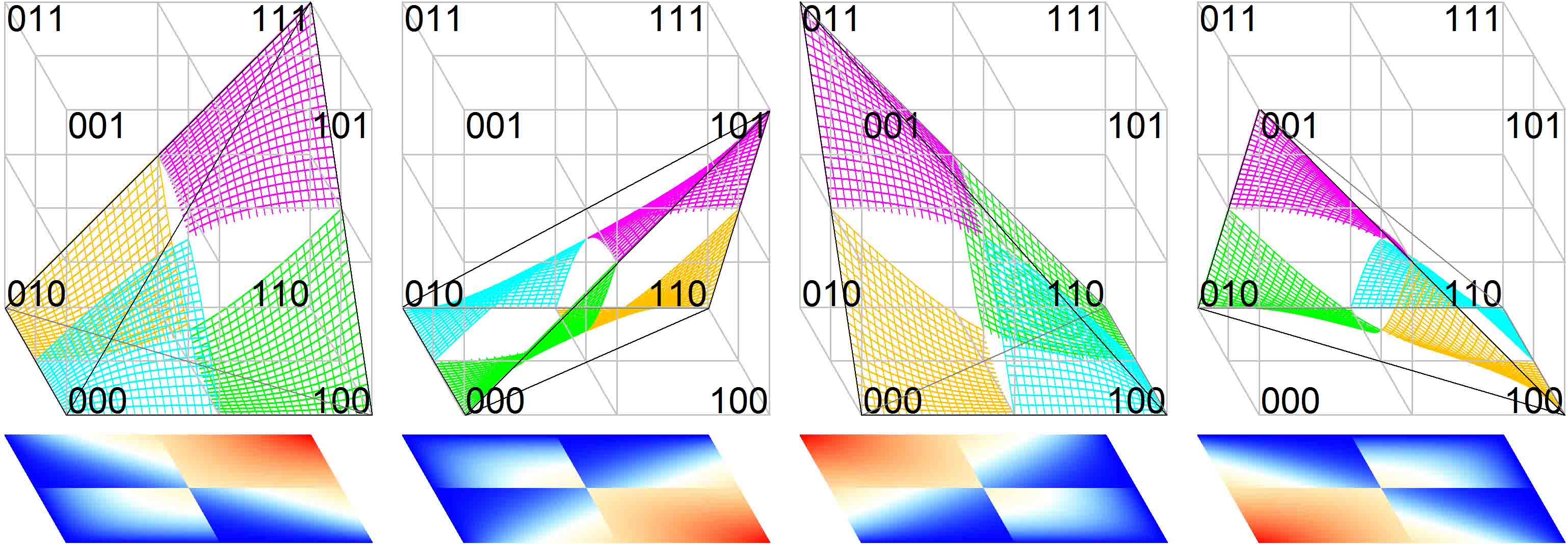}

\includegraphics[width=3.3in]{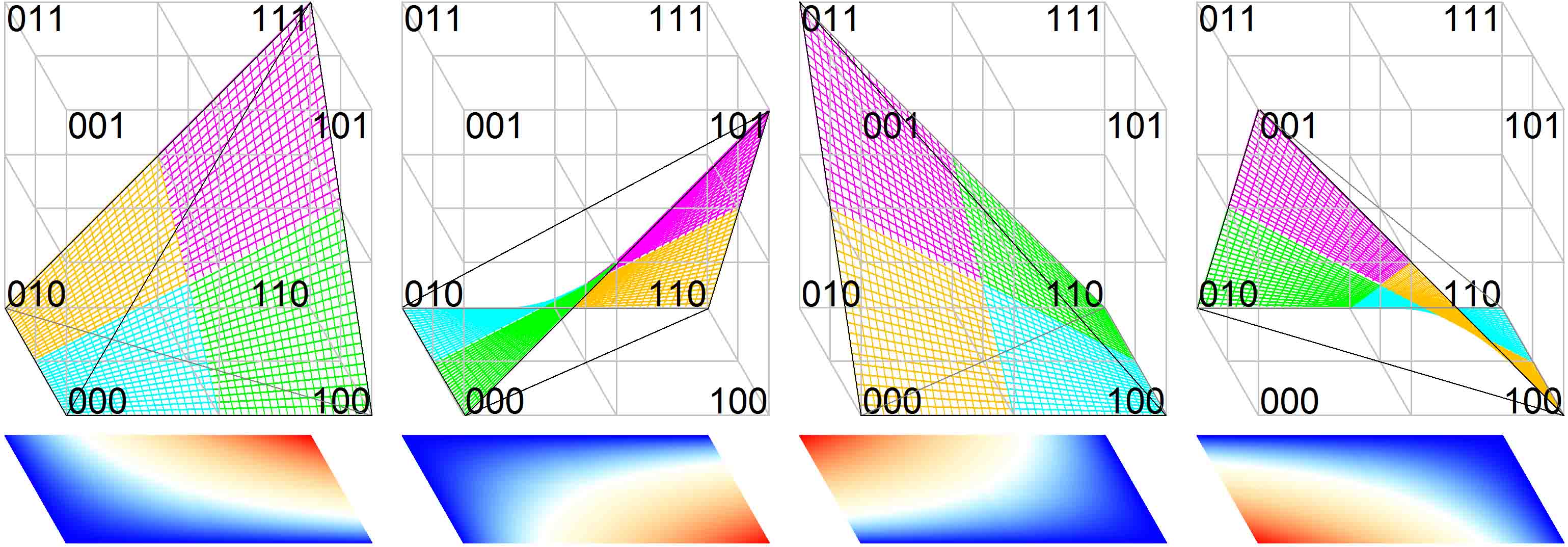}

\includegraphics[width=3.3in]{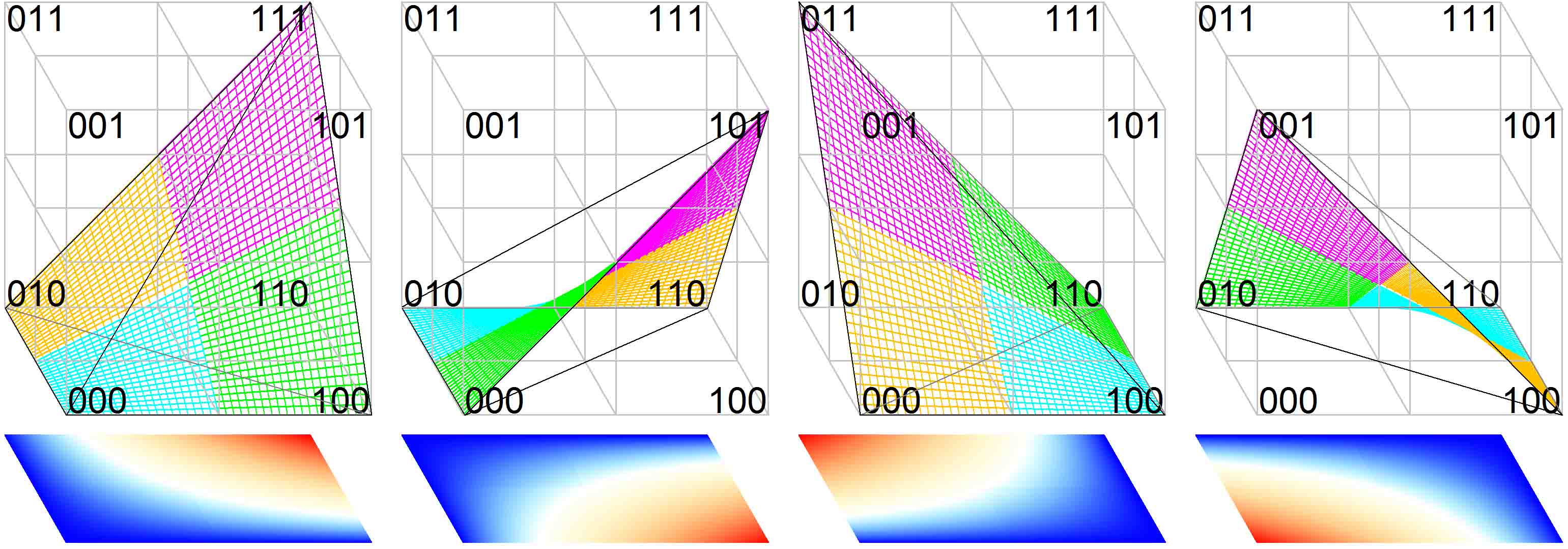}

\includegraphics[width=3.3in]{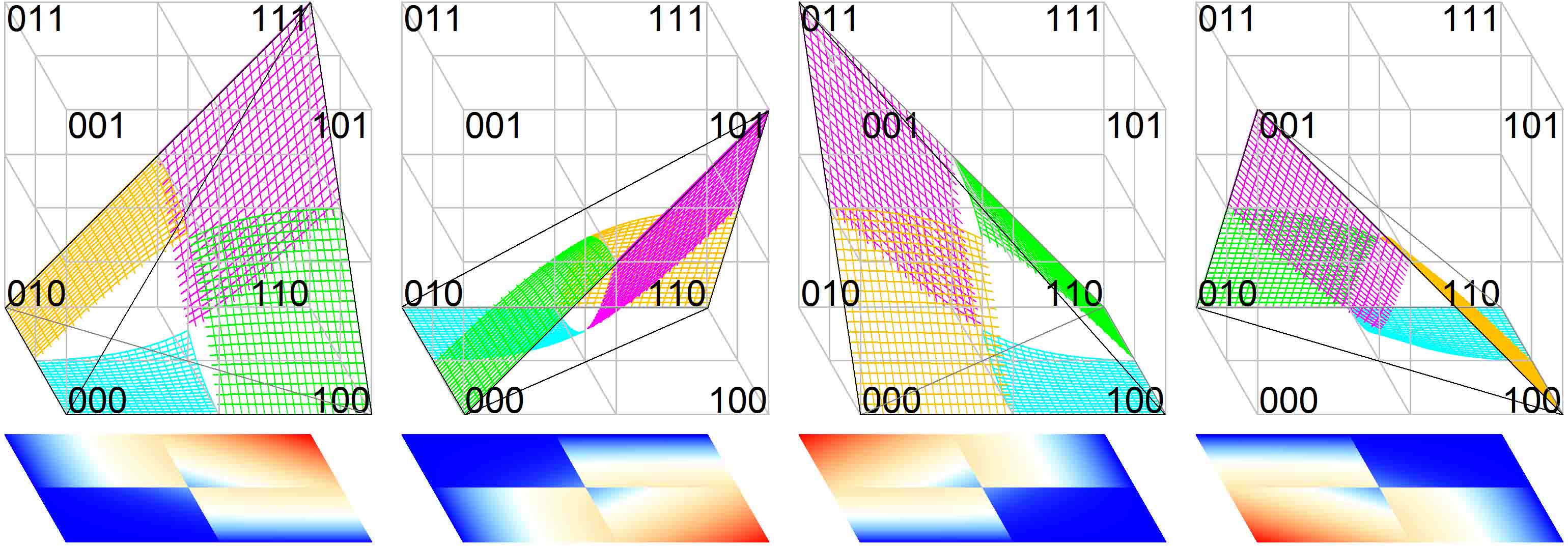}

\caption{Cartesian representations of 2-Kopulas of doublets of half-rare events $\frak{X}=\{x,y\}$ and its set-phenomena $\frak{X}^{(c|x)}$, $\frak{X}^{(c|y)}$ и $\frak{X}^{(c|xy)}$ corresponding to Frank Kopula (from up to down): from near-upper  ($\theta=6.5$) around pinked-independent ($\theta=0.1, \, -0.1$) to near-lower ($\theta=-6.5$).} %
\label{fig-Kopulas-FRANK}
\end{figure}

\begin{figure}[h!]
\centering
\includegraphics[width=3.3in]{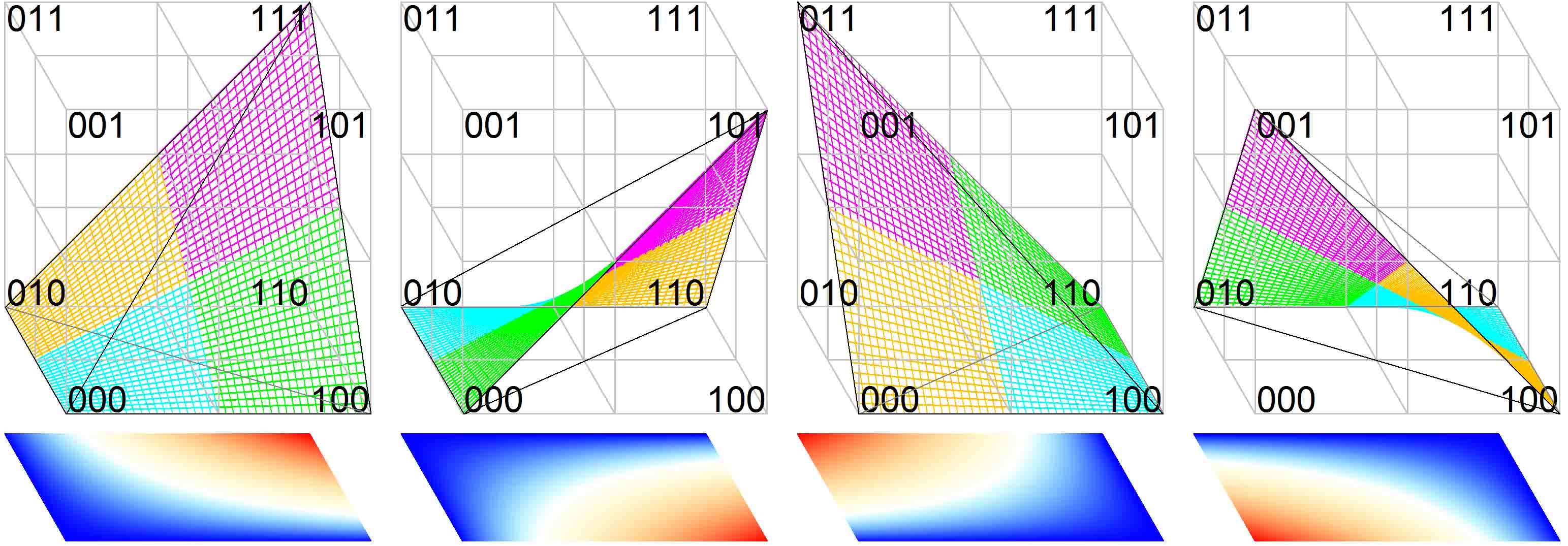}

\includegraphics[width=3.3in]{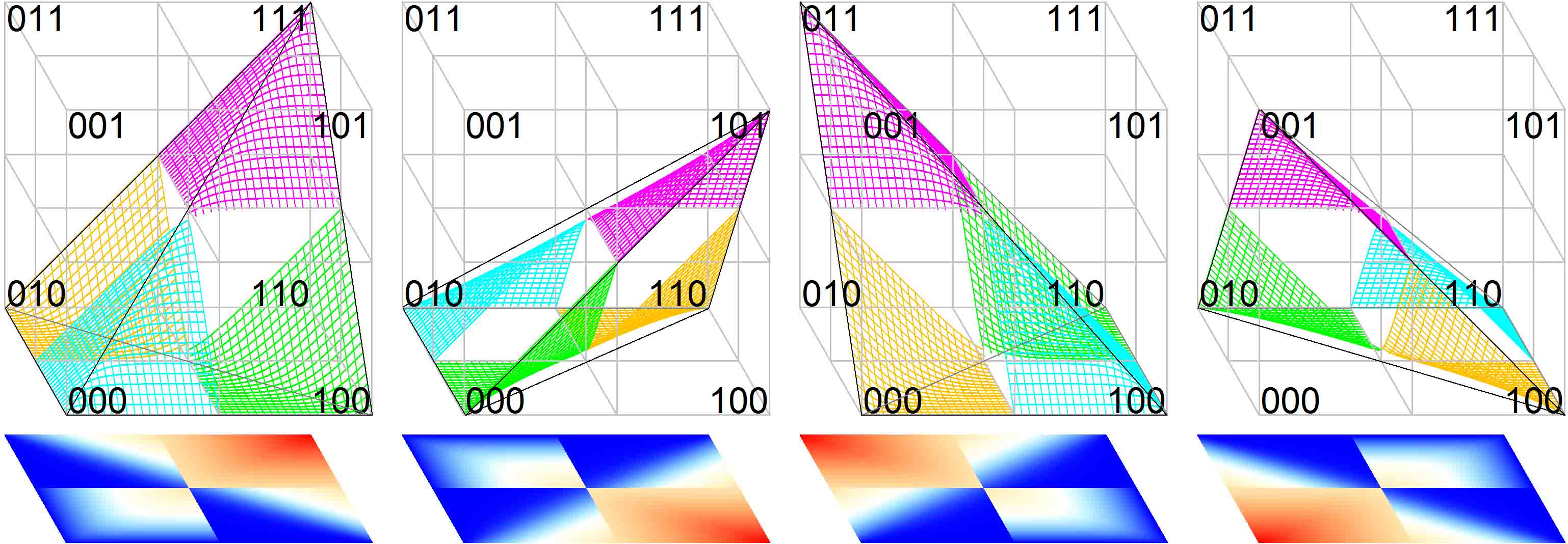}

\caption{Cartesian representations of 2-Kopulas of doublets of half-rare events $\frak{X}=\{x,y\}$ and its set-phenomena $\frak{X}^{(c|x)}$, $\frak{X}^{(c|y)}$ и $\frak{X}^{(c|xy)}$ corresponding to Gumbel Kopula (from up to down): from inndependent  ($\theta=1.0$) to near-lower ($\theta=6.5$).} %
\label{fig-Kopulas-GUMBEL}
\end{figure}

In Fig.s \ref{fig-Kopulas-AMH}, \ref{fig-Kopulas-CLAYTON}, \ref{fig-Kopulas-FRANK}, \ref{fig-Kopulas-GUMBEL}, and \ref{fig-Kopulas-JOE} it is shown 2-Kopulas of doublets of half-rare events $\frak{X}=\{x,y\}$ and its set-phenomena $\frak{X}^{(c|x)}$, $\frak{X}^{(c|y)}$, and $\frak{X}^{(c|xy)}$, corresponding to some classical copulas.

2-Kopula of Ali-Mikhail-Haq, $\theta \in [-1,1)$:
\begin{equation}\label{2-Kopula-AMH}
\begin{split}
&p_{xy}(w_x,w_y) =\\
&= \frac{w_x w_y}{1-\theta(1-w_x)(1-w_y)}.
\end{split}
\end{equation}

2-Kopula of Clayton, $\theta \in [-1,\infty)-\{0\}$:
\begin{equation}\label{2-Kopula-CLAYTON}
\begin{split}
&p_{xy}(w_x,w_y) =\\
&= \left[ \max\left\{ w_x^{-\theta} + w_y^{-\theta} -1 ; \, 0 \right\} \right]^{-1/\theta}.
\end{split}
\end{equation}

2-Kopula of Joe, $\theta \in [1,\infty)$:
\begin{equation}\label{2-Kopula-JOE}
\small
\begin{split}
&p_{xy}(w_x,w_y) =\\
&= 1\!-\!\left[ (1\!-\!w_x)^\theta\!\!+\!(1\!-\!w_y)^\theta\!\!-\!(1\!-\!w_x)^\theta(1\!-\!w_y)^\theta \right]^{1/\theta}\!\!.
\end{split}
\end{equation}

2-Kopula of Frank, $\theta \in \mathbb{R}-\{0\}$:
\begin{equation}\label{2-Kopula-FRANK}
\small
\begin{split}
&p_{xy}(w_x,w_y) =\\
&= -\frac{1}{\theta} \log\!\left[ 1+\frac{(\exp(-\theta w_x)-1)(\exp(-\theta w_y)-1)}{\exp(-\theta)-1} \right].
\end{split}
\end{equation}

2-Kopula of Gumbel, $\theta \in [1,\infty)$:
\begin{equation}\label{2-Kopula-GUMBEL}
\begin{split}
&p_{xy}(w_x,w_y) =\\
&= \exp\!\left[ -\left( (-\log(w_x))^\theta + (-\log(w_y))^\theta \right)^{1/\theta} \right].
\end{split}
\end{equation}

\subsection{Examples of 3-Kopulas, functional parameters of which serve Fr\'echet-correlations of events}

\begin{figure}[ht!]
\vspace*{-10pt}
\centering
\includegraphics[width=3.3in]{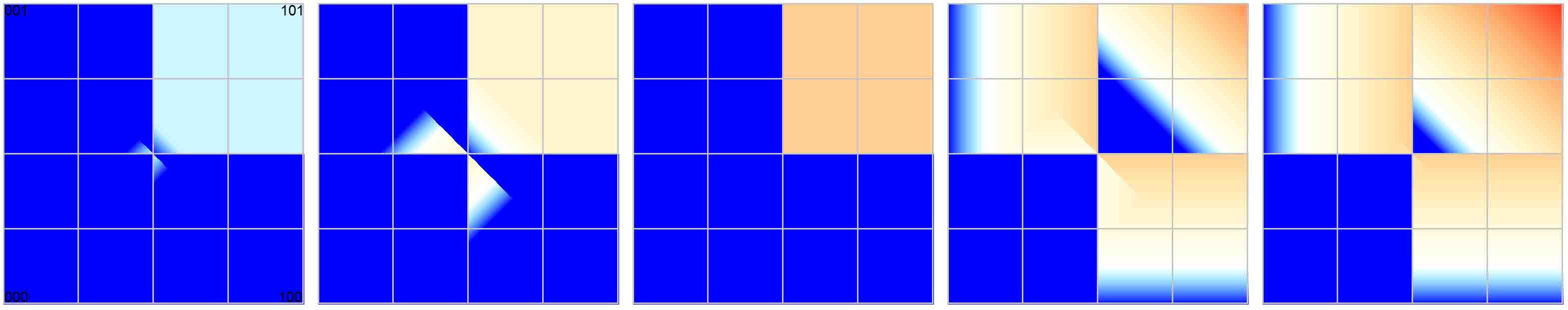}\\[-5pt]
$\textsf{Kor}=-1$\\[0pt]

\includegraphics[width=3.3in]{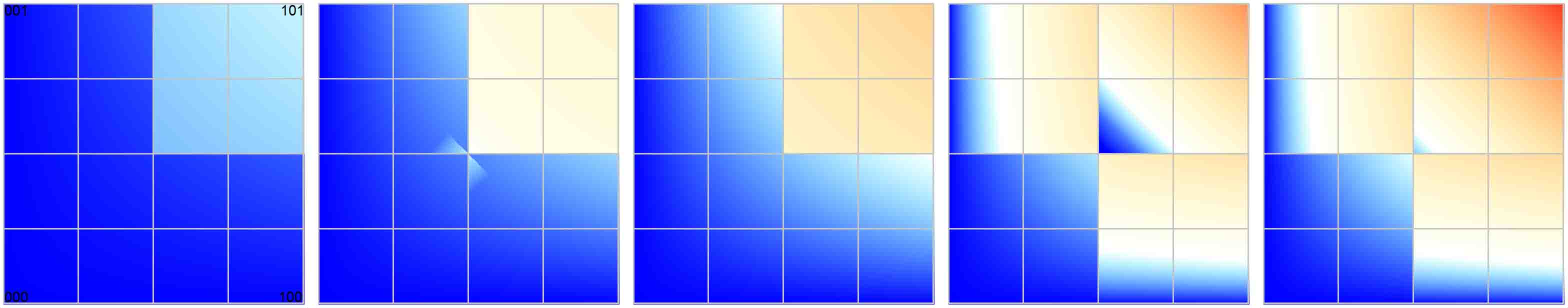}\\[-5pt]
$\textsf{Kor}=-0.5$\\[0pt]

\includegraphics[width=3.3in]{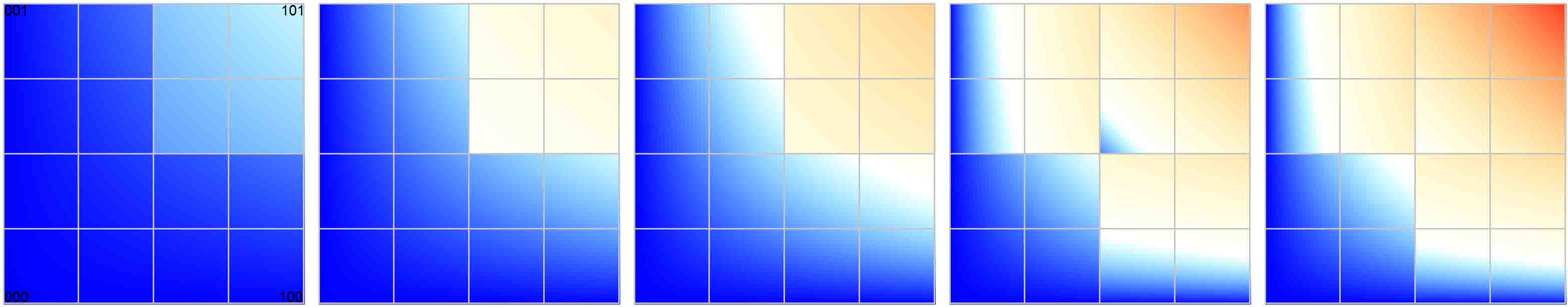}\\[-5pt]
$\textsf{Kor}=-0.3$\\[0pt]

\includegraphics[width=3.3in]{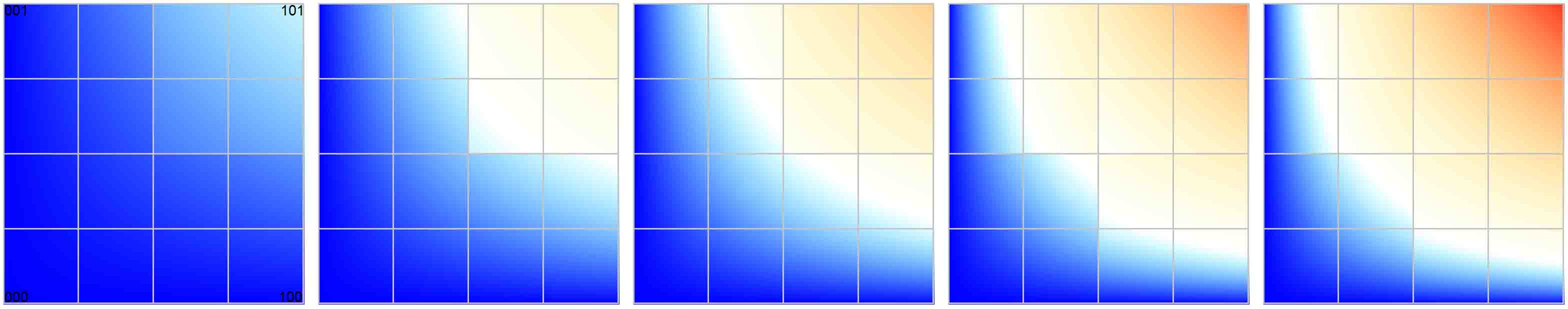}\\[-5pt]
$\textsf{Kor}=-0.1$\\[0pt]

\includegraphics[width=3.3in]{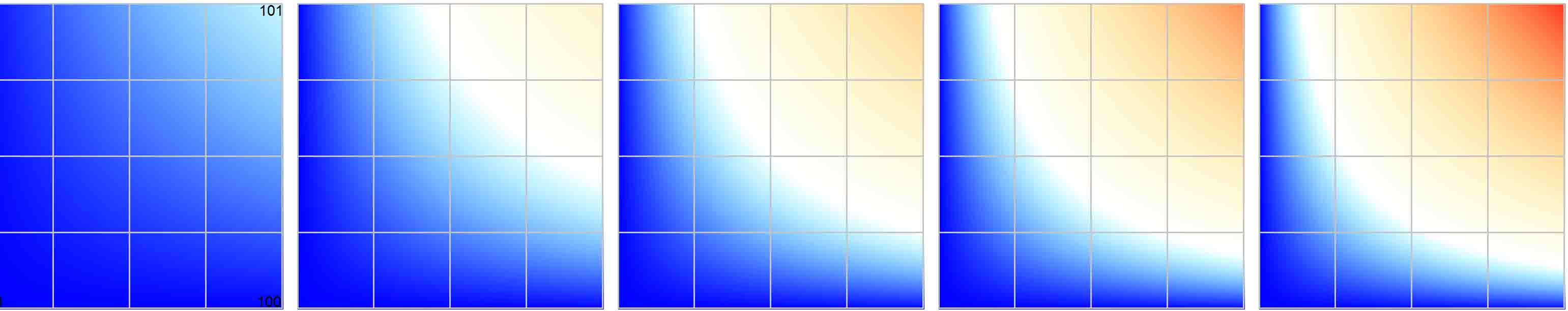}\\[-5pt]
$\textsf{Kor}=0$\\[0pt]

\includegraphics[width=3.3in]{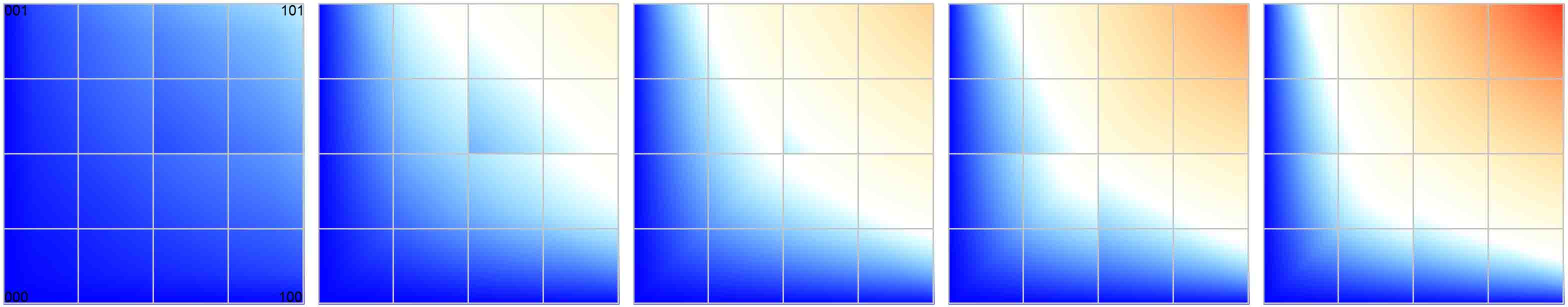}\\[-5pt]
$\textsf{Kor}=0.1$\\[0pt]

\includegraphics[width=3.3in]{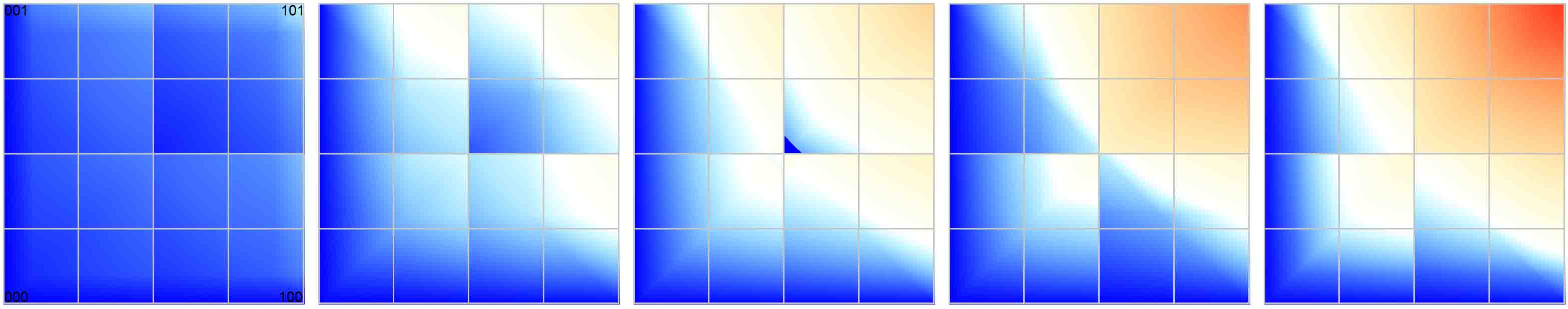}\\[-5pt]
$\textsf{Kor}=0.3$\\[0pt]

\includegraphics[width=3.3in]{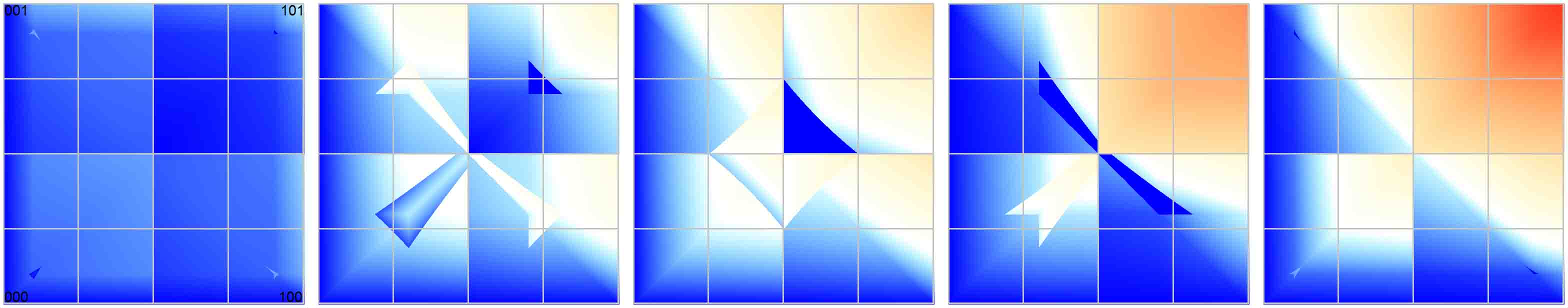}\\[-5pt]
$\textsf{Kor}=0.5$\\[0pt]

\includegraphics[width=3.3in]{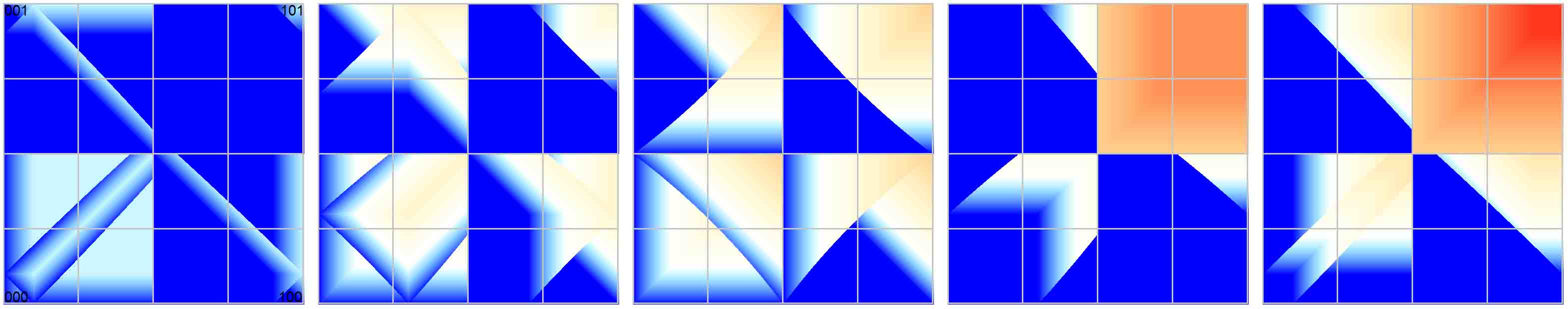}\\[-5pt]
$\textsf{Kor}=1$\\[0pt]

\vspace{-13pt}

\caption{The first modification of definitions. Cartesian representations of 3-Kopulas of triplets of half-rare events $\frak{X}=\{x,y,z\}$, constructed by the frame method (\ref{epd-triplet-8-4}) with non-negative values of a single parameter (from up to down) $\textsf{Kor}=-1,-0.5,-0.3,-0.1,0,0.1,0.3,0.5,1$, to which all four inserted \emph{Frechet-correlations} are equal  (see paragraph \ref{3Kopulas-parameters}). The independent 3-Kopula is obtained for $\textsf{Kor}=0$.} %
\label{fig-Kop3}
\end{figure}

\begin{figure}[ht!]
\vspace*{-10pt}
\centering
\includegraphics[width=3.3in]{Kop3_-1.jpg}\\[-5pt]
$\textsf{Kor}=-1$\\[0pt]

\includegraphics[width=3.3in]{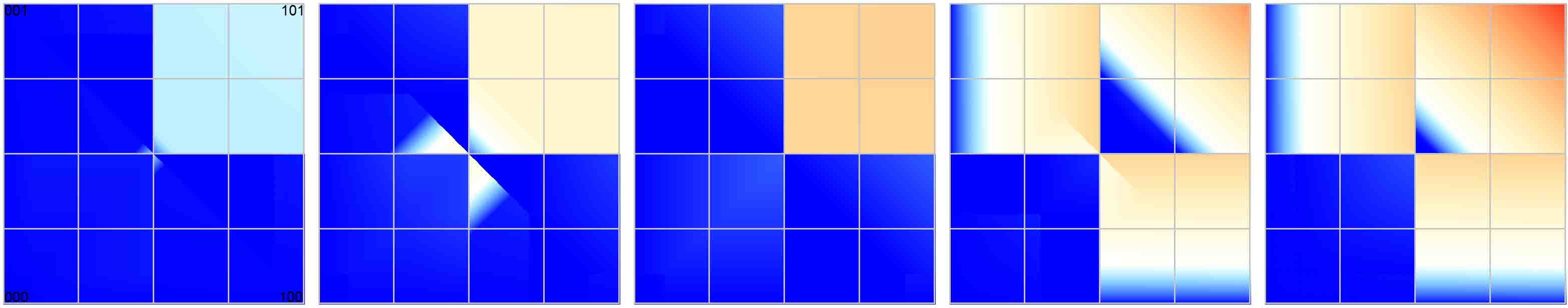}\\[-5pt]
$\textsf{Kor}=-0.9$\\[0pt]

\includegraphics[width=3.3in]{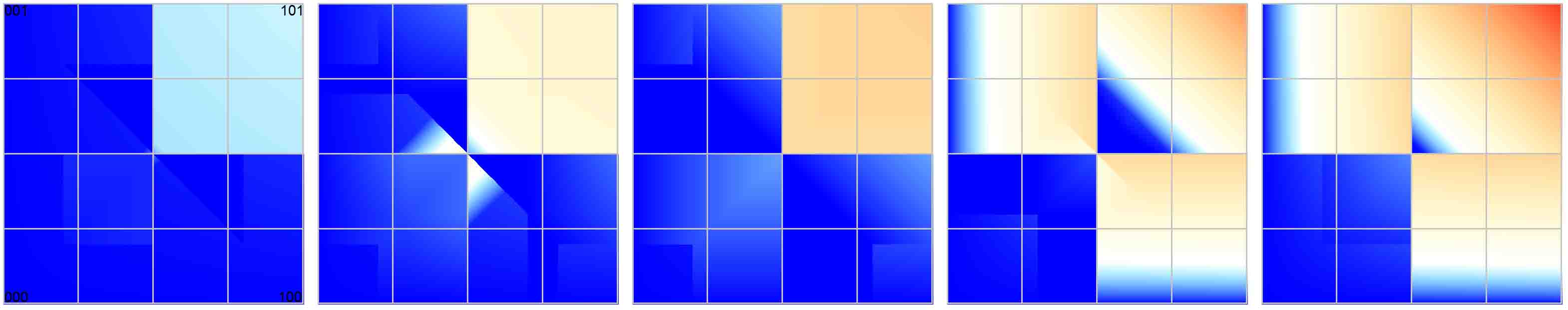}\\[-5pt]
$\textsf{Kor}=-0.8$\\[0pt]

\includegraphics[width=3.3in]{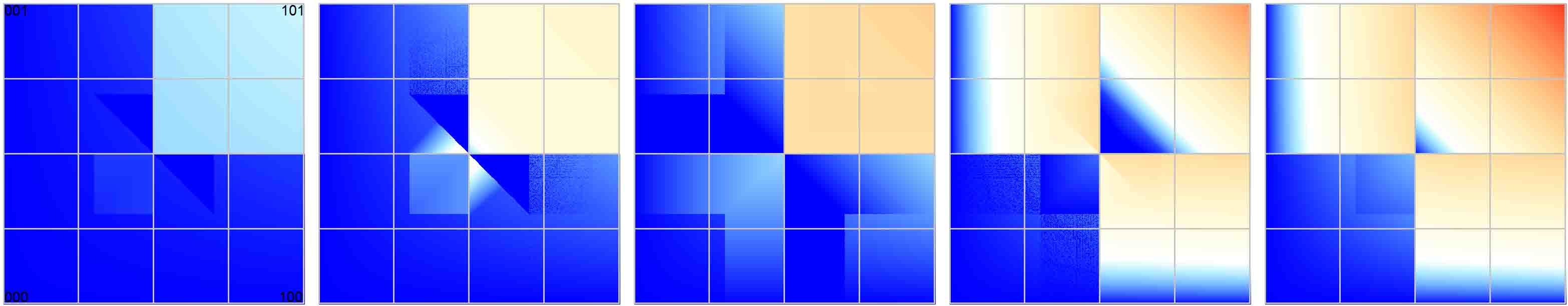}\\[-5pt]
$\textsf{Kor}=-0.7$\\[0pt]

\includegraphics[width=3.3in]{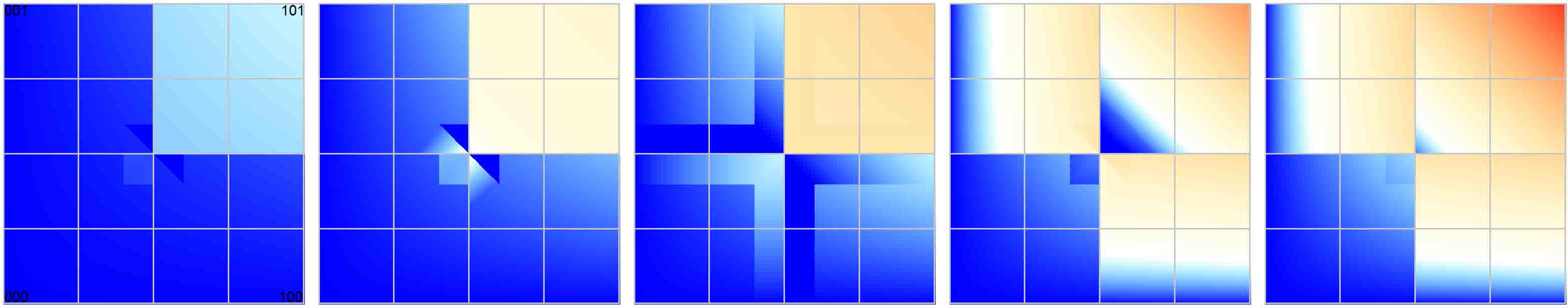}\\[-5pt]
$\textsf{Kor}=-0.6$\\[0pt]

\includegraphics[width=3.3in]{Kop3_-05.jpg}\\[-5pt]
$\textsf{Kor}=-0.5$\\[0pt]

\includegraphics[width=3.3in]{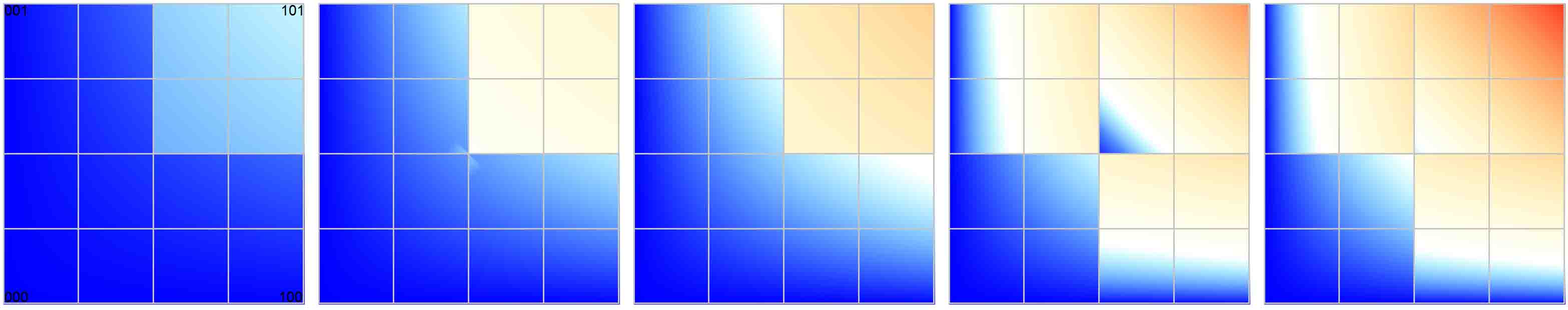}\\[-5pt]
$\textsf{Kor}=-0.4$\\[0pt]

\includegraphics[width=3.3in]{Kop3_-03.jpg}\\[-5pt]
$\textsf{Kor}=-0.3$\\[0pt]

\includegraphics[width=3.3in]{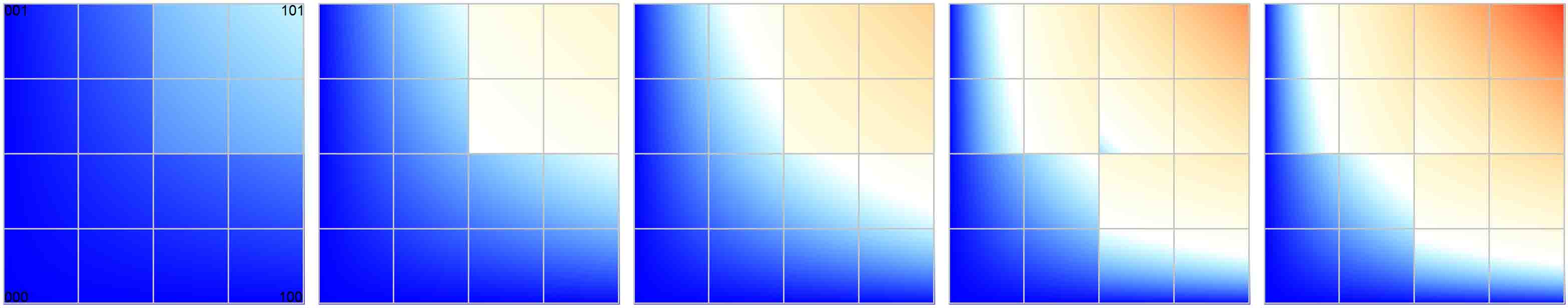}\\[-5pt]
$\textsf{Kor}=-0.2$\\[0pt]

\includegraphics[width=3.3in]{Kop3_-01.jpg}\\[-5pt]
$\textsf{Kor}=-0.1$\\[0pt]

\includegraphics[width=3.3in]{Kop31_ind.jpg}\\[-5pt]
$\textsf{Kor}=0$\\[0pt]

\vspace{-13pt}

\caption{The first modification of definitions. Cartesian representations of 3-Kopulas of triplets of half-rare events $\frak{X}=\{x,y,z\}$, constructed by the frame method (\ref{epd-triplet-8-4}) with non-positive values of the parameter $\textsf{Kor}=-1,-0.9,...,-0.2,-0.1,0$ (from up to down), to which inserted \emph{Frechet-correlations} are equal  (see paragraph \ref{3Kopulas-parameters}). The independent 3-Kopula is obtained for $\textsf{Kor}=0$.} %
\label{fig-Kop3_minus}
\end{figure}

\begin{figure}[ht!]
\vspace*{-10pt}
\centering
\includegraphics[width=3.3in]{Kop31_ind.jpg}\\[-5pt]
$\textsf{Kor}=0$\\[1.0pt]

\includegraphics[width=3.3in]{Kop3_01.jpg}\\[-5pt]
$\textsf{Kor}=0.1$\\[1.0pt]

\includegraphics[width=3.3in]{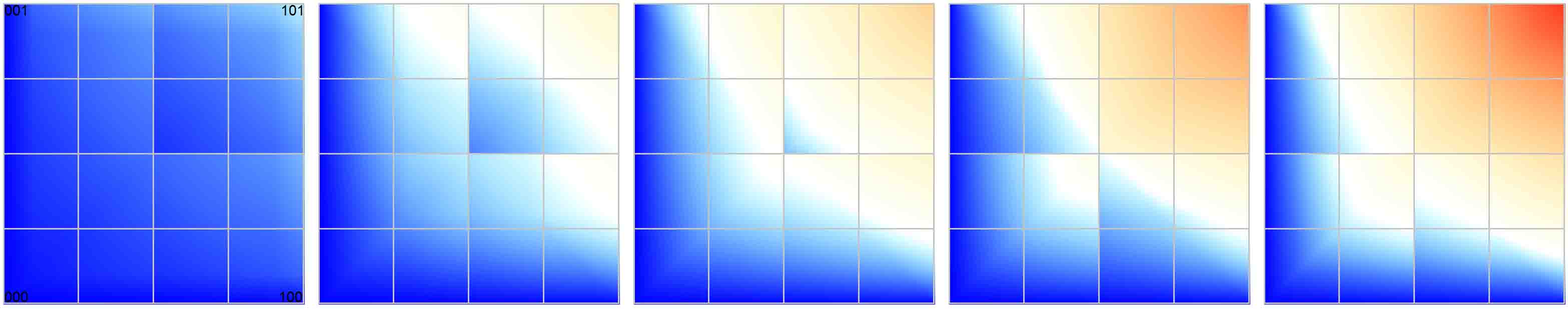}\\[-5pt]
$\textsf{Kor}=0.2$\\[1.0pt]

\includegraphics[width=3.3in]{Kop3_03.jpg}\\[-5pt]
$\textsf{Kor}=0.3$\\[1.0pt]

\includegraphics[width=3.3in]{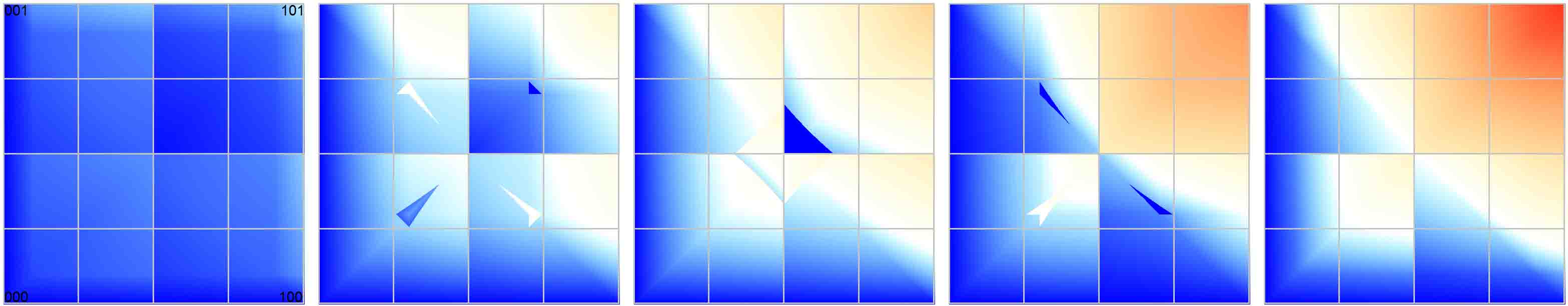}\\[-5pt]
$\textsf{Kor}=0.4$\\[1.0pt]

\includegraphics[width=3.3in]{Kop3_05.jpg}\\[-5pt]
$\textsf{Kor}=0.5$\\[1.0pt]

\includegraphics[width=3.3in]{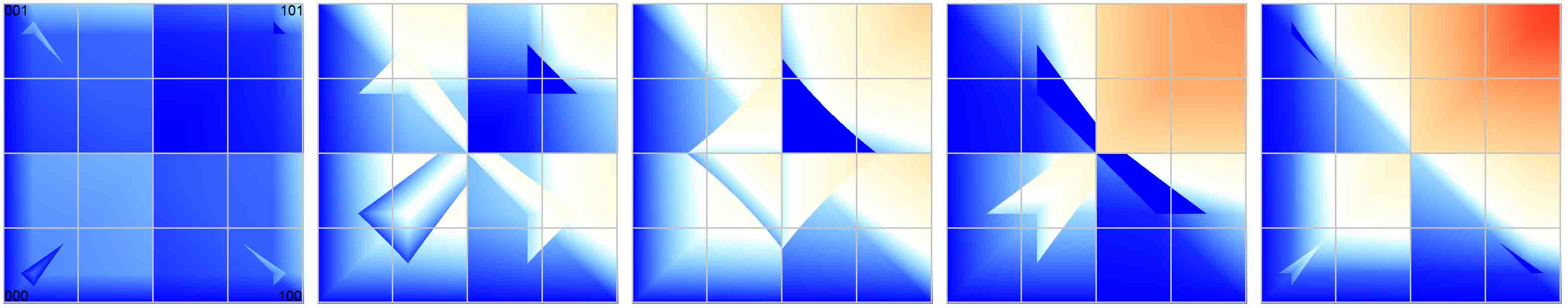}\\[-5pt]
$\textsf{Kor}=0.6$\\[1.0pt]

\includegraphics[width=3.3in]{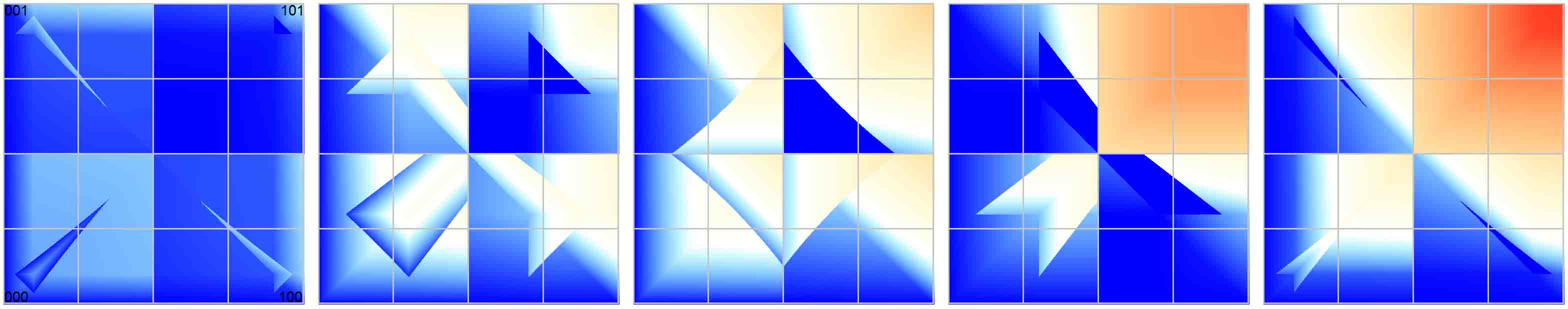}\\[-5pt]
$\textsf{Kor}=0.7$\\[1.0pt]

\includegraphics[width=3.3in]{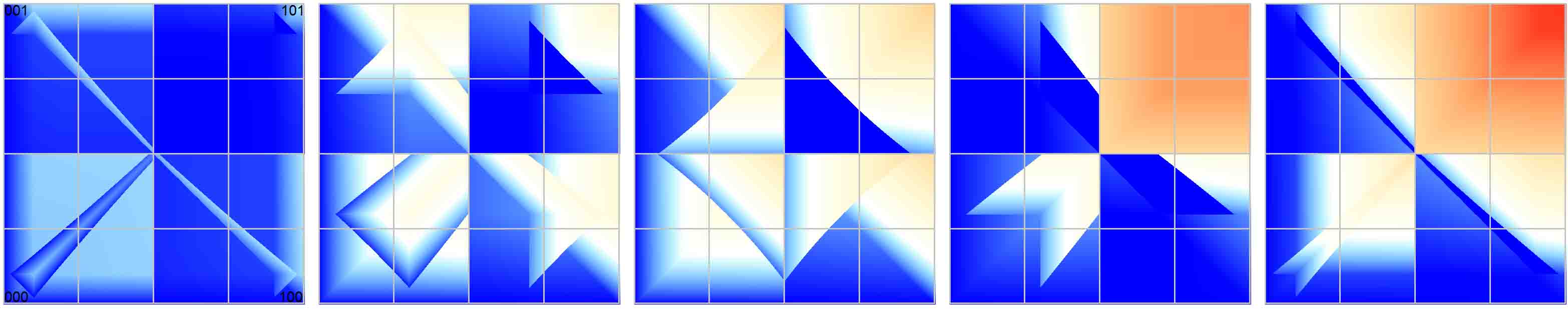}\\[-5pt]
$\textsf{Kor}=0.8$\\[1.0pt]

\includegraphics[width=3.3in]{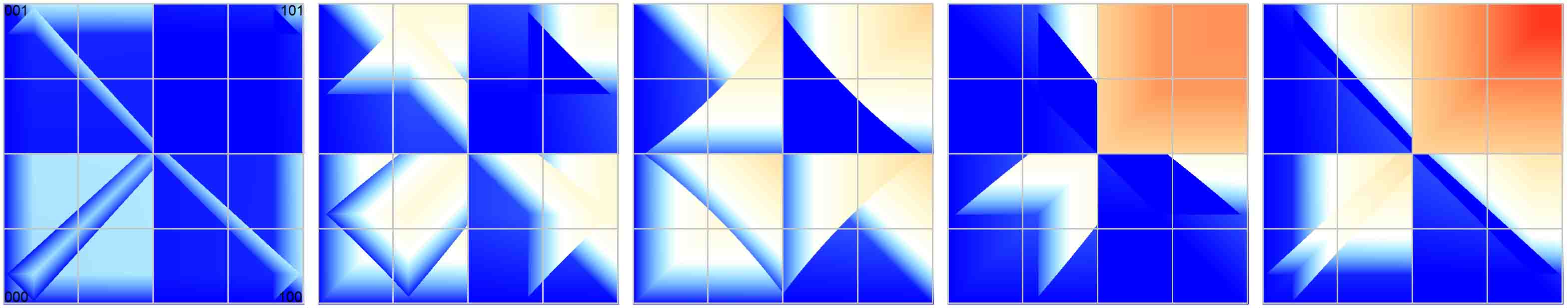}\\[-5pt]
$\textsf{Kor}=0.9$\\[1.0pt]

\includegraphics[width=3.3in]{Kop3_1.jpg}\\[-5pt]
$\textsf{Kor}=1$\\[1.0pt]

\vspace{-13pt}

\caption{The first modification of definitions. Cartesian representations of 3-Kopulas of triplets of half-rare events $\frak{X}=\{x,y,z\}$, constructed by the frame method (\ref{epd-triplet-8-4}) with non-negative values of the parameter $\textsf{Kor}=0,0.1,0.2,...,0.9,1$ (from up to down), to which inserted \emph{Frechet-correlations} are equal  (see paragraph \ref{3Kopulas-parameters}). The independent 3-Kopula is obtained for $\textsf{Kor}=0$.} %
\label{fig-Kop3_plus}
\end{figure}

In Fig.'s\footnote{Each figure shows maps of these 3-copulas on a cube in conditional colors, where the white color corresponds to the points at which the probabilities of terraced events are 1/8.} \ref{fig-Kop3}, \ref{fig-Kop3_minus}, and \ref{fig-Kop3_plus}
 it is shown 3-Kopulas of triplets of half-rare events $\frak{X}=\{x,y,z\}$, functional parameters of which serve Fr\'echet-correlations in the first modification of definitions
(see paragraph \ref{Kovariance3inserting}).

\section{Appendix\label{appendix13}}

\subsection{Abbreviations in the Kopula theory}

Consider the \emph{universal probability space} $(\Omega, \mathcal{A}^\mho, \mathbf{P})$  and one of its \emph{subject-name realizations}, a partial probability space $(\Omega, \mathcal{A}, \mathbf{P})$. The elements of the sigma-algebra $\mathcal{A}^\mho$ are the \emph{universal Kolmogorov events} $x^\mho \in \mathcal{A}^\mho$, and the elements of the sigma-algebra $\mathcal{A}$ --- \emph{events} $x \in \mathcal{A}$, which serve as \emph{names of universal Kolmogorov events} $x^\mho$ (see in details \cite{Vorobyev2014famems9}).

The notions that are relevant to a s.e. $\frak{X} \subseteq \mathcal{A}$ for which it is convenient to use the following abbreviations:

$\frak{X}=\{x : x \in \frak{X}\}$\\
\hspace*{6mm}{\scriptsize --- a set of events (s.e.);}\\[7pt]
$\breve{p}=\{p_x, x \in \frak{X}\}$\\
\hspace*{6mm}{\scriptsize --- an $\frak{X}$-set of probabilities of events from $\frak{X}$;}\\[7pt]
$\frak{X}^{(c|X)}=\frak{X}^{(c|X/\!\!/\frak{X})}=\{x:x \in X\}+\{x^c:x \in \frak{X}-X\}$\\
\hspace*{6mm}{\scriptsize --- an $X$-phenomenon of $\frak{X}, \ X\subseteq\frak X$;}\\[7pt]
$\frak{X}^{(c|\frak{X})}=\frak{X}^{(c|\frak{X}/\!\!/\frak{X})}=\frak{X}$\\
\hspace*{6mm}{\scriptsize --- an $\frak{X}$-phenomenon of $\frak{X}$ equal to $\frak{X}$;}\\[7pt]
$\breve{p}^{(c|X/\!\!/\frak{X})} = \{p_x, x \in X\}+\{1\!-\!p_x, x \in \frak{X}\!-\!X\}$\\
\hspace*{6mm}{\scriptsize --- an $\frak{X}$-set of probabilities of events from $\frak{X}^{(c|X)}, \ X \subseteq \frak{X}$;}\\[7pt]
$\breve{p}^{(c|\frak{X}/\!\!/\frak{X})}=\breve{p}$\\
\hspace*{6mm}{\scriptsize --- an $\frak{X}$-set of probabilities of events from $\frak{X}^{(c|\frak{X})}$ equal to $\breve{p}$;}\\[7pt]
$p(X /\!\!/ \frak{X})$\\
\hspace*{6mm}{\scriptsize --- a value of e.p.d. of the 1st kind of $\frak{X}$ for $X \subseteq \frak{X}$;}\\[7pt]
${\mbox{\boldmath$\mathscr{K}$}}\left(\breve{p}^{(c|X/\!\!/\frak{X})}\right)$\\
\hspace*{6mm}{\scriptsize --- a value of the Kopula of e.p.d. of the 1st kind of $\frak{X}$ for $X\subseteq\frak{X}$;}\\[7pt]
$p(X /\!\!/ \frak{X}) ={\mbox{\boldmath$\mathscr{K}$}}\left(\breve{p}^{(c|X/\!\!/\frak{X})}\right)$\\
\hspace*{6mm}{\scriptsize --- the definition of e.p.d. of the 1st kind of $\frak{X}$ by its Kopula, $X \subseteq \frak{X}$;}\\[7pt]
$\breve{s}=\{s_x, x \in \frak{X} \} \in [0,1/2]^{\otimes \frak{X}}$\\
\hspace*{6mm}{\scriptsize --- an $\frak{X}$-set of half-rare variables;}\\[7pt]
$\breve{w}=\{w_x, x \in \frak{X}\} \in [0,1]^{\otimes \frak{X}}$\\
\hspace*{6mm}{\scriptsize --- an $\frak{X}$-set of free variables.}

\subsection{Set-phenomenon renumbering a e.p.d. of the 1st kind and its Kopulas}

\texttt{Lemma \!\refstepcounter{ctrlem}\arabic{ctrlem}\,\label{lem-setPH-re-epd-Kopula}\itshape\enskip\scriptsize (Set-phenomenon renumbering a e.p.d. of the 1st kind and its Kopulas).} \emph{E.p.d. of the 1st kind and Kopulas of the s.e. $\frak{X}$ and of its $S$-phenomena $\frak{X}^{(c|S)}$ are connected by formulas of mutually inversion set-phenomenon renumbering for $X \subseteq \frak{X}$ и $S \subseteq \frak{X}$:}
\begin{equation}\label{setPH-re-epd}
\hspace{-3pt}\begin{split}
&p\!\left(X^{(c| S\cap X)} /\!\!/ \frak{X}^{(c|S)}\right) = p((S \!\Delta X)^c /\!\!/ \frak{X}),\\
&p(X /\!\!/ \frak{X})=p\!\left(\left((S\!\Delta X)^c\right)^{(c| S\cap (S\!\Delta X)^c)} /\!\!/ \frak{X}^{(c|S)}\right);
\end{split}
\end{equation}\\[-16pt]
\begin{equation}\label{setPH-re-Kopula}
\hspace{-3pt}\begin{split}
&{\mbox{\boldmath$\mathscr{K}$}}\!\left(\breve{p}^{\left(c|X^{(c| S\cap X)} /\!\!/ \frak{X}^{(c|S)}\right)}\right)=
{\mbox{\boldmath$\mathscr{K}$}}\!\left(\breve{p}^{(c|(S \!\Delta X)^c /\!\!/ \frak{X})}\right),\\
&\\[-8pt]
&{\mbox{\boldmath$\mathscr{K}$}}\!\left(\breve{p}^{(c|X/\!\!/ \frak{X})}\right)={\mbox{\boldmath$\mathscr{K}$}}\!\left(\breve{p}^{\left(c|\left((S\!\Delta X)^c\right)^{(c| S\cap (S\!\Delta X)^c)} /\!\!/ \frak{X}^{(c|S)}\right)}\right).
\end{split}
\hspace{-30pt}
\end{equation}
\texttt{Proof} follows immediately from the formulas of the set-phenomenon renumbering the terraced events of the 1st kind of the s.e. and of its set-phenomena proved in \cite{Vorobyev2015famems12}.

\subsection{Useful denotations for a doublet of events that are invariant relative to the $\breve{p}$-order\label{2-denotations}}

The following special denotations for a doublet of events $\frak{X}=\{x,y\}$ and the $\frak{X}$-set of marginal probabilities $\breve{p}=\{p_x,p_y\}$ that are invariant relative to the $\breve{p}$-order, are useful.

\begin{equation}\label{2-denotations}
\begin{split}
\hspace*{-50pt}\frak{X}&=\{x,y\}=\{x^{\uparrow}, x^{\downarrow}\},\\
\hspace*{-50pt}\breve{p}&=\{p_x,p_y\}=\{p^{\uparrow}, p^{\downarrow}\}, \ \ \
1/2 \geqslant p^{\uparrow} \geqslant p^{\downarrow},\\
\hspace*{-50pt}\breve{w}&=\{w_x,w_y\}=\{w^{\uparrow}, w^{\downarrow}\}, \ \ \
1 \geqslant w^{\uparrow} \geqslant w^{\downarrow},\\
\end{split}\hspace{-30pt}
\end{equation}
\begin{equation}\label{2-denotations1}
\begin{split}
\hspace*{-50pt}\{x^{\uparrow}\}&=\max\{\frak{X}\}=
\begin{cases}
\!\{x\}, &\!\!\!\! p_x \!>\! p_y,\\
\!\{y\}, &\!\!\!\! \mbox{иначе};
\end{cases}\\
\hspace*{-50pt}\{x^{\downarrow}\}&=\min\{\frak{X}\}=
\begin{cases}
\!\{x\}, &\!\!\!\! p_x \!\leqslant\! p_y,\\
\!\{y\}, &\!\!\!\! \mbox{иначе};
\end{cases}\\
\end{split}\hspace{-30pt}
\end{equation}
\begin{equation}\label{2-denotations1}
\begin{split}
\hspace*{-50pt}w^{\uparrow} &=\max\{\breve{w}\}\!=\!\!
\begin{cases}
\!w_x, &\!\!\!\! w_x \!>\! w_y,\\
\!w_y, &\!\!\!\! \mbox{иначе};
\end{cases}\\
\hspace*{-50pt}w^{\downarrow}  &= \min\{\breve{w}\}\!=\!\!
\begin{cases}
\!w_x, &\!\!\!\! w_x \!\leqslant\! w_y,\\
\!w_y, &\!\!\!\! \mbox{иначе};
\end{cases}\\
\end{split}\hspace{-30pt}
\end{equation}
\begin{equation}\label{2-denotations1}
\begin{split}
\hspace*{-50pt}p^{\uparrow} &=\max\{\breve{p}\}\!=\!\!
\begin{cases}
\!p_x, &\!\!\!\! p_x \!>\! p_y,\\
\!p_y, &\!\!\!\! \mbox{иначе};
\end{cases}\\
&= \max \Big\{\min\{w_x,1-w_x\}, \ \min\{w_y,1-w_y\}\Big\},\\
&\\
\hspace*{-50pt}p^{\downarrow}  &= \min\{\breve{p}\}\!=\!\!
\begin{cases}
\!p_x, &\!\!\!\! p_x \!\leqslant\! p_y,\\
\!p_y, &\!\!\!\! \mbox{иначе};
\end{cases}\\
&= \min \Big\{\min\{w_x,1-w_x\}, \ \min\{w_y,1-w_y\}\Big\}.
\end{split}\hspace{-30pt}
\end{equation}

In such invariant denotations, it is not difficult to write down the general recurrence formula for the half-rare 2-Kopula of the doublet $\frak{X}$, united combining both orders:
\begin{equation}\label{inserting-recurrent}
\hspace{-5pt}\begin{split}
&{\mbox{\boldmath$\mathscr{K}$}}\!\left(\breve{p}^{(c|X/\!\!/\frak{X})}\right) ={\mbox{\boldmath$\mathscr{K}$}}^{(c|\frak{X}/\!\!/\frak{X})}\!\left(\breve{p}^{(c|X/\!\!/\frak{X})}\right) =\\
&=\begin{cases}
{\mbox{\boldmath$\mathscr{K}$}}'\left(p_{xy}(\breve{p})\right),
&X=\frak{X},\\
{\mbox{\boldmath$\mathscr{K}$}}''\left(p^{\downarrow}-p_{xy}(\breve{p})\right), &X=\{x^{\downarrow}\},\\
{\mbox{\boldmath$\mathscr{K}$}}'\left(1-p_{xy}(\breve{p})\right)-1+p^{\uparrow}, &X=\{x^{\uparrow}\},\\
{\mbox{\boldmath$\mathscr{K}$}}''\left(1-p^{\downarrow}+p_{xy}(\breve{p})\right)-p^{\uparrow}, &X=\emptyset
\end{cases}
\end{split}\hspace{-13pt}
\end{equation}
where by Definition (\ref{2-denotations})
\begin{equation}\nonumber
\begin{split}
p^\uparrow= \max \Big\{\min\{w_x,1-w_x\}, \ \min\{w_y,1-w_y\}\Big\},\\
p^\downarrow=\min \Big\{\min\{w_x,1-w_x\}, \ \min\{w_y,1-w_y\}\Big\}.
\end{split}
\end{equation}
The mutual set-phenomenon inversion of 2-Kopulas of half-rare $\breve{p}$ and free $\breve{w}$ marginal probabilities has the form:
\begin{equation}\label{hrare-free-transforming}
\hspace{-5pt}\begin{split}
&{\mbox{\boldmath$\mathscr{K}$}}\!\left(\breve{p}^{(c|X/\!\!/\frak{X})}\right) =
{\mbox{\boldmath$\mathscr{K}$}}^{(c|X/\!\!/\frak{X})}\!\left(\breve{w}\right),\\
&{\mbox{\boldmath$\mathscr{K}$}}^{(c|X/\!\!/\frak{X})}\!\left(\breve{p}\right) =
{\mbox{\boldmath$\mathscr{K}$}}\!\left(\breve{w}^{(c|X/\!\!/\frak{X})}\right);
\end{split}\hspace{-13pt}
\end{equation}
\begin{equation}\label{hrare-free-transforming}
\hspace{-5pt}\begin{split}
{\mbox{\boldmath$\mathscr{K}$}}^{(c|S/\!\!/\frak{X})}\!\left(\breve{p}^{(c|X/\!\!/\frak{X})}\right)
&= {\mbox{\boldmath$\mathscr{K}$}}^{(c|\frak{X}/\!\!/\frak{X})}\!\left(\breve{p}^{(c|(S^c\Delta X)^c/\!\!/\frak{X})}\!\right)\\
&= {\mbox{\boldmath$\mathscr{K}$}}\!\left(\breve{p}^{(c|(S^c\Delta X)^c/\!\!/\frak{X})}\!\right).
\end{split}\hspace{-13pt}
\end{equation}
For example, for $X \subseteq \frak{X}=\{x,y\}$
\begin{equation}\label{hrare-free-transforming}
\hspace{-5pt}\begin{split}
{\mbox{\boldmath$\mathscr{K}$}}^{(c|\frak{X}/\!\!/\frak{X})}\!\left(\breve{p}^{(c|X/\!\!/\frak{X})}\right)
&= {\mbox{\boldmath$\mathscr{K}$}}\!\left(\breve{p}^{(c|X/\!\!/\frak{X})}\!\right),\\
{\mbox{\boldmath$\mathscr{K}$}}^{(c|\{x\}/\!\!/\frak{X})}\!\left(\breve{p}^{(c|X/\!\!/\frak{X})}\right)
&= {\mbox{\boldmath$\mathscr{K}$}}\!\left(\breve{p}^{(c|(\{y\}\Delta X)^c/\!\!/\frak{X})}\!\right),\\
{\mbox{\boldmath$\mathscr{K}$}}^{(c|\{y\}/\!\!/\frak{X})}\!\left(\breve{p}^{(c|X/\!\!/\frak{X})}\right)
&= {\mbox{\boldmath$\mathscr{K}$}}\!\left(\breve{p}^{(c|(\{x\}\Delta X)^c/\!\!/\frak{X})}\!\right),\\
{\mbox{\boldmath$\mathscr{K}$}}^{(c|\emptyset/\!\!/\frak{X})}\!\left(\breve{p}^{(c|X/\!\!/\frak{X})}\right)
&= {\mbox{\boldmath$\mathscr{K}$}}\!\left(\breve{p}^{(c|X^c/\!\!/\frak{X})}\!\right).
\end{split}\hspace{-13pt}
\end{equation}

\subsection{Recurrent properties of the $\breve{p}$-ordering a half-rare s.e.\label{non-order}}

\subsubsection{Recurrent properties of the $\breve{p}$-ordering a half-rare doublets of events\label{non-order2}}

Let us explain the role of $\breve{p}$-ordering in the frame method using the example of constructing a 2-Kopula of the $\breve{p}$-ordered half-rare events $\frak{X}=\{x,y\}$ with $\frak{X}$-set of marginal probabilities of events $\breve{p}=\{p_x,p_y\}$, that is, $1/2 \geqslant p_x \geqslant p_y$:
\begin{equation}\label{inserting-monoplets-formulas}
\begin{split}
\hspace*{-5pt}&{\mbox{\boldmath$\mathscr{K}$}}'\left(\breve{p}^{(c|X/\!\!/\{x,y\})}\right)=
\begin{cases}
p_{xy},     & \!\!\!\!\!X\!=\!\{x,y\},\\
p_{x}\!-\!p_{xy}, & \!\!\!\!\!X\!=\!\{x\},\\
p_{y}\!-\!p_{xy},     & \!\!\!\!\!X\!=\!\{y\},\\
1\!-\!p_{x}\!-\!p_{y}\!+\!p_{xy},     & \!\!\!\!\!X\!=\!\emptyset,
\end{cases}
\end{split}\hspace{-10pt}
\end{equation}
where, when selected as a function parameter $p_{xy}$ of the 1-Kopulas of inserted half-rare monoplates $\mathcal{X}'=\{s'\}=\{x\cap y\}$ and $\mathcal{X}'\!'=\{s'\!'\}=\{x^c\cap y\}$, are equal, respectively:
\begin{equation}\label{inserting-monoplets}
\hspace{-3pt}\begin{split}
&{\mbox{\boldmath$\mathscr{K}$}}'\left(\breve{p}^{(c|S/\!\!/\mathcal{X}')}\right)=
\begin{cases}
p_{s't'},     & S\!=\!\{s',t'\},\\
p_{s'}\!-\!p_{s't'}, & S\!=\!\{s'\},\\
p_{t'}\!-\!p_{s't'},     & S\!=\!\{t'\},\\
1\!-\!p_{s'}\!-\!p_{t'}\!+\!p_{s't'},     & S\!=\!\emptyset,
\end{cases}\\
&\\
&{\mbox{\boldmath$\mathscr{K}$}}'\!'\left(\breve{p}^{(c|S/\!\!/\mathcal{X}'\!')}\right)=\\
&=\begin{cases}
p_{s'\!'t'\!'},     & S\!=\!\{s'\!',t'\!'\},\\
p_y-p_{s'}\!-\!p_{s'\!'t'\!'}, & S\!=\!\{s'\!'\},\\
p_z-p_{t'}\!-\!p_{s'\!'t'\!'},     & S\!=\!\{t'\!'\},\\
1\!-\!p_y\!-\!p_z\!+\!p_{s'}\!+\!p_{t'}\!+\!p_{s'\!'t'\!'},     & S\!=\!\emptyset,
\end{cases}
\end{split}\hspace{-25pt}
\end{equation}
under the assumption that inserted half-rare monoplets have ``equally direct'' $\breve{p}$-orders:
\begin{equation}\label{2-order-0}
\begin{split}
p_y \geqslant p_{s'} \geqslant p_{t'},\\
p_z \geqslant p_{s'\!'} \geqslant p_{t'\!'}.
\end{split}
\end{equation}
However, nothing prevents the emergence of two more ``opposite $\breve{p}$ orders'' on the inserted half-rare monoplets:
\begin{equation}\label{2-order-1}
\begin{split}
p_y \geqslant p_{t'} \geqslant p_{s'},\\
p_z \geqslant p_{s'\!'} \geqslant p_{t'\!'}.
\end{split}
\end{equation}
\begin{equation}\label{2-order-2}
\begin{split}
p_y \geqslant p_{s'} \geqslant p_{t'},\\
p_z \geqslant p_{t'\!'} \geqslant p_{s'\!'};
\end{split}
\end{equation}
except for the ``equally inverse'' $\breve{p}$-order
\begin{equation}\label{2-order-3}
\begin{split}
p_y \geqslant p_{t'} \geqslant p_{s'},\\
p_z \geqslant p_{t'\!'} \geqslant p_{s'\!'},
\end{split}
\end{equation}
which can not be due to the consistency of the functional parameters, i.e., because
\begin{equation}\label{doublet-5}
\begin{split}
p_y=p_{s'} + p_{s'\!'} \geqslant p_{t'} + p_{t'\!'} = p_z.
\end{split}
\end{equation}

\subsubsection{Recurrent properties of $\breve{p}$-ordering the half-rare triplets of events\label{non-order3}}

Let us explain the role of $\breve{p}$-ordering in the frame method using the example of constructing a 3-Kopula of the $\breve{p}$-ordered half-rare events $\frak{X}=\{x,y,z\}$ with $\frak{X}$-set of marginal probabilities of events $\breve{p}=\{p_x,p_y,p_z\}$, that is, $1/2 \geqslant p_x \geqslant p_y\geqslant p_z$:
\begin{equation}\label{3-Kopula-recurrent}
\hspace{-0pt}\begin{split}
&{\mbox{\boldmath$\mathscr{K}$}}\left(\breve{p}^{(c|X/\!\!/\{x,y,z\})}\right)=\\
&\!\!=\!\!\begin{cases}
\!{\mbox{\boldmath$\mathscr{K}$}}'\left(p_{s'},p_{t'}\right),                         &\!\!\!\!X\!=\!\{x,y,z\},\\
\!{\mbox{\boldmath$\mathscr{K}$}}'\left(p_{s'},1\!-\!p_{t'}\right),              &\!\!\!\!X\!=\!\{x,y\},\\
\!{\mbox{\boldmath$\mathscr{K}$}}'\left(1\!-\!p_{s'},p_{t'}\right),              &\!\!\!\!X\!=\!\{x,z\},\\
\!{\mbox{\boldmath$\mathscr{K}$}}'\left(1\!-\!p_{s'},1\!-\!p_{t'}\right)\!-\!1\!+\!p_x,
                                  &\!\!\!\!X\!=\!\{x\},\\
\!{\mbox{\boldmath$\mathscr{K}$}}'\!'\left(p_y\!-\!p_{s'},p_z\!-\!p_{t'}\right),                   &\!\!\!\!X\!=\!\{y,z\},\\
\!{\mbox{\boldmath$\mathscr{K}$}}'\!'\left(p_y\!-\!p_{s'},1\!-\!p_z\!+\!p_{t'}\right),    &\!\!\!\!X\!=\!\{y\},\\
\!{\mbox{\boldmath$\mathscr{K}$}}'\!'\left(1\!-\!p_y\!+\!p_{s'},p_z\!-\!p_{t'}\right),    &\!\!\!\!X\!=\!\{z\},\\
\!{\mbox{\boldmath$\mathscr{K}$}}'\!'\left(1\!-\!p_y\!+\!p_{s'},1\!-\!p_z\!+\!p_{t'}\right)\!-\!p_x,
                                  &\!\!\!\!X\!=\!\emptyset,
\end{cases}
\end{split}\hspace{-32pt}
\end{equation}
where, when selected as function parameters $p_{s'},p_{t'},p_{s't'}$ and $p_{s'\!'t'\!'}$ and despite the fact that $p_{s'\!'}=p_y-p_{s'}, p_{t'\!'}=p_z-p_{t'}$, the 2-Kopulas of inserted half-rare doublets $\mathcal{X}'=\{s',t'\}=\{x\cap y,x\cap z\}$ and $\mathcal{X}'\!'=\{s'\!',t'\!'\}=\{x^c\cap y,z^c\cap z\}$ are equal, respectively:
\begin{equation}\label{inserting-doublets-formulas}
\hspace{-3pt}\begin{split}
&{\mbox{\boldmath$\mathscr{K}$}}'\left(\breve{p}^{(c|S/\!\!/\mathcal{X}')}\right)=
\begin{cases}
p_{s't'},     & S\!=\!\{s',t'\},\\
p_{s'}\!-\!p_{s't'}, & S\!=\!\{s'\},\\
p_{t'}\!-\!p_{s't'},     & S\!=\!\{t'\},\\
1\!-\!p_{s'}\!-\!p_{t'}\!+\!p_{s't'},     & S\!=\!\emptyset,
\end{cases}\\
&\\
&{\mbox{\boldmath$\mathscr{K}$}}'\!'\left(\breve{p}^{(c|S/\!\!/\mathcal{X}'\!')}\right)=\\
&=\begin{cases}
p_{s'\!'t'\!'},     & S\!=\!\{s'\!',t'\!'\},\\
p_y-p_{s'}\!-\!p_{s'\!'t'\!'}, & S\!=\!\{s'\!'\},\\
p_z-p_{t'}\!-\!p_{s'\!'t'\!'},     & S\!=\!\{t'\!'\},\\
1\!-\!p_y\!-\!p_z\!+\!p_{s'}\!+\!p_{t'}\!+\!p_{s'\!'t'\!'},     & S\!=\!\emptyset,
\end{cases}
\end{split}\hspace{-25pt}
\end{equation}
under the assumption that the inserted half-rare doublets have ``equally direct'' $\breve{p}$-orders:
\begin{equation}\label{2-order-0}
\begin{split}
p_y \geqslant p_{s'} \geqslant p_{t'},\\
p_z \geqslant p_{s'\!'} \geqslant p_{t'\!'}.
\end{split}
\end{equation}
However, nothing prevents the emergence of two more ``opposite $\breve{p}$ orders'' on the inserted half-rare doublets:
\begin{equation}\label{2-order-1}
\begin{split}
p_y \geqslant p_{t'} \geqslant p_{s'},\\
p_z \geqslant p_{s'\!'} \geqslant p_{t'\!'}.
\end{split}
\end{equation}
\begin{equation}\label{2-order-2}
\begin{split}
p_y \geqslant p_{s'} \geqslant p_{t'},\\
p_z \geqslant p_{t'\!'} \geqslant p_{s'\!'};
\end{split}
\end{equation}
except for the ``equally inverse'' $\breve{p}$-order
\begin{equation}\label{2-order-3}
\begin{split}
p_y \geqslant p_{t'} \geqslant p_{s'},\\
p_z \geqslant p_{t'\!'} \geqslant p_{s'\!'},
\end{split}
\end{equation}
which can not be due to the consistency of the functional parameters, i.e., because
\begin{equation}\label{doublet-5}
\begin{split}
p_y=p_{s'} + p_{s'\!'} \geqslant p_{t'} + p_{t'\!'} = p_z.
\end{split}
\end{equation}

\subsubsection{Extending the frame method to $\breve{p}$-non-ordered half-rare s.e.'s\label{non-order23}}

Above we outlined the frame method for constructing $N$-Kopulas of $\breve{p}$-\uwave{ordered} half-rare $N$-s.e.'s. It remains to extend it to construct $N$-Kopulas of $\breve{p}$-\uwave{disordered} half-rare $N$-s.e.'s using the following technique, based on the obvious invariance property of permutations of events in s.e.: ``as events from some s.e. do not order, the s.e. will not change "; and very useful in practical calculations.

We denote by
\begin{equation}\label{Norder}
\begin{split}
\frak{X}^* = \{x_0^*,x_1^*,...,x_{N-1}^*\},
\end{split}
\end{equation}
--- the $\breve{p}$-\uwave{ordered} \emph{half-rare $N$-s.e.},
which consists from the same events, that an ``arbitrary'' $\breve{p}$-\uwave{non-ordered} \emph{half-rare $N$-s.e.}
\begin{equation}\label{Nnonorder}
\begin{split}
\frak{X} = \{x_0,x_1,...,x_{N-1}\},
\end{split}
\end{equation}
i.e.,
\begin{equation}\label{Xequality}
\begin{split}
\frak{X}^*\!=\!\{x_0^*,x_1^*,...,x_{N\!-\!1}^*\}\!=\! \{x_0,x_1,...,x_{N\!-\!1}\}\!=\!\frak{X},
\end{split}
\end{equation}
but arranged in descending order of their probabilities. In other words, the $\frak{X}^*$-set of marginal probabilities
\begin{equation}\label{Pmarginals}
\begin{split}
\breve{p}^* = \{p_0^*,p_1^*,...,p_{N-1}^*\},
\end{split}
\end{equation}
is such that
\begin{equation}\label{Porder}
\begin{split}
1/2 \geqslant p_0^* \geqslant p_1^* \geqslant ... \geqslant p_{N-1}^*
\end{split}
\end{equation}
where
\begin{equation}\label{Pdef}
\begin{split}
&p_0^* = \max \{ p_x : x \in \frak{X} \},\\
&p_1^* = \max \{ p_x : x \in \frak{X}-\{x_0^*\} \},\\
&...\\
&p_{n+1}^* = \max \{ p_x : x \in \frak{X}-\{x_1^*,...,x_n^*\} \},\\
&...\\
&p_{N}^* = \max \{ p_x : x \in \frak{X}-\{x_1^*,...,x_{N-1}^*\} \}.
\end{split}
\end{equation}
Consequently, $\frak{X}^*$-set of marginal probabilities $\breve{p}^*$ which consists of the same probabilities that $\frak{X}$-set of marginal probabilities $\breve{p}$, i.e.,
\begin{equation}\label{Pequality}
\hspace*{-0pt}
\begin{split}
\breve{p}^*\!=\!\{p_0^*,p_1^*,...,p_{N-1}^*\}\!=\!\{p_0,p_1,...,p_{N-1}\}\!=\!\breve{p},
\end{split}\hspace*{0pt}
\end{equation}
but arranged in descending order.

Now, to construct the $N$-Kopulas of the $\breve{p}$-\uwave{disordered} $N$-s.e $\frak{X}$ by the frame method it is sufficient to construct this $N$-Kopula of the $\breve{p}$-\uwave{ordered} $N$-s.e $\frak {X}^* = \frak{X}$ by this method, reasoning by (\ref{Xequality}) and (\ref{Pequality}) reasoning that

Теперь для построения рамочным методом $N$-Копулы $\breve{p}$-\uwave{неупорядоченного} $N$-s.e. $\frak{X}$ достаточно построить этим методом $N$-Копулу $\breve{p}$-\uwave{упорядоченного} $N$-s.e. $\frak{X}^*=\frak{X}$,
reasoning by virtue of (\ref{Xequality}) and (\ref{Pequality}), that
\begin{equation}\label{NKopula-expanding}
\begin{split}
{\mbox{\boldmath$\mathscr{K}$}}\left(\breve{p}\right)=
{\mbox{\boldmath$\mathscr{K}$}}\left(\breve{p}^*\right),
\end{split}
\end{equation}
i.e., for $X \subseteq \frak{X}$
\begin{equation}\label{NKopula-expanding1}
\begin{split}
{\mbox{\boldmath$\mathscr{K}$}}\left(\breve{p}^{(c|X/\!\!/\frak{X})}\right)=
{\mbox{\boldmath$\mathscr{K}$}}\left(\breve{p}^{*(c|X^*/\!\!/\frak{X}^*)}\right)
\end{split}
\end{equation}
where
\begin{equation}\label{subset-order}
\begin{split}
X^* = \{x^* : x \in X\} \subseteq \frak{X}^*
\end{split}
\end{equation}
are subsets of the $\breve{p}$-\uwave{ordered} $(N\!-\!1)$-s.e. $\frak{X}^*$.

\texttt{Note \!\refstepcounter{ctrnot}\arabic{ctrnot}\,\label{not-order}\itshape\enskip\scriptsize (properties of functions of an unordered set of arguments).} Equations (\ref{NKopula-expanding}) and (\ref{NKopula-expanding1}) should not be regarded as a unique property of the Kopula invariance with respect to permutations of its arguments. This property is possessed by any Kopula, since it is a function of an unordered set of arguments. Therefore it is quite natural that the Kopula is invariant under permutations of the arguments, like any other such function. This property must be remembered only in practical calculations, when we volence-nolens must introduce an arbitrary order on a disordered set in order to be able to perform calculations.

Consider the examples of Kopulas of arbitrary, i.e., $\breve{p}$-disordered, s.e.'s
$$
\frak{X} = \{x_0,x_1,...,x_{N-1}\}
$$
in the notation just introduced, assuming that we have available Kopulas of the $\breve{p}$-ordered $N$-s.e's
$$
\frak{X}^* = \{x_0^*,x_1^*,...,x_{N-1}^*\} = \frak{X}
$$
for $N=1,2$.

\texttt{Example \!\refstepcounter{ctrexa}\arabic{ctrexa}\,\label{exa-2expanding}\itshape\enskip\scriptsize (invariant formula for the 2-Kopula of a half-rare doublet of events).} Let $\frak{X}=\{x_0,x_1\}$ be the $\breve{p}$-non-ordered half-rare doublet of events. Then its 2-Kopula is calculated at each point $\breve{p}^{(c|X/\!\!/\{x_{0},x_{1}\})} \in [0,1]^{\otimes \frak{X}}$ by the following formulas:
\begin{equation}\label{2-inserting-recurrent}
\hspace*{-0pt}
\begin{split}
&{\mbox{\boldmath$\mathscr{K}$}}\left(\breve{p}^{(c|X/\!\!/\{x_{0},x_{1}\})}\right) ={\mbox{\boldmath$\mathscr{K}$}}\left(\breve{p}^{(c|X^*/\!\!/\{x_{0}^*,x_{1}^*\})}\right) =\\
&=\begin{cases}
{\mbox{\boldmath$\mathscr{K}$}}'\left(p_{s'}\right), & X^*=\{x_{0}^*,x_{1}^*\},\\
{\mbox{\boldmath$\mathscr{K}$}}'\!'\left(p_{1}-p_{s'}\right), & X^*=\{x_{1}^*\},\\
{\mbox{\boldmath$\mathscr{K}$}}'\left(1-p_{s'}\right)-1+p_{0}, & X^*=\{x_{0}^*\},\\
{\mbox{\boldmath$\mathscr{K}$}}'\!'\left(1-p_{1}+p_{s'}\right)-p_{0}, & X^*=\emptyset,
\end{cases}
\end{split}\hspace*{-0pt}
\end{equation}
where
\begin{equation}\label{2-inserting-monoplets-yx}
\begin{split}
\mathcal{X}' &=\{s'\}=\{x_{0}^* \cap x_{1}^*\},\\
\mathcal{X}'\!' &=\{s''\}=\{(x_{0}^*)^c \cap x_{1}^*\}.
\end{split}
\end{equation}

\texttt{Example \!\refstepcounter{ctrexa}\arabic{ctrexa}\,\label{exa-3expanding}\itshape\enskip\scriptsize (invariant formula for the 3-Kopula of a half-rare triplet of events).} Let $\frak{X}=\{x_0,x_1,x_2\}$ be the $\breve{p}$-non-ordered  half-rare triplet of events. Then its 3-Kopula is calculated at each point $\breve{p}^{(c|X/\!\!/\{x_{0},x_{1},x_{2}\})} \in [0,1]^{\otimes \frak{X}}$ by the following formulas:
\begin{equation}\label{3-Kopula-recurrent}
\begin{split}
\hspace*{-15pt}&{\mbox{\boldmath$\mathscr{K}$}}\!\left(\breve{p}^{(c|X/\!\!/\{x_0,x_1,x_2\})}\right) =
{\mbox{\boldmath$\mathscr{K}$}}\!\left(\breve{p}^{(c|X^*/\!\!/\{x_0^*,x_1^*,x_2^*\})}\right) =\\
\hspace*{-15pt}&\!\!=\!\!\begin{cases}
\!{\mbox{\boldmath$\mathscr{K}$}}'\!\left(p_{s'},p_{t'}\right),                         &\!\!\!\!\!X^*\!=\!\!\{x_0^*,x_1^*,x_2^*\}\!,\\
\!{\mbox{\boldmath$\mathscr{K}$}}'\!\left(p_{s'},1\!-\!p_{t'}\right),              &\!\!\!\!\!X^*\!=\!\!\{x_0^*,x_1^*\}\!,\\
\!{\mbox{\boldmath$\mathscr{K}$}}'\!\left(1\!-\!p_{s'},p_{t'}\right),              &\!\!\!\!\!X^*\!=\!\{x_0^*,x_2^*\}\!,\\
\!{\mbox{\boldmath$\mathscr{K}$}}'\!\left(1\!-\!p_{s'},1\!-\!p_{t'}\right)\!-\!1\!+\!p_0,
                                  &\!\!\!\!\!X^*\!=\!\!\{x_0^*\},\\
\!{\mbox{\boldmath$\mathscr{K}$}}'\!'\!\left(p_1\!-\!p_{s'},p_2\!-\!p_{t'}\right),                   &\!\!\!\!\!X^*\!=\!\!\{x_1^*,x_2^*\}\!,\\
\!{\mbox{\boldmath$\mathscr{K}$}}'\!'\!\left(1\!-\!p_1\!+\!p_{s'},p_2\!-\!p_{t'}\right),    &\!\!\!\!\!X^*\!=\!\!\{x_2^*\},\\
\!{\mbox{\boldmath$\mathscr{K}$}}'\!'\!\left(p_1\!-\!p_{s'},1\!-\!p_2\!+\!p_{t'}\right),    &\!\!\!\!\!X^*\!=\!\!\{x_1^*\},\\
\!{\mbox{\boldmath$\mathscr{K}$}}'\!'\!\left(1\!-\!p_1\!+\!p_{s'},1\!-\!p_2\!+\!p_{t'}\right)\!-\!p_0,
                                  &\!\!\!\!\!X^*\!=\!\emptyset,
\end{cases}
\end{split}\hspace*{-26pt}
\end{equation}
where when selecting as function parameters $p_{s'},p_{t'},p_{s't'}$ и $p_{s'\!'t'\!'}$ and despite the fact that $p_{s'\!'}=p_1-p_{s'}, p_{t'\!'}=p_2-p_{t'}$,  2-Kopulas of the inserted half-rare doublets $\mathcal{X}'=\{s',t'\}=\{x_0^*\cap x_1^*,x_0^*\cap x_2^*\}$ and $\mathcal{X}'\!'=\{s'\!',t'\!'\}=\{(x_0^*)^c\cap x_1^*,(x_0^*)^c\cap x_2^*\}$ are equal respectively:
\begin{equation}\label{inserting-doublets-formulas}
\hspace{-3pt}\begin{split}
&{\mbox{\boldmath$\mathscr{K}$}}'\left(\breve{p}^{(c|S/\!\!/\mathcal{X}')}\right)=
\begin{cases}
p_{s't'},     & S\!=\!\{s',t'\},\\
p_{s'}\!-\!p_{s't'}, & S\!=\!\{s'\},\\
p_{t'}\!-\!p_{s't'},     & S\!=\!\{t'\},\\
1\!-\!p_{s'}\!-\!p_{t'}\!+\!p_{s't'},     & S\!=\!\emptyset,
\end{cases}\\
&\\
&{\mbox{\boldmath$\mathscr{K}$}}'\!'\left(\breve{p}^{(c|S/\!\!/\mathcal{X}'\!')}\right)=\\
&=\begin{cases}
p_{s'\!'t'\!'},     & S\!=\!\{s'\!',t'\!'\},\\
p_1-p_{s'}\!-\!p_{s'\!'t'\!'}, & S\!=\!\{s'\!'\},\\
p_2-p_{t'}\!-\!p_{s'\!'t'\!'},     & S\!=\!\{t'\!'\},\\
1\!-\!p_1\!-\!p_2\!+\!p_{s'}\!+\!p_{t'}\!+\!p_{s'\!'t'\!'},     & S\!=\!\emptyset.
\end{cases}
\end{split}\hspace{-25pt}
\end{equation}

\subsection{Geometric interpretation of set-phenomenon renumberings}

For a subset of events $V \subseteq \frak{X}$ $V$-phenomenon renumbering of the terrace events, generated by $(N\!-\!1)$-s.h-r.e. $\frak{X}$, is based on the replacement of events from the subset $V^c=\frak{X}-V$ by their complements:
\begin{equation}\label{...}
\begin{split}
\frak{X}^{(c|V)} &= V + (V^c)^{(c)} =\\
&=\{ x, x \in V \} + \{ x^c, x \in V^c \},
\end{split}
\end{equation}
from which the mutually inverse set-phenomenon renumbering formulas follow:
\begin{equation}\label{setPH-ter}
\hspace{-3pt}\begin{split}
&\textsf{ter}\!\!\left(\!X^{(c| V\cap X)} \!/\!\!/ \frak{X}^{(c|V)}\!\right)\!=\!
\textsf{ter}((V \!\Delta X)^c \!/\!\!/ \frak{X}),\\
&\textsf{ter}(X \!/\!\!/ \frak{X})\!=\!\textsf{ter}\!\!\left(\!\left((V\!\Delta X)^c\right)^{(c| V\cap (V\!\Delta X)^c)} \!\!/\!\!/ \frak{X}^{(c|V)}\!\right)
\end{split}
\hspace{-8pt}
\end{equation}
for $V \subseteq \frak{X}$ and $X \subseteq \frak{X}$.

\begin{figure}[ht!]
\centering
\includegraphics[width=1.6in]{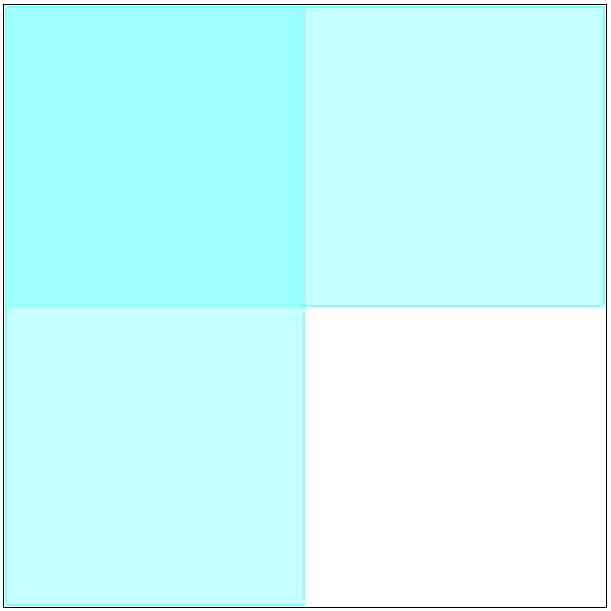}\hspace{-3pt}
\includegraphics[width=1.6in]{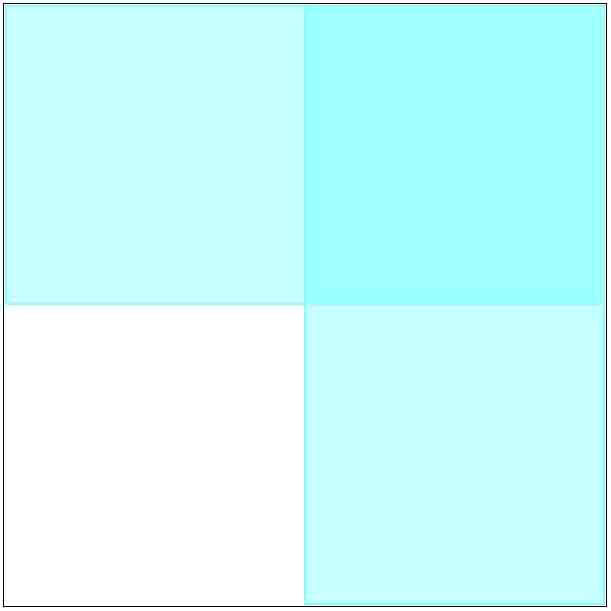}

\vspace{-95pt}

$\{x,y^c\}$ \hspace{27pt} $\{y^c\}$ \hspace{27pt} $\{y^c\}$ \hspace{27pt} $\{x^c,y^c\}$

\vspace{17pt}

$\frak{X}^{(c|x)}$ \hspace{85pt} $\frak{X}^{(c|\emptyset)}$

\vspace{17pt}

$\{x\}$ \hspace{42pt} $\emptyset$ \hspace{42pt} $\emptyset$ \hspace{42pt} $\{x^c\}$

\vspace{22pt}

\includegraphics[width=1.6in]{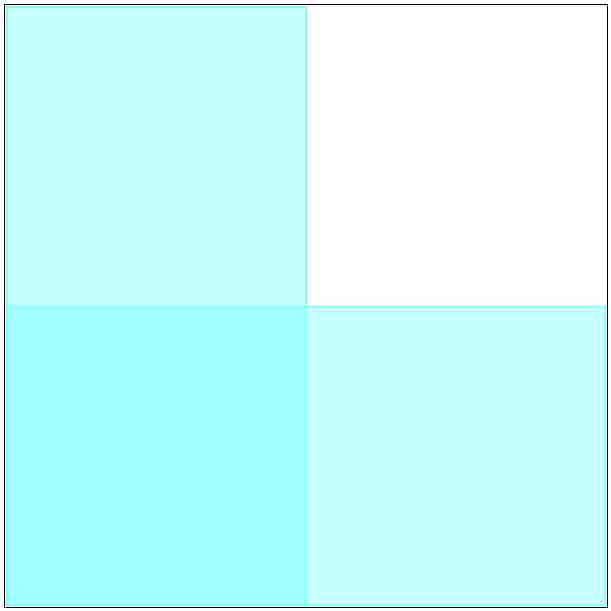}\hspace{-3pt}
\includegraphics[width=1.6in]{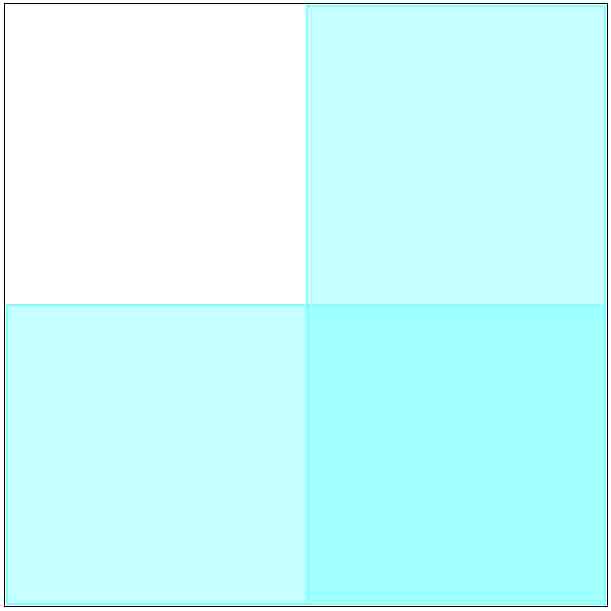}

\vspace{-95pt}

$\{x\}$ \hspace{42pt} $\emptyset$ \hspace{42pt} $\emptyset$ \hspace{42pt} $\{x^c\}$

\vspace{17pt}

$\frak{X}^{(c|xy)}$ \hspace{85pt} $\frak{X}^{(c|y)}$

\vspace{17pt}

$\{x,y\}$ \hspace{32pt} $\{y\}$ \hspace{32pt} $\{y\}$ \hspace{32pt} $\{x^c,y\}$

\vspace{22pt}

\caption{Geometric interpretation of a set-phenomenon renumbering the terraced events, generated by the \emph{doublet of half-rare events} $\frak{X}=\{x,y\}=\{x,y\}^{(c|xy)}$, by reflections with respect to straight lines orthogonal to the coordinate axes and intersecting them at the points 1/2. In the form of unit squares (with the origin in the bottom left corner of each square), the four Venn diagrams of \emph{doublets of half-rare events} $\frak{X}=\frak{X}^{(c|xy)}$ and its set-phenomena $\frak{X}^{(c|x)}=\{x,y^c\},\frak{X}^{(c|y)}=\{x^c,y\}$ and $\frak{X}^{(c|\emptyset)}=\{x^c,y^c\}$ are shown; the terrace events are marked with subsets of the \emph{doublet of half-rare events} $\frak{X}$ or its set-phenomena, consisting of both \emph{half-rare events} and its complements;
on each diagram pairs of events, from which the \emph{doublet of half-rare events} or its set-phenomena consists, are shaded (aqua).\label{fig-BlueBlue}}
\end{figure}

Therefore, the $V$-phenomenon renumbering of the terrace events, generated by $(N \! - \! 1)$-s.h-r.e. $\frak{X}$, by the formulas (\ref{setPH-ter}) is geometrically interpreted on the $(N \! - \! 1)$-dimensional Venn diagram of this s.e. as a reflection of the $\frak{X}$-hypercube relative to those hyperplanes that are orthogonal to the $x$-axes numbered by the events $x\in V^c \subseteq \frak {X}$ (see Fig. \ref{fig-BlueBlue} for the doublet of events).

\subsection{Projection of the $2^\frak{X}$-simplex on the $\frak{X}$-hypercube\label{simplex-cube-projection}}

Take an arbitrary $(N\!-\!1)$-s.e. $\frak{X} \subseteq \mathcal{A}$ with e.p.d. of the 1st kind, which, as is known \cite{Vorobyev2007}, is defined as the $2^\frak{X}$-set of probabilities of terrace events of the 1st kind
\begin{equation}\label{def-epd}
\begin{split}
\{p(X /\!\!/ \frak{X}), X \subseteq \frak{X}\}.
\end{split}
\end{equation}
Look at the $\frak{X}$-set (\ref{def-epd}) as a $2^\frak {X}$-hyper-point from a half-rare $2^N$-vertex simplex
\begin{equation}\label{def-simplex}
\hspace*{-3pt}
\begin{split}
&\mathscr{S}_{2^\frak{X}}\!=\!\\
&\!\!=\!\!\left\{\!\! \{p(X \!/\!\!/\! \frak{X}), X\!\subseteq\!\frak{X}\} : p(X \!/\!\!/\! \frak{X})\!\geqslant\!0, \!\!\sum_{X\!\subseteq\!\frak{X}} \!p(X \!/\!\!/\! \frak{X})\!\!=\!\!1
\!\!\right\}\!\!,
\end{split}\hspace*{-10pt}
\end{equation}
to each vertex of which the \emph{degenerate e.p.d.} corresponds. In this e.p.d., as is known, only one of the 1st kind of probability, equal to one, is different from zero. Number the vertex $2^\frak {X}$ of the simplex $\mathscr{S}_{2^\frak{X}}$ by the subset $X\subseteq \frak {X}$.
The \emph{degenerate e.p.d.} of the 1st kind with $p (X /\!\!/ \frak {X}) = 1$ corresponds to this vertex. And associate the vertex $ \ frak {X} $ with the hypercube $ [0,1] ^ {\ otimes \ frak { X}} $, numbered by the $ \ frak {X} $ -set:

каждой вершине которого, соответствует \emph{вырожденное e.p.d.}, у которого, как известно, лишь одна вероятность of the 1st kind, равная единице, отлична от нуля. Занумеруйте подмножеством $X \subseteq \frak{X}$ вершину $2^\frak{X}$-симплекса $\mathscr{S}_{2^\frak{X}}$, которой соответствует \emph{вырожденное e.p.d.} of the 1st kind с $p(X /\!\!/ \frak{X})=1$. And associate with it the vertex of $\frak{X}$-hypercube $[0,1]^{\otimes \frak{X}}$, numbering by the following $\frak{X}$-set:
\begin{equation}\label{def-indicator}
\begin{split}
\{ \Upsilon_{X /\!\!/ \frak{X}}(x), x \in \frak{X}  \}
\end{split}
\end{equation}
where
\begin{equation}\label{def-indicator1}
\begin{split}
\Upsilon_{X /\!\!/ \frak{X}}(x) =
\begin{cases}
1, &x \in X,\\
0, &\mbox{иначе},
\end{cases}
\end{split}
\end{equation}
are values of the indicator of subset $X \subseteq \frak{X}$ on events $x \in \frak{X}$.
Define a prohection
$$
\textsf{pr} : \mathscr{S}_{2^\frak{X}} \to [0,1]^{\otimes \frak{X}}
$$
of the $2^\frak{X}$-simplex $\mathscr{S}_{2^\frak{X}}$ on $\frak{X}$-hypercube $[0,1]^{\otimes \frak{X}}$ by the following formula:
\begin{equation}\label{def-projection}
\begin{split}
\textsf{pr}(\{p(X /\!\!/ \frak{X}), X \subseteq \frak{X}\}) =
\{p_x, x \in \frak{X}\}
\end{split}
\end{equation}
where
\begin{equation}\label{def-projection1}
\begin{split}
p_x &= \sum_{X \subseteq \frak{X}} p(X /\!\!/ \frak{X}) \Upsilon_{X /\!\!/ \frak{X}}(x)=\\
&= \sum_{x \in X \subseteq \frak{X}} p(X /\!\!/ \frak{X})
\end{split}
\end{equation}
is a convex combination of hypercube vertices, which,
as known \cite{Vorobyev2007},
is intepreted as the probability of event $x \in \frak{X}$.

With projection (\ref{def-projection}) vertices of the $2^\frak{X}$-simplex maps to vertices of the $\frak{X}$-hypercube, and edges map to its edges or diagonals (see \cite{Vorobyev2010b}, \cite{Vorobyev2014famems10} and Fig. \ref{fig-cubexyPR}).

\texttt{Example \!\refstepcounter{ctrexa}\arabic{ctrexa}\,\label{exa-projection}\itshape\enskip\scriptsize (projections of vertices of a $2^\frak{X}$-simplex).}
For example, the vertex of the $2^\frak{X}$-simplex enumerated by the subset $X_0 \subseteq \frak{X}$ corresponds to the degenerate e.p.d. of the 1st kind with probabilities
$$
p(X /\!\!/ \frak{X}) =
\begin{cases}
1, & X=X_0,\\
0, & X_0 \ne X \subseteq \frak{X}.
\end{cases}
$$
From (\ref{def-projection1}) you obtain that
\begin{equation}\nonumber
\begin{split}
p_x =\Upsilon_{X_0 /\!\!/ \frak{X}}(x) =
\begin{cases}
1, & x \in X_0,\\
0, & x \in \frak{X}-X_0.
\end{cases}
\end{split}
\end{equation}
Therefore, by (\ref{def-indicator})
$$
\{p_x, x \in \frak{X}\} = \{\Upsilon_{X_0 /\!\!/ \frak{X}}(x), x \in \frak{X}\}
$$
is a vertex of the $\frak{X}$-hypercube.

In particular, the $\emptyset$-vertex of the $2^\frak{X}$-simplex, i.e., the vertex numbered by the subset $X_0=\emptyset$ is projected into the $\frak{X}$-set $\{0,...,0\}$, consisting of the zero probabilities of marginal events, in other words, projected into the $\emptyset$-vertex of the $\frak{X}$-hypercube, i.e., to the vertex located at the beginning coordinates:
\begin{equation}\label{0-vertex}
\begin{split}
\{0,...,0\} \sim p(X /\!\!/ \frak{X}) =
\begin{cases}
1, & X=\emptyset,\\
0, & \emptyset \ne X \subseteq \frak{X};
\end{cases}
\end{split}
\end{equation}
and the $\frak{X}$-vertex of the $2^\frak{X}$-simplex, i.e., the vertex enumerated by the subset $X_0=\frak{X}$ is projected into the $\frak{X}$-set $\{1,...,1\}$, consisting of the unit probabilities of marginal events, in other words, projected into the $\frak{X}$-vertex of the $\frak{X}$-hypercube, i.e., to the vertex opposite to the origin:
\begin{equation}\label{1-vertex}
\begin{split}
\{1,...,1\} \sim p(X /\!\!/ \frak{X}) =
\begin{cases}
1, & X=\frak{X},\\
0, & \frak{X} \ne X \subseteq \frak{X}.
\end{cases}
\end{split}
\end{equation}

In general, due to the linearity of the projection (\ref{def-projection}), the set of such points of the $2^\frak {X}$-simplex that project into the same point of the $\frak {X}$-hypercube is convex and forms a sub-simplex of smaller dimension.

\subsection{Half-rare events on Venn $(N\!-\!1)$-diagram\label{N-Venn-diagram}}

We will figure out how a Venn $(N\!-\!1)$-diagram of an arbitrary $(N\!-\!1)$-s.e. is constructed on the basis of the projection (\ref{def-projection}), in which the role of the space of universal elementary events $\Omega$ is played by the \emph{unit $(N\!-\!1)$--dimensional hypercube}. Such a Venn $(N\!-\!1)$-diagram puts terraced hypercubes generated by dividing a unit hypercube in half orthogonal to each of the $N$ axes into a one-to-one correspondence with the terraced events generated by the given $(N\!-\!1)$-s.e.

Take first $(N\!-\!1)$-s.h-r.e. $\frak{X}$ and represent its Venn $(N\!-\!1)$-diagram\footnote{see the Venn 2-diagram \emph{doublet of half-rare events} in Fig. \ref{fig-cubexyPR}.} On which $\Omega$ is represented by a \emph{unit $(N\!-\!1)$-dimensional hypercube} that serves as \emph{ordered\footnote{The role of the order of events in s.e. when working with their images in $\mathbb{R}^N$ is discussed in \cite{vorobyev2015famems12}.} image} of the $\frak{X}$-hypercube
\begin{equation}\label{X-hypercube}
\begin{split}
[0,1]^{\otimes \frak{X}} = \bigotimes_{x \in \frak{X}} [0,1]^{x},
\end{split}
\end{equation}
broken by hyperplanes orthogonal to $x$-axis and intersecting them at points $1/2$ into $2^N$ $X$-terraced hypercubes for $X \subseteq \frak{X}$
\begin{equation}\label{X-terrace-hypercube}
\begin{split}
[0,1]^{\otimes \, \textsf{ter}(X /\!\!/ \frak{X})} = \bigotimes_{x \in X} [0,1/2]^{x} \bigotimes_{x \in \frak{X}-X} (1/2,1]^{x},
\end{split}
\end{equation}
where each marginal \emph{half-rare event} $x \in \frak{X}$ is represented as a \emph{$x$-half of $\frak{X}$-hypercube containing the origin}:
\begin{equation}\label{x-hypercube}
\begin{split}
[0,1/2]^{x} \otimes [0,1]^{\otimes (\frak{X}-\{x\})},
\end{split}
\end{equation}
its complement $x^c = \Omega-x$ is represented in the form of another \emph{$x$-half of $\frak{X}$-hypercube that does not contain the origin}:
\begin{equation}\label{x-hypercube}
\begin{split}
(1/2,1]^{x} \otimes [0,1]^{\otimes (\frak{X}-\{x\})},
\end{split}
\end{equation}
and the $X$-terraced event $\textsf{ter}(X /\!\!/ \frak{X})$ --- as a \emph{$X$-terraced hypercube} (\ref{X-terrace-hypercube}):
\begin{equation}\label{Venn-terrace-image}
\begin{split}
\textsf{ter}(X /\!\!/ \frak{X}) \sim [0,1]^{\otimes \, \textsf{ter}(X /\!\!/ \frak{X})}.
\end{split}
\end{equation}
The formula (\ref{Venn-terrace-image}) once again points to a one-to-one correspondence between the $2^N$-space of terraced hypercubes (\ref{X-terrace-hypercube}) from the Venn $(N\!-\!1)$-diagram of $(N\!-\!1)$-s.h-r.e. $\frak{X}$ and $2^N$-totality of terraced events, generated by $\frak{X}$.

If the correspondence between the terraced hypercubes and the terraced events looks natural, then for the sets of \emph{half-rare} events $\frak{X}$ the correspondence between the terraced hypercubes and the numbering of the vertices of the $2^\frak{X}$-simplex projected into the corresponding vertices of the $\frak{X}$-hypercube under the projection (\ref{def-projection}) is defined by the operation of the complement and requires a special

\texttt{Note \!\refstepcounter{ctrnot}\arabic{ctrnot}\,\label{not-projection-numbering}\itshape\enskip\scriptsize (correspondence between the numbering of terraced hypercubes and vertices of $2^\frak{X}$-simplex of e.p.d.'s of the 1st kind of $(N\!-\!1)$-s.h-r.e. $\frak{X}$ on its Venn $(N\!-\!1)$-diagram).} On the Venn $(N\!-\!1)$-diagram of $(N\!-\!1)$-set of \emph{half-rare} events $\frak{X}$, every $X^c$-terraced hypercube $[0,1]^{\otimes \, \textsf{ter}(X^c /\!\!/ \frak{X})}$ contains the $X$-vertex of $\frak{X}$-hypercube, into which corresponding $X$-vertex of $2^\frak{X}$-simplex $\mathscr{S}_{2^{\frak{X}}}$ of e.p.d.'s of the 1st kind of $(N\!-\!1)$-s.h-r.e. $\frak{X}$ for $X \subseteq \frak{X}$ is projected:
\begin{equation}\label{projection-numbering}
\begin{split}
\{1, \, x \in X\}\!+\!\{0, \, x \in \frak{X}-X\} \in [0,1]^{\otimes \textsf{ter}(X^c /\!\!/ \frak{X})},
\end{split}\hspace{-7pt}
\end{equation}
in particular, for $\emptyset$-vertex and $\frak{X}$-vertex we have:
\begin{equation}\label{projection-numbering1}
\begin{split}
\{0, \, x \in \frak{X}\} &= \{0,...,0\} \in [0,1]^{\otimes \textsf{ter}(\frak{X} /\!\!/ \frak{X})},\\
\{1, \, x \in \frak{X}\} &=\{1,...,1\}\in [0,1]^{\otimes \textsf{ter}(\emptyset /\!\!/ \frak{X})}.
\end{split}
\end{equation}

Нетрудно догадаться, что $(N\!-\!1)$-диаграмма Венна произвольного сет-феномена м.пр.\'с. отличается от $(N\!-\!1)$-диаграммы Венна самого $\frak{X}$ лишь перенумерацией террасных events по формулам из \cite{Vorobyev2015famems12}. Сделаем

It is not difficult to guess that the Venn $(N\!-\!1)$-diagram of an arbitrary set-phenomenon of s.h-r.e. differs from the Venn $(N\!-\!1)$-diagram of $\frak{X}$ itself only by renumbering terraced events using formulas from \cite{Vorobyev2015famems12}. Let's do

\texttt{Note \!\refstepcounter{ctrnot}\arabic{ctrnot}\,\label{not-projection-setPH}\itshape\enskip\scriptsize (Venn $(N\!-\!1)$-diagram of  set-phenomena of a set of half-rare events).} On the Venn $(N\!-\!1)$-diagram of the $V$-phenomenon $\frak{X}^{(c|V)}$ of $(N\!-\!1)$-set of \emph{half-rare} events $\frak{X}$, every $V\!\Delta\!X$-terraced hypercube $[0,1]^{\otimes \, \textsf{ter}(V\!\Delta\!X /\!\!/ \frak{X})}$ contains the $X$-vertex of $\frak{X}$-hypercube, into which the corresponding $X$-vertex of $2^\frak{X}$-simplex $\mathscr{S}_{2^{\frak{X}}}$ of e.p.d. of the 1st kind of $(N\!-\!1)$-s.h-r.e. $\frak{X}$ for $X \subseteq \frak{X}$ and $V \subseteq \frak{X}$ is projected:
\begin{equation}\label{projection-numbering}
\hspace{-0pt}\begin{split}
\{1, \, x \in X\}\!+\!\{0, \, x \in \frak{X}-X\} \in [0,1]^{\otimes \textsf{ter}(V\!\Delta\!X /\!\!/ \frak{X})}\!,
\end{split}\hspace{-7pt}
\end{equation}
in particular, for the $\emptyset$-phenomenon $\frak{X}^{(c|\emptyset)} = \frak{X}^{(c)}$ and for $\emptyset$-vertex and $\frak{X}$-vertex we have:
\begin{equation}\label{projection-numbering1}
\begin{split}
\{0, \, x \in \frak{X}\} &= \{0,...,0\} \in [0,1]^{\otimes \textsf{ter}(\emptyset /\!\!/ \frak{X}^{(c)})},\\
\{1, \, x \in \frak{X}\} &=\{1,...,1\}\in [0,1]^{\otimes \textsf{ter}(\frak{X}^{(c)} /\!\!/ \frak{X}^{(c)})}.
\end{split}
\end{equation}

\begin{figure}[h!]
\vspace*{8pt}
\large
\centering
  \includegraphics[width=3.33in]{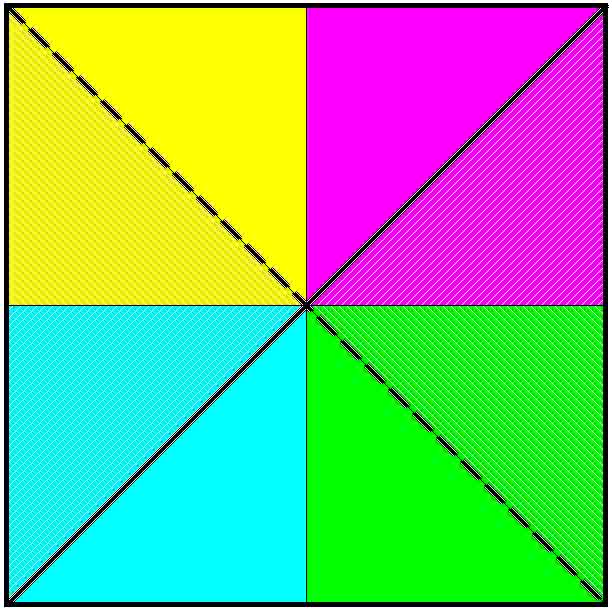}
  \vspace{-272pt}

\hspace{-6pt}  $\{y\} \hspace{193pt} \{x,y\}$

\vspace{4pt}

\hspace{30pt}  $p_x \geqslant 1-p_y \hspace{33pt} 1-p_x \geqslant 1-p_y$

\vspace{85pt}

\hspace{10pt}  $1-p_y > p_x \hspace{55pt} 1-p_y > 1-p_x$

\vspace{5pt}

\hspace{29pt}  $p_y > p_x \hspace{77pt} p_y > 1-p_x$

\vspace{84pt}

\hspace{10pt}  $p_x \geqslant p_y \hspace{55pt} 1-p_x \geqslant p_y$

\vspace{5pt}

\hspace{-2pt}  $\emptyset \hspace{215pt} \{x\}$

\vspace{-228pt}

\hspace{-4pt}  \textcolor{white}{\Large{\mbox{\boldmath$\mbox{\textsf{ter}}(x/\!\!/xy) \hspace{17pt} \mbox{\textsf{ter}}(\emptyset/\!\!/xy)$}}}

\vspace{45pt}

\hspace{-4pt}  \textcolor{white}{\Large{\mbox{\boldmath$\mbox{\textsf{ter}}(y/\!\!/yx) \hspace{82pt} \mbox{\textsf{ter}}(\emptyset/\!\!/yx)$}}}

\vspace{40pt}

\hspace{-3pt}  \textcolor{white}{\Large{\mbox{\boldmath$\mbox{\textsf{ter}}(yx/\!\!/yx) \hspace{80pt} \mbox{\textsf{ter}}(x/\!\!/yx)$}}}

\vspace{45pt}

\hspace{-3pt}  \textcolor{white}{\Large{\mbox{\boldmath$\mbox{\textsf{ter}}(xy/\!\!/xy) \hspace{15pt} \mbox{\textsf{ter}}(y/\!\!/xy)$}}}

\vspace{25pt}

\caption{\scriptsize The projection of a simplex (tetrahedron) of doublets events $\mathscr{S}_{2^{\{x,y\}}}$ on a unit $\{x,y\}$-square $[0,1]^{\otimes \{x,y\}}$ of its marginal probabilities $\breve{p}=\{p_x,p_y\}$. The $X$-вершины of this simplex are projected in corresponding $X$-vertices of $\{x,y\}$-square, $X \subseteq \{x,y\}$, and the all of e.p.d.'s of doublets of events with given $\{x,y\}$-set of probabilities of marginal events $\breve{p}=\{p_x, p_y\}$ are projected in each point $\breve{p} \in [0,1]^{\otimes \{x,y\}}$. In the left down quadrant (aqua) e.p.d.'s of the all of \emph{doublets of half-rare events} of two kind are projected: $1/2 \geqslant p_x \geqslant p_y$ (unshaded) and $p_x < p_y \leqslant 1/2$ (shaded triangle); in the remaining 3 quadrants e.p.d.'s of $\emptyset$-phenomena, $\{y\}$-phenomena and $\{x\}$-phenomena of \emph{doublets of half-rare events} are projected. The half-rare doublets of the second kind: $p_x < p_y \leqslant 1/2$, are projected in the shaded triangle of left down quadrant, and its set-phenomena --- in shaded triangles of corresponding quadrants. The terraced events, generated by \emph{doublets of half-rare events} $\{x,y\}$, are marked by the white formulas.}
\label{fig-cubexyPR}
\end{figure}

\begin{figure}[h!]
\vspace*{8pt}
\large
\centering
  \includegraphics[width=3.33in]{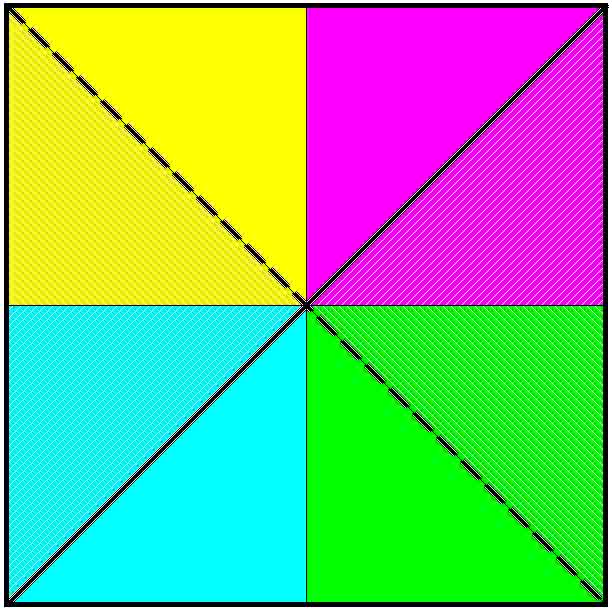}
  \vspace{-272pt}

\hspace{-7pt}  $\{y\} \hspace{190pt} \{x,y\}$

\vspace{4pt}

\hspace{30pt}  $p_x \geqslant 1-p_y \hspace{33pt} 1-p_x \geqslant 1-p_y$

\vspace{85pt}

\hspace{10pt}  $1-p_y > p_x \hspace{55pt} 1-p_y > 1-p_x$

\vspace{5pt}

\hspace{29pt}  $p_y > p_x \hspace{77pt} p_y > 1-p_x$

\vspace{84pt}

\hspace{10pt}  $p_x \geqslant p_y \hspace{55pt} 1-p_x \geqslant p_y$

\vspace{5pt}

\hspace{-3pt}  $\emptyset \hspace{213pt} \{x\}$

\vspace{-233pt}

\hspace{-14pt}  \textcolor{white}{\Huge{\mbox{\boldmath$\{x\} \hspace{42pt} \emptyset$}}}

\vspace{25pt}

\hspace{-14pt}  \textcolor{white}{\Huge{\mbox{\boldmath$\{y\} \hspace{142pt} \emptyset$}}}

\vspace{40pt}

\hspace{-21pt}  \textcolor{white}{\Huge{\mbox{\boldmath$\{y,x\} \hspace{95pt} \{x\}$}}}

\vspace{30pt}

\hspace{-20pt}  \textcolor{white}{\Huge{\mbox{\boldmath$\{x,y\} \hspace{20pt} \{y\}$}}}

\vspace{22pt}

\caption{\scriptsize The projection of a simplex (tetrahedron) of doublets events $\mathscr{S}_{2^{\{x,y\}}}$ on a unit $\{x,y\}$-square $[0,1]^{\otimes \{x,y\}}$ of its marginal probabilities $\breve{p}=\{p_x,p_y\}$. The $X$-вершины of this simplex are projected in corresponding $X$-vertices of $\{x,y\}$-square, $X \subseteq \{x,y\}$, and the all of e.p.d.'s of doublets of events with given $\{x,y\}$-set of probabilities of marginal events $\breve{p}=\{p_x, p_y\}$ are projected in each point $\breve{p} \in [0,1]^{\otimes \{x,y\}}$. In the left down quadrant (aqua) e.p.d.'s of the all of \emph{doublets of half-rare events} of two kind are projected: $1/2 \geqslant p_x \geqslant p_y$ (unshaded) and $p_x < p_y \leqslant 1/2$ (shaded triangle); in the remaining 3 quadrants e.p.d.'s of $\emptyset$-phenomena, $\{y\}$-phenomena and $\{x\}$-phenomena of \emph{doublets of half-rare events} are projected. The half-rare doublets of the second kind: $p_x < p_y \leqslant 1/2$, are projected in the shaded triangle of left down quadrant, and its set-phenomena --- in shaded triangles of corresponding quadrants. The terraced events, generated by \emph{doublets of half-rare events} $\{x,y\}$, are marked by the white formulas.}
  \label{fig-cubexyPR}
\end{figure}

\subsection{Set-phenomenon spectrum of functions on the $\frak{X}$-hypercube}

\texttt{Definition \!\refstepcounter{ctrdef}\arabic{ctrdef}\,\label{spectrum}({\itshape\footnotesize set-phenomenon spectrum of functions on the $\frak{X}$-hypercube}\,).}
With each function $\psi \in \Psi_\frak{X}$, defined on the $\frak{X}$-hypercube, the $2^N$ functions are connected. These functions are defined on the  $\frak{X}$-hypercube by formulas:
$$
\psi_X\left(\breve{w}\right) = \psi\left(\breve{w}^{(c|X/\!\!/\frak{X})}\right)
$$
for $X \subseteq \frak{X}$. The family of the all such functions
$$
\{ \psi_X : X \subseteq \frak{X} \}
$$
is called the \emph{set-phenomenon $\frak{X}$-spectrum} of the function $\psi$.

Let's define for each $X \subseteq \frak{X}$ the \emph{terraced $X$-hypercube}
$$
\textsf{ter}^\otimes(X /\!\!/ \frak{X}) = \bigotimes_{x \in X} \left[\frac{1}{2}, 1\right] \bigotimes_{x \in \frak{X}-X} \left[0,\frac{1}{2}\right),
$$
from which the $\frak{X}$-hypercube is composed:
$$
[0,1]^{\otimes \frak{X}} = \sum_{X \subseteq \frak{X}} \textsf{ter}^\otimes(X /\!\!/ \frak{X}).
$$

\texttt{Lemma \!\refstepcounter{ctrlem}\arabic{ctrlem}\,\label{normalizing}({\itshape\footnotesize on a set-phenomenon $\frak{X}$-spectrum of normalized function}\,).}
In order that the family of functions $\{ \theta_X : X \subseteq \frak{X}  \}$ из $\Psi_\frak{X}$
is a set-phenomenon $\frak{X}$-spectrum of some function normalized on the $\frak{X}$-hypercube, it is necessary and sufficient that
\begin{equation}\label{intro.2}
\begin{split}
\sum_{X \subseteq \frak{X}} \theta_X\left(\breve{w}\right) = 1
\end{split}
\end{equation}
for all $\breve{w} \in [0,1]^{\otimes \frak{X}}$.

\texttt{Proof.} 1) If the family $\{ \theta_X : X \subseteq \frak{X}  \}$ is a set-phenomenon $\frak{X}$-spectrum of some normalized function, then by Definition \ref{spectrum} the equality (\ref{intro.2}) is satisfied.
2) Let now the equality (\ref{intro.2}) is satisfied. Construct the function $\psi$ on the $\frak{X}$-hypercube by the following way
$$
\psi\left(\breve{w}\right) =
\begin{cases}
\theta_\emptyset\left(\breve{w}^{(\!c|\!\emptyset\!)}\right), & \breve{w} \in [0,1/2)^{\otimes \frak{X}},\\[-5pt]
\ldots,&\cr
\theta_X\left(\breve{w}^{(c|X/\!\!/\frak{X})}\right),& \breve{w} \in \textsf{ter}^\otimes(X /\!\!/ \frak{X}),\\[-5pt]
\ldots,&\cr
\theta_\frak{X}\left(\breve{w}^{(c|\frak{X})}\right),& \breve{w} \in [1/2,1]^{\otimes \frak{X}}
\end{cases}
$$
and show that the $\psi$ is normalized on the $\frak{X}$-hypercube. Indeed, noting that for an arbitrary $X \subseteq \frak{X}$ the equality
$
\psi\left(\breve{w}\right) = \theta_X\left(\breve{w}^{(c|X/\!\!/\frak{X})}\right)
$
is equivalent to the equality
$
\psi\left(\breve{w}^{(c|X/\!\!/\frak{X})}\right) = \theta_X\left(\breve{w}\right),
$
we obtain the required:
$$
\sum_{X \subseteq \frak{X}} \psi\left(\breve{w}^{(c|X/\!\!/\frak{X})}\right) = \sum_{X \subseteq \frak{X}} \theta_X\left(\breve{w}\right) = 1.
$$

\texttt{Lemma \!\refstepcounter{ctrlem}\arabic{ctrlem}\,\label{1-function}({\itshape\footnotesize on a set-phenomenon $\frak{X}$-spectrum of the  1-function}\,).}
In order that the family of functions $\{ \theta_X : X \subseteq \frak{X}  \}$ из $\Psi_\frak{X}$
is a set-pehomenon $\frak{X}$-spectrum of some 1-function on the $\frak{X}$-hypercube, it is necessary and sufficient that for each $x \in \frak{X}$
\begin{equation}\label{intro.4}
\begin{split}
\sum_{x \in X \subseteq \frak{X}} \theta_X\left(\breve{w}\right) = w_x
\end{split}
\end{equation}
for all $\breve{w} \in [0,1]^{\otimes \frak{X}}$.

\texttt{Proof.} 1) If the family (\ref{intro}.3) is a set-phenomenon $\frak{X}$-spectrum of some 1-function, then partial sums of functions from the family at $x \in X \subseteq \frak{X}$ равны $w_x$:
$$
\sum_{x \in X \subseteq \frak{X}} \psi_X\left(\breve{w}\right) = w_x
$$
for each $\breve{w} \in [0,1]^{\otimes \frak{X}}$ by Definition \ref{one}. 2) Let now the equalities (\ref{intro}.4) are satisfied. Let's construct the function $\psi$ on $\frak{X}$-hypercube by the following way
$$
\psi\left(\breve{w}\right) =
\begin{cases}
\theta_\emptyset\left(\breve{w}^{(\!c|\!\emptyset\!)}\right), & \breve{w} \in [0,1/2)^{\otimes \frak{X}},\\[-5pt]
\ldots,&\cr
\theta_X\left(\breve{w}^{(c|X/\!\!/\frak{X})}\right),& \breve{w} \in \textsf{ter}^\otimes(X /\!\!/ \frak{X}),\\[-5pt]
\ldots,&\cr
\theta_\frak{X}\left(\breve{w}^{(c|\frak{X})}\right),& \breve{w} \in [1/2,1]^{\otimes \frak{X}}
\end{cases}
$$
and show that $\psi$ is a 1-function on the $\frak{X}$-hypercube. Indeed, noting that for an arbitrary $X \subseteq \frak{X}$ the equality
$
\psi\left(\breve{w}\right) = \theta_X\left(\breve{w}^{(c|X/\!\!/\frak{X})}\right)
$
is equivalent to the equality
$
\psi\left(\breve{w}^{(c|X/\!\!/\frak{X})}\right) = \theta_X\left(\breve{w}\right),
$
we obtain the required:
$$
\sum_{x \in X \subseteq \frak{X}} \psi\left(\breve{w}^{(c|X/\!\!/\frak{X})}\right) = \sum_{x \in X \subseteq \frak{X}} \theta_X\left(\breve{w}\right) = w_x.
$$

\begin{figure}[h!]
\vspace*{8pt}
\large
\centering
  \includegraphics[width=3.33in]{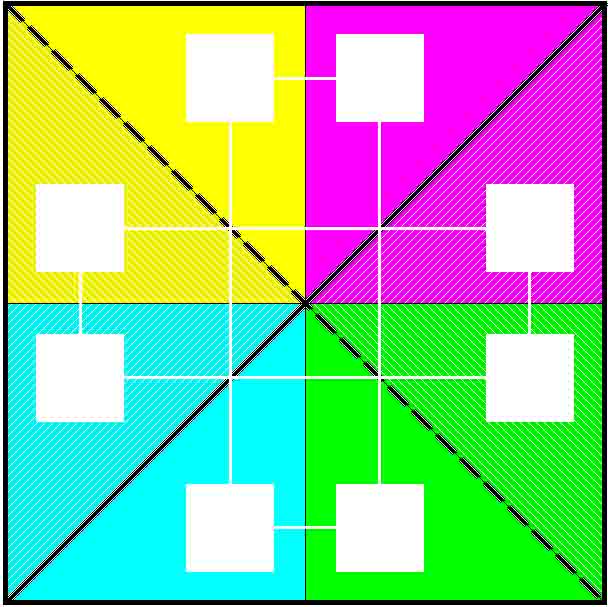}
\vspace{-272pt}

{$\{0,1\}$ \hspace{173pt} $\{1,1\}$}

                     \vspace{18pt}

\hspace{-4pt} {\small $xy$ \hspace{39pt} $xy$}

\vspace{-4pt}

\hspace{-4pt} {\small $10$ \hspace{39pt} $00$}

                     \vspace{36pt}

\hspace{-4pt} {\small $yx$ \hspace{157pt} $yx$}

\vspace{-4pt}

\hspace{-4pt} {\small $01$ \hspace{157pt} $00$}

                     \vspace{35pt}

\hspace{-4pt} {\small $yx$ \hspace{157pt} $yx$}

\vspace{-4pt}

\hspace{-4pt} {\small $11$ \hspace{157pt} $10$}

                     \vspace{36pt}

\hspace{-4pt} {\small $xy$ \hspace{39pt} $xy$}

\vspace{-5pt}

\hspace{-4pt} {\small $11$ \hspace{39pt} $01$}

\vspace{21pt}

\hspace{-5pt}  $\{p_x,p_y\}=\{0,0\} \hspace{127pt} \{1,0\}$

\vspace{-250pt}

\hspace{-10pt}  \textcolor{white}{\Huge{\mbox{\boldmath$\{\!x\!\} \hspace{182pt} \emptyset$}}}

\vspace{180pt}

\hspace{-3pt}  \textcolor{white}{\Huge{\mbox{\boldmath$\{\!x,\!y\!\} \hspace{137pt} \{\!y\!\}$}}}

\vspace{15pt}

\caption{\scriptsize The projection of $2^2$-vertices simplex on a square, on which the scheme is superimposed, illustrating a connection of two permutations of events in a half-rare doublet with the $2^2$ set-phenomena.}
\label{fig-cubeXY}
\end{figure}

\begin{figure}[h!]
\vspace*{8pt}
\large
\centering
\includegraphics[width=3.33in]{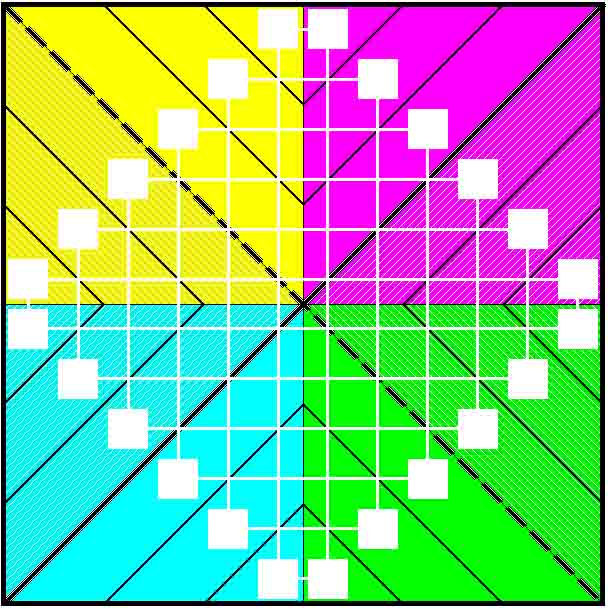}
\vspace{-272pt}

\hspace{-6pt} {$\{0,1,1\}$ \hspace{152pt} $\{1,1,1\}$}

   \vspace{-2pt}

\hspace{-4pt} {\small $xyz$ \hspace{-1pt} $xyz$}

\vspace{-5pt}

\hspace{-4pt} {\small $100$ \hspace{-1pt} $000$}

   \vspace{-4pt}

\hspace{-4pt} {\small $xzy$ \hspace{39pt} $xzy$}

\vspace{-4pt}

\hspace{-4pt} {\small $100$ \hspace{39pt} $000$}

   \vspace{-4pt}

\hspace{-4pt} {\small $yxz$ \hspace{78pt} $yxz$}

\vspace{-4pt}

\hspace{-4pt} {\small $010$ \hspace{78pt} $000$}

   \vspace{-4pt}

\hspace{-4pt} {\small $zxy$ \hspace{118pt} $zxy$}

\vspace{-4pt}

\hspace{-4pt} {\small $010$ \hspace{118pt} $000$}

   \vspace{-5pt}

\hspace{-4pt} {\small $yzx$ \hspace{157pt} $yzx$}

\vspace{-4pt}

\hspace{-4pt} {\small $001$ \hspace{157pt} $000$}

   \vspace{-4pt}

\hspace{-4pt} {\small $zyx$ \hspace{197pt} $zyx$}

\vspace{-4pt}

\hspace{-4pt} {\small $001$ \hspace{197pt} $000$}

   \vspace{-4pt}

\hspace{-4pt} {\small $zyx$ \hspace{197pt} $zyx$}

\vspace{-4pt}

\hspace{-4pt} {\small $011$ \hspace{197pt} $010$}

   \vspace{-4pt}

\hspace{-4pt} {\small $yzx$ \hspace{157pt} $yzx$}

\vspace{-4pt}

\hspace{-4pt} {\small $101$ \hspace{157pt} $100$}

   \vspace{-4pt}

\hspace{-4pt} {\small $zxy$ \hspace{118pt} $zxy$}

\vspace{-4pt}

\hspace{-4pt} {\small $011$ \hspace{118pt} $001$}

   \vspace{-4pt}

\hspace{-4pt} {\small $yxz$ \hspace{78pt} $yxz$}

\vspace{-4pt}

\hspace{-4pt} {\small $110$ \hspace{78pt} $100$}

   \vspace{-4pt}

\hspace{-4pt} {\small $xzy$ \hspace{39pt} $xzy$}

\vspace{-5pt}

\hspace{-4pt} {\small $101$ \hspace{39pt} $001$}

   \vspace{-4pt}

\hspace{-4pt} {\small $xyz$ \hspace{-1pt} $xyz$}

\vspace{-4pt}

\hspace{-4pt} {\small $110$ \hspace{-1pt} $010$}

\vspace{0pt}

\hspace{-5pt}  $\{p_x,p_y,p_z\}=\{0,0,1\} \hspace{90pt} \{1,0,1\}$

\vspace{-250pt}

\hspace{-10pt}  \textcolor{white}{\Huge{\mbox{\boldmath$\{\!x\!\} \hspace{182pt} \emptyset$}}}

\vspace{180pt}

\hspace{-3pt}  \textcolor{white}{\Huge{\mbox{\boldmath$\{\!x,\!y\!\} \hspace{140pt} \{\!y\!\}$}}}

\vspace*{35pt}

\includegraphics[width=3.33in]{cubeXYZ.jpg}
\vspace{-272pt}

\hspace{-6pt} {$\{0,1,0\}$ \hspace{152pt} $\{1,1,0\}$}

   \vspace{-2pt}

\hspace{-4pt} {\small $xyz$ \hspace{-1pt} $xyz$}

\vspace{-5pt}

\hspace{-4pt} {\small $101$ \hspace{-1pt} $001$}

   \vspace{-4pt}

\hspace{-4pt} {\small $xzy$ \hspace{39pt} $xzy$}

\vspace{-4pt}

\hspace{-4pt} {\small $110$ \hspace{39pt} $010$}

   \vspace{-4pt}

\hspace{-4pt} {\small $yxz$ \hspace{78pt} $yxz$}

\vspace{-4pt}

\hspace{-4pt} {\small $011$ \hspace{78pt} $001$}

   \vspace{-4pt}

\hspace{-4pt} {\small $zxy$ \hspace{118pt} $zxy$}

\vspace{-4pt}

\hspace{-4pt} {\small $110$ \hspace{118pt} $100$}

   \vspace{-5pt}

\hspace{-4pt} {\small $yzx$ \hspace{157pt} $yzx$}

\vspace{-4pt}

\hspace{-4pt} {\small $011$ \hspace{157pt} $010$}

   \vspace{-4pt}

\hspace{-4pt} {\small $zyx$ \hspace{197pt} $zyx$}

\vspace{-4pt}

\hspace{-4pt} {\small $101$ \hspace{197pt} $100$}

   \vspace{-4pt}

\hspace{-4pt} {\small $zyx$ \hspace{197pt} $zyx$}

\vspace{-4pt}

\hspace{-4pt} {\small $111$ \hspace{197pt} $110$}

   \vspace{-4pt}

\hspace{-4pt} {\small $yzx$ \hspace{157pt} $yzx$}

\vspace{-4pt}

\hspace{-4pt} {\small $111$ \hspace{157pt} $110$}

   \vspace{-4pt}

\hspace{-4pt} {\small $zxy$ \hspace{118pt} $zxy$}

\vspace{-4pt}

\hspace{-4pt} {\small $111$ \hspace{118pt} $101$}

   \vspace{-4pt}

\hspace{-4pt} {\small $yxz$ \hspace{78pt} $yxz$}

\vspace{-4pt}

\hspace{-4pt} {\small $111$ \hspace{78pt} $101$}

   \vspace{-4pt}

\hspace{-4pt} {\small $xzy$ \hspace{39pt} $xzy$}

\vspace{-5pt}

\hspace{-4pt} {\small $111$ \hspace{39pt} $011$}

   \vspace{-4pt}

\hspace{-4pt} {\small $xyz$ \hspace{-1pt} $xyz$}

\vspace{-4pt}

\hspace{-4pt} {\small $111$ \hspace{-1pt} $011$}

\vspace{0pt}

\hspace{-5pt}  $\{p_x,p_y,p_z\}=\{0,0,0\} \hspace{90pt} \{1,0,0\}$

\vspace{-250pt}

\hspace{-7pt}  \textcolor{white}{\Huge{\mbox{\boldmath$\{\!x,\!z\!\} \hspace{147pt} \{\!z\!\}$}}}

\vspace{180pt}

\hspace{-3pt}  \textcolor{white}{\Huge{\mbox{\boldmath$\{\!x,\!y,\!z\!\} \hspace{100pt} \{\!y,\!z\!\}$}}}

\hspace{15pt}

\caption{\scriptsize These are not geometrical projections of $2^3$-vertices simplex on a cube, but two conditional schemes of these projections, which illustrate a connection of six permutations of events in a half-rare triplet of events with its $2^3$ set-phenomena. The conditional scheme of the projection on the upper half of the cube is shown at the top, on the lower half --- at the bottom. In the Venn diagram of half-rare events: $x$ is the left, $y$ is the right, and $z$ is the lower half of the cube.}
\label{fig-cubeXYZ}
\end{figure}

\section{Remaining behind the scenes}

In the text and, in particular, in the Appendix, the value of the \emph{$\breve{p}$-ordering condition} of the set of events is specified, which complicates the computational implementation of the above algorithms in the frame method of constructing Kopulas as set functions of the set of marginal probabilities. The reason for this complication lies in the properties of the set-functions, i.e., functions of a set that differ from the properties of arbitrary functions of several variables. The point is that the set-function of the set of marginal probabilities is necessarily a symmetric function of the marginal probability vector (\emph{Cartesian representation of Kopula}, see Prolegomenon \ref{pro-9}), to determine which it is sufficient to specify its values only on those vectors whose components are ordered, for example, in descending order, so that the remaining values can be determined by the appropriate permutations of the arguments. For example, the \emph{Cartesian representation of an $N$ -Kopula in $\mathbb{R}^N$} is sufficient to define on the $1/N!$ part of the unit $N$-hypercube so that this representation becomes definite on the whole hypercube by continuing permutations of arguments.

Although this task is purely technical, but its solution opens the way for the application of the proposed Kopula (eventological copula) theory to the construction of the eventological theory of ordinary copulas that determine the joint distribution of a given set of marginal distributions.
The author encountered this when developing the program code, which calculated all the illustrations for the Kopula examples. The problem is solved programmatically, but requires a detailed description of this solution (see Fig. \ref{fig-cubeXY} and \ref{fig-cubeXYZ}), which, of course, together with the eventological theory of copula deserves a separate publication.

In conclusion, I can not resist the temptation to quote the formulation of the tenth Prolegomenon of the Kopula theory, which reveals the content of these my next publications.

\texttt{Prolegomenon \!\refstepcounter{ctrPRO}\arabic{ctrPRO}\,\label{pro-10}\itshape\footnotesize (Cartesian representation of the $N$-Kopula defines $2^N$ classical copulas of $N$ marginal uniform distributions on $[0,1]$).}

\section{On the inevitable development of language}

This first work on the theory of the eventological copula is over at the end of July 2015. It sums up the work on the eventological theory of probabilities, raising the theory of Kopula to its apex. The work is written in a mathematical language, in which the state of the eventological theory was reflected precisely at the time when the author unexpectedly, but by the way, got a brilliant example of two statisticians from sociology and ecology, who immediately forced him to postpone polishing of the Kopula theory for almost a year in order to immediately immerse themselves in the destructive creation of a new unifying \emph{eventological theory of experience and chance} by the agonizing fusion of two dual theories: the eventological theory of believabilities and the eventological theory of probabilities.

Because of this, the mathematical language of this work is just a pretension to the eventological probability theory, which does not yet know that there is a very close twin that exists --- the eventological theory of believabilities. Therefore, in the terminology of this work, those crucial changes in the basic concepts and notations that were invented to construct a unifying eventological theory did not find any worthy reflection. Of course, the new unifying theory suggests the development of the original mathematical language of dual Kopulas, one of which hosts the eventological probability theory, and the other --- in the eventological believability theory.

$\bigstar$ The English version of this article was published on November 12, 2017. Therefore, my later works \cite{Vorobyev2015famems13,Vorobyev2016famems3,Vorobyev2016famems2,Vorobyev2016famems1}, which expand the themes of this work, are added to the list of references. Due to the arXiv.org limitation on the volume of publication, the work is reduced by removing some illustrations. In full, the work is available at: {\tiny\url{http://www.academia.edu/35218637/}}.

{\footnotesize
\bibliography{vorobyev6}

\begin{thebibliography}{10}

\bibitem{Vorobyev2007}
O.~Yu. Vorobyev.
\newblock {\em Eventology}.
\newblock Siberian Federal University, Krasnoyarsk, Russia, 435p.,
  \mbox{\tiny\url{https://www.academia.edu/179393/}}, 2007.

\bibitem{Sklar1959}
A.~Sklar.
\newblock Fonctions de r\'epartition \'a $n$ dimensions et leurs marges.
\newblock {\em Publ. Inst. Statist. Univ. Paris}, 8:229--231, 1959.

\bibitem{Nelsen1999}
R.B. Nelsen.
\newblock {\em An introducvtion to copulas}.
\newblock Springer, New York, 1999.

\bibitem{Nelsen2003}
R.B. Nelsen.
\newblock Properties and applications of copulas: A brief survey.
\newblock {\em Proceedings of the First Brazilian Conference on Statistical
  Modelling in Insurance and Finance (J. Dhaene, N. Kolev, and P. Morettin,
  eds.), Institute of Mathematics and Statistics, University of Sao Paulo},
  pages 10--28, 2003.

\bibitem{Alsina2006}
C.~Alsina, M.J. Frank, and B.~Schweizer.
\newblock {\em Assocative functions: Triangular Norms and Copulas}.
\newblock World Scientific Publishing Co. Pte. Ltd., Singapore, 2006.

\bibitem{Vorobyev2015famems12}
O.~Yu. Vorobyev.
\newblock Ordered sets of half-rare events and its set-penomena.
\newblock {\em \emph{In.} Proc. of the XIV Intern. FAMEMS Conf. on Financial
  and Actuarial Mathematics and Eventology of Multivariate Statistics,
  Krasnoyarsk, SFU (Oleg Vorobyev ed.)}, pages 52--59, ISBN
  978--5--9903358--5--1, \mbox{\tiny\url{https://www.academia.edu/24351285}},
  2015.

\bibitem{Vorobyev2006aa}
O.~Yu. Vorobyev and G.~M. Boldyr.
\newblock On a new notion of conditional event and its application in
  eventological analysis.
\newblock {\em Notes of Krasnoyarsk State University, Phys. Math. Series},
  1:152--159, 2006.

\bibitem{Vorobyev2014famems9}
O.~Yu. Vorobyev.
\newblock On the foundations of the eventological method.
\newblock {\em \emph{In.} Proc. of the XIII Intern. FAMEMS Conf. on Financial
  and Actuarial Mathematics and Eventology of Multivariate Statistics,
  Krasnoyarsk, SFU (Oleg Vorobyev ed.)}, pages 112--121, ISBN
  978--5--9903358--4--4, 2014.

\bibitem{Vorobyev2010b}
O.~Yu. Vorobyev.
\newblock Fr\'echet boundary eventological distributions and its applications.
\newblock {\em \emph{In.} Proc. of the XV Intern. EM Conf. on {E}ventological
  {M}athematics. Krasnoyarsk, KSU (Oleg Vorobyev ed.)}, 1:57--69, 2010.

\bibitem{Vorobyev2014famems10}
O.~Yu. Vorobyev.
\newblock Eventologial generalization of {F}r\'echet bounds for a set of
  events.
\newblock {\em \emph{In.} Proc. of the XIII Intern. FAMEMS Conf. on Financial
  and Actuarial Mathematics and Eventology of Multivariate Statistics,
  Krasnoyarsk, SFU (Oleg Vorobyev ed.)}, pages 122--130, ISBN
  978--5--9903358--4--4, 2014.

\bibitem{Vorobyev2015famems13}
O.~Yu. Vorobyev.
\newblock Elements of the kopula (eventological copula) theory.
\newblock {\em \emph{In.} Proc. of the XIV Intern. FAMEMS Conf. on Financial
  and Actuarial Mathematics and Eventology of Multivariate Statistics,
  Krasnoyarsk, SFU (Oleg Vorobyev ed.)}, pages 78--125, ISBN
  978--5--9903358--5--1, \mbox{\tiny\url{https://www.academia.edu/35218637}},
  2015.

\bibitem{Vorobyev2016famems3}
O.~Yu. Vorobyev.
\newblock Theory of dual co$\sim$event means.
\newblock {\em \emph{In.} Proc. of the XV Intern. FAMEMS Conf. on Financial and
  Actuarial Mathematics and Eventology of Multivariate Statistics \& the
  Workshop on Hilbert's Sixth Problem; Krasnoyarsk, SFU (Oleg Vorobyev ed.)},
  pages 44--93, ISBN 978--5--9903358--6--8,
  \mbox{\tiny\url{https://www.academia.edu/34357251}}, 2016.

\bibitem{Vorobyev2016famems2}
O.~Yu. Vorobyev.
\newblock Postulating the theory of experience and of chance as a theory of
  co$\sim$events (co$\sim$beings).
\newblock {\em \emph{In.} Proc. of the XV Intern. FAMEMS Conf. on Financial and
  Actuarial Mathematics and Eventology of Multivariate Statistics \& the
  Workshop on Hilbert's Sixth Problem; Krasnoyarsk, SFU (Oleg Vorobyev ed.)},
  pages 25--43, ISBN 978--5--9903358--6--8,
  \mbox{\tiny\url{https://www.academia.edu/34373279}}, 2016.

\bibitem{Vorobyev2016famems1}
O.~Yu. Vorobyev.
\newblock An element-set labelling a {C}artesian product by measurable binary
  relations which leads to postulates of the theory of experience and chance as
  a theory of co$\sim$events.
\newblock {\em \emph{In.} Proc. of the XV Intern. FAMEMS Conf. on Financial and
  Actuarial Mathematics and Eventology of Multivariate Statistics \& the
  Workshop on Hilbert's Sixth Problem; Krasnoyarsk, SFU (Oleg Vorobyev ed.)},
  pages 9--24, ISBN 978--5--9903358--6--8,
  \mbox{\tiny\url{https://www.academia.edu/34390291}}, 2016.

\end{thebibliography}
}

\end{document}